THE OBJECTIVE FUNCTION:

SCIENCE AND SOCIETY IN THE AGE OF MACHINE INTELLIGENCE

by

EMANUEL D. MOSS

A dissertation submitted to the Graduate Faculty in Anthropology in partial fulfillment of the requirements for the degree of Doctor of Philosophy, The City University of New York

2021

The Objective Function:

Science and Society in the Age of Machine Intelligence

by

Emanuel D. Moss

This manuscript has been read and accepted for the Graduate Faculty in Anthropology in satisfaction of the dissertation requirement for the degree of Doctor of Philosophy.

______________________________          ______________________________
Date                                                              Karen Strassler

                                                                  Chair of Examining Committee

______________________________          ______________________________
Date                                                              Jeff Maskovsky

                                                                  Executive Officer

Supervisory Committee:

Mandana Limbert

Dána-Ain Davis

danah boyd

THE CITY UNIVERSITY OF NEW YORK

*Abstract*

The Objective Function: Science and Society in the Age of Machine Intelligence

by

Emanuel D. Moss

Advisor: Karen Strassler

Machine intelligence, or the use of complex computational and statistical practices to make predictions and classifications based on data representations of phenomena, has been applied to domains as disparate as criminal justice, commerce, medicine, media and the arts, mechanical engineering, among others. How has machine intelligence become able to glide so freely across, and to make such waves for, these domains? In this dissertation, I take up that question by ethnographically engaging with how the authority of machine learning has been constructed such that it can influence so many domains, and I investigate what the consequences are of it being able to do so. By examining the workplace practices of the applied machine learning researchers who produce machine intelligence, those they work with, and the artifacts they produce—algorithmic systems, public demonstrations of machine intelligence, academic research articles, and conference presentations—a wider set of implications about the legacies of positivism and objectivity, the construction of expertise, and the exercise of power takes shape.

The dissertation begins by arguing that machine intelligence proceeds from a "naïve" form of empiricism with ties to positivist intellectual traditions of the 17th and 18th centuries. This naïve empiricism eschews other forms of knowledge and theory formation in order for applied machine learning researchers to enact *data performances* that bring objects of analysis into existence as entities capable of being subjected to machine intelligence. By "data performances," I mean generative enactments which bring into existence that which machine intelligence purports to analyze or describe. The enactment of data performances is analyzed as an agential cut into a representational field that produces both stable claims about the world and the interpretive frame in which those claims can hold true. The dissertation also examines how machine intelligence depends upon a range of accommodations from other institutions and organizations, from data collection and processing to organizational commitments to support the work of applied machine learning researchers. Throughout the dissertation, methods are developed for analyzing the expert practices of machine learning researchers to transform situated, positional knowledge into machine intelligence and re-present it as objective knowledge. These methods trace the chains of dependencies between data collection, processing, and analysis to reveal where and how hidden assumptions about the phenomena being analyzed are advanced.

The second half of the dissertation focuses on how the authority of machine intelligence to control or ensure compliance is developed. This authority rests not only on applications of machine intelligence which constrain the freedom of others to act in accordance with their own desires, but also on the ways in which attempts to critique or curtail the authority of machine intelligence are assimilated into the logics and practices of machine intelligence itself. Attempts to limit the authority of machine intelligence, particularly *AI ethics* and *algorithmic fairness* are explored ethnographically to conclude that even in recognizing and attempting to take responsibility for the harms it risks producing in the world, machine intelligence nevertheless remains resistant to forms of accountability that are external to its own practices. This ensures that machine intelligence remains a deeply conservative project, contrary to its presentation as futuristic or transformative, that conserves the power of those who already wield it.

# ACKNOWLEDGEMENTS

While this dissertation is presented as a work of sole authorship, it has not come into being through my efforts alone. Rather, it exists because of those who have supported, worked with, and humored me over the past many years.

I first thank my advisor, Karen Strassler, for her unenumerable contributions and unwavering support. I also thank the other members of my committee, Dàna-Ain Davis and Mandana Limbert, for their encouragement and support throughout this process. I also thank Ellen DeRiso for her indefatigable assistance in navigating the doctoral program, and Katherine Verdery for her early and sustained support of my doctoral studies. I am also indebted to danah boyd for the profound contributions she has made to my intellectual development and the formulation of this project from its earliest incarnations. I have been incredibly fortunate to work closely with Jacob Metcalf in producing scholarship and developing ideas which ramify throughout this dissertation. I have benefited from, and am grateful for, opportunities to develop my ideas through participation in the Algorithmic Fairness and Opacity Group at UC-Berkeley led by Deirdre Mulligan and Jenna Burrell, the UC-Irvine Tech, Law, and Society Summer Institute directed by Bill Maurer, the Stanford Law Clinic on the Future of Algorithms led by Bryan Casey and Philip Malone, the Harvard STS Summer School directed by Sheila Jasanoff, and the REALML workshop directed by William Isaac.

I have also benefited tremendously from the Interuniversity Graduate School Consortium, which afforded me the opportunity to engage in coursework with and receive valuable mentorship from Emily Martin, Richard Rottenburg, and Zoë Crossland, whom I thank. This dissertation would not exist without the mentorship of Elizabeth Ferry, Janet McIntosh, Jonathan Anjaria, Sarah Lamb, Richard Parmentier, and Chandler Rosenberger, who encouraged me to pursue doctoral studies and equipped me to be able to do so. I also wish to thank Lily Hu, Alexa Hagerty, Lev Manovich, Alondra Nelson, Talal Asad, Katie Shilton, Casey Fiesler, Michael Zimmer, Matthew Bietz, Jessica Vitak, Andrew Smart, Deb Raji, Joshua Kroll, Morgan Ames, Abigail Jacobs, Hanna Wallach, Christopher Roosevelt, Christina Luke, Bradley Sekedat, Ben Roome, and especially Friederike Schüür, for the crucial conversations that—whether they knew it or not—have shaped this project and the trajectory of my career.

This research has been enabled by generous support from the Wenner Gren Foundation, the National Science Foundation (NSF Awards # 1704369 and # 1633400), the CUNY Graduate Center Early Research Initiative Pre-Dissertation Fellowship, the CUNY Advanced Research Collaborative (ARC), and the CUNY Graduate Center Fellowship program. I extend my gratitude to the boards and administrators of these programs for making my work possible.

Tremendous thanks are also due to the Data & Society Research Institute and the community of camaraderie and scholarship it anchors. The leadership team and the AI on the Ground Initiative, in particular, have encouraged my research every step of the way and have opened countless doors for my scholarship and career along the way. I benefited immensely from research collaborations with danah boyd and Jacob Metcalf, as well as Madeleine Clare Elish, Elizabeth Anne Watkins, and Ranjit Singh. I also benefitted from the generous mentorship and thought partnership I have received from Sareeta Amrute, Andrew Selbst, Mark Latonero, Alex Rosenblat, Kadija Ferryman, Jessie Daniels, and others. The winter intensive studies Janet Vertesi led were also invaluable to this dissertation, as were the daily writing calls with Robyn Caplan and Kinjal Dave.

Portions of work represented in Chapter 2 has been published in an article titled "How Modes of Myth-Making Affect the Particulars of DS/ML Adoption in Industry" in the Proceedings of the 2018 Ethnographic Praxis in Industry Conference (EPIC), which was co-authored with Friederike Schüür. Portions of work represented in Chapter 4 has been published in an article titled "Owning Ethics: Corporate Logics, Silicon Valley, and the Institutionalization of Ethics" in Social Research, co-authored with Jacob Metcalf and danah boyd, and in a research report for Data & Society Research Institute titled "Ethics Owners: A New Model of Organizational Responsibility in Data-driven Technology Companies" co-authored with Jacob Metcalf. The ideas and work presented in those chapters of this dissertation are my own, except where explicitly cited.

Thank you to my parents, Ron and Fran, for their loving support. Thank you to Adam and David for believing in me. Thank you to my dearest friends, near and—all too often—far. Thank you to Igor, Grace, and Cousin Tony for making my time away from home feel more like home. And thank you, Mona, for helping me see this through.

Finally, bottomless thanks are reserved for my many interlocutors across my field sites with whom I developed this research, who must remain anonymous at this time but whose contributions to this scholarship can be found on every page of this dissertation.

All errors, oversights, and omissions are, of course, my own.

*T*ABLE OF *C*ONTENTS

*T*ABLE OF *F*IGURES

## INTRODUCTION: BETTER LIVING THROUGH ALGORITHMS

In the spring of 2018, I was attending an all-hands meeting at an applied machine learning research lab called Over the Horizon in New York City. The meeting was dedicated to deciding what the lab's next project should be. It was focused on how best to demonstrate the capabilities of a new technique in machine learning called multi-task learning. The team was tossing around ideas about what kind of objective they could give to a product demonstrating this new technique that would best illustrate its capabilities. Markus, one of the senior researchers, spoke up, saying: "your objective function is subjective. The more fully you can describe your objective, the easier it is to automate everything else." Markus made this statement unironically. For him, and for other practitioners of machine intelligence, the objective function—the mathematical expression that acts as the fulcrum of machine learning—can be anything they want it to be. In practice, it is a piece of code that specifies their goal, the objective they are pursuing. But it is also a means through which they can make claims about the objectivity with which they pursue that goal. In other words, it is both how they accomplish their work, and how they present their work as authoritative.

Entitled "The Objective Function: Science and Society in the Age of Machine Intelligence," my dissertation explores how applied machine learning researchers construct the authority of machine intelligence. Ultimately, the authority of machine intelligence is derived from its conservatism. Contrary to the trappings of technological advancement and futuristic automation, it conserves the power of already-existing institutions and it reinscribes social relations of the past into the present and future. Machine intelligence may, on the surface, threaten the authority of powerful actors like judges, doctors, or loan officers by way of automation. But, in reality, machine intelligence preserves the power of the institutions — the courts, hospitals, and banks — that confer authority upon those actors.

The work presented here represents ethnographic fieldwork conducted over nearly two years, traversing an applied machine learning research lab in New York, two data science bootcamps, several machine learning conferences and artificial intelligence workshops, online discussion boards, interviews with practitioners in New York and Silicon Valley, and the artifacts produced by applied machine learning researchers—datasets, models, conference papers, and various other publications. The dissertation is roughly divided into two registers. One register examines how authority is constructed through the making of machine intelligence, as undertaken by applied machine learning researchers. The other register expands the aperture of study to examine how authority is constructed through the institutions and organizations it is developed for and within. It begins by arguing that machine intelligence proceeds from a "naïve" form of empiricism closely tied to positivist intellectual traditions of the 17th and 18th centuries. This naive empiricism eschews theory-laden approaches to data analysis, enabling what I call data performances. Through the statistical analysis of large datasets, these data performances enact the objects that applied machine learning researchers analyze as particular kinds of entities capable of being subjected to machine intelligence. The dissertation also examines other performances conducted by applied machine learning researchers to garner excitement about machine intelligence, as well as organizational commitments to support their work.

I draw upon Karen Barad's notion of agential realism to analyze the practices of machine intelligence as making an agential cut into unfolding, indeterminate fields of meaning and social relations to enact and stabilize specific configurations of them through the apparatus of machine intelligence. A machine intelligence analysis of sentiment using natural language processing—for example classifying whether a tweet about a particular name-brand product is positive or negative—performs sentiment as something inhering in text, situated across a specific gamut of possibilities, and independent of its context of use. It also performs positive-ness or negative-ness, as dimensions of sentiment, by mapping those sentiments onto statistical attributes of a body of text.

Over the course of the dissertation, I develop a method for tracing the secret life of data through tracing how the expert practices of applied researchers operate across datasets, algorithmic models, and software products. I use this method to reveal how situated, positional knowledge is transformed and re-presented as objective knowledge produced through machine intelligence. The later part of the dissertation focuses on how the authority of machine intelligence is actually extended through attempts to limit it. I discuss how AI ethics and algorithmic fairness assimilate outside critiques and concerns into the logics and practices of machine intelligence itself. Machine intelligence practitioners convert concerns about contested moral concepts like ethics and fairness into the practices of machine intelligence development. I show how practices of machine intelligence exert a form of deontic power that, in shaping what kinds of things its objects of analysis are, creates obligations for others to treat them as such. By this, I mean that the logics of machine intelligence are enfolded within a longer history of measurement, quantification, and accountability that acts as a check on other forms of expertise. This can be the case whether a judge is obligated to follow legislative sentencing guidelines or the outputs of an algorithmic recidivism prediction model. The dissertation ends with a demonstration of how, even through the pursuit of algorithmic fairness, machine intelligence re-entrenches not only the power of the institutions that wield it, but also reinscribes essentialist understandings of social categories like race and gender as the object of positivist knowledge, rather than as social constructs.

The authority of machine intelligence is constructed through data performances that enact its objects of analysis, through workplace performances that orchestrate organizational commitments to practices of machine intelligence, through institutional practices that extend bureaucratic power to machine intelligence, and through a wide range of practices that co-opt efforts to hold machine intelligence accountable. Nevertheless, my research also points toward ways to construct accountability regimes that avoid co-optation. Such regimes would involve practices that see the methods used to evaluate algorithmic impacts as also making agential cuts which stabilize particular understandings of algorithmic harm. Doing so would carry with it an imperative to include multiple forms of expertise—including that of technologists, social scientists, and those with lived experience as subjects of machine intelligence—in the construction of more robust forms of accountability.

Machine learning is difficult to define, even for experts. Differing definitions exist in textbooks, and practitioners will shift their working definitions to emphasize some aspects over others, in relation to their own interests or the audience they are speaking with. Briefly, machine learning refers to a set of computational techniques that use digital data to produce classifications that are somehow useful for making sense of or acting on phenomena that are linked to that data. As such, it is closely related to other knowledge practices like "data science" and "artificial intelligence". Machine learning can be understood as a particular approach to artificial intelligence, as a means for conducting data science, as a technical statistical practice, an approach to software engineering, and as its own area of applied and academic research. Rather than rigidly delineate these different registers of and relations to machine learning, I will refer to these practices collectively as *machine intelligence*. Before I shift my discussion to the slightly more expansive term, I will first introduce machine learning as a specific computational and statistical technique. Throughout, I will generally refer to my interlocutors as *applied machine learning researchers*; while I also engaged with a few individuals who referred to themselves as "data scientists," the vast majority of those interlocutors were trained in and employed to research and apply machine learning techniques.

The computational techniques machine learning uses are algorithmic, they are a rule-governed procedures for transforming inputs into outputs in ostensibly predictable, mechanical ways. Producing useful classifications, as machine learning does, has long been a power-laden exercise of institutions (Foucault [1966] 1994; Bowker and Star 1999; Hacking 1986), but over the first decades of the 21$^{st}$ century machine learning has rapidly become insinuated in the classificatory practices of government agencies, corporations, and disciplinary academic knowledge production. Machine learning researchers have developed applications of these techniques in domains as disparate as medicine (Cleophas and Zwinderman 2015), finance (Dixon, Halperin, and Bilokon 2020), policing (Ferguson 2017), business and marketing (Ammerman 2019), and music composition (Dong et al. 2018). Machine learning has skipped across the placid surface of vast and deep lakes of data, applying itself to certain well-posed problems in one domain, hopping on to other problems in other domains, and leaving expanding ripples its wake.

How has machine learning become able to glide so freely across, and to make such waves for, matters as important as criminal justice, commerce, medicine, media and the arts, mechanical engineering, and a seemingly endless list of other domains? In this dissertation, I take up that question by ethnographically engaging with how the authority of machine learning has been constructed such that it can influence so many domains, and I investigate what the consequences are of it being able to do so. In doing so, I move from the passive voice of so much popular discourse on machine learning, in which machine learning "has become a key technique for solving problems" (Mathworks 2020) or "has become far more accessible to non-experts" (Schwartz 2016) toward the active voice of applied machine learning researchers. In doing so I resurface the labor of such researchers, whose own contributions are structurally effaced from view in the final presentation of knowledge produced through machine intelligence. I examine applied machine learning researchers' training, work practices, public and private forms of discourse, and their corporate interlocutors to understand how the authority of machine learning is co-produced (Jasanoff 2004) with and through the knowledge it gives rise to. At the same time, I examine how this authority is co-constructed (Bijker, Hughes, and Pinch 1987) alongside the development of machine learning techniques, the growing availability of data, increased financing, and computational capacity—all of which are also within the scope of applied machine intelligence researchers' work lives.

## SITES AND METHODS

The dissertation is based on ethnographic fieldwork conducted over twenty-one months between 2017 and 2019 in several locations across the United States and at industry conferences held internationally. My primary field site was an applied machine learning research lab in Brooklyn, NY where I worked for twelve months as a technical writer and research team member. I also conducted participant-observation (Bernard 2006) at data science and machine learning bootcamps and meetups in New York City and San Francisco, where I also conducted an extensive series of semi-structured interviews, and I regularly attended machine learning conferences across the United States, Europe, and Canada. I closely read and analyzed the proceedings of these conferences, as well as other academic machine learning research papers. Additionally, I made extensive use of online discussion boards for machine learning researchers and practitioners, and I consider the public discourse about the growing influence of machine learning over society as a key component of my ethnographic analysis. Throughout this dissertation, I develop a set of theoretical and methodological lenses that can traverse the broad set of practices—professional, inter-personal, discursive, textual, and material—that comprise these sites, artifacts, and interactions. I draw heavily upon Karen Barad's notions of post-human performativity, intra-action, and enactment (Barad 2003; 2007) to understand how machine intelligence constitutes what I call *data performances* that reshape the onto-epistemological status of its objects of analysis (Chapter 1). I also draw heavily upon what Michael Silverstein and Greg Urban call "the natural histories of discourse" (Silverstein and Urban 1996a) in developing a method for understanding how data and machine learning come into intra-action with the phenomena they purport to represent, and how machine learning systems discursively construct analyses that come to be seen as meaningful and authoritative (Chapter 3).

The overlapping practices that constitute the authority of machine intelligence were visible to me in part because of my outsider status—I was able to "play the stranger" to machine intelligence without being the stranger, as Shapin and Shaffer enrolled themselves in their study of 17th century debates between Boyle and Hobbes about the experimental method (1985, 20). The role of professional stranger is, of course, familiar territory for anthropology (Powdermaker 1966) and has been a mainstay of laboratory studies approaches to STS (Latour and Woolgar 1979; Traweek 1988). I often had to make myself a stranger across the many sites I discuss in this dissertation, even though I entered those sites as someone positioned with the same race, gender, and class privileges to be there unquestioningly. I had to perform my own ignorance and naivety in order to ask the questions I wanted, or to receive the kinds of demonstrations and explanations that helped me document what applied machine learning researchers understand for themselves as the key practices of machine intelligence.

I crashed open houses for companies launching their machine intelligence products, I showed up for recruiting events at "bootcamps" offering training in machine intelligence to prospective students, I lurked on message boards hosting machine learning "competitions" and other discussions, and I applied for jobs I was only marginally qualified for at companies that were actively working in machine intelligence. One of these job applications actually resulted in an offer to work as a "technical writing intern" at the pseudonymous Over the Horizon (OTH), an applied machine learning research lab in the greater New York City area. It was there that I spent more than a year working with applied machine learning researchers, although I was often at pains to understand what distinguished "applied" research from "basic" research.

Over the Horizon called itself a "laboratory". It was founded and led by "Hanah," one of the first women to hold the title of "data scientist". Hanah had then worked as a consultant, crafting "data strategies" for large well-capitalized U.S. companies, and helped bring data science and machine learning to prominence through engaging and persuasive keynote talks at industry conferences. She founded OTH with "Markus," a PhD astronomer turned machine learning researcher. They quickly hired two other applied machine learning researchers full-time—"Frankie" and "Matthieu," and another two researchers part-time. The lab also employed "Graham"—a graphic designer and front-end coder, "Brendan"—a sales lead who had previously worked as a quasi-venture capitalist in a research and development agency within the U.S. federal government, "Roger"—a lawyer who served as chief legal counsel and was teaching himself machine learning in his downtime, "Daveeda"—the OTH administrative assistant, and several "interns" including myself. Other employees would come and go during my tenure at OTH. Matthieu relocated to Los Angeles and began working remotely six months after I joined OTH. Two other applied machine learning researchers joined the laboratory towards the end of my tenure there, and several sales and marketing specialists joined for short periods of time. Before I concluded my fieldwork, OTH merged with "Nebulosa," a large publicly-traded cloud services company, but OTH maintained its own Brooklyn workspace instead of moving into Nebulosa's Manhattan offices. Hanah led business development, bringing in potential clients through one-on-one meetings and by giving keynotes at industry conferences attended by potential clients, and only worked "hands on" with a few select clients. Hanah handed most potential client leads off either to Brendan, the sales lead, or directly to Markus or another applied machine learning researcher.

In addition to publishing research reports and developing prototypes to demonstrate "new and newly possible machine learning techniques" for their clients (see Chapter 1), OTH also served as a consultancy to large companies that had small data science and machine learning teams but lacked the resources to conduct their own applied research. Overall, OTH was a crucial node in an effort that spanned the entire tech industry to build a market (Strasser 2004) for machine intelligence itself. In my capacity as a "technical writing intern," I would draft short synopses of recent events in machine learning for the OTH newsletter, which was mailed to subscribers on a weekly basis. I would also write longer, 800-1200 word blog posts for the OTH website, as well as summaries of academic machine learning research articles that had recently been posted to the pre-print server arXiv[1] or in the proceedings of machine learning conferences. As I became more involved in writing research reports, I would interview practitioners and developers of the techniques which were the subjects of reports. I would research application of those techniques and try to identify the business case for using them as well as the ethical implications of doings so. Eventually, I would write chapters of the reports myself, under the supervision and direction of the applied machine learning researcher who was acting as lead author of the report. I would also sit in on meetings with clients and take notes, occasionally asking probing or clarifying questions. As time went on, I would help prepare workshop presentation materials, help conduct data strategy workshops, and help write the workshop reports for clients to use as they integrated OTH recommendations.

---

[1] https://arxiv.org/

These consulting interactions—particularly the data strategy workshops—are largely the focus of Chapter 2, but working at this lab connected me to the several other sites that comprised my field site(s) and gave me a modicum of legitimacy with which to gain entrée. Lab researchers also connected me to the sites where machine intelligence methods were legitimated: industry conferences, research conferences, salons, and online discussion boards where machine intelligence techniques are commonly shared and discussed. While working at the lab, I learned which conferences are seen as trustworthy for the quality of research presented there, which are seen as useful for networking and understanding business needs for and applications of machine intelligence, and how applied machine learning researchers navigate posts on discussion boards. I used these para-sites to understand how techniques gain prominence as acceptable best practices for applying machine intelligence, and how what machine intelligence *is* changes as they move beyond the laboratory—where it is an indeterminate product open to contestation—to the conferences, boardrooms, and applications where its authority is enacted.

The researchers at the lab also had strong connections to data science and machine learning "bootcamps" where machine intelligence skills were taught outside of a traditional university setting. Some of the lab researchers were themselves former graduates of or staff at these bootcamps, and I benefited from introductions made on my behalf that connected me to current teachers and recent graduates, beyond what I would have been able to negotiate for myself by merely attending open houses and recruiting events. These forms of entrée, as well as a research collaboration with Data & Society Research Institute that gave me access to an even broader network of applied machine learning researchers and others within the technology industry, were useful when I moved to San Francisco, CA for a second phase of ethnographic fieldwork. There, I continued attending machine learning events—meetups, conferences, and workshops—as I conducted long-form, semi-structured interviews with researchers. These interviews also addressed the emerging sets of practices within and adjacent to machine intelligence that were concerned with what was becoming known as "AI ethics" (Chapter 4) and "algorithmic fairness" (Chapter 5).

Throughout the fieldwork, I regularly attended academic and industry-focused machine intelligence conferences. Academic machine intelligence conferences are fertile sites for ethnographic study, as they present formal descriptions of academic work alongside discussions of this works' history and underlying motivations, as well as reflections on how work has been taken up by the rest of the field. Formal presentations—delivery of conference papers and demonstrations—are accompanied by informal mixing, unscripted discussions of resolved and unresolved challenges, and narrations of self in relation to various machine intelligence undertakings. Two of these conference papers are analyzed in detail below (Chapters 3 and 5). I attended several academic machine intelligence conferences throughout my fieldwork: the 2018 International Conference on Machine Learning (ICML) in Stockholm, Sweden and the 2018 Neural Information Processing Systems Conference (NeurIPS) in Montreal, Quebec. These conferences are massive, in terms of attendance and square footage of convention space. Post-fieldwork, when I attended the 2019 NeurIPS Conference in Vancouver, British Columbia, it was the highest attendance they had ever had at their (capacious) waterfront convention center. I also attended several smaller "special interest" machine intelligence conferences during my fieldwork. These smaller conferences are usually held under the aegis of a larger professional society; in my fieldwork these societies were the Association of Computing Machines (ACM) and the Institute of Electrical and Electronics Engineers (IEEE).

## An Epistemological Fetish

Throughout this fieldwork, and over the course of this dissertation, my aim is to expand the lens through which the practices of machine learning researchers and their interlocutors can be seen as contributing to the authority of machine learning in everyday life. Generally, my goal is to focus on how machine learning foregrounds its "automatic" and even "magic" (Elish and boyd 2017) aspects while obscuring the economic, political, social, and labor relations that undergird it. These relations, which are enacted by the practices of those who work in the field of machine learning, also connect a range of other actors—"data subjects" who produce detailed data traces with their everyday existence in the world, private corporations that collect real-time data through networked devices like smartphones and networked transport vehicles, bureaucracies that publish datasets, digital infrastructures of healthcare and media, platformed gig-work laborers, advocates of digital rights, emerging data policies and regulations, "e-commerce," basic and applied research funding from both corporate and government sources, and the powerful graphical processing unites (GPUs) housed in nebulous server farms where data is processed by machine learning algorithm. I hope to show, as these relations rapidly fade from view through the intentional, purposive effort of machine learning engineers, how the products of machine learning come to instead appear as separate from their labor—an *epistemological fetish* like Marx's commodity (K. Marx [1867] 1990, 165), not only automatic or magic, but also natural and inevitable.

The primary contribution of this dissertation is to show how machine learning has gained a great deal of its authority at the same time that it has become a field on which long-standing debates play out; machine learning has become a focal point in debates over what distinguishes science and engineering, the legitimacy of private enterprise, the ethics of prediction, the validity of social categories, the limits of data's ability to mediate real-world phenomena, and the responsibilities of both private and public institutions. Where does the production of knowledge stop, and the application of knowledge begin? Machine learning further blurs the line between science and engineering as applied machine learning researchers continually ask themselves, and each other, whether they are generating novel insights into real-world phenomena or merely making data accessible through complex data engineering and descriptive statistics. What role do new technologies play in how organizations change (Volkoff, Strong, and Elmes 2007)? Machine learning finds itself at the center of prior tensions and conflicts within organizations as they make recommendations for how decision-making power can best be allocated to machine learning tools, and for how organizations can reallocate their data resources productively (Vedres and Stark 2010). How are racism, misogyny, prejudice, and other forms of social bias perpetuated through mundane bureaucratic practices of accounting? Machine learning finds itself subject to trenchant critiques of automated bias (Angwin et al. 2016; Noble 2018; Barocas and Selbst 2016; Tu, Nelson, and Hines 2001) and its practitioners position themselves as well-suited to provide technical fixes to these long standing issues (Metcalf, Moss, and Boyd 2019; Selbst et al. 2019). And, what is the best way to act toward others? Machine learning has found itself at the center of centuries-old debates of ethics and moral philosophy in which the stakes of virtuous personal behavior seem incredibly high, the role of government in regulating the use of data about its citizens' every action becomes a major question, and calculating the balance between the greatest good for the greatest number suddenly seems possible.

As applied machine learning researchers have found themselves at the center of long-standing debates, the challenges they face and the responses they mount are, in many ways, continuous with how those in prior iterations of these debates have acted. But in other ways these debates, when framed around machine learning, present several discontinuities with the past. This is not to say that the technical aspects of machine learning are in any way determinative (cf. Winner 1986), but rather that machine learning connotes a range of sociomaterial practices (Mol 2002; Barad 2003; L. Suchman 2007), i.e. "the constitutive entanglement of the social and the material in everyday organizational life" (Orlikowski 2007, 1437), that have altered the terrain upon which science, engineering, entrepreneurship, social science, ethics, and governance occur.

## WHAT IS MACHINE INTELLIGENCE?

This dissertation takes machine *intelligence* as its object of analysis, ethnographically accessed through participant observation with those who practice machine *learning*—applied machine learning researchers and other technologists, as well as discursive analysis of public writing and messaging about machine learning. Apart from any singular textbook definition, I take machine intelligence to encompass a range of practices, technologies, analytical lenses, and imaginaries that substantially overlap with "big data," "data science," and "artificial intelligence".[2] By doing so, I extend the lines of critical questioning that have been levied at big data (boyd and Crawford 2012) and artificial intelligence (Forsythe 2001) to machine learning. These lines of questioning probe the inherently *social* character of machine learning by asking how machine learning practices participate in the production of knowledge, leverage that knowledge for the accumulation of power and capital, and "delete the social" (Star 1991; cited in Forsythe 2001, 53)—both in terms of the social relations that structure the work of machine learning engineering and the socially-constructed components of what machine learning takes as the given inputs to its data processing. The many varieties, techniques, and methods that comprise machine intelligence are both complicated technical concepts requiring detailed explanation and complex sociomaterial practices requiring ethnographic analysis situated within some degree of social, economic, and political context. I take machine intelligence as an expansive set of technical, interpersonal, economic, discursive, and cultural practices that are themselves sociomaterial and operative across scales. Occasionally, I discuss *algorithms* more specifically as the rule-governed technical components of a machine learning system. It will from time to time become necessary to pause the narrative and undertake a detailed technical explanation of a particular term or method before being able to continue the analysis, as I do throughout the dissertation.

---

[2] Exceedingly briefly, big data refers to a set of computational capabilities of, complemented by social and organizational practices that contribute to, recording and storing massive amounts of information about a wide range of phenomena digitally, in ways that support its retrieval and analysis (see Mayer-Schönberger and Cukier 2013; Podesta 2014). Artificial intelligence refers to the ability of computational systems to perform "human-like" tasks ranging from simplistic game play to reading and writing to robotic surgery to imperceptibly replicating the entire repertoire of human behavior. Artificial intelligence research on so-called "expert systems" (see Buchanan and Shortliffe 1984; Forsythe 2001) predates developments in big data and machine learning that have undergirded the current "second wave" of big data- and machine learning-fueled artificial intelligence (see Hurlburt 2017).

While some applications of machine intelligence might seem like frivolous novelties, or of niche interest to specialists, machine intelligence is also deeply implicated in sociotechnical systems of sweeping reach into the daily life of ordinary human subjects living in the digitally-mediated contemporary world. These systems use machine intelligence to generate product suggestions for e-commerce sites used by hundreds of millions of shoppers, to dynamically place advertisements on websites tailored to the predicted interests of website visitors, to rank and sort news stories seen on social media sites, to inform judges about the likelihood a defendant might violate the terms of pretrial release, to allocate social services targeted for high-risk children, and to accomplish thousands of other tasks on a daily basis. Indeed, machine intelligence is increasingly implicated in how knowledge is produced and how powerful institutions—markets, governments, and corporations—are configured to make use of knowledge, allocate resources, and exercise power.

Rather than asking how machine intelligence has reshaped these institutions or how the needs of these institutions have led the state of machine learning to its present configuration, this dissertation asks how machine learning and the social uses to which it is put are socially constructed (Bijker, Hughes, and Pinch 1987) through the practices of applied machine learning researchers and those they interface with. Machine learning is only one node in the meshwork of people, organizations, practices, and technical artifacts from which these systems emerge, but in the 21$^{st}$ century it has become an increasingly important node. This importance cannot be neatly explained by any steady improvement in machine intelligence as a technology *per se*, but instead will be explored by interrogating how machine intelligence has come to be seen by various groups as a solution to their problems *as they perceive them*. These groups reorganize their practices—of data collection and decision making in particular—to make use of machine learning, and in solving these groups' problems, machine intelligence also produces knowledge about organizations, social interactions, and the physical world. This dissertation also adopts a co-productionist approach that sees the practices of machine learning and the knowledge produced by machine learning "as being produced together" (Jasanoff 2004, 2). Doing so opens up the ethnographic study of machine learning across scales—from the bits of computer code written in a cramped meeting room to sweeping reevaluations of scientific knowledge in disparate domains, and from academic conference papers to the predictive systems that fade in the background while still exerting high degrees of control over flows of people, information, goods, and capital.

As with other new technologies, the role humans play in, at turns, making machine intelligence visible or invisible is of crucial importance for this dissertation. Importantly, the dissertation will argue that the authority of machine learning is in large part produced through how it both erupts into view, at times dominating or making highly public contributions to conversations about matters of significant public concern, but also recedes from view, at times operating behind the scenes as the inputs to other systems, or under the cover of other identities like "artificial intelligence," "chatbots," "analytics," or any number of other aliases. In part, this dissertation answers the call of Susan Leigh Star "to study boring things" (Star 1999b, 377). But beyond this, it studies how machine learning has been *made* boring; how it has taken on the properties of infrastructure that Star enumerates, particularly its *embeddedness* within other structures, the *transparency* with which it is integrated into daily life, its *reach* beyond local contexts, its reliance on an *installed base* of prior technologies, and the ways in which it becomes *visible upon breakdown*. Using these properties as a rubric for studying machine learning points towards and organizes my analysis of the practices of the applied machine learning researchers I studied amongst. They labored to embed their work into other structures, to connect their code to already installed bases, to extend its reach and scope, to make it transparent to their clients and other end users, to return it to invisibility following breakdown, and so on.

In the sense that machine intelligence is not hidden in the background for those I studied with, this dissertation is only truly an "infrastructural inversion" (Bowker and Star 1999, 34) when it focuses on that which has been made infrastructural for machine intelligence, particularly when exploring that which is infrastructural to my interlocutors. Exploring datasets like MNIST and the statistical packages that are ready-to-hand as the "background for other forms of work" (Star 1999b, 380) in the way they are embedded into software programming packages is a form of infrastructural inversion. These explorations delve into how machine intelligence has been made possible by prior labor dedicated to making the world representable as data and justifying the need to do so, as well as laying the groundwork for the application of machine intelligence through audit and accountability practices that lend themselves to automated decision-making systems that can embed machine intelligence.

Machine learning, then, is not merely a node that lies betwixt data collection practices, information infrastructures, and social practices. Although they work within a liminal space (V. W. Turner 1979) in which machine learning models are trained and then deployed as objects that tend toward closure (Latour 2003), machine learning engineers and researchers are active participants not only in how knowledge is collected, analyzed, and operationalized to predict and classify, but also in how those predictions and classifications are made to matter for others. This "mattering" may take on many forms, but the role researchers play in making machine learning matter has as much to do with stabilizing the truth value of their analyses as it does with justifying the financial returns on any investments in their work or building an aura of magic and enchantment around the technological objects they construct (Gell 1996; Nye 1996; Elish and boyd 2017). Human society has long endeavored to make predictions through oracles (Evans-Pritchard 1976; Flad 2008), prophesy (Guyer 2007), or the scientific method (Popper [1935] 2002). Each of these techniques participate in, depend upon, and help to build authoritative narratives about the nature of reality. For machine learning, the particular nature of reality implicated in how its techniques produce predictions and classifications is comprised of several interrelated trends in the production of knowledge through which its claims are made meaningful and actionable.

While "trends" is perhaps too small a word for such sweeping approaches to the production of knowledge, machine learning nevertheless depends on several distinct, but overlapping, historically situated developments in knowledge production: positivism, objectivity, quantification, and datafication. While these developments are often discussed as if they were synonymous with each other, and with modernity writ large, I will accept the challenge offered by Lorraine Daston and Peter Galison to see these developments as "distinct epistemic virtues" that "may sometimes converge … but … may also diverge" (Daston and Galison 2010, 4). In doing so, I posit that machine learning, artificial intelligence, big data, and data science represent a specific mobilization of these epistemic virtues, which I refer to as *machine intelligence*, that has its own historical, geographic, and political trajectory and which will be explored throughout this dissertation.

I use "machine intelligence" not to posit a particular type of "intelligence" that can be compared to other types of intelligence, or to posit a definition of intelligence that would allow for comparison between humans and machines (Legg and Hutter 2007). Rather, I use "machine intelligence" to refer to the broad class of sociotechnical systems built around computerized predictions and classifications, and to transcend debates about fine-grained distinctions between "data science," "machine learning," "deep learning," and "artificial intelligence". These terms act as boundary objects (Star 1989) that have different, but significant, meanings within various communities of practice across the broad domain of machine intelligence. While particular meanings are important for specific communities of practice, and will be discussed with the specificity required to distinguish between different sets of practices and organizational solidarities, "machine intelligence" will be used when discussing the set of capabilities that allow for, and buttress the authority of, computational prediction and classification.

## STUDYING MACHINE INTELLIGENCE AS SCIENCE, AS TECHNOLOGY, AND AS CULTURE

In studying machine intelligence through an ethnographic investigation of the practices of applied machine researchers, my goal is not to prove or disprove the claims of applied machine learning researchers on their own terms. Rather, I seek to understand how applied machine learning researchers contribute to the idea that numbers can speak for themselves. This requires asking how applied machine learning researchers make numbers speak, how machine intelligence is constructed as having validity, how it exists alongside (and often in tension with) other ways of knowing the world, how it participates in the authority granted to machine learning practices over a wide range of domains, and how applied machine learning researchers grapple with the "theory-free" claims of machine intelligence as they are integrated into broader social networks of which they are only a small but powerful element. These epistemological and ontological questions—how machine intelligence is able to make knowledge claims and the kinds of things machine intelligence is able to make knowledge claims about—speak directly to the core concerns of science and technology studies (STS) and the anthropology of science and technology. Here, I use ethnographic methods to empirically demonstrate how the knowledge claims of applied machine learning researchers attains the status of authoritative truth through the application of "machine intelligence". These claims in turn stabilize and are stabilized by the technological artifacts, machine learning practices, and objects of analysis that applied machine learning researchers engage with.

To do so, I draw heavily from across the range of theoretical lenses STS and anthropology offer, seeking insight into the consequences of machine intelligence and how it is made to travel far beyond the laboratory. I make no attempt to police a disciplinary divide between STS, anthropology, sociology, media and information studies, or other related disciplines that are relevant to my own work. Nor do I make any claim that the use of ethnographic or other methodological approaches can neatly distinguish any of these disciplines from the others. Rather, I draw liberally from across these intellectual traditions by acknowledging the decades-long interchange between of ideas between these literatures and the commitment to ethnographic methods (Hess 2001) within these disciplines (see, e.g. Martin [1987] 2001; Gusterson 1998; Kelty 2008; boyd 2014; Vertesi 2015). That said, it is worthwhile to trace the intellectual trajectories of science and technology studies, as well as anthropological approaches to the study of science, which emerged from the philosophy of science (Popper [1935] 2002; Fleck [1935] 1979) and the sociology of knowledge (Merton 1937) that were active areas of inquiry during the inter-war and post-WWII periods. These incipient fields tended to focus on the exceptionalism of science as a mode of knowledge production and the conditions that made science possible. At around the same time, historians of science and technology were writing triumphalist and "great man" narratives (Boring 1950) celebrating unidirectional progress and techno-utopian futures (Segal 2005; F. Turner 2006), while Marxist approaches to the sociology of technology were invested in determinist accounts that privileged the role of technology in producing particular social orders (Shaw 1979; Thompson 2016).

Such determinist narratives, as told by many applied machine learning researchers and their interlocutors, play a significant role in constructing the authority of machine intelligence by presenting it as an irresistible force that will reshape every field of human endeavor and even replace human physical, intellectual, and emotional labor in the near future (Singh 2019; Schwab 2015; Manyika and Sneader 2018). Boosters of machine intelligence craft seductive depictions of technological objects that will, one day soon, embody machine intelligence and operate in the world autonomously at a degree of performance that rivals or exceeds human analogues in ways that can nevertheless be seamlessly integrated into already-existing roles. These claims are evident in breathless reports of the coming wave of AI replacing truck drivers (Linder 2019; cf. Levy 2015), making doctors obsolete (Simonite 2019; cf. Elish and Watkins 2020), and even spelling the end of capitalism (Marr 2018). Triumphalist narratives about the inevitable progress of machine intelligence are not immune from the tendency to engage in "great man" hagiography, either, as computer scientists who conducted some of the foundational basic research underlying machine learning techniques—Turing Award winners Yann LeCun, Yoshua Bengio, and Geoff Hinton—have been elevated to the to the pantheon of "tech saviors" alongside Silicon Valley entrepreneurs like Elon Musk, Mark Zuckerberg, Jeff Bezos, and Bill Gates (Wiggers 2019). The lenses of STS provide ways of challenging such determinist narratives through a consideration of technology as socially constructed (Bijker, Hughes, and Pinch 1987) and by moving toward sociotechnical frames of analysis (Bijker and Law 1992). Importantly for questions addressing the authority of machine intelligence, STS challenges technological determinism by grappling with the political forces that shape technology and the political imperatives that technology serves (Winner 1980; Joerges 1999), the sociotechnical imaginaries that lend validity to technological interventions (Hecht 2009; Medina 2011; Halpern 2015), as well as the ways in which technological systems, objects of knowledge, and social orders co-produce each other through discourses, institutions, representations, and subjectivities (Mol 2002; Jasanoff 2004; Barad 2007).

I also draw heavily on theoretical lenses from the anthropology of science and technology, particularly those which overlap with, inform, and are informed by approaches from STS. Anthropological work that I draw upon has focused on the laboratory as a site for ethnographic investigation into knowledge production (Latour and Woolgar 1979; Traweek 1988; Roeppstorff 2002). Lab studies were part of a broader turn within social sciences, and particularly anthropology, away from structuralist-functionalist approaches and toward Marxian frames of analysis that attend to modes of knowledge production and subjecting those with power to the ethnographic lens (Nader 1969). Laboratory studies can be seen as, in part, transporting the "culture concept" (Geertz 1983; Boas 1940; Benedict [1934] 1959) of anthropology to science studies (Knorr-Cetina 1999), but beyond that laboratory studies also were concerned with explaining how the apparent stability of facts was constructed. Laboratory studies paid attention to how knowledge was assembled, how controversies became settled as facts, and how laboratory practices acted as sites of labor through which raw materials were transformed into finished products (Roeppstorff 2002; Lynch 1985; Traweek 1988; Latour and Woolgar 1979). Portions of this ethnography fit nicely within the laboratory studies model, as analysis of the practices of applied machine learning researchers within the lab, and their role in transforming the world beyond the lab (Latour 1983), are crucial for understanding how the authority of machine intelligence is produced.

My ethnographic field site begins within, but extends beyond the laboratory—it both is and is not a laboratory study. It addresses how the applied machine learning research lab I worked in is a site that produces knowledge about machine intelligence, machine intelligence products, and the authority of machine intelligence from its inputs—clients' data, researchers' labor, other researchers' published papers. But I also study how those artifacts that flow into and out of the lab. I examine how research papers, as a mode of scholarly communication closely tied to the experimental method (Shapin and Schaffer 1985), make epistemological and ontological claims about the world. I examine where the expertise of lab researchers comes from, by visiting machine learning bootcamps and engaging in close readings of instructional texts. And I examine how machine intelligence products—algorithmic systems, statistical models, and software services—intersect with organizational practices outside the lab, particularly with respect to how those inside Silicon Valley tech companies must grapple with the ethical implications of machine intelligence. Many of these ethical implications arise from how the authority of machine intelligence is implicated in transforming the world beyond the laboratory (Latour 1983). Therefore, I also draw upon ethnographic work that explores how scientific metaphors produce meaning within and beyond the laboratory (Martin [1987] 2001; Gusterson 1998) and can transform relations of power organized through the management of scientific systems (Helmreich 2009; Caduff 2012). However, machine intelligence depends upon ways of structuring data production and collection, as well as non-computational forms algorithmic decision making and knowledge production, that dissolve boundaries that might cleanly divide what can be described as "inside" or "outside" the "lab". I also draw upon ethnographies of technology that examine how the design and affordances of the built, and digital, environment shape—and are shaped by—social action (Mcintosh 2010; Coleman 2010; Schüll 2012; Dumit 2012). Additionally, prior work exploring the epistemological and ontological implications of expert ways of seeing (Dumit 2004; Saunders 2008) informs my own.

Data science, machine learning, and artificial intelligence are relatively recent topics of study for social science. Few laboratory ethnographies of machine intelligence have been conducted, and little of this work has yet been published (see Passi and Jackson 2018). The earliest social scientific work on artificial intelligence can be traced to Margaret Mead's involvement in the Macy Conferences on cybernetics (Pias and Vogl 2003) and Diana Forsythe's groundbreaking work on "expert systems" approaches to artificial intelligence (Forsythe 2001). More recent related social scientific work has interrogated the epistemology and claims to objectivity of data (boyd and Crawford 2012; Gitelman 2013), the methodological challenges of studying algorithms ethnographically (Dourish 2016; Seaver 2017), and the role of machine learning in producing knowledge and shaping subjectivities (Cheney-Lippold 2011; Mackenzie 2017). Ethnographies of the working world of software development, computers, and computing (Kelty 2008; Coleman 2013; Amrute 2016) have focused on computer work as labor, but also as a grounds for ideological contestation and the formation of race and class identities (see Tu, Nelson, and Hines 2001; Noble 2018; Benjamin 2019b). My own work builds on this prior work by tracing the authority of machine intelligence across several sites of its production and elaboration. I interrogate how machine intelligence produces knowledge and trace how it does so by shifting the ontological status of its objects of knowledge, with particularly drastic consequences for conceptualizations of human difference. I extend methodological approaches to the study of algorithmic systems attending to how machine intelligence is enacted through data performances, in the laboratory, within the tech industry, and across the academic literature of machine intelligence.

I define "authority" as both the power to influence, grounded in specialized forms of expertise, and the power to control, here grounded in automated systems that predict and classify in ways that can be used to determine access to information, resources, and spaces. These two registers through which to understand the authority of machine intelligence roughly map onto Deleuze's distinction between "disciplinary societies" that shape categories through which knowledge is made, and "societies of control" that shape how access to spaces and resources in managed (Deleuze 1992), albeit with the kinds of significant overlaps that characterize Deleuze's own distinctions between these two "societies". My investigation of machine intelligence's authority as an influential power grounded in expertise proceeds from prior work on how science and statistics have been implicated in the production of trust in individuals and institutions. Datafication, as discussed above, has played an important role in shifting legitimate authority from that of the sovereign to that of bureaucratic institutions. This authority has depended upon the capacity for statistical representations of the world to be verified through replicable, transparent mechanisms (Porter 1995). These "rituals of verification" are scripted interactions that require expertise to navigate and in so doing demonstrates the trustworthiness of officials for skeptical publics (Power 1997). Similarly for scientific knowledge production, scientific authority depends on the ability of experimenters not only to persuasively document and disseminate experimental findings over time and distance but also to be seen as trustworthy documentarians of those proceedings (Shapin and Schaffer 1985). For practitioners of machine intelligence to make authoritative claims of expertise across many the many domains of knowledge production that they do, they develop their own "rituals of verification"—namely, the optimization of an objective function. They also engage in practices that construct their authoritative expertise across time and distance through the range of discursive activities—workshops, conference presentations, academic papers, and spectacular public demonstrations—that I document throughout this dissertation.

I also examine how the authority of machine intelligence based on claims to expertise, depends on bringing the objects of its expertise into existence. This becomes legible by applying feminist STS framings of performativity that approach objects of knowledge as *hybrid*; as material-semiotic entities whose agencies and meanings must be enacted through skilled human and non-human work. Early work in feminist STS pointed to the embodied qualities of knowledge and called for the study of women's work, women's bodies, and other viewpoints that had previously been ignored or overlooked by male scholars, both in STS but also in the sciences themselves (Martin [1987] 2001; L. Suchman 2008; L. A. Suchman and Wynn 1984; L. A. Suchman 1980; Rapp 2005). Additionally, feminist STS, drawing heavily from Marxist feminism, studies the invisibilized forms of labor that undergird hegemonic power structures (Rubin 1975) but outright rejects the "othering" dualisms of alternate approaches to STS. Beyond sex and gender binaries, feminist STS breaks down the nature/culture, science/society, representation/reality, epistemology/ontology, and object/subject dualisms that characterize approaches to STS to instead provide a new vocabulary around performativity and the social production of such distinctions (Haraway [1985] 2004; Barad 2007; Mol 2002). This is not to say that these distinctions are not real, but that they must actively be brought into existence through enrollment, enactment, and performance. I focus in particular on Karen Barad's "agential realism" to show how machine intelligence ought to be thought of not so much as analyzing statistical representations of real-world phenomena or revealing hidden truths about reality, but instead as an agentive practice that cuts into an indeterminate social world to enact the objects of its analysis.

Focusing on the agentive aspects of machine intelligence brings into sharp relief the multiplicity of standpoints and perspectives it brings into an assemblage (G. E. Marcus and Saka 2006). Applied machine learning researchers have their own standpoints, as do the corporate entities the work for and within, which generally correspond with the white, male, upper-middle class subject positions that predominate in Silicon Valley, the tech industry more broadly, and academic computer science during the time period of my fieldwork. So too do the developers of machine learning packages they employ, as well as the datasets they work with, which were collected within particular contexts for specific purposes (Rosenberg 2013) that often differ from the contexts in which they are incorporated into a machine intelligence project. Feminist approaches to STS see these forms of knowledge as *situated*; what we know depends on who we are and how we come to know things (Haraway 1988; Harding 1991). The knowledge produced by machine intelligence, then, is enacted by and through these variously-positioned standpoints.

I use actor-network theory (ANT) to make these semiotic-material enactments are more legible (Latour 2005), to help discern how these situated knowledges are assembled together with the artifacts—algorithms, software, datasets, and computer servers—of machine intelligence. Actor-network theory draws attention to the mundane objects—Latour's seat belts, door grooms, and speed bumps (Latour 1992)—that act in social life, and how they come to have meanings through which they are able to act agentially. Such a framing is useful even if many of the non-human actors of machine intelligence do not present themselves as nearly so mundane. Examples of spectacular new technologies that act "autonomously"–from self-driving cars to script-writing robots (Sharp 2016)–abounded during my fieldwork, and represented a significant component of how the authority of applied machine learning researchers was produced. Nevertheless, the idea that inanimate (or even autonomously animate) objects are embedded within social networks, and are themselves social actors, is fertile ground for understanding how machine intelligence's predictions and classifications shape and reshape sociocultural understandings.

Attending to machine intelligence as an actor in social life also points toward the controlling form of authority it holds. Incorporated into automated decision-making systems, machine intelligence is a set of practices that produce disciplinary effects through classification—determining what is normal and what is abnormal, or what is acceptable within a given threshold, for someone to receive a loan, or parole, or a medical treatment, or even an online advertisement. But in enforcing that threshold machine intelligence constitutes what John Searle calls "deontic power," which induces people to act in ways other than they might desire (Searle 2006). This constitutes a form of control that in turn shapes and reshapes what it means for a subject to be loan-worthy, an exemplary candidate for parole, or a viable candidate for a medical treatment, but also the degrees of freedom a loan officer, a parole board member, or a clinical research board might have when they are expected or required to "automated decision-making systems" that operationalize machine intelligence.

The authority granted to machine intelligence to control, here, answers Langdon Winner's question, "do artifacts have politics?" in the affirmative (Winner 1980). Machine intelligence is increasingly the means through which political goals are accomplished, not just in the distribution of government services (Eubanks 2018), but also in how political, technological, and scientific orders are intertwined (Jasanoff 2004). Automated decision making systems which operationalize machine intelligence enact political agendas by controlling access to information and resources (Pasquale 2015) at the same time that "the faith we put in the power of technology shields algorithmic systems from critical interrogation" (Waldman 2019, 614). Narratives about technological artifacts—their efficiency, tirelessness, and equanimity—serve political agendas (Joerges 1999) at the same time that these artifacts have their own politics. In this dissertation, the technopolitics of machine intelligence is primarily focused on how indeterminate social concepts, particularly race and gender, but also justice and ethics, are made to seem like calculable objects of knowledge, and the attendant consequences of this process. This "ethico-onto-epistemology" (Barad 2007, 90) is a particularly useful lens for studying how applied machine learning researchers, critical scholars, community advocates, and others who hold stakes in how machine intelligence is constructed understand the ways they are entangled with the risks and harms of algorithmic systems that affect the world in serious, but sometimes unpredictable ways (Metcalf, Moss, and boyd 2019), particularly given the entanglements between machine intelligence, venture capitalism, and neoliberal imperatives toward efficiency and corporatization (W. Brown 2015)—particularly in the North American context I studied.

## The Working World of Machine Intelligence

Mid-morning inside an anonymous convention center that could have been in Miami just as easily as it could have been in Montreal, following opening remarks and a panel discussion but prior to the much-needed catered coffee break, a critical scholar of data began her keynote by polling the assembled audience by asking, "How many of you are workers?" In an audience of several hundred, less than two dozen hands went up. She followed up with, "How many of you are employed by a tech company?" and almost half the audience raised their hands. "Everyone who has raised your hands, keep them up. One last question," she went on. "How many of you work for an academic research lab?" Most of the rest of the audience's hands went up. "Now why didn't you all raise your hands to the first questions!" she finished with a rhetorical flourish. Her point, which was not lost on this particular audience, was that no matter highly valued, rarified, or elite the work of applied machine learning research is seen as, it is still nevertheless *work*. This message was being delivered in early 2018, at a time when applied machine learning research was being written about in the popular press as one of the most lucrative jobs available in North America (Rayome 2018).

Machine learning, at that time, was in its full fluorescence. It had been applied to problems that previously seemed intractable, like complex image recognition (He et al. 2016), and was regularly completing tasks that had been set up as "grand challenges," like besting human competitors in multiplayer video games (Vinyals et al. 2019). And those applied machine learning researchers who were accomplishing these feats were being portrayed in hagiographic terms, given prestigious computing awards, able to command huge salaries and massive grants to lead research teams, and could move from company to company or research lab to research lab (Vincent 2019). To some extent, it seemed like every applied machine learning researcher saw themselves as potential Turing Award winners, capable of steering the field as they might like, working on what they might like, when they might like. But for the vast majority of such folks, they worked for a startup, a corporate research lab, an academic research incubator, or on a small data science team inside a company that might not even realize they *had* a data science team.

For those who did work in tech-centric companies, the worker-like quality of their careers could be hard to discern. The applied machine learning research lab where I conducted my Brooklyn, NY fieldwork lacked the foosball table, the climbing wall, the videogame consoles, the catered lunches, and the kombucha bars that could be found at some of the Silicon Valley tech company headquarters I also visited. But it did have a fully stocked fridge and, after being acquired by a much larger Palo Alto, CA based company, an entire "snack station" with individually wrapped energy bars that would have cost up to $3 at the bodega around the corner, single-serve breakfast cereals, and more beef jerky than seemed economically or environmentally sustainable. It also had a hammock, a bottle or two of bourbon on the bookcase next to some boardgames, and ample indoor bicycle parking. All of these amenities, and the ability to work from home on any given day, made it somewhat easy to believe that the office was just another place to hang out for a bit, and not a workplace.

The other implication of what that critical data scholar was trying to get across to her audience was that each person sitting in front of her ultimately played a small part in larger projects, the goals of which were determined at levels beyond their control, and that any agency they held over their own working lives was bound by that of the salaried labor relationship—no matter how much their time at the office resembled play more than it resembled work. It was around the same time of this conference that a number of stories began appearing in the popular press that even more viscerally reminded everyone that tech work was, in fact, *work*. The full sweep of the consequences of these stories for how applied machine learning researchers approach their work is told in Chapter 4, but one thing that the Cambridge Analytica revelations about the use of machine intelligence in manipulating the U.K. Brexit vote and the 2016 U.S. election (Cadwalladr and Graham-Harrison 2018), the Project Maven contract between Google and the U.S. Department of Defense to build drone strike targeting applications with machine intelligence (Deahl 2018), and especially the Google Walkout over the company's handling of sexual harassment allegations (Wakabayashi et al. 2018) made clear was the fact that applied machine learning research was a *job*, and that applied machine learning researchers were *workers*. And, as workers, they did not always get to set the terms of their employment or set individual limitations on the uses to which their research labor was put. And for many it was becoming increasingly difficult to maintain a personal narrative about the independence, objectivity, or purity of their research.

As much as this dissertation explores epistemologies and ontologies of machine intelligence—how machine intelligence produces knowledge and stabilizes itself and the objects of the knowledge it produces—it also is about the people who "do" machine intelligence and the "things" they do machine learning to and with. Their identities, the discourses they engage in, the institutions they comprise, and the way they represent themselves and their work to others are all co-produced together with machine intelligence (Jasanoff 2004). While machine intelligence appears autonomous, automatic, or mechanical, it is the product of labor invested by applied machine learning engineers who work in labs, communicate with each other at conferences and in virtual spaces, derive meaning from their work, build careers based on their interpersonal networks, accumulate status, invest their work with significance, and often reflect on their position in the world with respect to the products of their labor.

From the earliest days of its modern incarnation, social science has developed well-established methodological and theoretical frameworks to address work and labor (Durkheim [1933] 1997; Weber [1921] 2019; K. Marx [1867] 1990). And from the mid-20th century those frameworks have been applied to, and refined within, professional and elite settings like those in which machine intelligence is developed (DiMaggio and Powell 1983; Friedland 2012; Kunda 2006). Anthropological approaches to work and labor have proceeded from an early preoccupation with peasant labor and subsistence economies, to work and labor under socialism, to development economics, globalization, and labor migration (Hann 2018). Ethnographic work in high-status, corporate settings (Ho 2009) have been less common but has generally proceeded from the "studying up" (Nader 1969) moment of the late 1960s and late 1970s. This dissertation extends such work into the working world of machine intelligence by attending to the workplace practices of applied machine learning researchers, practices they engage in for career advancement and personal improvement, techniques they employ to build interpersonal networks for career advancement, and how they incorporate their working personas into their personal identities.

Machine intelligence cannot, however, only be analyzed in terms of the practices of applied machine learning researchers and their collaborators. It is also very much a function of the algorithmic techniques, the data those techniques are applied to, the mechanisms through which that data is collected, and the subsequent tools, products, and automated decisions algorithmic outputs feed into. Social science, and particularly anthropology, have only more recently turned to computer technologies of objects of ethnographic study themselves. Methodological and theoretical lenses from STS, particularly ANT and technopolitical approaches, are well-attuned to the ethnographic potential of analyzing the various entanglements of the social and the technical. However algorithmic systems represent, for many, an exceptional instance of such entanglement. Merging, as they do, mechanical rigidity in the step-wise, rule-following actions of computational algorithms with the "messy," context-laden stochasticity of data drawn from the social and natural world, algorithmic systems have received their own methodological and theoretical attention. This specific scrutiny is called for, scholars argue, because of the range of attunements that are necessary for an ethnography of algorithms. Here, they deserve specific scrutiny because throughout my fieldwork they are indexically linked to the objectivity, power, and authority of machine learning practices writ large.

Nick Seaver proposes that algorithms, as sociotechnical systems, be studied both *in* culture and *as* culture, and suggests a number of methodological approaches for doing so, drawing on the work of Paul Dourish and Annemarie Mol (Seaver 2017). Algorithms *in* culture take on different meanings and are variably defined from one social context to the next (Dourish 2016). Algorithms *as* culture do not so much act upon culture, inflecting it in this way or that, but through their operations bring culture into existence in a multiplicity of ways (Mol 2002). Studying algorithms *in* culture leads towards questions of the social significance granted to algorithmic systems, the power they hold in the popular imagination, and which individuals and groups are acknowledged as having the legitimacy to accept, limit, or regulate their development and operation. Recent scholarship in this direction has focused on how algorithms have been mythologized in popular culture and in policy (Elish and boyd 2017; Ames 2018), how they are incorporated into existing cultural practices as they are racialized (Cave and Dihal 2020) or metaphorically transposed into more familiar objects (Stark and Hoffmann 2019), and how people go about living their lives in interaction with algorithmic systems. This latter set of studies typically addresses how individuals and communities adjust their practices in full or partial awareness of how an algorithmic system is intervening into those practices, as with automated content moderation (Geiger 2017), or how algorithmic systems produce harmful impacts to individual communities.

Studying algorithms in culture can also lead towards questions of how algorithms intervene in and mediate social relations. Research along these lines has focused on the institutional functions algorithms play in criminal justice (Barocas and Selbst 2016; Citron 2008; Wexler 2018) or social services (Eubanks 2018), on how algorithmic systems intervene into and amplify already-existing racial, gender, and class disparities in society (Sweeney 2013; Noble 2018), or on how they reproduce carceral systems and racial logics across society (Benjamin 2019b; 2019a). In practice, however, many of these research directions blur the boundary between an algorithms-in-culture and the algorithms-as-culture approach for two reasons. One, as people self-consciously ascribe intention and meaning to the outputs of an algorithmic system, they become enrolled in enactments of new social relations between themselves, institutions, and algorithms. In my fieldwork, this was discussed as a concern about people learning to "game" an algorithm; if they knew the decision rules of an algorithmic system, users might change their behavior to get the decision they wanted. This was a problem for the applied machine learning researchers I worked with, as such "gaming" would interfere with the normative goals of the algorithmic system to allocate resources appropriately. But "gaming" is common in the "real world," and demonstrates how the meanings algorithmic systems have *in* culture lead to social enactments *as* culture, too (Geiger 2017). Two, algorithms act within existing social relations to reproduce culture. As algorithmic systems become the means through which institutions, both public and corporate, exert control over the flow of information, resources, and power (Deleuze 1992), they increasingly create the cultural significance of that which they exert their authority over. The work algorithmic systems do to predict and classify becomes self-fulfilling prophecies, enacting the cultural meaning of creditworthiness, sickness and health, race and gender, and anything else that can be crafted as an objective function.

In my fieldwork, I was more likely to find Seaver's algorithms *as* culture take shape in the pursuit amongst applied machine learning research to build algorithmic representations *of* culture. The idea of "representations" was a key concept in this fieldwork, and is worth exploring here. Machine intelligence depends upon learning a mathematical "representation" of all the datapoints it has access to. This representation is more abstract than the full dataset, but consists of statistical patterns that, more often than not, correspond to the distribution of variables within that dataset.

The fully-trained machine learning model, then, is an algorithmically-constructed representation of all the numerals the model has been trained on. And for practical purposes, a model that has been sufficiently trained to reach an adequate degree of accuracy when used to predict new examples of handwritten numerals can be said to be a representation of all handwritten Arabic numerals. It is a representation, although not the reality (Seaver 2015), of the culture of numeral-writing, as instantiated in the dataset from which MNIST was drawn. Similar representations can be constructed for an entire corpus of language using machine learning techniques, as has been done with all user-generated content posted to the social media site Reddit, or to Wikipedia, or to Twitter (Goldberg and Levy 2014). Even relatively simple algorithmic models, like a grading rubric used to evaluate student essays, can be thought of as a representation of all the ways of responding to an essay prompt within the context of that course.

As I argue throughout this dissertation, algorithmic representations are always already cultural. They are at once the product of the culturally-bound understandings the applied machine learning researchers that develop them bring to their framing of the projects they engage in, the culturally-contingent practices that structure data collection and classification, the cultural particularity of the phenomena represented as data, and cultural practices that are enacted through the operation of the algorithmic system. Tracing the practices through which algorithms embed aspects of culture, how those aspects are elided from professional and popular understandings of algorithms *in* culture, and the cultural conventions that support the operation of algorithms *as* culture are key components to understanding the authority machine intelligence has been granted in the production of knowledge across society.

## CHAPTER 1: DATA DANCES ON THE GRAVE OF THEORY

**"OUT WITH EVERY THEORY"**

It was the summer of 2014 when I first heard the word "algorithm" spoken of with a sense of reverence and possibility. I was in a dusty tractor shed that had been converted to a temporary archaeology lab where ceramic sherds were being sorted, photographed, and labeled for later analysis. I had been volunteering as a 3D modeling specialist on an archaeological excavation of a Middle Bronze Age citadel site in western Turkey (Roosevelt et al. 2015), which meant working closely with the project's database administrator, who was also finishing his doctoral studies on Anatolian ceramics. While he was supervising undergraduate student volunteers diligently documenting artifacts, we were discussing how to accommodate 3D objects within his database design. As our conversation shifted into a discussion of the general categories of data the database was designed to handle, he dropped an offhand comment that seemed entirely disconnected to the pot sherds, soil samples, and mud brick fragments we were surrounded by. He said, "I'm just trying to build something everyone working in the eastern Mediterranean could use so that we could run algorithms over it and figure out what people in the past were really up to." Having worked with him for several seasons, I knew he had a wry sense of humor but also that he was being sincere about his belief in the power of algorithms. I also knew the type of "naïve empiricism" this archaeologist was articulating—"the belief that the data speak for themselves"—was common in archaeology (M. H. Johnson 2011), despite having been compellingly rebutted in a series of critical interventions (Wylie 2002). So I was still taken aback to hear him defer to algorithms as capable of standing for the academic archaeologists whose ranks he was training to join. What was it about algorithms that they could usurp even this dedicated researcher's burgeoning expertise?

When I returned to New York City at the end of that summer, I began to see that "naïve empiricism" was not the province of lapsed processual archaeologists alone. This idea that "data could speak for itself" had recently come back into vogue, most visibly in a special issue of Wired Magazine, hugely influential on popular understandings of technology development and entrepreneurship, dedicated to "the petabyte age". It bears repeating that this new era was one in which machine intelligence would enable "massive amounts of data and applied mathematics replace every other tool that might be brought to bear. *Out with every theory of human behavior, from linguistics to sociology. Forget taxonomy, ontology, and psychology.* Who knows why people do what they do? The point is they do it, and we can track and measure it with unprecedented fidelity. With enough data, the numbers speak for themselves" (Anderson 2008, emphasis added). And speaking, they were. In criminal justice, data was speaking for itself when arrest and re-arrest data was being used to classify defendants who should receive pretrial release and those who should not (Angwin et al. 2016). In medicine, data was speaking for itself when electronic health records were being used to predict patient outcomes (Miotto et al. 2016). And all across the Silicon Valley startup ecosystem, data was being coaxed into speaking for itself through products and services that would replace call centers with chatbots (Sangroya, Saini, and Anantaram 2017), recommend products that online shoppers wouldn't be able to resist (Gomes 2017), or even master tasks thought to be squarely within the domain of human intelligence like strategic gameplay (Silver et al. 2017).

Anderson's formulation of how data comes to speak seems to eschew any theoretical basis for knowledge in favor of whatever knowledge applied mathematics can coax from very large data sets. It also eschews any consideration for the contents of datasets, the conditions under which they were compiled, or what relationship they hold with the phenomena they purport to represent. Anderson's formulation also stands in stark contrast to how many applied mathematicians, machine intelligence practitioners like the applied machine learning engineers, and specifically data scientists, were describing their emergent field at that same time. In 2008, DJ Patil and Jeff Hammerbacher held meetings that would later be celebrated as founding the job of data scientist, during which they determined that "the dominant trait among data scientists is an intense curiosity—a desire to go beneath the surface of a problem, find the questions at its heart, and distill them into *a very clear set of hypotheses that can be tested*" (Davenport and Patil 2012, emphasis added). Here, Patil's formulation of how data comes to speak closely resembles the theory-guided testing of carefully formulated hypotheses against empirical data that Anderson dispenses with entirely.

In this chapter, I argue that the naïve empiricism of Anderson and the hypothesis-driven data science of Patil are ultimately irreconcilable with each other, even as they both circulate widely within the same discursive space. Though irreconcilable, these two framings play off of each other to constitute the authority and legitimacy of machine intelligence. This interplay bestows upon machine intelligence an aura of validity from the hypothesis-driven scientific method (Hepburn and Andersen 2021) commonly taught in high school classrooms, as well as the impersonal distance of naïve empiricism that removes the experimenter from the picture and grounds findings directly "in the data" itself. My intention in making this observation is not to then say that one of these framings is science and therefore the other is not. Rather, my intention is to show that agreements about the legitimate means through which knowledge is constructed are themselves the product of labor exerted to construct knowledge, establish conventions, and forge consensus around the use of particular techniques of machine intelligence.

In this, I follow Shapin and Shaffer (1985) in trying to see the authority of machine intelligence as products of social and historical practices "integrated into patterns of activity," which they refer to as Wittgensteinian "forms of life" (1985, 29). To do so, I explore how the development of applied mathematical and statistical techniques that are now central to machine intelligence, many of which are incredibly complex manipulations of data but packaged as subroutines that can be called by a single line of code and linked together in chains of subroutines, contributed to naïve empiricism on the part of practitioners in the "era of big data". I then document and analyze how these framings simultaneously circulate throughout the training and validation of practitioners, at data science "bootcamps" and in online machine learning "competitions". In doing this, I examine how applied machine learning researchers navigate these irreconcilable frames when faced directly with their contradictions. This reveals how the actual labor applied machine learning researchers exert in navigating these frames is erased from view, sealed within "black boxes" that appear to mechanically transform inputs to outputs. These transformations, rather than being more abstract (and therefore useful) representations of more complex datasets, are enactments—data performances—of new objects of knowledge that arise from the "intra-actions" (Barad 2003, 803) of data, researcher, and algorithm.

## A Brief Lesson on Machine Learning

Without looking up from my laptop, I heard a door open and two pairs of footfalls softly treading down the stairs from the loft. The loft held three cramped conference rooms—named "Gaussian," "Binomial," and "Poisson" after different types of statistical distributions—which had been constructed sometime between when the building's life began as a brewery and its current home as headquarters to an applied machine learning research lab. The footsteps were accompanied by soft conversation. "It would be great to get our hands on their telemetry data… there's huge cost savings for the airline industry in predictive maintenance!" Markus and Matthieu were discussing a prospective client call with a turbine parts manufacturer that they had just concluded, and they were excited about the cost savings they could find by predicting which specific aircraft engines' parts were likely to fail and should therefore be inspected and replaced. Traditionally, aircraft components are inspected, repaired, and replaced at regular, recommended maintenance intervals provided by manufacturers. But instead, these two guys thought that statistical predictions about each component's likelihood to fail in the near future could be inferred from huge rafts of data manufacturers kept about things like flight hours, rotational velocity, ambient temperature, and the performance characteristics of related components for every engine in their product line. Using these predictions, maintenance crews could spend their time working on components that were most likely to fail, while leaving parts that had more life in them in place for a bit longer.

The two guys loping down the stairs after their sales call had not worked a day in the aerospace industry (as far as I knew). They didn't work for an airline or an aircraft manufacturer. They weren't mechanical engineers. But they were so confident in their algorithmic tools that they thought they could achieve tasks impossible for aerospace or mechanical engineers to accomplish. They were applied machine learning researchers, and they spent their workdays—which sometimes started late in the day but lasted deep into the evening—developing ways to use the emerging techniques of machine learning to solve all sorts of problems and adapting those problems into the kinds of problems machine learning can solve.

Because machine learning's social character is not always immediately obvious as conveyed by the excitement of applied machine learning engineers about being able to predict the imminent failure of airplane turbine components through machine learning it is worth beginning with an exploration of some of the more technical definitions of machine learning, as the set of machine intelligence practices most central to this dissertation. In doing so, it is possible to interrogate how machine learning researchers themselves implicate the social in the descriptions of machine learning that they themselves offer. One of the earliest such descriptions is found in a 1968 issue of the journal *Nature*, which describes the hypothesis that "computers could learn from experience and thus automatically improve the efficiency of their own programs during execution" (Michie 1968, 19). Fifty years on, descriptions of machine learning remain relatively unchanged in machine learning textbooks:

> "Machine learning is programming computers to optimize a performance criterion using example data or past experience. We have a model defined up to some parameters, and learning is the execution of a computer program to optimize the parameters of the model using the training data or past experience. The model may be predictive to make predictions in the future, or descriptive to gain knowledge from data, or both" (Alpaydin 2014, 3).

In this more recent description, the modal verb "could" has been replaced with the present tense "is," signaling the accomplishment of what had once been mere possibility. But importantly, the words "data," "prediction," and "knowledge" also appear. In adding these three nouns to his description, Alpaydin shifts machine learning from the register of well-bounded computer science—in which a computer adapts itself based on conditions internal to the computational system—to situate machine learning in more social registers where the social uses of machine learning techniques predominate. This shift follows the trajectory of computer science's gradual shift toward more "applied" domains.[1] Other widely circulated definitions of machine learning similarly foreground the social uses of machine learning as well. Tech journalist Karen Hao, writing for the MIT Technology Review describes machine learning as using "statistics to find patterns in massive amounts of data" (Hao 2018) and frames both "data" and "patterns" within the context of online social media and e-commerce applications. This framing is fair, as Silicon Valley companies engaged in these domains are some of the most significant actors in applying machine learning techniques and in undertaking the basic research developing the scope of statistical and computational tools for machine learning.

But machine learning is a far more general-purpose a tool than its most visible applications would seem to imply, in that it can be deployed in a vast range of contexts that make its applications seem almost incommensurable. It is difficult to compare, for example, machine learning used to correct for atmospheric distortion in astronomical observatories with machine learning used to suggest the next book a fan of science fiction might be interested in reading or with that used to identify the identity of someone in a photograph, or with that used to analyze the sentimental valence of a text message. The definition offered by Alpaydin above is open enough to allow for these many uses, but in the interest of bringing some specificity to the sociomaterial practices that comprise machine learning in the field sites I studied, a "toy" example may be helpful. Such toy examples are common in popular articles explaining what machine learning "is" for lay audiences, critical scholars, or beginners in the field of machine learning. My own toy example goes something like this:

> Let's say I want to teach a computer to achieve a goal. Because it's a computer, I have to be extremely explicit about what I tell it to do, because it can only follow directions it's given and can't supply any of the context it might need to be able to follow vague or insufficiently well-defined instructions. Let's say the goal I want the computer to achieve is to be able to distinguish handwritten Arabic numerals, zero (0) through nine (9), each from the other. I might do this in any number of ways. I might craft a description of each numeral—zero is round and consists of a single looping line that doesn't intersect itself, in contrast to eight (8) which is round but consists of two intersecting loops. One (1) is a straight line, six (6) is a loop and a line with the loop at the bottom while nine (9) is a loop and a line with the loop at the top, etc. But I have already gotten myself into some difficulty here; handwriting is highly variable, and we all have seen one (1) written both with and without the flag coming off the top of it. Sometimes the flag is so pronounced it looks almost like a seven (7), so much so that some people write their sevens with a crossbar across its vertical line. Some people draw curvy sixes and nines, others use straight lines. Some strike through their zeros, and some people leave a tail at the top of their eights. I will have to explicitly tell the computer to look out for these exceptions, or else it will make a lot of mistakes! My list of rules for the computer to follow could continue to grow with each exception I encounter, and I would have to be quite the expert in handwriting to be able to enumerate every single exception with sufficient rigor for the computer to follow my directions well enough to make heads or tails of anyone's checkbook.

---

[1] There is a small irony, here, given that computer science was once thought of as "applied mathematics" an epithet computer scientists had to work against as they professionalized their discipline (Knuth 1972). A consequence of this professionalization was also the rapid masculinization of the all computing professions during this same period (Hicks 2017).

But there are other ways to teach a computer how to accomplish such a task. I could gather as many examples of each handwritten numeral as I can find and then explicitly tell the computer which numeral is represented by each image. I can then instruct the computer to guess which numeral is depicted in each of the images, based on the patterns of light and dark zones in the image. If the computer guesses correctly, it retains a sense of which pattern of light and dark zones matches which numeral. Over successive correct guesses, it updates that pattern until it is able to make correct guesses often enough to meet an accuracy threshold I tell it to try to reach. For high stakes applications, like reading numbers off handwritten checks, I might want to tell the computer to keep guessing and updating its pattern until it can guess correctly at least 97% of the time. If I give the computer enough examples of handwritten numerals, and let it guess often enough, the computer can "learn" a pattern of light and dark zones that allow it to "guess" the correct numeral nearly every time. With additional computational tools, it can learn to guess correctly regardless of how the numeral is oriented within an image (it could be upside down or sideways), what color it is drawn in, what the background looks like, etc. And if I translate those patterns of light and dark into—numbers representing pixel values, say—and try to classify patterns based on their quantified representations, then I'm doing machine learning.

This toy example contrasts two possible methods of producing computational behaviors that resemble what is otherwise thought of as the distinctly "human" intelligence required to make sense of handwritten numerals. Both of these methods are decidedly *algorithmic*—they lay out a rule-governed set of steps that accomplishes a goal. While more precise definitions of an algorithm have proven elusive for computer science (Blass and Gurevich 2003), algorithms are directly implicated in all computation, even those forms of computation that predate electronic computing machines. Etymologically derived from the name of the 9$^{th}$ century Persian mathematician Abdallah Mohamed Ibn Musa al-Khwarizmi al-Magusi (Ausiello and Petreschi 2013), algorithms can refer to any step-wise process for solving a problem. In computer science, algorithms are used for basic functionality like simple arithmetic, checking data integrity, and alphabetization, as well as far more complex functionality like that employed in machine learning.

The two methods for identifying handwritten characters outlined above also typify two approaches—one which predates machine learning and another that depends on it—to *artificial intelligence*, an interdisciplinary field of study that can be traced back to the early 1950's in which the founders of computer science began to question whether or not computers can be said to "think" (Turing 1950). The first method lays out an early approach to artificial intelligence in which so-called "expert systems" (Buchanan and Shortliffe 1984) were built that could replicate human intelligence by programming them with a sufficiently specific set of rules to replicate human decision-making around tasks far more complicated than recognizing handwritten numerals, including medical diagnosis and logistical supply chain management (Forsythe 2001). The second method lays out "machine learning" in a way that conforms with the definitions given above. The computer is "programmed to optimize a performance

criterion"—it is programmed to pursue optimum accuracy for guessing handwritten numerals correctly—using a large number of "example data"—the labelled images (Alpaydin 2014, 3).

Importantly, this method can be used *predictively* or *descriptively*, depending on how the objective is specified. In the example above, I have asked the computer to *predict* which numeral a handwritten image is most likely to represent, based on example images labeled with their corresponding numeral. However, I could have also given the same set of images to the computer without giving it any labels, asked it to *describe* those images, and it would have used related statistical techniques to report that there were 10 more-or-less distinct groups of images, one group for each numeral, although the computer would not have had much to say about which group corresponded with each numeral. In machine learning, when labels are supplied by the engineer, the computer is conducting "supervised" machine learning—it pursues an objective under the supervision of the assigned labels. In contrast, "unsupervised" machine learning performs analysis without the benefit of such supplied information.

"Prediction" in particular, is a heavily trafficked term in machine learning. That is to say, a machine learning application that generates natural-seeming language is predictive in the sense that the application "predicts" each word that should follow the next (Wen et al. 2015). Prediction is also operative when asking a supervised machine learning system to engage in *classification* (by predicting what class a data point belongs to) or to engage in *regression* (which involves predicting an output value for a given set of inputs). Descriptive machine learning techniques also serve crucial functions in both classification and regression, as will also be expanded upon below.

Narratively, the toy example given above is similar to many descriptions of machine learning written for lay audiences. These descriptions might charge the reader to imagine teaching a computer how to make appetizing meals for the author's children (O'Neal 2016), or teaching it how to distinguish pictures of cats from pictures of dogs (A. Liu 2019), or whether an image has a face in it or not (Meserole 2018). During my fieldwork, several applied machine learning researchers and I sat down and engaged in the exercise outlined above in an effort to teach ourselves TensorFlow, a popular machine learning package open-sourced by Google during my fieldwork. The package itself contained MNIST (Figure 1), a standardized dataset containing 70,000 examples of handwritten numerals (LeCun, Cortes, and Burges 2010).

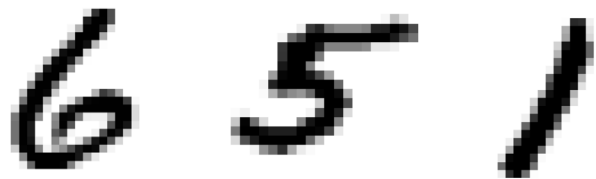

Working from a textbook (Géron 2017) but conferring among ourselves whenever we reached a step we didn't fully understand, we followed each step, sometimes even copying and pasting lines of code directly from a website into our programming interfaces, until we had "trained" the computer to accurately classify the numerals. It was actually viscerally exciting… sitting around a small table, each looking at our own screens, we watched the computer's accuracy at guessing numerals improve—in fits and starts—as each training "epoch" passed, until the rate of improvement slowed to almost a standstill. What we were watching was Alpaydin's machine learning definition in action, but it was also the realization of centuries-old dreams of transforming the messy world of human behavior and social interaction into the object of certain knowledge. If handwriting could be made legible to a computer, then what couldn't be made legible in such a way? The applied machine learning researchers I was sitting with had other ways of describing the "optimization" of Alpaydin's definition. To them we were "minimizing the loss function" or "pursuing the objective function," which are both ways of describing the statistical, mathematical, and computational techniques that allow machine learning to engage in goal-oriented behavior. Put simply, the objective function is a mathematical representation of a problem that can be used to calculate a numerical value that a machine learning system is trying to either maximize or minimize (a loss function refers to a value that the system is trying to minimize).

Objective functions are also closely linked to the "predictive" role of machine learning. The objective function, also referred to as a "risk function" or a "loss function" (Alpaydin 2014) is the component of machine learning that orients the computer program toward its goal (Breiman 2001b). The loss, risk, or objective, of these functions is to minimize the risk of a wrong prediction, or of error in predictions. Objective functions may be crafted in a variety of ways by applied machine learning researchers, for example by calculating the overall cost of errors (in which predictions that are far off the actual value "cost more" than near misses) or by tallying the total number of mistakes (in which all misses "cost the same"). In discussions among researchers, the objective function is often pointed to as the fulcrum on which a successful machine learning project might rest. When discussing a machine learning application, they try to refine the problem to which machine learning is being applied from what they see as an abstract, fuzzy challenge like how to prioritize patients needing a scarce medical treatment to be—as one applied machine learning researcher put it—"more precise, more precise, until you get to the objective function".  This quest for precision transforms the original goal into one that can be pursued through statistical means, and in the process necessarily eschews many of the contextual, qualitative considerations that might be brought to bear on that question. The objective function represents both the goal of the machine learning application and, once trained, the optimal solution to the problem as posed by the researchers.

For our MNIST project, we were trying to minimize the number of incorrect "guesses" the computer was making about which numeral a given image represented. Without providing the computer with any information about what each numeral looks like, or what numerals represent, but merely by providing it with an association between images of numerals and their label the computer was able to iteratively improve its guesswork through the relentless pursuit of a minimal loss function. It required almost 60,000 examples for the computer to get good enough to be able to recognize images of numerals it hadn't seen before, however. And of course, those images had to come from somewhere. They also had to be organized into a format in which the image and the correct label could be ingested by the computer. For MNIST, the images were collected by the National Institute of Standards from handwritten entries made by U.S. Census workers (National Institute of Standards and Technology 1994)—literally professional writers of numerals—and were modified for machine learning applications by an academic machine learning research lab (LeCun, Cortes, and Burges 2010). But the implication of such "toy" exercises is clear: with a sufficiently large set of data, an adequately designed objective function, and enough time or processing power to train a computer, it can "learn" how to make accurate classifications or predictions without having been told much else about the phenomena it was learning about. This is why the applied machine learning researchers thought they could predict component failures in aircraft engines; if they could just get their hands on enough data about components that had failed, the data could instruct them when and where to look for subsequent failures. In this way, machine learning has been applied to translating extinct languages (Connor-Simmons 2020), discovering antibiotics (Sample 2020), locating terrorists (Tarantola 2017), and almost every other problem for which there is data and that can be posed as a machine learning problem.

**FROM NAÏVE EMPIRICISM TO DATA SCIENCE**

By referring to Chris Anderson's 2008 formulation of how machine intelligence lets "the numbers speak for themselves" as "naïve empiricism," I am excavating not only mid-20th century debates within archaeological theory, but also a contentious epistemological debate from the 19th century. This debate counterposed the positivism of August Comte with a neo-Kantian emphasis on the importance of *a priori* knowledge. Comte extended Galileo's view of mathematics as the language of nature to construct knowledge about the world from empirical observations presumed to be independent of human subjectivity, while neo-Kantians acknowledged the role prior knowledge plays in the collection of empirical observations. It was these neo-Kantians who derided Comte's "theory free" empirical programme as naïve, and contestations over the appropriate role of prior knowledge and the "theory-ladenness" of data collection would characterize debates within the philosophy of science through the 20th century (T. Shapiro 2017). I argue here that this debate has been reconstituted in the 21st century under the guise of novel technological approaches through contestations over machine intelligences practices that were visible throughout my fieldwork.

Engaging with the naïve empiricism of Comte is appropriate in the context of machine intelligence because machine intelligence practitioners explicitly link their work to of Comte. Alex "Sandy" Pentland—a computer scientist and cognitive psychologist who has been foundational for the field of "computational social science" (Lazer et al. 2009) and the MIT Media Lab's influential work in machine intelligence—re-coined Comte's term "social physics" to describe his project of seeking to make social graph data speak for itself by deriving laws of human social behavior from large datasets (Pentland 2015). Pentland lauds Comte in the introduction to his book by observing that "his theories were in many ways too simplistic, but he was going in the right direction … modern mathematics and big data produce a description of humanity that is driven not by fixed ideas, but by the flow of ideas between people" (Pentland 2015, ix), by which he meant the datafied patterns of exchange. Revisiting the historical connections between positivism, objectivity, and practices of quantification and datafication as antecedents and epistemological premises of machine intelligence is instructive here:

## POSITIVISM

Positivist approaches hold that knowledge about natural phenomena is produced through observations about the world, and statements about the natural world can be verified by similar means. Positivism is often attributed to the early-19th century work of Auguste Comte and his predecessor Claude-Henri de Saint-Simon (J. H. Turner, Beeghley, and Powers 2012), who are now seen as foundational thinkers for the social sciences but who were writing at a time before clean distinctions had been drawn between the social and the natural sciences. Comte and Saint-Simon saw positivism as the telos of all knowledge production, with fields of inquiry progressing through earlier stages of polytheism, then theism, and finally positivism. At a time when the authority of the church and the role it played in the feudal order across Europe was waning, positivism was a rejection of metaphysical explanation and a turn toward knowledge that rested on that which can be rendered observable, either through direct sensory experience or through instruments and experimental apparatuses (Shapin and Schaffer 1985). The "stages of development" for individual fields of inquiry allowed Comte and Saint-Simon to construct a hierarchy of sciences, with those that reached the positivist stage earliest—astronomy, physics, and chemistry—at the top, above fields like biology, physiology and sociology. While all sciences ought to be positivist, implicit in this hierarchy is the idea that some sciences are more amenable to positivism than others, whether because of the difficulty in making positivist observations or the overall complexity of the phenomena themselves. For Comte, this hierarchy was inverted; the complexity of a positivist social science made it the "queen science" from which an ameliorative program could ensue—a "positive philosophy" that formulated "the laws of the social universe" from which to construct a better society (J. H. Turner, Beeghley, and Powers 2012, 40).

Positivism, and by extension the logical positivism of the Weimar-era Vienna Circle that admitted logical propositions alongside empirical observations (Popper [1935] 2002), can also be traced back to even earlier attempts to represent the world—natural phenomena, human behavior, social relations—in ways that allow such things to be classified and subjected to predictions. These ways of representing the world can be traced back to 17$^{th}$ century efforts to leverage the power of direct observation in service of more generalizable laws about the universe. Although couched in much older Platonist debates (Gorham et al. 2016), the idea that direct observation and measurement can be leveraged to extrapolate laws of nature finds its most vivid representation in Galileo Galilei's assertion in 1623 that the "all-encompassing book that is constantly open before our eyes, that is the universe … is written in mathematical language" (Galilei 2008, 183). While the formal, often mathematical legibility of all things in the universe, including human society, may at times elude human mathematical techniques, for Galileo it was nevertheless an article of faith that the universe was mathematical whether human knowledge could represent it as such or not. Indeed, while Galileo might admit that many natural phenomena that were sufficiently "mathematically complex" might remain unknowable (Gorham et al. 2016), machine intelligence often seems to take up the implicit challenge Galileo offered.

## OBJECTIVITY

From Galileo's assertion that math is the language of nature to the elucidation of mathematical laws of nature by the likes of Isaac Newton a few short decades later (Newton [1687] 1999), the positivist impulse to "let nature speak for itself" strengthened into the dominant concern of 19th century scientific thought. Positivist approaches to the natural world led to an orientation toward the organization of experience into standardized units, of experience as well as of measurement. Scientific endeavors of this era sought to remove the subjective impressions of the scientist as much as possible from their efforts to observe and record nature directly (Daston and Galison 2010), or to attempt to account for the bodily production of sense impressions (Crary 1990). The forms of "mechanical objectivity" this impulse gave rise to increasingly relied on instruments to aid scientists in faithfully reproducing nature, but also on scientists to become more machine-like themselves—subverting their own idiosyncratic impulses, aesthetic judgements, and mortal failings—in their pursuit of positivist knowledge. They sought to transform subjective impressions into externalized objects that could be the target of objective validation, often by a scientific community that has been organized in order to do so. Daston and Gallison point out the close coupling of objectivity and subjectivity over time, "the negative character of all forms of objectivity" revealing "as wax is to seal" the "bolder and more solid features of subjectivity" (Daston and Galison 1992, 82). What Porter calls "the taming of human subjectivity" extended beyond science to order all sorts of social processes, particularly bureaucratic administration, economics, and public health through quantification (Porter 1995, 21).

Mechanical objectivity, as the inverse of personal subjectivity, relies on rule-following—rigorous compliance with proper procedures for achieving a verifiable, replicable result—but it does not replace personal expertise. Rather, mechanical objectivity reframes expertise to include "an ability to follow the rules" (Porter 1995, 4) even if trained judgement (Daston and Galison 2010), personal discretion, and practical experience (J. C. Scott 1998) can never be fully eschewed. Algorithms, however, give the appearance of doing just that. They are automated rule-following routines which appear to operate entirely apart from human influence once set upon their course to inevitably arrive at an objective outcome. To do so, they must be provided with inputs, the ingredients for the recipe they tirelessly and exactingly follow, that precisely conform to their requirements. While the demand for algorithms that can handle less precise inputs, that are "robust to noise," has resulted in algorithmic preprocessing techniques that can be deployed "with the turn of a key". Most algorithmic approaches require improvisation and cunning on the part of computer scientists and machine learning engineers. Often, these workers must cobble together and patch programming packages, sometimes transforming data, estimating missing values, truncating (in)significant digits, or inserting intermediate statistical techniques to suit the requirements.

Data collection practices (Gitelman 2013), categorizations enacted through data collection (Bowker and Star 1999), engineering practices like "feature engineering" (which will be explored later), and the ways problems are formulated as addressable by machine intelligence (Selbst et al. 2019) all form the initial conditions from which algorithms proceed "objectively". Machine intelligence maintains its aura of objectivity by continuing to shuffle the shells under which more subjective practices are hidden or by reframing those practices as objects of positivist claims. Alternately, machine intelligence can restore an aura of objectivity to more subjective practices by incorporating into algorithmic subprocesses, as has recently been done through "automated feature engineering" that replaces human discretion with a statistical method (Kaul, Maheshwary, and Pudi 2017).

**QUANTIFICATION**

What Ian Hacking calls the "avalanche of printed numbers" made exercising power into a positivist endeavor (Hacking 1982), particularly as civil servants in 19[th] century Europe replaced the aristocracy in the administration of state power. The authority with which the aristocracy wielded that power, by virtue of their association with the sovereign, came to be replaced by the trust in quantified, auditable accountings of their activities (Porter 1995; Strathern 2000). Since the 19th century, quantitative techniques have been motivated by positivist claims about the measurability of the world. This has led, in turn, to the development of techniques like probability calculus and statistics which have been deployed across the sciences—especially in physics, biomedicine, economics, political science, sociology, and psychology but also in accounting, audit, and administration—in ways that have over time been validated within the disciplinary norms of each of these areas of knowledge production. Applied machine learning research has taken the face validity of these techniques and sought to further develop them through the practices of machine learning. In doing so, it separates the epistemological claims of quantificatory reasoning from the disciplinary theories that quantification has been developed within in the past. This allows machine learning researchers to make claims that, to their way of thinking, do not require the types of causal claims that traditionally hold within individual disciplinary norms. This separation between purely quantificatory reasoning and modes of disciplinary knowledge production has been articulated most clearly in a 2008 editorial by Chris Anderson, then-editor-in-chief of Wired Magazine:

> "This is a world where massive amounts of data and applied mathematics replace every other tool that might be brought to bear. Out with every theory of human behavior, from linguistics to sociology. Forget taxonomy, ontology, and psychology. Who knows why people do what they do? The point is they do it, and we can track and measure it with unprecedented fidelity. *With enough data, the numbers speak for themselves*" (Anderson 2008, emphasis added).

Making "the numbers speak for themselves" is the key work of applied machine learning researchers, and is in some way the inevitable conclusion of Comte, Galileo, and Newton's metaphysic. But the idea that numbers *can* speak for themselves rests on the confluence of positivism, objectivity, quantification, and datafication that constitutes *machine intelligence*. Machine intelligence is the set of computational practices that engage seemingly autonomous software techniques to construct positivist claims based on predictions and classifications, derived from numerical representations of the world, and for enabling those claims to be seen as valid.

## DATAFICATION

The quantified, numerical world of machine learning is one in which numbers have long been made to matter because they can travel in ways the phenomena they represent cannot, because they have come to be seen as more trustworthy than the fallible human actors that produce them, and because they can render dissimilar phenomena commensurable and comparable (Porter 1995). The pervasive production of digital data–what Kenneth Cukier calls "datafication" because of its seemingly-inexhaustible ability "to render into data many aspects of the world that have never been quantified before" (Cukier and Mayer-Schönberger 2014, 29)– is the apotheosis of quantification. Enumerating these aspects of the world that have been datafied, Cukier points to how location has become data thanks to the GPS capabilities of smartphones and personal vehicles (A. Shapiro 2018), health has become data thanks to medical records infrastructure (Ruckenstein and Schüll 2017), and even starlight and moonbeams have become data thanks to the CCD sensors now mounted at the rear of refracting telescopes (Kremer et al. 2017).

Machine learning researchers are both voracious consumers of datafied phenomena and prolific datafiers themselves. Any properly structured dataset affords applied machine learning researchers an opportunity to demonstrate the capabilities of machine learning–perhaps through a spectacular or "toy" example," or to construct a product that could be the jumping-off point for a potentially profitable startup company. At the applied machine learning research lab where I conducted my fieldwork, an email newsletter that highlighted newly available public datasets was eagerly awaited each week because of how useful the datasets it linked to could be for trying out new techniques or building a demonstration project for clients. The datasets highlighted in this newsletter tell a near-real-time story about the state of datafication, demonstrating how the banal bureaucratic practices of the municipal governments, the rise of data journalism (Fink and Anderson 2014), and the increasingly networked digital media, among other trends, translate into a panoply of datasets for applied machine learning researchers to explore. In just one week, this newsletter highlighted a dataset containing the name of every dog wearing a license tag from the New York City Department of Health, every flight a presidential candidate took prior to taking office in 2017, the count of *E. coli* bacteria on San Francisco beaches, and the migratory patterns of research scientists who use ORCID to track their citations (Singer-Vine 2017). From these datasets we can learn that there are 227 dogs named "Duke," 154 named "Nala," and 1 named "Marx" registered in New York City (NYC Department of Health 2017), that one candidate's presidential campaign had a carbon footprint equivalent to the yearly carbon consumption of 500 U.S. citizens (Bump 2017), and that genome scientist Rimantas Kodzius is the world's most peripatetic scientist (Bohannon 2017). But in the hands of applied machine learning researchers, these datasets can form the heart of a potentially saleable product–like the "Beach Poo Bot" built to alert surfers of when it was microbially safe to head down to Bay Area beaches (Brandon 2017).

Any limits to datafication present themselves as an addressable challenge to applied machine learning researchers, as the scope of their work includes developing datafication techniques for any phenomena they are interested in but for which data do not already exist. Thus they find themselves managing datasets for their own work, whether that be by allocating resources for microtask laborers (see Gray and Suri 2019) to "hand-label" datasets, spending time themselves "cleaning" or "ingesting" data, or by deploying complex machine learning tools to gather well-structured data for training machine learning models (Roh, Heo, and Whang 2019). Applied machine learning researchers, as producers and consumers of datafied phenomena, exist within a much longer history of quantificatory practices, not only in that they use data but also in *how* they use data. The methods and practices of machine learning are contiguous with many of the statistical techniques developed "for inventing, constructing, and proving scientific facts, both in the natural and social sciences" (Desrosières 1998, 3).

"Facts," as cultural objects, have their own social and historical trajectories (Daston 1994; Poovey 1998). Prior work has demonstrated how that which can be taken as facts emerge not simply as given or universally accessible truths about the world of phenomenological experience, but are constructed through socially situated practices of expert witnessing (Shapin and Schaffer 1985), cultural conventions of subjectivity and objectivity (Daston and Galison 1992; Harding 1992a), historical shifts in power and authority over who gets to speak to whom about the nature of reality (Porter 1995; Power 1997; Strathern 2000), and the ways such shifts in power and authority are operationalized as practices of administration, bureaucratic control, and commercial regulation (Poovey 1998; Desrosières 1998). Similarly, "quantification," as a set of sociomaterial practices contiguous with machine learning and datafication, has developed along socially and historically situated trajectories. Foucault cautions that for the natural histories produced in this time period to appear "it was necessary … for History to become Natural" (Foucault [1966] 1994, 128), which is to say that the objects of scientific study had first to become stabilized as objects capable of being studied through historically contingent knowledge practices.

Foucault points to how Linnean classification relied on "systematic seeing" of variables that could be represented numerically as crucial for the stabilization of objects of study. Bracketing out sense impressions that could not be rendered comparable through precise numerical or geometrical language manages sources of uncertainty that lie with the observer, and establishes a schematic (what Foucault refers to as "structural") validity for collapsing numerous scientific observations onto a single class of phenomena—a plant's genus and species, for example. For botanical taxonomies, a natural historian might determine that flowers whose various measurements fall into the same proportion to each other belong to the same species. Regardless of subjective impressions of various botanists, the unity of a species can be established if the proportion of petal width to petal length and sepal length to sepal width is uniform.

It is worth noting here that precisely this Linnean taxonomic practice that caught Foucault's attention is entirely reproduced within the practices of machine intelligence. In the same text that the applied machine learning researchers and I were using to teach ourselves TensorFlow (Géron 2017), we worked through an exercise that used a statistical technique of "logistic regression" to classify hundreds of measurements from individual iris specimens into three iris species (*Iris setosa, Iris versicolor,* and *Iris virginica*). This exercise is common for students of machine learning, whether self-taught or formally trained, and is demonstrated using a dataset containing iris flower measurements built into the `scikit-learn` machine learning package. This dataset itself was first published in the *Annals of Eugenics* in 1936 (Fisher 1936), around the same time that logistic regression techniques were being developed. While Linnaeus was writing about a century before the techniques of regression were developed (Cramer 2002), his insistence on establishing classifications of species based on attributes that can be measured (Linneaus 2003, 71) served as one of the primary application for which regression techniques were developed. The word "regression" persists as a label for this suite of statistical techniques from early work by Francis Galton, who observed that the full-grown height of children of tall parts demonstrate "filial regression toward mediocrity," or the arithmetic mean of a population (Galton 1886, 247). As will be discussed at greater length in Chapter 1, the links between the development of quantificatory practices in the sciences, scientistic management of people and bodies and machine intelligence are strong.

## THE "TWO CULTURES" OF LEARNING FROM DATA

The rhetorical connection between machine intelligence and naïve empiricism is similarly explicit in a 2011 keynote presented by John Rauser at Strata, an influential data science industry convention that I regularly attended throughout my fieldwork. In his keynote, he argued that 18th century astronomer Tobias Mayer (b. 1723) should be seen by the machine intelligence community as the first data scientist because of how he calculated the libration of the moon (the wobble that can be perceived in the side of the moon that always faces earth). Mayer used a surfeit of astronomical observations (the "big data" of the time in Rauser's telling) and a data analysis technique that relied only on the distribution of that observational data (without any explanatory recourse to explicit guiding theory to interpret that distribution) to make his calculation (Rauser 2011). What makes Mayer's work an exemplar of data science, for Rauser, is precisely its naïve empiricism. Mayer collects a series of astronomical observations, examines how they are distributed, and then devises an algebraic technique to calculate lunar motion using all his data points. By including more data points (n=27) in his calculation than the three that are geometrically necessary, Mayer's was able to produce a solution with three times higher accuracy than would otherwise have been possible. It is crucial to note that this higher accuracy comes not from paying attention to how the astronomical observations were collected, the observational apparatus, or the skill of the observer, but rather from the empirical data itself which *purposefully excludes* prior interpretive frames, observational effects, or subjectivities.

Despite rhetorical tendrils that reach back to the positivist, naïve empiricism of the 17th and 18th centuries, the 21st century practices of machine intelligence have their most direct roots in the statistical mathematics of the 20th century. The development of what we can now call machine intelligence—techniques that can be used to transform data into classifications or predictions—is the apotheosis of a move that had been building in mathematical statistics since the 1960s. Writing in 1962, statistician John Tukey began to argue that existing approaches to statistics, which aided in the production of generalized mathematical models against which explanatory scientific theories could be tested, were being subsumed by a set of practices that were claiming the mantle of "science" by virtue of their ability to predict phenomena (Tukey 1962). Coined by Tukey as "data analysis,"[2] he posited a hierarchy of sciences nearly identical to Auguste Comte's 19th century hierarchy and then claimed a spot atop that hierarchy, above even the social and behavioral sciences: "In the sense that biology is more complex than physics, and the behavioral sciences are more complex than either, it is likely that the general problems of *data analysis* are more complex than those of all three" (Tukey 1986, emphasis added).

Passing a judgement on Tukey's arguments that "data analysis" constitutes a true "science" in ways that pure mathematics could never be is less relevant than identifying where and how he locates the science of data analysis in relation to other scientific practices. Tukey posits a science of large datasets, one that enables data analysts (read, "data scientists") to learn from data (Tukey 1962). In doing so, Tukey operationalizes naïve empiricism, effectively constructing a scientific research trajectory into how to derive empirical knowledge from data while eliding the interpretive frames through which data was collected, observational effects from the data collection process, or subjectivities at play throughout the complex entanglement of phenomena and their datafied representations. Tukey himself went to great lengths to point out that his science of data analysis was by no means "theory-free," but he always subsumes theory to the data analyst's (read: data scientist's) trained judgement to interpret the data directly (read: naïvely). He repeatedly cautions his readers that "if data analysis is to be well done, much of it must be a matter of judgment, and "theory," whether statistical or non-statistical, will have to guide, not command" (Tukey 1962, 10). Tukey's programme for data analysis has laid the groundwork for naïve empirical approaches that still characterize widespread machine intelligence practices today, as is made clear in a 50-year retrospective on the paper in which he laid out data analysis as a novel science (Donoho 2017).

---

[2] The original text uses "data analysis", the practices of which as described in that text are in line with the practices associated with "machine intelligence" adopted throughout this analysis.

For the purposes of this chapter, the emphasis Tukey placed on the "judgement" of those who work on machine intelligence is crucial for identifying moments within which those who work on machine intelligence shift between naïve empiricist and hypothesis-informed frames. It is these work practices and subjectivities that comprise the authority of machine intelligence writ large. It is this very judgement that structures the "epistemic virtues" at stake in the work of machine intelligence (Daston and Galison 2010, 4). For Tukey, and as we will also see for those who work in machine intelligence, explicit acknowledgement of the role analysts' judgements play in data analysis is a fraught affair. They must employ the kind of *mētis*, or contextual improvisation and ingenuity, that James Scott counterposes to *technè*, or technical exactitude (J. C. Scott 1998), not only to get the algorithm churning in the first place, but to maintain its operation over time.

The *mētis* involved in constructing and maintaining an algorithm is far from the only way in which algorithmic objectivity merely plays a shell game with human subjectivity, which still infuses the entire "stack" of decisions that are compiled to produce knowledge through machine intelligence (Passi 2018). On one hand, data analysis should lead inexorably toward an optimal result with mathematical precision and certainty, which Tukey articulates under the widespread sentiment that "statistics are optimization" and that "data analysis should not *appear* to be a matter of judgement" (Tukey 1962, 9; emphasis in the original). On the other hand, Tukey points out that *appearances* to the contrary, the results of data analysis do indeed "depend upon the assumptions and criteria [of the data analyst], whose selection may, perhaps, even be admitted to involve judgment" (*Ibid.*).

Over forty years, the techniques available for Tukey's nascent field of "data analysis" expanded to include many of the machine intelligence techniques like "decision trees" and "neural nets" that form the basis for the technical practices of machine learning, data science, and artificial intelligence today.

Between 1967 and 2001, computer scientists and statisticians developed and operationalized these techniques, which Leo Breiman reflected upon in his observations about the state of statistics in a 2001 article. In this article, which has been hugely influential within machine intelligence circles, Breiman posits that there are "two cultures" of statistics: a traditional approach to statistics, and a newer approach that comprises what we can now call machine intelligence. The older approach, for Breiman, constitutes "the data modeling culture," which makes assumptions about how and why data are distributed as they are and tries to update a model to fit the data. He calls the newer, machine intelligence approach "the algorithmic modeling culture," which views the distribution of data as fundamentally unknown and tries to construct a model that can predict where in the distribution any given data point might fall (Breiman 2001, 199).

Both 'cultures' are concerned with developing *models*—a concept which Philip Agre points out as having "no single technical specification" (Agre 1997, 143) but which is also worth pausing to reflect upon momentarily. 'Models' are necessarily simplified representations of phenomena that contain the characteristics relevant to understanding those phenomena. A mathematical equation, like

$$y = mx + b$$

can model the relationship between two variables, $x$ and $y$, provided the characteristics one is interested in are relevant to those variables' relationship on a cartesian plane. A chemical formula, like

$$6CO_2 + 6H_2O \xrightarrow{\Delta} C_6H_{12}O_6 + O_2$$

can model the inputs and outputs of photosynthesis that are relevant to chemistry. And a concrete model of the Mississippi River can model the hydraulic characteristics of the entire watershed (Cheramie 2011). Each of these 'models' is an abstraction of a much more complex phenomenon that isolates the relevant characteristics of that phenomenon for a particular purpose, while ignoring other aspects that are entangled with those other characteristics. Identifying a trend in the relationship between two variables does not allow one to grapple with any causal relationship between those two variables. Understanding how photosynthesis transforms carbon dioxide and water into oxygen and sugar elides important details that are necessary for maintaining a garden or grappling with the economic drivers of deforestation. A model that aids in managing the flood control infrastructure of a river system is not as useful for understanding water pollution. Statistical modeling has a long history in the political economic exercise of power, which from the 19th century has hinged on "the modern concept of representativeness ... and the centralization of national statistics" stratified "according to *a priori* divisions of the (national) population" (Asad 1994, 74). These divisions and claims to representativeness are necessarily incomplete and reductive; a model that captures *all* the complexity of the system it represents is as useful as a Borgesian map (Borges and Hurley 1998); models must reduce this complexity to the elements about which analysts are concerned, and thus choices about what is and is not modeled are a function of analysts' position, interests, and assumptions (Harding 1992a).

Breiman's "data modeling culture" constructs statistical models about what is known about a phenomenon; it is a hypothesis based on a theory about what characteristics are relevant to a problem, and is subsequently tested and updated. The "algorithmic modeling culture," on the other hand, is naïve to the phenomenon being modeled in that it requires no prior knowledge about the phenomena represented by the dataset. From the dataset, it is possible to construct a statistical representation—a simplified model that captures the relationship between aspects of the dataset—that can be used to make predictions about that dataset. Breiman declares himself as clearly on the side of the algorithmic modeling culture, which he describes as leading to the use of more suitable models (for the specific problems statisticians work on) and more exciting methods (that draw from an expanded palette of statistical techniques), but also as less prone to "irrelevant theory and questionable scientific conclusions," noting that "the theory in this field shifts focus from data models to the properties of algorithms ... and what gives them good predictive accuracy" (Breiman 2001, 205).

The field of applied machine learning research, as I show below, has enacted this shift by developing highly visible practices for presenting and elaborating the algorithmic components of machine intelligence while downplaying the role of prior knowledge and theoretical understandings of the phenomena that are represented as data. In doing so, applied machine learning research participates in the 19th century imperative to "let numbers speak for themselves" while embedding a naïve form of empiricism into a set of practices that can be presented as scientific and deeply engaged with theory, even if that theoretical engagement applies only to algorithmic approaches to statistics and data, and not to the phenomena about which machine intelligence is making claims. The relationship between data and what the data purport to represent remains unexamined. The naïve empiricism of Comte has been brought into academic statistics by John Tukey, further elaborated by academic statisticians like Breiman, and operationalized by applied machine learning researchers who declare themselves in alignment with "the algorithmic modeling culture" of statistics. In training, applied machine learning researchers learn not only that the data can speak for itself, and their task is to coax it into doing so, but also that the data must be seen as speaking for itself. They learn to constrain how they describe their own agency and influence over the outputs of machine intelligence so that it is always the data speaking, not themseves.

Practitioners of machine intelligence make these declarations explicitly, claiming the prior work of Tukey, Breiman, and others as part of their own intellectual, disciplinary, and professional lineages in public presentations about the development of machine intelligence techniques. I repeatedly observed reprises of the celebration of 18th century astronomer Tobias Meyer at Strata, as applied machine learning researchers indexed proponents of naïve empiricism in their public talks. In 2018, I attended the Neural Information Processing

Conference (NeurIPS), one of three annual marquee machine intelligence conferences[3] widely attended by academic and industry practitioners. At an invited talk opening the final day of that conference, the speaker set the entire field of machine intelligence, including its claims to be legitimate producers of knowledge, within a lineage that tied back directly to Leo Breiman by using Breiman's Google Scholar search results page as the first slide of his presentation (Figure 2). This was a powerful rhetorical device because some of the most innovative and pervasive research techniques that had defined machine intelligence in the past year were revealed to be the titles of Breiman's scholarship dating back to the 1980s and 1990s, accompanied by citation counts in the tens of thousands (Spiegelhalter 2018). The sentiment behind this slide was replicated in conversations I had with applied machine learning researchers as I began my fieldwork; practitioners I worked with sent me copies of Breiman's article on the "two cultures" of statistics. Perhaps this was because they thought that I, as an anthropologist, would be interested in the "cultural" framing of statistics (they weren't wrong), but given the author's polemical stance about the incompatibility of these two cultures, I also read the article as a statement of these practitioners' affinity for one culture ("algorithmic modeling") over the other.

---

[3] The ACM International Conference on Machine Learning (ICML) and the American Association for the Advancement of Artificial Intelligence (AAAI) conference are the other two marquee conferences.

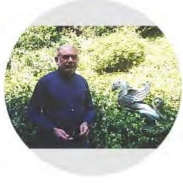

**Leo Breiman 1928-2005**  
Professor of Statistics, UC Berkeley  
Verified email at stat.berkeley.edu - Homepage  
Data Analysis   Statistics   Machine Learning

FOLLOW

| TITLE | CITED BY | YEAR |
|---|---|---|
| **Random forests**<br>L Breiman<br>Machine learning 45 (1), 5-32 | 41246 | 2001 |
| **Classification and Regression Trees**<br>L Breiman, JH Friedman, RA Olshen, CJ Stone<br>CRC Press, New York | 39082 * | 1999 |
| **Classification and regression trees**<br>L Breiman<br>Chapman & Hall/CRC | 39082 * | 1984 |
| **Bagging predictors**<br>L Breiman<br>Machine learning 24 (2), 123-140 | 19724 | 1996 |

As academic statistics has moved, slowly and contentiously, toward data-centric approaches founded on the "algorithmic modeling culture" and away from theoretical framings from other domains, this movement has been replicated at a tremendous power of magnification in applied machine learning research. This has been accomplished through professional practices that foreground algorithmic techniques, and that obscure the role of situated judgement on the part of the analyst, techniques which have developed apace. The result of this is that practitioners of machine intelligence are always at pains to remove any evidence from an analysis that might index their own hand at work, to always point to the mechanism that produced the analysis and not their own effort in bringing that mechanism to bear on the data. In practice, this occurs through the redefinition of any problem—classifying criminal defendants, predicting online shopping behavior, or generating natural-seeming language—into an "optimization problem". By selecting an attribute or feature of a dataset to optimize for—say, rearrest of a parolee—and linking that feature to other features for that

parolee—age, gender, zip code, number of close relatives, years of unemployment—the statistical relationship between the objective and other attributes can be pursued mathematically. Risk of reoffence become a function of other attributes within the dataset when applied across a large enough sample of individuals. From that point, the machine intelligence takes over, and no human judgement needs to appear.

**INTUITIONS AND OPTIMIZATIONS**

On the ground, it was often difficult to disambiguate the many kinds of judgements applied machine learning researchers were making. Practitioners of machine intelligence, I would observe, could see themselves as generating and testing hypotheses at the same time that they saw themselves as "letting the data speak for itself". They would say "let's see what the data says if we assume …" and then they would update their assumptions and ask, "what does the data say now?" But the assumptions they articulated about the phenomena represented by data were not generally seen as "hypotheses" in any formal sense. Hypotheses were reserved for more thoroughly elaborated statistical and algorithmic theories about the technical processes they were implementing, whether they be "random forests," "support vector machines," or "multi-task learning" for any task that could be formulated as an *optimization problem*. An optimization problem can be thought of as a prediction or classification problem for which errors can be measured by the machine intelligence system and minimized as the system 'learns'. The identification of handwritten numerals discussed above is an optimization problem, because the task is set up in a way that the computer program is tasked with 'guessing' the correct numeral represented in an image, is told when it is correct or incorrect, and can update its method for guessing as it minimizes the errors in its guesswork.

Optimization problems are sometimes called "minimization problems" because they are set up to minimize error across all the data a machine learning program is trained on, and the task the computer program is optimizing for, or minimizing error for, is called the *objective function*. Crafting an objective function—a well-specified task for which error can be minimized—is a crucial task for applied machine learning researchers, but it is a task that is constrained by the goals they are seeking to achieve. These goals are often set by product managers, designers, and others within their organization who do not have expertise in machine intelligence. Applied machine learning researchers must navigate the goals of their coworkers, shifting and shaping those goals until they can be posed as an optimization problem. Faced with the goal of classifying potential parolees' danger to a community should they be granted parole, they must transform that goal into a question that can be answered by "the data itself". In the absence of data about, or even meaningful definitions of, "danger to a community," they must resort to data they *do* have—re-arrest data—in crafting that objective function. While the directive they have been given is clear—find a way to predict danger and rank parolees according to that danger—many other things remain opaque to these researchers: what constitutes "danger" for community members, what constitutes "danger" for their managers and their clients, what the relationship between offenses for which one might be re-arrested and any "danger" those offenses pose for the community, and how the likelihood of rearrest is skewed by race. Also opaque are the motivations that might lie behind the setting of thresholds that separate those on a ranked list into those deserving of or too risky for parole. Those motivations might be to variously keep people in prison, speed their release, provide a statistical rationale for parole boards' decision-making process, etc.

The centrality of optimization problems to the work of applied machine learning researchers is perhaps best explored through how the "common task framework" became the most important way of demonstrating the proficiency of both applied machine learning researchers and machine intelligence itself. The common task framework can be traced back to its origins at Bell Labs the 1960s (Liberman 2010) and through its popularity at the National Institutes of Standards and Technology (see our earlier discussion of MNIST) and the Defense Advanced Research Projects Agency (DARPA) before becoming the common currency of machine intelligence through Kaggle-style competitions, which began in 2010 (Donoho 2019). Kaggle, Inc. is a private, well-funded company that operates a platform hosting data science competitions. These competitions address academic research and business problems, and have been used as both recruitment tools for companies seeking machine intelligence talent and for applied machine learning researchers to demonstrate their skills (Bojer and Meldgaard 2021). Companies that share data as part of a Kaggle competition benefit from hundreds of researchers applying their skill to business problems and often offer cash prizes (in addition of fees provided to Kaggle) for successful competitors. These competitions provide a dataset and a goal to competitors who vie with each other to develop a technique that accomplishes that goal with as high an accuracy rate as possible. The common task framework can be applied to a seemingly endless array of well-posed problems (i.e., problems that can be re-framed as optimization problems), given sufficient quantities of data.[4] It can accommodate challenges to predict the presence of cancer in lung scans, to predict stock market movements, to identify highway obstacles in dashcam footage, etc. By producing "winners" for each of these contests the goals of these tasks appear to have been achieved, but these challenges demonstrate that some machine intelligence approaches to each challenge *are better than others*, not that the mission of object detection or stock market prediction has been accomplished.

---

[4] "Kaggle-style competitions" refers to the website https://www.kaggle.com/, launched in 2010, to host machine intelligence competitions by providing a dataset and a prediction task, and then numerically scoring attempts to "solve" that task, with the winner determined by the highest accuracy rate (according to a pre-established accuracy metric) achieved on that task.

In one famous Kaggle competition, for example, Netflix (an online movie streaming platform) provided a $1 million prize for any team that could predict how specific viewers in a "qualifying dataset" would rate films, ostensibly based on the "training dataset" of ratings they had given other films and the ratings that film had been given by other viewers (Figure 3). The winner was judged based on the predictive accuracy of their model, and the award ultimately went to a team from AT&T Labs (formerly Bell Labs) who called themselves 'Pragmatic Chaos' (Kaggle 2017). Notably, the data Netflix provided (600 MB worth of data) was stripped of almost all context; individual movies were represented by one set of numerical codes (although they could be matched to their actual titles by joining to a different data table), users by another, and the only additional information contestants were given was the date on which a movie was given a rating by a user, and the date of the movie's initial release. The overall competition was then judged based on the average error for each competitor, calculated by squaring the error for each prediction, averaging that figure, and then taking the square root of that average to generate a single error metric, the root mean squared error (RMSE), that treated both over- and under-estimations similarly. The winning team had an astonishingly miniscule RMSE: they accurately predicted a great many of the ratings users wound up actually assigning to movies.

```
For example, if the qualifying dataset looked like:

111:
3245,2005-12-19
5666,2005-12-23
6789,2005-03-14
```

```
225:
1234,2005-05-26
3456,2005-11-07
```

then a prediction file should look something like:

```
111:
3.0
3.4
4.0
225:
1.0
2.0
```

which predicts that customer 3245 would have rated movie 111 3.0 stars on the 19th of Decemeber, 2005, that customer 5666 would have rated it slightly higher at 3.4 stars on the 23rd of Decemeber, 2005, etc.

Figure 3. Text from the README file of the Kaggle competition Netflix Prize.

Their application of machine intelligence to the provided data clearly worked very well for predicting the missing values in the qualifying dataset, but it is less clear whether what this team accomplished can be seen as wielding any explanatory power over why users gave a certain rating or why movies received a certain ranking. This explanatory power would look something like a generalizable model about the relationships between movies, users, and ratings, that could be abstracted from the data of the Netflix Prize competition and used to inform similar questions in other contexts. But for Kaggle-style competitions that typify the Common Task Framework, generalizable models are not the goal. They merely need to make predictions that are valid within the bounds of the provided dataset. Predictive accuracy is the goal. And these two goals are what separate, in the words of statistician Leo Breiman, the "two cultures" of statistics (Breiman 2001a). Furthermore, the Common Task Framework delimits an arena of machine intelligence practice focused on optimization—selecting algorithmic techniques, performing 'feature engineering', tuning 'hyperparameters'—that is structurally and temporally distinct from the problem formulation itself, and severs machine intelligence practices that operationalize naïve empiricism from practices that engage with the context of a machine intelligence problem.

Once they have an optimization problem at hand, they must exercise professional judgement as they develop an objective function and a machine intelligence technique to perform that optimization. This task falls squarely within the "algorithmic modeling culture" of Brieman and the "data analysis" of Tukey. Applied machine learning researchers have a repertoire of algorithmic techniques, parameterizations, hyperparameter tunings, "feature engineering" techniques, and "loss functions" that they can draw upon, modify, and customize based on the objective function and the data they are working with, and these techniques are the basis for an overwhelming majority of conversations and representations made about machine intelligence in trainings, online discussion forums, and conference presentations. But in workplace conversations, these techniques overlap with discussions about applied machine learning researchers' "intuitions" and "assumptions" about the "real-world" phenomena represented by the datasets they were working with.

'Intuition', as it was mobilized rhetorically, often conflated disciplinary expertise and a preternatural instinctive grasp of data. In conversation, applied machine learning researchers would talk about how, in order to succeed, one would "need some data intuition" and to also have shown an ability to "go deep in a technical area like math, physics or electrical engineering". Developing technical skills in one of these areas was seen as training a practitioner's general intuitions for data, regardless of domain. These "intuitions" were sometimes as thoroughly elaborated as the statistical theories they engaged with—many applied machine learning researchers I worked with had completed doctoral training in other fields and would mobilize theoretical perspectives from those fields. But other times these intuitions were just that: intuitions. After following an intuition that entropy models from astrophysics might be used to predict the costs to financial services customers failing to follow financial advisors' advice, or that chatbots would be more engaging (according to a discrete metric for "engagement") if they could vary their scripted responses, these researchers would sometimes have to conclude for themselves that they had been following a dead end.

Intuition is crucial for the production of knowledge through machine intelligence. Without *some* intuitions about what a dataset contains, how it is related to real-world phenomena, and what can be *done* with it, data remain mute and any one application of machine intelligence cannot be evaluated meaningfully against another. My interlocutors would often press each other on why a particular decision was made by asking questions like, "What is your intuition as to why these recall metrics are so low?" and would mobilize intuition as a hypothesis to be tested. The "rigor" of machine intelligence was often counterposed to the generative ideation of "intuition" regardless of its source. During one presentation of a data science team working on operationalizing personality "types" drawn from psychology research, the applied machine learning researcher on the team described his work "proving the intuitions" of the psychologists. Applied machine learning researchers I worked closely with were equally pleased when their intuitions were disproven, taking this as demonstrative of the "rigor" of their methods.

Intuition was not merely the obverse of "rigor," the "trained judgement" applied machine learning researchers develop from interacting with datasets, algorithms, and their peers over the course of their careers (Traweek 1988; Daston and Galison 2010). "Intuition" was sometimes presented as the background for all work on machine intelligence, with one applied machine learning researcher observing, that "the intuitions we bring to the table shape the work that we do, so it matters what our background is". Nevermind that the researcher was standing on stage at an "AI for Good" event (see Chapter 3) next to acclaimed cellist Yo-Yo Ma when he said this, drawing an unspoken parallel between the intuitions shaping artistic, creative expression and those shaping complex sociotechnical systems. In conversations with each other, applied machine learning researchers could move between all these framings of intuition fluidly, sometimes within the space of a single sentence. An intuition could be an explanation as to why something worked or did not work. It could be a partially articulated analysis of the data or code they were working with. It could be a hypothesis that needed to be tested. And it could be something that is taken for established truth within the discipline in which they were trained, as when an applied machine learning researcher compared her intuitions about "ablation" techniques within a neural network to ablation studies in the cognitive neuroscience discipline she was trained in.

Intuition often sat easily next to and within algorithmic theories that were used to construct the complex machine learning systems my interlocutors developed, and both intuition and algorithmic theory played significant roles in making sense of the outputs of these systems. At times shifts between intuition and theory happened unreflexively, as part of agreed-upon practices for doing machine intelligence work. But at other times, practitioners questioned what kind of knowledge production they were engaged in—were they engaging theory to produce empirically valid and externally verifiable results, or were they externalizing their own intuitions by creating a statistical representation that merely mapped their prior knowledge? This would happen when comparing and combining machine intelligence techniques.

One applied machine learning researcher I interviewed was working with medical imaging technologies, developing a technique that used machine intelligence to predict "regions of interest," learned from radiologists, at the same time that it classified an image as "normal" or "abnormal". The question he struggled with was whether or not his assumption was valid that the regions radiologists pay attention to matters for the diagnostic task (the classification between normal and abnormal images). The intuition he voiced was that, "of course it matters, radiologists know where to look to make a diagnosis," but he was also committed to a naïve empirical approach that would imply classification could be done from the overall image, without recourse to medical doctors' "theories" about where abnormality was located. This researcher resolved this concern for himself by reducing his problem to what he called "an empirical question": by measuring the size of the contribution from each of these tasks—localization and classification—to the overall accuracy, the strength of his assumptions could be tested. The fact that these tasks inform each other, and not that one task or the other was entirely responsible for the accuracy his system was achieving, indicated for him that his intuitions were on the right track—doctors' trained judgements matter, but do not entirely explain the classifications machine intelligence was capable of making.

## BASIC TRAINING

My first foray as a "professional stranger" (Shapin and Schaffer 1985, 20) found me walking into the lobby of a building near Madison Square Park that I had long admired as I passed by on my way to teach undergrads "Introduction to Anthropology" at a nearby CUNY campus. I had signed up online for an open house hosted by a "data science bootcamp" looking to recruit new students, which I announced to the doorman before being allowed to take the elevator up to a set of floors leased by a well-known co-working space operator. After mingling and being offered the requisite pizza, barely flavored seltzer, and bottled mid-range beer, the other attendees and I were ushered into a conference room that was the impermanent home of a private continuing-education company that referred to itself as a "coding academy". Bootcamps and coding academies like the one I visited on that day had sprouted up starting in 2012, and by the time I started pre-dissertation exploratory fieldwork in 2016 there were dozens of companies running these bootcamps, often with training locations spread across multiple cities, in the United States and sometimes also abroad (Schaffer 2015).

Bootcamps were designed with many different types of students in mind—some are oriented toward mid-career managers hoping to supplement their resumes with more technical skills, some toward "blue-collar" workers hoping to be re-skilled into the "white-collar" workforce, others toward recent doctoral students hoping to find lucrative employment after deciding to leave (or being unable to secure) a post-doctoral academic position. Hundreds of these types of programs currently exist in the United States, thanks to policy priorities that support bootcamps as alternative pathways to training a high-skilled workforce outside of the traditional educational system (Arbeit et al. 2019). At the first open house I attended, I was given an overview of a single course focused on data visualization that promised to help students "add data to their storytelling" no matter what field they were in. Using data science and making data central to the rhetorical heft of arguments one might make, was presented as a crucial job skill that could lead to promotions, raises, or other forms of career advancement. I would go on to attend similar open houses, with attendance facilitated by online event management companies like Meetup and Eventbrite, throughout my fieldwork. While I would later attend instructional sessions at one of the bootcamps, early on I attended open houses in an attempt to map the field, trying to get a sense of the distinction between different curricula and different machine intelligence techniques, and to identify potential field sites and interlocutors.

For prospective trainees, these distinctions might matter a great deal for their careers: fine-grained differences between machine intelligence practices could be drawn from the curricula of these training bootcamps. Prerequisites like proficiency in specific coding languages were more common for machine learning and artificial intelligence training programs, prior study of statistics was recommended for those interested in data science and machine learning, linear algebra was a prerequisite for some machine learning and artificial intelligence programs, data science training programs were pitched towards those who wanted to pursue careers inside of big companies, and artificial intelligence training was pitched toward those seeking careers in venture capital-backed start-ups. These open houses were a crucial site for entering the field, but also for the production of machine intelligence as a field. Bootcamp open houses were sales and recruitment events, they were free to attend, but those deemed most likely to spend $15-25,000 received the most attention from staff.

The open house I attended that first night was different in form and scope, but not in aspiration from the data science bootcamp I joined a year later as a participant-observer. This bootcamp was tailored specifically to recent doctoral scholars and former postdocs who were leaving the academic track. In interviews with bootcamp faculty and administrators, they candidly characterized the pool they recruit from as those squeezed out by an academic job that overproduces PhDs, particularly in STEM fields, and under-employs recent graduates. Consequently, careers in data science, machine learning, and artificial intelligence were presented as fallback options for students who already had robust statistical training, rigorous instruction in the scientific method, and well-developed analytical skills. Given these students' prior training, they were seen as already having "some intuition with data" but requiring rigor in machine intelligence techniques. Additionally, at least according to advertising copy from the bootcamp, such individuals needed help adapting their scientific skills to industry settings where machine intelligence applications were in high demand. This meant being able to package their work in short presentations that telegraph the business problem they are addressing and how they have solved it by applying the techniques of machine intelligence.

From the curricula offered at this bootcamp, and from the presentations and feedback sessions I observed and participated in, adapting these newly-minted Ph.D.'s to successful job candidates in the tech industry consisted of supplementing their hard-earned scientific skills with what they lacked. The bootcamp supplemented already-existing scientific skills with coding lessons, an overview of how standard machine intelligence software packages can be applied and integrated into a data product, and instruction in product design. The entire bootcamp, which was promoted as "an intensive 7-week post-doctoral training fellowship bridging the gap between academia & data science," offers an inversion, of sorts, of the "deficit model of education" with respect to science and technology (Bak 2001). Within the context of this bootcamp, post-doctoral trainees come in with their "science" skills firmly in hand, but having a deficit in "data" and "business" skills that the bootcamp is intended to rectify. They were likely to have good data "intuitions," but they needed to adapt the rigor of their doctoral training to the rigor demanded for practicing machine intelligence. What this meant in practice was that bootcamp trainees were being taught to shift the scope of their professional judgement from their disciplinary perspectives, in which problems were framed as hypotheses to be disproven, to the practices of naïve empiricism, in which problems were framed as questions that could be answered by applying machine intelligence techniques directly to data.

This point bears some unpacking, as it seemingly sets up a mutually exclusive dichotomy between knowledge production practices characterized by logical positivism (Popper [1935] 2002) and those characterized by naïve empiricism. This dichotomy is helpful analytically, but knowledge has long been produced through practices that blend and overlap logical positivism and naïve empiricism in machine intelligence, as well as in the STEM fields that bootcamp trainees were already experts in. This is evident in the history of archaeological theory discussed above, as well as in the 20th century elaboration of machine intelligence techniques by statisticians like Tukey and Breiman. But it is also evident in the adoption of machine intelligence techniques within STEM fields themselves. This is unsurprising, as many of the machine intelligence techniques that have since been applied across domains of inquiry were first developed to solve statistical problems in nuclear physics (Kahn 1954) and before that in correcting for observational error through probability calculus (Sekula 1986). Machine intelligence techniques have been adopted by myriad other STEM fields including optics (Fol et al. 2019), astronomy (Kremer et al. 2017), and genomics (Zou et al. 2019), and trainees in this particular bootcamp often entered their re-training with more than a passing familiarity with applications of machine intelligence in their fields of doctoral study. Within these domains, machine intelligence techniques have served a crucial role in the production of knowledge, lending the techniques themselves tremendous legitimacy as scientific tools. The work of applied machine intelligence researchers, however, is focused on applying the techniques of machine intelligence outside the context of scientific domains, unmoored from other scientific practices that stabilize machine intelligence products as facts.

For trainees at this particular bootcamp, then, their challenge is to shift the emphasis of their professional judgement onto machine intelligence techniques and away from the scientific norms of hypothesis generation, testing, and proof from their prior training. This is not to say that hypotheses or theory are absent from the work they produced in the bootcamp because they were tasked with constructing problems that could be solved by machine intelligence, which is formally identical to hypothesis generation. Rather, it is to say that it was more important for them to learn how to narrate their claims through the authority of machine intelligence. By justifying the suitability of their machine intelligence techniques to the data and the problem at hand, the particular modifications they made of parameters, and the resulting accuracy metrics, trainees became skilled manipulators of machine intelligence and datasets who could find answers to any problem—not just those they had earned a doctoral degree to address. In the context of a bootcamp that was retraining PhD scientists to enter industry roles, this shift makes sense. Their new professional competency was not to be in pursuing open-ended scientific inquiries but in providing concrete answers to questions posed by others in their organization, using whatever data was available that could be brought to bear on those questions.

    The bootcamp was structured around a "portfolio project" that trainees would build over the course of their seven weeks, scoped to address a business need and to demonstrate their ability to apply machine intelligence to that business need. The bootcamp culminated in a "pitch" event in which trainees' portfolio projects were presented to potential employers. The bootcamp itself had ongoing relationships with these potential employers; the tuition-free bootcamp was only possible because the bootcamp itself was sustained by the recruitment fees paid by companies when they hired trainees out of the pitch meetings. These ongoing relationships also manifested in materials trainees had access to during the bootcamp. Potential employers offered their companies' proprietary data, upon which trainees could build their projects. These employers visited the bootcamp in the early weeks of each session to present their business needs and generalized descriptions of the problems they face. Trainees, then, between classroom instruction on machine intelligence methods and skill-building exercises, were expected to develop their projects by taking provided data (or sourcing other publicly available data), applying a machine intelligence technique taught in class to that data, thereby to address some business problem they inferred from potential employers' early

presentations. The result of this process, in terms of how trainees presented their portfolio projects as machine intelligence projects, was that machine intelligence was treated like a set of *a la carte* menu options that could be applied to an available dataset, the outputs of which were then treated as a worthwhile or important finding. As training exercises, not all projects were necessarily worthwhile or important. Trainees often inadvertently replicated existing analyses, or were given more limited datasets than they would have access to as full time employees. But the goal of this exercise was not to produce groundbreaking or novel insights. It was to develop the skills need to formulate a problem, choose an appropriate machine intelligence technique, conduct an analysis, and pitch the results of that to other stakeholders.

On one of the days that I visited the bootcamp, trainees were presenting early versions of their pitches for practice and instructor feedback, in a cramped, couch-lined room with curtained windows looking out over Union Square in Manhattan. Each trainee would deliver their pitch, aided by a slide deck, and then open the floor for questions and feedback. These early iterations of their presentations were based on projects that were still very much works-in-progress, and so for some trainees these feedback sessions were mostly useful for getting experience with public speaking. They would frame the problem they were trying to solve, the data source(s) that they chose as relevant for the problem, the machine intelligence techniques they were in the process of applying to that data, and would conclude with a results slide highlighting the significance of what they had done. Often these later slides would be placeholders, or would be narrated by saying "I haven't gotten this to work yet, but my intuition is…". They had picked up the importance of intuition early in their bootcamp, and were learning to pair it with the techniques they had learned and language that indexed that rigor by standing up to the scrutiny that followed each presentation. During post-presentation Q&A sessions, first their fellow trainees would ask questions like, "why did you use a minmax function there?" or "how did you scrape that data set?" or "did you consider retraining with different holdout data?" Instructors would then mix questions with suggestions like, "you could also try the same analysis but with gradient boosted decision trees," or "you need to deliver this pitch with confidence... you have solved a problem they didn't even know they have." Representatives of potential employers, who were not always members of data teams, would generally have the last word, offering statements of encouragement or clarifying how they think about their business model.

Trainees presented a heterogenous set of projects. The goal of each was to demonstrate the ability to translate the type of scientific thinking they had developed in their doctoral training into business applications, while also showing some form of mastery over at least one machine intelligence technique. This resulted in presentations about projects ranging from detecting the type of crops being grown in satellite imagery, suggesting potentially desirable items to online shoppers based on items they had already placed in their shopping cart, and analyzing the text of twitter messages to discern the approximate location of the sender. In speaking with trainees about their projects after the feedback session, it was clear that many of the "problems" they had chosen to solve were based more on the fit between the machine intelligence technique they were interested in, or understood well enough, or had enough confidence in to try to apply, and less to do with any sort of deep engagement with a potential employer's business model, market analysis, or problem formation. One trainee told me that they liked the in-class exercise on gradient-boosted decision trees, a machine intelligence approach packaged as a module called "xgboost" so it could be neatly used inside the programming package "R" (Chen, Benetsy, and Khotilovich 2021), and that they looked for an available dataset that they could use "xgboost" on for their portfolio project. Another trainee, who had presented a project on abandoned shopping cart items, had been impressed by the quality and "cleanness" of a shopping cart dataset provided by a potential employer, and built their project around that data to avoid any additional "data cleaning," during which she would have had to spend precious bootcamp time meticulously looking for missing values or changes in data collection practices that might invalidate large portions of the dataset.

While it might seem strange that trainees would be encouraged to work on projects with such seemingly arbitrary motivations, or that trainees' personal whims would shape the projects they spent the better part of their seven-week training pursuing, the project development process within the bootcamp had a great deal of resonance with how I saw applied machine learning research being carried out in other contexts. Throughout my fieldwork—in public meetups, discussion boards for practicing applied machine learning researchers, and particularly in mediatized representations of machine intelligence—machine intelligence was presented as a "black box" technology that takes data as an input and produces useful answers as an output. The naïve empiricism of machine intelligence supports this approach; if the data can speak for itself, the practitioner's task is to let it do so. The appropriate machine intelligence technique, applied to an appropriate data set in a rigorous way will allow the data to speak. Conversely, if an inappropriate technique is applied, or an inapt dataset is chosen, the data will remain mute. For these trainees, learning to apply these black box machine intelligence techniques in ways that get the data to speak was far more important than their motivations for doing so. The pragmatic result of "data speaking" outweighed any other considerations.

That said, choosing the *right* black box, for the right *reasons,* would be the crucial locus of professional rigor for these incipient data scientists as their careers progressed. But online trainings I observed, textbooks I studied, and public talks about implementing various techniques often bracketed out the details that make one machine intelligence technique appropriate for one set of tasks and not for others, leaving the details of those implementations beyond public view, and even beyond the scope of work for apprentice practitioners.

## DIPPING INTO THE BLACK BOX

The "black box" concept is a crucial component for understanding how science and social practices are deeply intertwined. For Latour, black boxes are the *ne plus ultra* of scientific activity—a phenomenon understood well enough that it can becomes a stable piece of knowledge that others can build upon. The work of scientists is to enclose phenomena in black boxes, and scientific knowledge progresses by building upon black boxed knowledge that remains unperturbed by subsequent experience (Latour 2003). Latour draws the language of black boxes from the terminology of science and engineering, attributing his usage to the field of cybernetics, which reached its fullest fluorescence around the Macy Conferences of 1946-1953, which also produced foundational work on the possibility of artificial intelligence (Pias and Vogl 2003). Latour uses the black box concept to analyze science in the making, and for him telling the story of how black boxes are "closed" reveals the social context and contingency of the content of scientific knowledge. But for applied machine learning researchers, *the* enclosing of machine intelligence within black boxes serves to simultaneously buttress and undermine their authority as they derive useful answers from data. To explore what I mean by that, it is useful to consider how the concept of the black box works within science and engineering.

Scientists and engineers describe black boxes in pragmatic terms. Particularly in electrical engineering and computer science, black boxes are ways to isolate subprocesses or to reduce complexity by bracketing off more complex phenomena and only addressing how those phenomena are inputs or outputs for other parts of a system. These inputs and outputs could be as simple as the amount of current going into a transistor and the amount of current flowing out of it (Belevitch 1962), or as complex as the pixels of a photograph going into an algorithmic system and a real person's name being output from it. Assembling black-boxed knowledge can look like taking something that is known from prior scientific experimentation, like the double-helix shape of DNA that Latour is fond of discussing, and deriving a new bit of knowledge from that established research, like the ribosomal assembly of a specific protein, without having to re-establish the earlier fact about DNA's shape. Or it can look like enclosing a complex chemical reaction in a sub-process of a larger industrial process.

In machine intelligence, the use of black boxes proceeds similarly. In epistemological terms, however, these techniques often embed assumptions about the nature of reality that go entirely unrecognized by applied machine learning researchers. These assumptions can be illustrated by considering how machine intelligence techniques are developed for one purpose and then adapted—or re-purposed—for another, metaphorically linking these purposes. Regression, a foundational technique for establishing a trend line or binary classification, can be calculated in a straightforward (if time-consuming) way by hand or in explicitly written lines of code. For a given dataset, linear regression entails using the datapoints in a set of data to calculate the slope (and y-intercept) of the trend line. The trend line can also be used to classify data into two groups, one of which falls above and one of which falls below that line. In computer code, this would take approximately 45 lines of code (Figure 4) and require a firm grasp of the "ordinary least squares" technique for calculating a regression line (Hutcheson and Sofroniou 2006).

```python
import numpy as np
import pandas as pd
from matplotlib import pyplot as plt

class Model():

    def __init__(self, x, y):
        self.x = x
        self.y = y

    def fit(self, intercept = False):
        self.intercept = intercept

        x = np.hstack([np.ones(len(self.x))[:, np.newaxis], self.x]) if self.intercept else self.x

        y = self.y
        self.betas = np.linalg.inv(x.T @ x) @ x.T @ self.y

    def predict(self):

        x = np.hstack([np.ones(len(self.x))[:, np.newaxis], self.x])\
            if self.intercept else self.x

        self.y_hat = x @ self.betas

        # If you want to see the predictions, you can 'return self.y_hat' here

    def plot_predictions(self):
        plt.scatter(self.x, self.y, c='orange', label = 'Observed values')
        plt.plot(self.x, self.y_hat, label = 'Fitted values')
        plt.legend()
        title_label = 'Fitted OLS Regression: intercept={}, slope={}'\
            .format(np.round(self.betas[0][0],2), np.round(self.betas[1][0],2))
        plt.title(label=title_label)
        plt.show()

if __name__ == '__main__':
    x = np.arange(50)[:, np.newaxis]
    y = np.array([i + np.random.rand() for i in range(50)])[:, np.newaxis]
    ols_test = Model(x, y)
    ols_test.fit(intercept=True)
    ols_test.predict()
    ols_test.plot_predictions()
```

Figure 4. Calculating a linear regression in python using ordinary least squares (Shemet 2020).

This technique of calculating linear regression would undoubtedly be time consuming for an applied machine learning researcher, is an opportunity to introduce error into an application if even a single character is typed incorrectly, and most practitioners would be hard-pressed to construct this procedure from memory. Instead, they use programming packages like `scikit-learn` that have enclosed the ordinary least squares technique inside a black box that can be called from a single line of code (Figure 5): `reg = LinearRegression().fit(X, y)` (Pedregosa et al. 2011). For applied machine learning researchers, the mathematical intricacies of many of the tools they employ are often black-boxed in precisely the same way as linear regression is. This affords them the opportunity to conduct fantastically powerful analyses of very large datasets with very little effort.

```
>>> import numpy as np
>>> from sklearn.linear_model import LinearRegression
>>> X = np.array([[1, 1], [1, 2], [2, 2], [2, 3]])
>>> # y = 1 * x_0 + 2 * x_1 + 3
>>> y = np.dot(X, np.array([1, 2])) + 3
>>> reg = LinearRegression().fit(X, y)    ←
>>> reg.score(X, y)
1.0
>>> reg.coef_
array([1., 2.])
>>> reg.intercept_
3.0000...
>>> reg.predict(np.array([[3, 5]]))
array([16.])
```

Figure 5. Snippet of python code implementing linear regression at arrow (Pedregosa et al. 2011).

But such affordances are only superficially effortless. For an applied machine learning researcher to call linear regression on a set of data, they must first ascertain that the data set is appropriate for linear regression—not all data sets are appropriate for linear regression because not all phenomena are linear. Knowing which packaged machine intelligence techniques are capable of handling which types of data, and for addressing specific questions about various types of data, is the main task of applying machine learning according to bootcamp instructors and the authors of any number of textbooks on machine learning (Alpaydin 2014; Bishop 2006). Additionally, black-boxing machine intelligence techniques within programing packages like `scikit-learn` renders opaque many of the underlying assumptions a technique makes about the data. In practical terms, this means that practitioners use default settings for complex techniques, rather than carefully adjusting parameters that might have drastic consequences for the outputs. For regression, there are many such parameters that affect the outcome: whether the phenomenon in question is linear or logarithmic, how smooth or jagged a regression line might be, etc. all affect the final output.

## BACK TO BOOTCAMP

For the trainees I visited with at the bootcamp, recognizing a potential fit between an available dataset and a machine intelligence technique they felt comfortable with was never couched in an awareness of such deeper logics, even if they did at times worry about how the techniques they were applying would compare (favorably or unfavorably) to notions of statistical validity within the disciplines in which they were trained. Finding an appropriate machine intelligence technique was instead merely a necessary early step toward completing their final project. For the student using xgboost to build a product recommendation application, they were already familiar with "decision trees" from their biomedical research, where it can be used to classify risk factors for disease in sets of health records (Karaolis et al. 2010). And they were excited about the application of "gradient boosting" to these decision trees offered by the xgboost software package.[5] But they were not optimistic or excited about any potential project they might have chosen to do until they realized that one of the potential employers participating in the bootcamp had provided a dataset about the shopping cart history of their e-commerce customers that was perfectly suited to the specific affordances of xgboost. Once they realized this, they were able to apply the "black box" of xgboost to that employer's dataset in a fairly straightforward way.

Calling that application straightforward elides and erases, of course, the actual labor it took to get their project to work, but that elision and erasure is crucial to how the work of applied machine learning research is black boxed at a much larger scale. Getting xgboost to work on that particular dataset took many, many steps. The data had to be transformed to fit the inputs expected by the xgboost software package. Parameters of that package had to be researched and chosen carefully, often through time-consuming and laborious trial and error. Computer code had to be written to automate some of that trial and error, at least according to the "best practices" taught by the bootcamp. And results had to be checked and double-checked, again to satisfy the bootcamp's (and the industry's) "best practices". But when it came time to present their work to the potential employer, the trainee, too, elided all the work they put in. Their presentation even featured a slide that showed an icon representing the data they worked with, with an arrow leading to a box (it was blue on the slide) labeled "xgboost," followed by pair of arrows pointing to the results: the three items most likely to result in a completed purchase if the customer added them to their shopping cart.

---

[5] The specifics of gradient-boosted decision trees as applied in xgboost are beyond the scope of this discussion, but see Chen, Benetsy, and Khotilovich 2021.

To say that much of the effort this trainee put into their applied machine learning project is made invisible or elided from the final presentation of that project is somewhat of a commonplace. Labor is often stripped of its visibility or its particular character as it is condensed into commodities (Roseberry 1997). The commodity produced by the trainee is both a product, i.e. a black box that can transform shopping cart history into product recommendations that will increase the likelihood of a completed order, and an "insight". In the world of applied machine intelligence, insights are presented as fungible commodities. What Marx calls "the fetishism which attaches itself to the products of labour, so soon as they are produced as commodities" (Marx [1867] 1990, 165) also applies to the insights produced by applied machine learning researchers' labor. Insights are severed from the activities of those who produced them to stand apart from machine intelligence. They are even made to seem as if they precede machine intelligence, having existed prior to any data collection or algorithmic analysis. At recruiting events and job fairs for data scientists, representatives of successful Silicon Valley data applied machine learning research teams would assert that "data pulled together into insights … is the voice of the customer at scale," and would cheer their role in delivering these precious insights for their employer.

In this framing, insights are the output of a well-developed machine intelligence project. They are arrived at through the specially-developed skills of applied machine learning researchers and the tools they have exclusive access to. But they are also made to seem immanent in the world, always there awaiting discovery by intrepid researchers. They are products, and while the practices that lead to "producing insights is more like making lures than pulling up fish—a matter of intentions and construction, not the retrieving of objective relations from an obliging sea of data" (Seaver 2015, 33), I suggest that insights are made to stand apart from this effort and take the form of objective truths that came to be represented as data. An applied machine learning research team, or a well-trained data scientist can be expected to regularly produce insights that can be treated as objective knowledge, but then instrumentalized, either by building them into a product (as the trainee did with their project), or by informing corporate strategy. In a visit to present their work to OTH, a potential client's machine learning team presented slides that displayed the work "insights" enclosed in a cog-shaped icon that neatly articulated with icons labeled "data" on one side and "action" on the other. Another potential client discussed setting up "a content studio based on data insights."

## FROM RESEARCH TO PRODUCT

For applied machine intelligence researchers, producing *insights* is of key importance. Often, these insights are taken as commodities which can travel far beyond the contexts that produced them. However, insights derive from a situated understanding of the world and the purposes to which they might be put. They are more like lures than fish, to reuse Seaver's analysis given above. As lures, they work in certain lakes and streams and not others, at particular times of year, and not at others. But, treated as fish, they can be flash frozen and shipped all over the world. Similarly, a machine intelligence product that is narrowly tuned to a specific purpose—detecting sepsis at a specific hospital in a discrete geographic location, for example—cannot easily be used for a different ailment, in a different city, or a different hospital (see Sendak et al. 2020). The machine intelligence product, then, is an assemblage of the data, the phenomena represented by the data, the context in which those phenomena occurred and could be captured as data, the algorithmic techniques applied to the data, the researchers applying those techniques, and the physical infrastructure making data storage, retrieval, and processing possible. Mistaking machine intelligence as a fish rather than a lure, i.e. as a context-independent portable commodity rather than a situated apparatus for producing a desired result, leads to a tension for applied machine learning researchers as to what the goal of their labor is, and how researchers navigate this tension has profound consequences for how they understand their work, how they wield their authority, and how the claims of machine intelligence come to be seen as authoritative by their coworkers, employers, and the broader public. If what they produce is seen as universal—able to travel without being transformed—their analyses can break down or produce harm when they are transported from one context to another. Misappropriating an earthquake prediction algorithm to serve as a component of a crime prediction tool led to over-policing of certain neighborhoods in major cities across the United States, for example (Haskins 2019). And yet the idea *that* their products can travel—that their work produces generalizable insights—is responsible for a portion of the authority granted to machine intelligence.

This tension arises in part because generating "insights" and building machine intelligence "products" are differentially rewarded depending on the organizational structures applied machine learning researchers work within. But tensions also arise when applied machine learning researchers begin to ask themselves which of these tasks is the "real" work that they are doing. Researchers would observe, particularly when stuck on a thorny passage of computer code, that they "feel like they are just messing around with the plumbing". At other times they would ask each other, "is this even science?" before launching into a lengthy debate about what was and what was not scientific about the particular problem they were grappling with. In trying to resolve the question of "is this even science?" for themselves, they would oscillate between Tukey's data analysis—building algorithmic models that had predictive accuracy within a dataset—and the hypothesis-driven deductive methods of their own training. For these researchers, the status of science and scientific work was deeply entwined with their sense of self. They held doctorates in science disciplines like cognitive neuroscience, astronomy, or physics but had transitioned into applied machine learning research roles (either through training bootcamps like the one described above or through autodidactic means) and balked at job titles that would identify them as engineers, even where engineering staff were situated at the top of their workplaces' corporate hierarchies.

Another part of the tension that arose for applied machine learning engineers derives from the split focus within statistics between Breiman's "data modeling" and "algorithmic modeling" cultures (Breiman 2001a), which could also be glossed as Tukey's turn toward "data analysis" (Tukey 1962). Recall that "data modeling" involves making assumptions about how and why data are distributed as they are and trying to describe that distribution mathematically in a way that "fits" the data. "Algorithmic modeling," in contrast, views the distribution of data as fundamentally unknown and tries to construct a model that can predict where in the distribution any given data point might fall. The tension between these two "cultures" for the applied machine learning engineers I worked with, then, was whether the assumptions they made about a dataset pertained to the data itself, or to the phenomenon represented by the data. This played out in dramatic relief at OTH throughout my fieldwork.

At this lab, applied machine learning research was marketed to corporate data science and machine learning teams through offering consulting hours and hosting workshops for corporate clients, but the greater balance of the effort in the lab was dedicated to producing quarterly research reports about a new or emerging technique in machine learning that would then be distributed to their clients on a subscription model. This research report was a multi-person endeavor, although one researcher led the overall effort. It involved an elaborate topic-selection process, a research period that involved a literature review and interviews with early adopters of the technique within industry, development of a prototype machine learning product that used the technique, and drafting of the report itself (complete with stylized illustrations and a letter-press binding). Developing a machine learning product for a new technique was the centerpiece of the research report effort. It required understanding the technique itself, in minute technical detail, and how to apply it to problems. But it also required constructing a problem that was well-suited to be addressed by that new technique.

In the sense that most work in machine intelligence involves applying a well-suited technique to a well-posed problem, the exercise these applied machine learning researchers undertook for their research report was merely a highly reflexive version of how machine intelligence is applied generally. Inspired by the horizontal organizational structures common in the technology industry (W. R. Scott 1975; Pfeffer and Leblebici 1977), researchers at OTH shared responsibilities and steered many business decisions through collaboration and consensus. The topic-selection process at the lab was emblematic of this. It was run by the researcher whose "turn" it was to write the next report and involved a series of whiteboard exercises in which all lab members would nominate candidates, research their potential as a topic, and then present their favored candidates to the entire team, who would then collectively winnow down the candidates to a few finalists. The report author would ultimately select the topic. The reflexivity of the process these applied machine learning researchers engaged in offers a valuable window into questions that go un-asked and assumptions that go un-examined in more mundane, less research-oriented applications of machine intelligence. The research report I participated in addressed a technique called "multi-task learning" that was first described in the late 1990s (Caruana 1997), but the technique rose to the top of the topic-selection process because it had recently found applications in industry (see McCann et al. 2018), and was more easily integrated with newer machine learning programming packages.

Multi-task learning is a machine intelligence technique that is designed to accomplish two or more tasks simultaneously. Extending the previously discussed example of handwritten numeral identification, a multi-task learning setup might simultaneously predict which numeral was represented in an image as well as, hypothetically, the unique individual whose handwriting produced that numeral. To do this, each image of a handwritten numeral would have to be associated with a label for the numeral (0-9) represented in the image and an additional label identifying the person who produced the drawing represented in the image. A standard approach to supervised learning could train an algorithm to predict either of these labels for a given image, but a multitask learning approach uses a single loss function to minimize the error for both tasks. The "exciting" thing about multi-task learning, as it was explained to me, is that "intuitively, focusing on a single task should produce the highest accuracy, but counter-intuitively, focusing on more than one task actually can improve the accuracy for each task compared to what can be achieved through single-task learning". This was explained, again with recourse to intuition, by observing that tasks that are somehow related can inform each other as they are learned by an algorithmic model. In her presentation arguing that multi-task learning ought to be the topic of the upcoming report, an applied machine learning researcher explained the technique in terms of "natural language processing," describing how language tasks are related. She said that an algorithm that can identify the part-of-speech of a word in a sentence (noun, verb, adjective, etc.) can do so with greater accuracy if it can borrow from statistical information it uses to identify the semantic role (agent, predicate, goal, etc.) of that sentence. The "intuition behind this," she explained, is that nouns are more likely to be agents, verbs are more likely to be predicates, and so on. She drew this intuition from her own experience with language, but also from the research she had been conducting across the computer science literature on multi-task learning (e.g. Strubell et al. 2018). Similar intuition could be applied to the handwritten numerals example, in that learning the stylistic quirks of individual numeral-writers could inform how the algorithm discerns between various examples of each numeral—perhaps one writer includes serifs or writes at a sharp angle—and these quirks could be used to help distinguish between person A's "7" and person B's "1".

The challenge for the applied machine learning researchers, having selected multi-task learning as the topic for their report, was to create a demonstration product for the technique. The "demo" was intended to be a "real" product that report subscribers could visit through a password-protected URL and interact with to examine the capabilities of multi-task learning. This lab had developed demos for prior reports that could summarize long text passages into single sentences, predict customers that were likely to "churn" or cancel their subscriptions to a service, or identify the objects contained in a photograph. Multi-task learning was discussed by applied machine learning researchers working on the report as more difficult to design a demo for, because it was "an approach, not a technology" and therefore had to demonstrate a way of thinking about problems in ways that would be novel for others working in machine intelligence. After several rounds of ideation, and a false start that would have 'learned' online shoppers' product reviews, the researchers landed on a concept for demonstrating multi-task learning that would predict two things simultaneously: what section (i.e., world news, sports, lifestyle, politics, etc.) a newspaper article belong in, and whether that article was from a 'tabloid' or a 'broadsheet'. Having chosen this product for the demo, the researchers were faced with a series of questions that characterize every approach to machine intelligence, but are seldom discussed as explicitly as they were in developing this product. These questions typically are along the lines of: What problem are we trying to solve? What data is relevant to this problem, and is this data available to us? What is the format of the data? What technique is most appropriate for prediction or classification based on this dataset? Is it structured properly for the technique we want to apply to it? What parameters do we need to use?

For the newspaper classification demo project, researchers quickly found themselves asking additional questions that were not part of the standard machine intelligence process, at least based on what I had been observing at data science bootcamps, machine learning conferences, and in the consulting engagements during which I accompanied these applied machine learning researchers. The questions they started asking themselves, and each other, were based on their encounter with the limits of the naïve empiricism that characterizes the algorithmic modeling culture of machine intelligence. These limits were the limits of what can be known from the data itself.

Some of these limits were wrapped up in categorical confusion. Other limits were wrapped up in how algorithmic modeling can be applied to data that is heavily laden with what I will, for the time being, call "cultural meaning". In evoking 'culture' and 'meaning', I am invoking a textual definition of culture as "a system of inherited conceptions expressed in symbolic forms" (Geertz 1983, 89). As definitions of culture go, Geertz's formulation leaves much to be desired, and has been rightfully critiqued (Roseberry 1994; K. E. Hoffman 2009; Ortner 1997). But in the context of machine intelligence, and particularly for machine intelligence approaches to language, it is a useful starting point for beginning to explore what it might mean for machines to be 'intelligent' at all, as what they are being tasked with 'learning' are—by any definition—cultural concepts. And what they are learning is how to manipulate, if not concepts themselves, then the symbolic encoding of those concepts. Anything that is deemed worth predicting or classifying—whether it be handwritten numerals, the expected failure of turbine parts, or movie ratings—is a cultural concept represented as a machine-readable symbol. So, 'tabloid', 'broadsheet', 'sports', 'business', etc. are labels for cultural concepts. Applied machine learning engineers are not trying to "teach" a machine what "tabloids" are, or what "sports" are. Rather, they are trying to get a machine "to pick up on those properties"—other symbols that have been associated with that concept—in the text of an article that can be used to distinguish a broadsheet article from a tabloid article, or a sports article from a news article.

So, what are the properties of a news article that a machine can "pick up on"? In the handwritten numerals example, the properties a machine can pick up on are the variously shaded pixels of each image, and how they are arranged in relation to each other. For text, a nearly identical machine learning setup can pick up on the presence of different words and their relationship to each other. Applied machine learning researchers could show the machine a large set of articles, each with labels for the section of a publication and the type of publication they are from, and the machine could "learn" to predict those labels by updating itself each time it makes a mistake, to minimize its overall error. The most common way of accomplishing this is through a technique called "deep learning" (Goodfellow, Bengio, and Courville 2016), which the applied machine learning engineers used in their multi-task learning demo.

## A Brief Plunge into Deep Learning

Deep learning has widely been credited for the resurgence of interest in artificial intelligence (G. Marcus 2018) following the long 'winter' that followed disillusionment in expert-systems, or rule-based, approaches from the 1970s (Forsythe 2001). Deep learning builds on a 'neural network' approach (Anthony and Bartlett 1999), which is itself an attempt to construct computational systems modeled after biological systems. 'Artificial neurons' are modeled after 'biological neurons'—in the textbook my coworkers and I used to teach ourselves `TensorFlow`, the resemblance between artificial and biological neurons was described thusly: "Biological neurons receive short electrical impulses called *signals* from other neurons via ... synapses. When a neuron receives a sufficient number of signals from other neurons within a few milliseconds, it fires its own signals. Thus, individual biological neurons seem to behave in a rather simple way, but they are organized in a vast network of billions of neurons, each neuron typically connected to thousands of other neurons. Highly complex computations can be performed by a vast network of fairly simple neurons, much like a complex anthill can emerge from the combined efforts of simple ants" (Géron 2017). Similarly, artificial neurons can be linked together and trained to pass 'signals' on to other neurons when they receive appropriate inputs. In deep learning, successive layers of neural networks are linked together to perform complex tasks. The validity of a deep learning approach to problems of machine intelligence was demonstrated in 2012 when it was applied to a large dataset of images to correctly predict the labels solely based on the content of those images (Krizhevsky, Sutskever, and Hinton 2012). Since then, it has been adapted and applied to a wide range of problems, enjoying a faddish popularity amongst applied machine researchers over the course of my fieldwork.

As a supervised learning approach, deep learning is seen by applied machine learning researchers as particularly well-suited to learning from data that do not have easily accessible 'features'. In contrast to, say, biometric descriptions of iris flowers, the features of which (sepal length, petal width, etc.) can be measured and subjected to regression analysis, digital images or large bodies of text do not have discrete features. Instead, they have clusters of pixels or arrangements of words. Deep learning is a technique that can examine the relationship between individual words or pixels and feed statistical information about those relationships into a subsequent 'layer' of analysis that can compare the relationship between clusters of words or pixels to identify structures within those relationships (e.g., edges or curves, syntactical regularities). It can then feed statistical information about the relationship between those structures into subsequent layers that can identify how those structures combine into more complex features (e.g., anatomical units, topic sentences). A final layer, then, can associate features with the training labels provided to the algorithm (e.g., '7', '1', 'sports', 'tabloid') to predict which label an input might belong to. Deep learning is 'deep' because it connects several layers of statistical representation together to construct successively more complex features from unstructured and (from the perspective of the learning algorithm) incoherent inputs (see Figure 6).

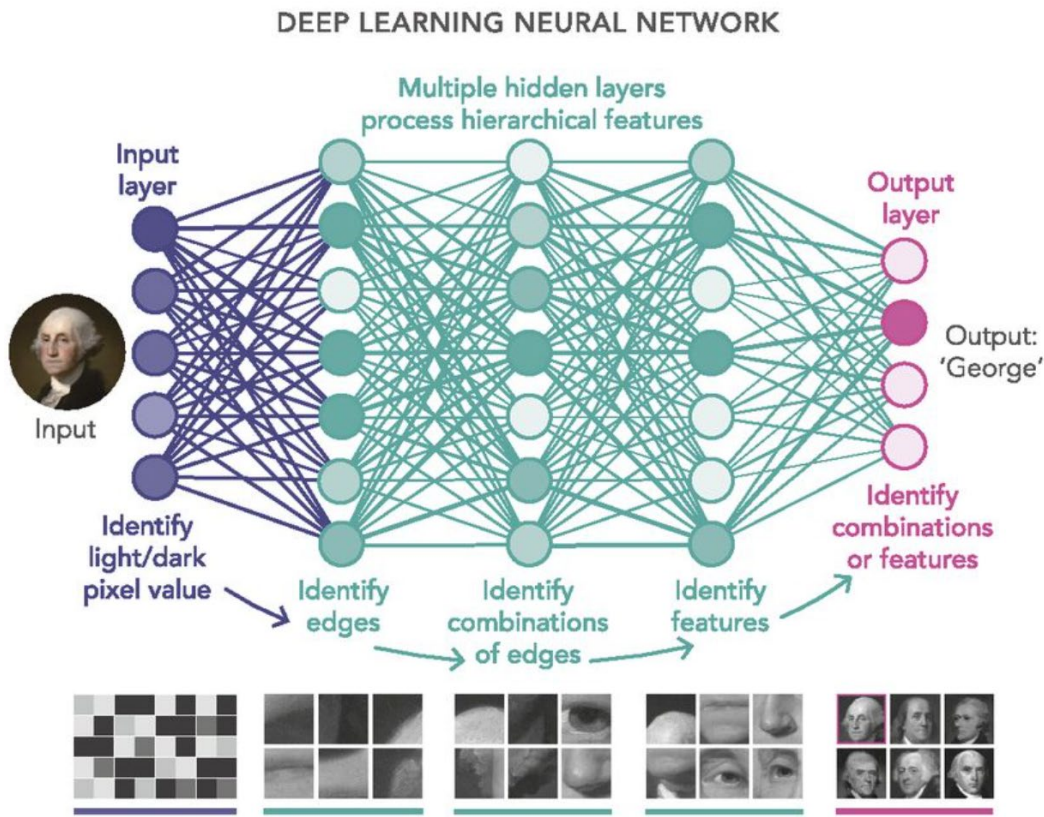

In doing so, the overall network is said to have learned a *representation* of the data it is trained on—the statistical information stored across the nodes, connections, and layers in a deep neural network *stands in for* the entire dataset the network was trained on. A deep neural network trained on human faces learns a "representation of human faces," i.e. it learns what human faces 'are' at least for the purpose of identifying faces within a photograph containing other objects or identifying individual faces from a "perpetual lineup" of other individuals' faces (Garvie et al. 2016). This speaks to applied machine learning researchers' "tendency to conflate representations with the things they represent" (Agre 1997, 143). Representationalism of this kind is not unique to machine intelligence. As Karen Barad points out, the belief that representations "mirror preexisting phenomena" supports both realist, empirical beliefs as well as social constructivist beliefs (Barad 2003, 802), as will be explored at greater length below.

For machine intelligence, this conflation of deep neural networks that are said to "learn a statistical representation" with the idea that they are learning something about the ontological status of what the objects in a dataset "are" reveals that "AI is a discursive practice" . Applied machine learning engineers use technical terms of art like 'learning' in ways that take on the "vernacular meaning of the word" (Agre 1997, 140). This simultaneously obscures the technical specificity of machine intelligence techniques and lends those techniques to broader analogies that are accessible to a lay public. This is why an algorithm that 'learns' can be said to 'understand', and why a deep neural network that has formed a representation of human faces can be said to 'recognize' or 'identify' those faces. These verbs—"learn," "recognize," "identify"—have vernacular meanings that are far more capacious and less precise than their technical referents, which are more narrowly bound by the material computational limits of machine intelligence. Nevertheless, this gap allows a lay public to imagine machine intelligence as capable of engaging in the full gamut of activities such verbs signify when in actuality they are far more limited. The power and authority of machine intelligence is, in this way, magnified by the gap between the technical denotation and vernacular connotation of what it claims to be able to accomplish.

## TABLOIDS AND BROADSHEETS

To demonstrate multi-task learning for their research report, the applied machine learning engineers I worked with used a deep learning approach to text classification that could pick up on statistical regularities associated with each 'label' assigned to a news article. These labels, functionally identical to the numeral associated with handwritten characters in previous examples, were the specific news category (politics, sports, lifestyle, etc.) and the type of publication (tabloid, broadsheet). These statistical regularities were informed by other approaches to "natural language processing" (Manning and Scheutze 1999) that are commonly used to perform classification and prediction on text. Semantic role labeling and part-of-speech tagging (as discussed above), as well as sentiment analysis and topic modeling (Wallach 2006), were incorporated into the demonstration project to provide additional features for the deep neural network to build representations with.

The data they used were tens of thousands of news articles they "scraped" (Seitz 2015) from newspaper websites. This dataset contained text of articles, the publication the article was scraped from, and the publication-specific news category the article was from. In designing the data collection strategy, researchers needed to ensure that they would wind up with enough points of data, from enough different categories of articles, from enough different types of publications. They needed to do this to make sure that they had a dataset that was representative of the categories they were interested in and was balanced between those categories. Where publications like *The Wall Street Journal* and *The Washington Post* might share news section labels like "Politics," "Business," "Sports," and "Tech," one might have a "Lifestyle" section where the other has a "Life & Arts" section. As the researchers scraped more articles from additional publications, this categorical confusion expanded. Was "World" in one paper the same thing as "International Politics" in another? Was "Entertainment" the same thing as "Gossip," or was it something entirely different? Perhaps most importantly, researchers asked themselves if the distinction between "tabloid" and "broadsheet" was "real, or just in our heads".

These were the questions that researchers discussed with each other at weekly project meetings, and the goal of these discussions was to ensure that data that was being scraped—the news articles being downloaded—would be able to "train" the multi-task learning algorithm the head author of the report was designing, in such a way that the capabilities of multi-task learning would be evident to their audience. This required them to think carefully about the how the deep neural networks they were training with their scraped data formed representations of that dataset. One dominant question was whether the distinction between 'broadsheet' and 'tabloid' was real. Were these terms of art in publishing, and if so, were they discrete categories that had material differences in terms of features in the structure of their text that a neural network could "pick up on"? For the first part of this question, the lead researcher recognized that "I think that maybe this decision shouldn't be made by us. But maybe we can find it in some external source like the Institute for Media Studies that considers this".[6] But for the second part of this question, machine intelligence supplies its own answer—if the algorithmic system can accurately predict which type of publication a given text is from, then there is, almost definitionally, a difference between the two categories. This difference is then reinforced by researchers' intuition, in this case found in claims that "the language that somebody uses, or like the writing style definitely betrays some things" that the deep neural network could pick up on.

---

[6] Ultimately, researchers would refer to a range of scholarship to inform their understanding of the distinctions between tabloids and broadsheets (see e.g. Nadkarni 2011).

Another dominant question was how the proportion and sources of training data could affect the outputs from the system they were building. The researchers' concern was that if they only had a few examples of articles from one category, say 'lifestyle," or if one category's articles tended to have comparatively fewer words per article, as 'obituaries' might, then the product might fail in one of two ways. One failure mode would be that under-represented categories would be under-represented in predictions about a text—the algorithmic system would predict the likelihood an article belonged to one category or another in proportion to how those categories were represented in the training data. The other failure mode would be that error rates would be higher for under-represented categories. If there were fewer 'lifestyle' articles in the training data, the algorithmic system might misclassify lifestyle articles more frequently. These concerns were pressing concerns for the applied machine learning researchers building this demonstration product, but they were discussing problems of dataset imbalance in ways that have become central to the emerging discourse about algorithmic fairness and racial bias in machine intelligence. Critical audits of facial recognition systems were, at around this very same time, revealing that facial recognition systems that did not sufficiently represent minority populations in their training data were more likely to misclassify minority group members' faces in real-world applications (Buolamwini and Gebru 2018; Jacob Snow 2018), as will be discussed at greater length in Chapter 4.

A final question that these applied machine learning researchers asked themselves was about *what* they were teaching the deep neural net. They were feeding it text belonging to different categorical labels, and the system was identifying features for itself—ultimately the researchers couldn't be certain if the system was picking up on stylistic differences, lexical differences, syntactic differences, or some other trace in the dataset that they themselves were not aware of, but that the neural network was using to minimize its error. Critics of machine intelligence refer to this as a problem of *opacity* (Burrell 2016) or an inability to fully explain or understand the inner workings of machine intelligence, while others see this as contributing to the "magic" workings of machine intelligence (Elish and boyd 2017). Particularly for deep neural networks that have hundreds of nodes and thousands of connections between them, it is often impossible for those who construct algorithmic systems to explain, or even understand, exactly how a prediction or classification is being made. Outside of research contexts, this problem of opacity is exacerbated by trade secrets or intellectual property protections that protect private developers and prevent independent auditors or critical scholars from accessing the system to even attempt to understand it.

But within OTH, researchers could be candid about how little could be known about the relationship between what was provided to the deep neural network and what its outputs were. They were keenly aware, in scraping news articles and assigning labels to it, that they were shaping what stylistic features would be available to the deep neural network for it to build its representation. They were shaping what the algorithmic system would think 'sports', 'news', broadsheets', or 'tabloids' *are*, at least in terms of what the prototype would learn that they are. And while they felt confident that what should be labeled a 'sports' article was obvious, labeling other categories required more subjective judgements. Across the working world of machine intelligence, researchers acknowledge that two activities require the most effort, and are the most important steps toward a final product: data cleaning and data labeling. Data cleaning (see Passi and Jackson 2017) is a pre-processing step that is crucial for ensuring that each entry in a dataset is complete, and compatible with other entries. This can be as simple as transforming some entries' values from percentages ("95%") to decimals ("0.95") or as complex as interpolating missing values based on other, similar entries. Data labeling is a pre-processing step that assigns an important attribute to each entry in a dataset. Sometimes attributes of a dataset are made to serve as labels—a healthcare dataset might use a specific diagnosis in medical records as label, and a machine intelligence system might be built to learn what other attributes are predictive of that diagnosis. Other datasets must be "hand labeled". Image recognition, for example, requires thousands of images to pass before human reviewers' eyes as they assign labels to the contents of those images (e.g., "dog," "dachshund," "human," "boy," "car," "red car," "red Toyota truck," etc.). This labor-intensive process captures the cultural concepts of human labelers and affixes them to images so that machine intelligence can "learn" the relationships of differently colored pixels that correspond with each label.

These labels are arbitrary, conventional symbols as far as machine intelligence is concerned. In other contexts, the relationship between data and their labels could be verified against some "ground truth"—an independent assessment of what a piece of data *was*, at least in the eye of the beholder. Its iconic resemblance to the label could be verified. One way of establishing ground truth could be polling online pieceworkers—often using Amazon's MechanicalTurk (mTurk) platform (Gray and Suri 2019)—about what was in an image. If a majority agreed that an image was "of" the same thing, if an image labeled as a 'hamburger' appeared to depict a hamburger, that label could be considered to have been "ground truthed". Another ground-truthing method, particularly for datasets where one attribute was being take for the label being "learned," is to spot-check the dataset by looking for associated documentation for a small sample of entries. In health data, individual patient records might be pulled to verify entries and confirm the various attributes in the dataset are clearly identified.

Even in sufficiently ground-truthed datasets, a machine intelligence system might pick up on features that have little bearing on the 'true' labels of an image. If the image of a hamburger depicts a scene in which the hamburger is accompanied by French fries or a soft drink, the system may 'learn' that those objects are 'hamburgers' too, since the label for the image does not specify which configuration of pixels pertain to the subject of the image and which pertain to the background. A system that outputs 'hamburger' for any image taken within a fast-food restaurant, then, might be accurate to the data provided to the system, but inaccurate to the purposes for which the system was developed. The validity of a machine intelligence system designed to identify the contents of images then, has to be considered as an assemblage of the tacit knowledge of mTurk workers, the statistical accuracy of the system itself, and the practical effect of the system's outputs.

This points to the involvement of many hands in "the creation and maintenance of complex classification as a work practice" (Bowker and Star 1999, 5), but also in the relationship between subjectivity and objectivity. Machine intelligence's naïve empiricism depends, for its naivety, upon the "given-ness" of the data it uses to produce knowledge. This "given-ness" derives from the earliest 17th century usage of the word as that which was "given by the conventions of argument … that category of facts and principles that were, by agreement, beyond argument" (Rosenberg 2013, 20), and which later came to be applied to "the modern sense of *fact*" (Poovey 1998, 90). But what distinguishes 'facts' and 'evidence' from 'data', Rosenberg argues, is that "the semantic function of data is *specifically* rhetorical" (Rosenberg 2013, 18) in that data can be 'taken' as 'given'. Data *declares* what is taken for the truth. For applied machine learning researchers, this is crucial—their work is to bring insights contained within data to the surface, to amplify and clarify these declarations, not to apply their own subjectivity to data. Any space left for the researcher's subjectivity, like making a judgement call about whether an image is a dog or a wolf, threatens to create "a gap between subjectivity and reality," as Joseph Dumit puts it, and "this gap becomes the structure of the researcher's desire, motivating him to eliminate the appearance of subjectivity through automating all the tasks he presently is involved in" (Dumit 2004).

As a theme that will recur throughout this ethnography of machine intelligence, "automating all the tasks" becomes a way of reinscribing specific subjectivities—of microtask workers, of institutions that produce data, and of researchers who curate and clean datasets—into analyses that are then made to seem objective through the practices of applied machine learning engineers. Accounting for these subjectivities could lead toward a stronger form of objectivity that considers these different subjectivities' positionality (Harding 1992b), but that possibility is foreclosed by researchers' packaging of machine intelligence as a singular technological object and not an assemblage of multiple perspectives, subsystems, and forms of labor. In the case of image labels, researchers do not risk their own objectivity by labeling the contents of images themselves. Researchers 'automate' the work of labeling by creating a task that can be completed by crowdworkers on platforms like mTurk. The mTurk platform automates the labeling process, randomizing images shown to crowdworkers, pooling results, and updating the dataset with workers' labels. Through the platform, the work of labeling is automated, the applied machine learning researchers' subjectivity never comes into play, but the subjectivity of the crowdworkers is sublimated into an objective set of labeled images. Ways of ensuring that images are labeled 'objectively' can be enforced through the platform—mTurk workers are routinely disciplined into a mechanical form of objectivity when payment for tasks performed 'incorrectly' is withheld, and the same task is often given to multiple workers and the results 'pooled' to create a normative (and therefore more objective) result. Thus, 'automation' and 'objectivity' are conflated, as are any means at applied machine learning researchers' disposal to offload subjectivity onto others. This is slightly different than the injection of external authority seen above, when what could be a subjective decision about what newspapers are indicative of 'tabloids' and 'broadsheets' was offloaded to an authority seen by researchers as more 'objective' on the issue, the Institute of Media Studies.

For the multi-task learning prototype system that predicted the publication section and type of newspaper articles, it would be impossible for researchers to 'ground truth' the data they were using to train their product themselves. Nor did they necessarily need to, as the scraping methodology supplied the labels they had for each article, which were assigned by editors at each respective publication. And the title of the publication was also known, and could be assigned to a publication type after outsourcing the decision to other sources—a well-reasoned, but ultimately subjective, determination made by consulting with media studies scholarship. This was ultimately suitable for the researchers' purposes, because "really, it's not a machine that is, in the end, consuming the predictions. It is a human, right? And so what's great about humans is that they can make sense of a lot of things. But as long as … it's a meaningful distinction for a person, where if they look at articles or … a bit of output and go 'oh yeah, interesting'. That's all we want, right?". The researchers understood themselves to be constructing a representation of news media that made distinguishing between various sources and topics possible, but would rely upon their audience to reconcile the distinctions the algorithmic system produced to the conceptualizations about news articles that they already held in mind.

The audience of the prototype was enrolled as such through an interactive 'front-end' for the algorithmic system that faced the viewer through a web browser. Visitors could view a selection of recent headlines to view how the algorithmic system classified that article: whether it was classified as a broadsheet or a tabloid, what news category the article was predicted to belong to, and the "degree of confidence" the algorithm had in that news category. Clicking on a headline would take visitors to an "article view" that showed "how the model arrives at the classification on a word-by-word level." This view displayed the text of the article and highlighted each word with a color that corresponded to how that word "nudged" the overall classification toward one category or another (Figure 7). Visitors could also "toggle" between a "broadsheet lens" and a "tabloid lens" to see how different words would nudge predictions toward or away from different sections, in a counterfactual world in which the article had been written in a different publication type.

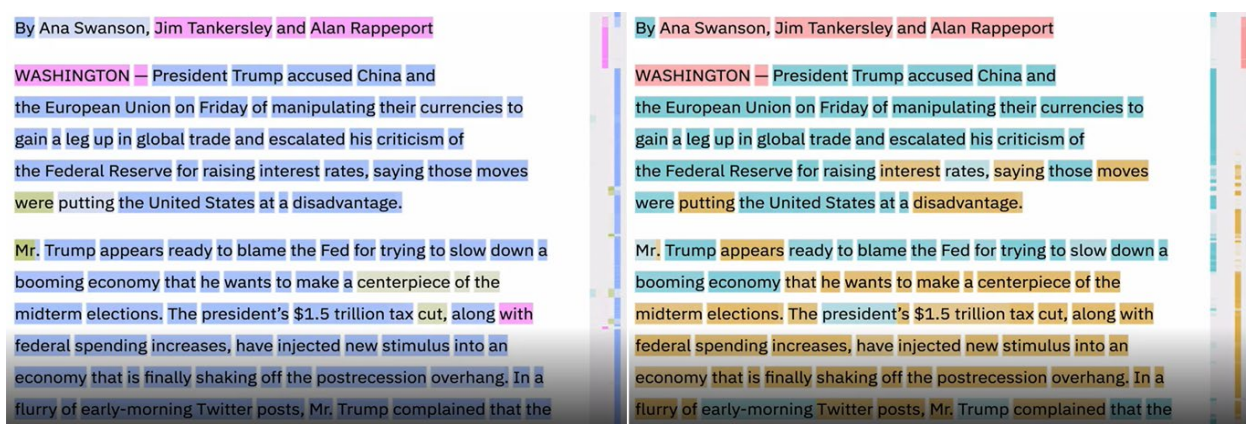

Figure 7. Example of the "article view" of the multi-task learning prototype

In the example shown, words like "centerpiece" and "cut" shift what is predominantly predicted to be a "politics" article toward the category of "business" when viewed through a broadsheet lens, but through a tabloid lens a "news" article gets nudged toward the category of "sports," even if the overall classification is not swayed by such nudges. The reasons why this might be the case are manifold—the 'sports' category might be overrepresented in the 'tabloid' dataset, financial figures might be more central to tabloid sports coverage but downplayed in broadsheet sports coverage, etc.—but they are not resolved through machine intelligence. The prototype's designers acknowledged the importance of context in the production of meaning from text, saying that "from surrounding sentences to place of publication, context changes word meaning," and the word-by-word highlighting indicates how the deep neural network they employed perceives these contextual changes. But there is always a contextual surplus that exceeds the algorithmic system's capacity, because the system cannot include its audience, but the audience is deeply enmeshed in its own interpretive contexts. The prototype uses a visual format to layer additional information onto text, enabling a form of play between text and audience in creating the meaning (Drucker 2014; Halpern 2015); the audience draws their own conclusions about what constitutes the categories of 'tabloid' or 'broadsheet', and with the aid of the "lenses" supplied by the applied machine learning researchers can draw conclusions about what distinguishes "tabloid news" from "broadsheet politics". While a longer, semiotic analysis of how meaning is constructed between data, algorithm, applied machine learning researcher, and audience through this prototype is beyond the scope of the current undertaking—in part because it was not designed as a reception study (Mankekar 1999; Radway 2008; Goldstein and Machor 2008) and in part because the construction of meaning is addressed in a subsequent chapter (Chapter 3)—this chapter will nevertheless conclude with a discussion of how these elements are brought into relation with one another through the practices of machine intelligence to produce the authority of machine intelligence.

## DATA PERFORMANCES

In learning how to enter the working world of machine intelligence, bootcamp trainees must engage in highly scripted *performances* before potential employers. They stand in front of slide presentations and display their competence and expertise. This performance is an important component of transforming their subject position into that of an applied machine learning researcher. Their role, their professional identity, "is itself a kind of becoming or activity" that, like gender, is brought into existence through the performativity of subjects within a discursive frame in which their subject position has meaning (Butler 1990, 112). But machine intelligence researchers are not the only entities brought into existence through performance (Austin 1962). Machine intelligence itself *enacts* the objects of its analysis. From the demonstration projects built by bootcamp trainees, to the winning entries to Kaggle competitions, to the prototypes built by researchers at OTH, each instantiation of machine intelligence discussed throughout this chapter constitutes a *data performance* of the conceptual categories to which techniques of machine intelligence were applied.

The applied machine learning researchers who prototyped an approach to multi-task learning were orchestrating a performance of 'tabloid'-ness and 'broadsheet'-ness for their audience. Similarly, the Kaggle competitors working on the Netflix prize produced a performance of the concept of 'movie ratings,' and the bootcamp trainees produced a performance of a particular form of online shopping behavior. The performative aspects of the newspaper prototype are most straightforward. Whatever existence, meaning, or cultural significance the concepts of 'broadsheet' or 'tabloid' might have—whether those be normative definitions, or implicit connotations that emerge through common usage—are ultimately as indeterminate as the quantum state of a subatomic particle. Like a probability cloud of potential meanings, they collapse into their particular significance when arrested within a specific context: a supermarket checkout aisle, an academic treatise on yellow journalism, or a dismissive comment about the veracity of a news source. As the object of a machine intelligence system, the meanings for 'tabloid' and 'broadsheet' are the result of how a technical apparatus—the multi-task learning system—intervenes into the many possible, indeterminate meanings for these terms. Machine intelligence *enacts* a particular meaning for otherwise flexible, indeterminate, and context-dependent objects through the "agential cut" it makes "intra-actively" with the datasets it operates upon and the statistical techniques that perform this statistical operation.

Karen Barad refers to this as "agential realism": the way "concepts obtain their meaning in relation to a particular physical apparatus, which marks the placement of a ... cut between the object and the agencies of observation, resolving ... semantic-ontic indeterminacy" (Barad 2007, 120). The objective function of machine intelligence, whether calculated on personal desktop computers or on distributed cloud servers, is precisely this type of apparatus which can resolve the meaning and existence of otherwise indeterminate concepts. Machine intelligence can make an agential cut through thousands of arrest records to resolve a specific meaning of "propensity to reoffend" (Corbett-Davies and Goel 2018) for a parole board hoping to bring the aura of objectivity to their deliberations. Machine intelligence can make an agential cut into thousands of newspaper articles to resolve a specific meaning of 'tabloid' and classify literary styles accordingly. And machine intelligence can make an agential cut into thousands of images of handwritten numerals and resolve which resemble a given number closely enough to pragmatically function as that number in a banking transaction.

These enactments are performances choreographed by applied machine intelligence researchers who operationalize the fact that "meaning is not a property of individual words or groups of words but an ongoing performance of the world in its differential dance of intelligibility and unintelligibility" (Barad 2007, 149). Like Barad, I distinguish between the pure representationalism that would render all things as linguistic phenomena and the material-discursive practices that "allows matter its due as an active participant in the world's becoming" (Barad 2003, 803). My contention is that the practices of machine intelligence are a set of material practices that *make* meaning through the material-discursive practices that are particular to applied machine learning researchers. They entail the rendering of material-discursive phenomena as machine-readable semantic representations, the manipulation of these representations through complex statistical techniques, and the generative enactment of specific material-discursive objects. These objects—with the material-discursive conditions of their production safely quarantined within the black boxes of machine intelligence—seemingly stand apart from data, analyst, and machine as autonomous and persuasive on their own, possessed of fetishistic truth-value and authority. Data analyses, algorithmic products, prototypes, and insights enter into a broader discourse as separable from the applied machine learning researchers themselves by virtue of the ways in which researchers' individual subjectivities, and research teams' corporate (or collective) subjectivities, are eliminated from view through black-boxing, automation, and a dogged commitment to naïve empiricism.

## Conclusion

This chapter has traced in both broad strokes and minute detail how these performances are orchestrated, but it is worth tying all this together explicitly. For machine intelligence, naïve empiricism, as refracted through the "data analysis" of Tukey and the "algorithmic modeling culture" of Breiman, can claim no prior knowledge of what the concepts catalogued within a dataset are. Where that prior knowledge is required—to distinguish between classes within that dataset, whether those classes be handwritten numerals, species of garden perennials, labels for images, or types of newspaper publications—applied machine learning researchers are at pains to absent themselves from subjective judgement by "automating" those determinations. This can happen through outsourcing subjective judgements to semi-automated platforms like Amazon's Mechanical Turk, automating the data collection process through techniques like scraping (which include attributes that can be used as labels for the basis of classification and prediction), or by taking the data as given (whether it is provided by boot camp patrons, open-source datasets, or other 'found' sources). Whatever way they obtain this data, the subjective judgements made within the dataset—"the day-to-day work of building classification systems" (Bowker and Star 1999, 10)—are kept at arm's length from the researchers themselves.

The products these researchers build are then presented to their users, who must incorporate the claims made by those products into their own understanding of the concepts addressed by that product. What a machine intelligence product 'says' a 'tabloid' actually 'is' may conflict or come into tension with understandings the recipients of this assertion might already hold. In some domains, this tension might be merely amusing. But in other contexts, this tension might only be resolved with dire consequences. A judge who must choose whether to accept or reject a machine intelligence judgement about a defendant's propensity to re-offend while awaiting trial might doubt their own discretion in the face of a precise numerical representation of the risk a defendant is calculated to pose (Jasanoff 1995; Wisser 2019). A doctor may face charges of negligence if she overrides or disregards the calculations of a machine intelligence system's sepsis warning.

As these outputs are incorporated into their targeted audience's prior understandings, machine intelligence creates a context in which its authoritative claims are made valid, and in which the concepts they manipulate *are* as they claim they are. The applied machine learning researchers' prototype says to its audience, "this is what a tabloid *is*, and this is what a broadsheet *is*. Look at the differences, see how some words are highlighted through one lens, and other words are highlighted through the other lens. There is a difference between these two categories. Look and see how confident we are in being able to distinguish between these concepts! Look at how accurate our predictions have been!". And the audience, seeing these differences, also sees the "whole complex of meanings" (Derrida 1973, 260) that make the distinctions between categories salient. The distinction between 'tabloid' and 'broadsheet' goes from being subjective, arbitrary, or deeply contextual (a tabloid in a U.S. supermarket is a different type of object than a tabloid in a British newsstand) to being concrete and objective. The concrete *meaning* of the category is performatively enacted through the way machine intelligence demonstrates the concept to its audience. Machine intelligence, then, is both an epistemological and an ontological project. It is a set of material-semiotic practices that constitutes knowledge, but also constitutes the *objects* of knowledge.

## CHAPTER 2: OPTIMIZING ORGANIZATIONS

**CONFERENCE TABLE PERFORMANCES**

If machine intelligence is capable of enacting data performances that produce and give meaning to its objects of knowledge (see Chapter 1), applied machine learning researchers engage in their own set of performances to make machine intelligence matter for the world. As we have already seen, formal, almost theatrical performances are crucial components for machine intelligence at various moments. Industry conference presentations, bootcamp trainee pitches, and even Kaggle competitions can be seen as performances orchestrated to promote individual applied machine learning researchers in their job search. But they are also used to promote specific machine intelligence techniques, and even machine intelligence writ large. Such theatrical performances are integral to the work of machine intelligence, as the authority of machine intelligence does not derive solely from the way it has been applied across various domains. Rather, practitioners of machine intelligence must enroll allies within domains in which they see potential applications of machine intelligence and persuade them of the benefits of applying machine intelligence techniques (Callon 1984; Latour 1988). Michel Callon has written persuasively about attending to this process of enrollment as an approach to studying the power of scientists and technology, pointing to how scientific researchers "become indispensable to the problems of other actors … by defining the nature of the problems of the latter and then suggesting that these would be resolved [through] the researchers' programme of investigation" (Callon 1984, 196). While applied machine learning researchers' enrolling performances are certainly legible to Callon's "sociology of translation," these performances also straddle a set of unresolvable contradictions that are inherent to how the authority of machine intelligence is produced, at several scales of this analysis.

For the applied machine learning researchers I worked with at Over the Horizon (OTH), performance was a central component of their work. Whether they were "building their personal brand" by writing blog posts about machine intelligence techniques that performed their expertise and mastery, or were sitting in on sales calls performing the high caliber of services the lab could provide potential clients, or were conducting workshops with a client to craft a "data strategy" that would orient the client's entire set of business practices around machine intelligence, applied machine learning researchers spent most of their days engaged in performances of one kind or another. These performances were both scripted, choreographed performances in front of boardroom audiences, as well as performative presentations of the self (Goffman 1956) mediated by sales and consulting calls, deliberate and incidental transgressions of workplace norms, idea pitches, and code reviews. The data strategy workshops were a peculiar genre of performance, which I encountered nowhere else in my fieldwork, and were one of the marquee consulting services offered by OTH to their clients. Workshops commanded a great deal of attention from applied machine learning researchers on staff, required a great deal of preparatory work, and carried with them a concomitantly large fee for the workshop itself. The workshops were intended to take a company—which may or may not already have a team of applied machine learning researchers on staff dedicated to implementing machine intelligence approaches into their everyday business practices—through the data strategy development process over as short a period of time as a single day. Those at OTH conceived the goal of a "data strategy" as a way to help companies realize the potential, and particularly the potential economic value, of the data they had access to.

A significant portion of the workshops, therefore, consisted of persuading variously-positioned stakeholders within the clients' organizational structure that there was indeed potential value in their company's data. The business world has long been convinced of the "data imperative" to collect data and use it to guide decision making (Fourcade and Healy 2017, 14), but most companies do not have a clearly articulated strategy to guide what data is collected, where it is stored, and how it is used to produce additional value for an organization through the predictive and classificatory capabilities of machine intelligence. This is why each data strategy workshop opened with a performance: the OTH consultants stood up in front of a conference room and narrated a brief history of machine intelligence, demonstrating what machine intelligence had accomplished in the world, why other actors in the audience's industry were adopting it, and why the people in the room should be excited about it. But this performance also had to walk a knife's edge to build excitement while keeping clients' expectations grounded. I was told by my interlocutors at OTH that "there's the hype, and then there's the hard decisions" that lead to incremental changes, modestly ambitious machine intelligence projects, and a reliable return on investments in applied machine learning research personnel.

Over the course of my time working with OTH, I attended several such workshops as a notetaker, and was occasionally asked to contribute with constructive questions or observations that took me beyond my own comfort zone occupying the role of workshop stenographer and forced me to acknowledge to myself that my role was not that of passive observer but active participant within my field site. I would ask a chatbot company's applied machine learning researcher who told OTH consultants "we're optimizing for good conversations" how they defined "good" and "conversation," opening up a discussion about how to practically bound a conversational interaction for the purposes of optimization, and how to develop metrics around the quality of an interaction with a chatbot. I would ask another company that proposed avoiding racial and gender discrimination by not collecting information about race or gender how they would know they had succeeded if they had no data about the race or gender of their users. And sometimes I would play up my outsider status as an anthropologist amongst technologists by asking for progressively simpler explanations of their data management processes until gaps in that process—which could be addressed as part of a data strategy—emerged into view.

Whether these questions were helpful in the moment or not, they often led to productive conversations with OTH applied machine learning researchers on the subway ride back to their office. These conversations ranged beyond what was directly pertinent to the data strategy workshop we had just completed; the post-workshop debriefs amongst OTH personnel were generally wide-ranging and delved into adjacent topics that could be turned into machine intelligence problems: are there any ways to measure what makes a conversation "good"? Is there a definition of "good" that could be borrowed from another discipline and turned into an objective function for machine intelligence? If so, where would you look for data to solve such problems? These conversations would sometimes swing wildly between dorm-room bull session, professional gossip, and practical ideation, but often found their way into the debriefing documents and after-action reports prepared by OTH staff for the clients to build upon. After attending a few of these workshops as a notetaker, and helping prepare workshop reports, I was eventually asked to help conduct a workshop led by Frankie, one of the first handful of employees to be hired on to OTH. Frankie was about my age, commuted by bicycle to work (like many OTH employees including myself), and had lived in a neighborhood I had once also lived in. She was trained in cognitive neuroscience and had worked in several European cities before coming to New York for a postdoc and eventually transitioning into machine intelligence through the bootcamp program discussed in Chapter 1, eventually becoming an instructor there prior to joining OTH. Frankie and I had had several opportunities to work together, including on the multi-task learning report she was the project lead for, and we had bonded over a week when we were two of only three or four staff members in the office and dared to play music from a shared playlist we had created—not on headphones, but over speakers, so the entire room could hear! This shockingly transgressive move was a marked contrast from the typical heads-down, headphones-on, screen-mediated interactions that normally characterized work hours at OTH.

At any rate, Frankie had drafted me into a slightly elevated role as her second-in-command on a data strategy workshop for Intricate Media, a digital media company focused on producing written articles and short videos about music, art, and fashion for a youthful audience. This is how I found myself standing in the front of a conference room in midtown Manhattan, narrating a series of slides for an audience that nodded along, polite enough to at least feign understanding. The slides and the patter were not really my own; they had been road tested and refined into a boilerplate that could be bolted onto the intro of any presentation. But even though most of the words were not my own, I moved through the narrative as deliberately and enthusiastically as possible—controlling my nerves enough to avoid talking too quickly to be understood and stealing glances at Frankie for a bit of encouragement. I was telling a story: artificial intelligence and the techniques of machine learning have accomplished so many amazing feats in the past few years, things that seemed impossible before they were done, but which I promised could be explicated over the course of the day-long workshop we had convened.

I ran through a litany of machine intelligence achievements: Google DeepMind's project AlphaGo that defeated human grandmasters at the notoriously complex board game Go (Silver et al. 2017), another project from Google that had 'solved' the problem of energy consumption used to cool its own data centers (Evans and Gao 2016), and Carnegie Mellon University's poker-playing machine intelligence project Liberatus AI (N. Brown and Sandholm 2018). I also screened a short film called *Sunspring* (Sharp and Goodwin 2016), reciting a silent prayer of thanks to the silicon gods that the video embedded in the slideshow presentation actually played when I clicked it with the unfamiliar remote control in my hand. *Sunspring* is a curious project that filmed, with live actors, a screenplay that had been 'written' by a neural network trained on sci-fi movie scripts and that used text 'prediction' to author a unique script, replete with dialogue and stage directions (Newitz 2016). Frankie liked using *Sunspring* as an example of an AI achievement because it was not at all flawless—the screenplay shows the algorithmic system caught in a loop, with characters successively voicing a lack of knowing—"I don't know," "I don't know what you're talking about," "What do you mean?"—over and over again, in a way that seems totally unnatural for spoken dialogue. For Frankie, this flaw plants the seed of an idea that beneath the spectacle and triumph of these achievements lies something—if not darker, then more mundane—that foreshadowed the work ahead of us all in that workshop.

The feats this litany of marvels accomplished seemed almost magical, and we presented their achievements in a way that was intended to inspire awe in the audience. But the ways these marvels had been achieved were presented as almost deceptively simple, at least at first glance: Get two computers to play Go games against each other thousands of times, learning from their mistakes each time until they surpass the skills of human grandmasters. Let a neural network train itself over hundreds of movie scripts until it can predict which word follows the next well enough to write one of its own. Develop precise mathematical notation for Texas hold 'em and its rules and optimize for a betting strategy that takes one's opponents' betting strategy into account. These examples were drawn from newspaper and magazine stories that many in the audience would likely have already encountered. And even if they had not, popular media coverage of machine intelligence was everywhere. Self-driving cars, facial recognition, and automated hiring could be found in the pages of the New York Times Magazine every week, and even though Frankie and I were presenting the awe-inspiring capabilities of machine intelligence in order to dismantle and de-mystify their construction, our task was made more difficult by the durability of media coverage that presented these applications as *fait accompli* of machine intelligence.

After I presented this hit parade of achievements, it was time for Frankie and me to try to turn the tide of our audiences thinking about the magic and mystification of machine intelligence that we had just raised across our audience's eyes. We explained that indeed Go is a phenomenally complex board game, with millions of possible moves and billions of possible board configurations. But just because the accomplishments of AlphaGo appear as evidence that AlphaGo must be very intelligent—since Go is commonly understood as a highly complex game that only the most intelligent humans can learn to play well—Frankie and I stressed the idea that "AlphaGo is a narrow AI system that can play Go and that's it". The point we wanted to get across to our audience at the start of the workshop was that few 'real-word' problems that can be addressed by machine intelligence, particularly the problems faced by those at Intricate Media, are shaped like the game of Go. In fact, as Karpathy concludes in his analysis of AlphaGo (from which we drew our presentation slides), AlphaGo demonstrates not so much a spectacular achievement of machine intelligence but rather a shrewd choice on the part of DeepMind in choosing a tractable yet impressive problem (Karpathy 2017a). Despite our rehearsal of the triumphant, magical applications of machine intelligence that had gotten us in the door that morning, Frankie and I were not trying to sell magic. Rather, we were trying to hold that narrative at the same time that we punctured the veil of magic and mystification around machine intelligence to persuade our audience that behind the magic of machine intelligence were a series of deliberate steps and fortuitous conditions—like DeepMind's choice of Go as a problem to try to solve with machine intelligence—and that similar steps could be taken and conditions could be sought through the "data strategy" we would help them elucidate over the course of the day's workshop. Or at least we thought we would.

Early on in the workshop, almost immediately following our initial presentation, our audience suggested a roster of potential machine intelligence projects that could have been drawn directly from the most breathless headlines about the spectacular capabilities of machine intelligence. Intricate Media produced original content—written and short-form video—for multiple different reader- and viewerships grouped under a variety of different branded media channels. And so, given their existing business model, Intricate Media staff were interested in a specific set of machine intelligence capabilities: automated content generation, AI-enabled editing of existing content to appeal to audiences of additional media channels, and an internal-facing chatbot that could be used to improve cross-team communications within Intricate (and which would be particularly desirable if it could automatically suggest "new old" or "evergreen" content to promote on various media channels). These potential projects were all well within the realm of possibility for machine intelligence writ large, but as Frankie put it Intricate Media was "miles away from being able to even think about being able to build such things".

For Intricate Media specifically, this distance was a function of how they managed the data they were currently collecting and how they saw the value of that data for building new products. But the distance between Intricate Media's ambitions and capabilities was common across the range of clients OTH worked with, and across the world of machine intelligence. Only a few very large or highly specialized technology firms were responsible for most of the more prolific applications of machine intelligence over the past few years. These firms were responsible not only for demonstration projects that raised the profile of machine intelligence and made it seem desirable for smaller firms to pursue, but also created the market for the services OTH, and dozens of other machine intelligence consultancies, were offering at that time. This distance between ambitions and capabilities also points to a contradiction that runs through the history of technology development, and also informs the authority of machine intelligence: new technologies are presented to the public as spectacles that induce sublime experiences in their audiences even as time, familiarity, and integration into social practices renders them mundane.

## THE ALGORITHMIC SUBLIME

Technology studies scholar David E. Nye argues that "the sublime" is a "a preferred American trope" (Nye 1996, xiv) that is continually redeployed as new developers introduce novel technologies to society. Technological objects are successively constructed as "new sources of popular wonder and amazement" (*Ibid.*, 8), with all the attendant expectations and promises of social renewal that their popularity entails. This particularly *American* form of the technological sublime has been analyzed by Nye as producing the collective effervescence (Durkheim [1912] 2001) that might bind together a de-sacralized, settler-colonial state lacking a longer shared history. Nye traces a history of such sublime spectacles that are intimately linked to the emergence of a national identity for the United States, beginning with the wagon trains that would venture to western Virginia to gaze upon Natural Bridge (now a National Historical Monument), the railroad lines that connected city dwellers to the awesome power of Niagara Falls, and the tourist pilgrimages to the natural wonders of the 'wild' West— Yellowstone, Yosemite, and the Grand Canyon. He continues his traversal of the American technological sublime with the triumphs of human ingenuity: the Boulder Dam, the Golden Gate Bridge, the Empire State Building, and eventually the temporary attractions that demonstrated the beauty (and safety) of electric light at turn-of-the century fairs and the expositions.

Machine intelligence, I argue here, is implicated in this process as it is developed for integration into society.

That the American technological sublime registered in both natural wonders—the Grand Canyon and Niagara Falls—and technological wonders—railroad engines, urban electrification, and space shuttle launches—is perhaps a distinction without a difference. Natural wonders were made accessible to the public through technological progress in advancing the frontier and were often augmented with technological apparatuses, as with spectacular light shows illuminating Niagara Falls or Natural Bridge National Historic Monument, just as technological marvels—bridges, skylines, and train trestles—are naturalized as indelible features of the American landscape. What these spectacles share is that, like the train whistle that interrupts Henry David Thoreau's reveries (L. Marx 1964), technological objects burst forth into consciousness, as if out of nowhere. The bright light of these spectacles casts the conditions of their production into shadow, and foregrounds how they *appear* to work (Debord 1967) rather than how they must be *made* to work.

For natural features, this tendency elides the forced removal of indigenous populations from Niagara, Natural Bridge, Starved Rock, Yellowstone, and Yosemite (Spence 1999) and the role of chattel slavery in clearing the frontier (F. J. Turner 2008). For technological objects, this tendency elides not only the extractive labor conditions that produced infrastructural projects like bridges and railroads (Shu 2020), but also the co-ordination of efforts that make these beacons of technological process possible. This co-ordination of effort involves social construction, certainly, in which engineers and architects interpret society and society interprets space through its lived experience in and through technological systems (Bijker and Law 1992). But it also includes the articulation of standards (Busch 2011) that make

interoperability—between, say, rail cars and railroad tracks—possible, as well as the installed base (Star 1999a) on which successively 'spectacular' technologies are constructed.

Machine intelligence, as Frankie and I learned during our visit to Intricate Media, can be skewered by the same fork as the American technological sublime. While spectacular demonstrations of machine intelligence like AlphaGo can evoke an *algorithmic sublime* that, while mediated by digital and virtual technologies compared to wonders of the immediacy of the natural world and of civil engineering, nevertheless also renders mute the rational functions of all those who gaze upon its material instantiations, mediated by screens and other computational devices. The algorithmic sublime also serves to obscure the labor and negotiation needed to construct machine intelligence products that function in line with their design specifications. Indeed, the design specifications of machine intelligence products themselves require the kind of negotiations with human and non-human systems, as well as the introduction and coordination of technical standards, that spectacular demonstration projects tend to blot from view. The sublime, in its Kantian, Enlightenment-era sense has both a 'mathematical' and 'dynamic' aspect. The mathematical sublime describes an experience of scale that produces wonder in a human subject. This experience can be mediated or immediate—standing directly in front of the Grand Canyon and watching video of a nuclear detonations mushroom cloud are both experiences of the mathematical sublime, which exists at scales that dwarf the normal realm of human experience and evoke the mathematical sublime. The dynamic sublime is more closely associated with a sense of terror, but encountered at a safe and secure remove—an experience of power which "has no dominion over us" (Kant 2007, 90)—as when standing against the guardrail at Niagara Falls or observing a passing thunderstorm from afar.

The *algorithmic sublime*, as I articulate it, shares in the "awe and terror that overrides rational thought"(Ames 2018, 1) which "so fills the mind that it cannot … apply reason to it" (Mosco 2004, 23). The suspension of reason around encounters with machine intelligence helps explain how it comes to be seen as possessing "mythical" and "magical" capabilities (Elish and boyd 2017), or as a "fetish … imbued with capabilities that are not inherently properties or functions of the object itself" (Thomas, Nafus, and Sherman 2018). While we have already discussed the data performances of machine intelligence as producing a statistical fetish that obscures and disguises the source of the knowledge it stands in for, David Graeber's assertion that "a fetish is a god under process of construction" (Graeber 2005, 427) is also instructive here. Discussing the development of trust in the colonial market encounter, he argued that the fetish was a kind of "trading zone" between magic and organized religion. Placing the fetish "precisely at the point where conventional distinctions between 'magic' and 'religion' become meaningless," Graeber implicates it in the evolution of "new social forms and institutional arrangements," specifically markets, that could achieve pragmatic ends. The spectacular demonstrations of machine intelligence that evoke the algorithmic sublime similarly sit astride, and were instrumentally deployed by Frankie and me, as a transition from machine intelligence projects that function purely as spectacular demos to a more integrated product development process that was routinized and could serve ongoing business needs at Intricate Media. Even so, the discourse around machine intelligence projects tends to foreground the awe and the magic while their role in social creativity and institution-building remains as elusive as in the "pidgin-markets" of Graeber's analysis. A documentary about the defeat of Go champion Lee Sedol by AlphaGo tells us that "a game of Go has more possible configurations than there are atoms in the universe" (Kohs 2020), turning the machine intelligence system into an iconic representation of a common source of the mathematical sublime. In coverage of other spectacular demonstrations of machine intelligence, the number of petaflops a computer is capable of, the number of cores and GPUs brought to bear on a computational problem, and the nearly infinite permutations of possible outcomes the system must navigated contribute to the production of a mathematical sublime.

These spectacular demonstrations produce a distant terror, as well. Even setting aside the many terrifying scenarios of an "AI apocalypse" in which machines actually attack humanity (see Dowd 2017), machine intelligence confronts its audience with a performance of the dynamic sublime in which humanity is rendered irrelevant, relegated to the sidelines in a world where automation performs all meaningful or productive work. Through self-driving cars, automated screenplay writers, and factory robots we glimpse a future—from the comfort of our own present—where, like Niagara Falls, the world will keep churning oblivious to our existence (Geraci 2010). Even AlphaGo offers a glimpse at this world: during the 2016 challenge to Lee Sedol, AlphaGo selected a move that commentators, the applied machine learning researchers who designed and built the machine intelligence system, and professional Go players alike described as "inhuman but beautiful … so beautiful" (Metz 2016).

The algorithmic sublime shares more than just its mathematical and dynamic facets with the concept of the technological sublime developed by David Nye. In his analysis, Nye focused particular attention on the "electrical sublime," which surrounded the development of electricity as an infrastructural technology. In its early days, electricity "was invoked as the panacea for every social ill and the key to a whole range of social and personal transformations that promised to lighten  the toil of workers and housewives, to provide faster and cleaner forms of transport, and to revolutionize the farm" (Nye 1996, 143) and was popularized through spectacular demonstrations including the illumination of the Brooklyn Bridge, nighttime lightshows bathing Niagara Falls in electric light, and the "Great White Way" at the 1893 Chicago World's Fair. These spectacular demonstrations of electric light made the invisible visible, and also "combined the mathematical and dynamic sublime as the spectator encountered both extreme magnitude and irresistible power" (*Ibid.*, 151). Additionally, the electrical sublime shared in and promulgated the nationalism of the American technological sublime that can be traced back to the earliest days of American Empire. Spectacular demonstrations of electricity at commemorations—Independence Day Celebrations and centennials—and the use of national founding myths in electrical diorama—depicting events like the signing of the Declaration of Independence—rhetorically linked electricity to national belonging and statehood. Machine intelligence, over the course of my fieldwork, recapitulated many of the early claims that surrounded electricity, including invocations as a panacea and revolutionary force in society (K. Johnson 2020; Jackie Snow 2019; Lant 2016).

But for the applied machine learning researchers I worked with, the algorithmic sublime was both a product of their work and a barrier to their work. As discussed in Chapter 1, creating demonstration projects that foreground the spectacular capabilities of machine intelligence was part and parcel of their work. And as discussed above, cultivating a persona through performances of fluency and capability with machine intelligence techniques was part of the professional responsibilities of applied machine learning researchers at OTH. They frequently authored blog posts that engaged with the most spectacular achievements of machine intelligence, leveraging the sublime effects of projects like AlphaGo and *Sunspring* for their own promotional purposes. At times they would contribute to the sublime aura of these projects by foregrounding the scale of these achievements, and at other times they would foreground their own expertise by deconstructing and demystifying these technological objects by penning "explainers" that denatured the sublime aspects of such projects. But in workshops, and in their consultations with other clients, the algorithmic sublime was an obstacle to the practical, mundane practices they were trying to promulgate, and which would lead to more lucrative engagements with their clients.

## THE NEW ELECTRICITY

Metaphors, as they are deployed around machine intelligence, are socio-material practices that matter for the adoption of emerging technologies (Sturken 2004). They describe the world at the same time that they have practical consequences for it (Lakoff and Johnson 1980); and it is therefore crucial not to take them as simply "being *about* the world in ways that make little of how they may be materially located *within* it" (Keane 2003, 410). The metaphors our audience was primed to engage with on the day of our workshop at Intricate Media were in part captured by the spectacular demonstrations of machine intelligence we presented. But our audience was also fully enrolled in AI as a "discursive practice" (Agre 1997, 140) having already been primed to accept metaphorical conflations of machine intelligence with human intelligence, machine learning with human learning, and generative applications of machine intelligence (like the screenplay-writing *Sunspring*) with human creativity. The task before us, then, was to shift the metaphorical underpinnings of machine intelligence to a different footing.

As Frankie took the clicker from my hand and advanced to the next section of the presentation, she began the work of replacing the metaphors I had just provided a tour of with a new one. The image that appeared on the next slide was of Andrew Ng (Figure 8), formerly the chief data scientist at the very large Chinese tech company Baidu. Ng was at that time the current Co-Chairman of Coursera, an online training portal for aspiring applied machine learning researchers, and an adjunct professor of computer science at Stanford University. A member in good standing of the pantheon of machine intelligence thinkers, the mention of his name alone was capable of hushing the crowd at any tech-centric event. Frankie narrated his most recent contribution to the discourse around machine intelligence: that "AI is the new electricity" (Ng 2017).

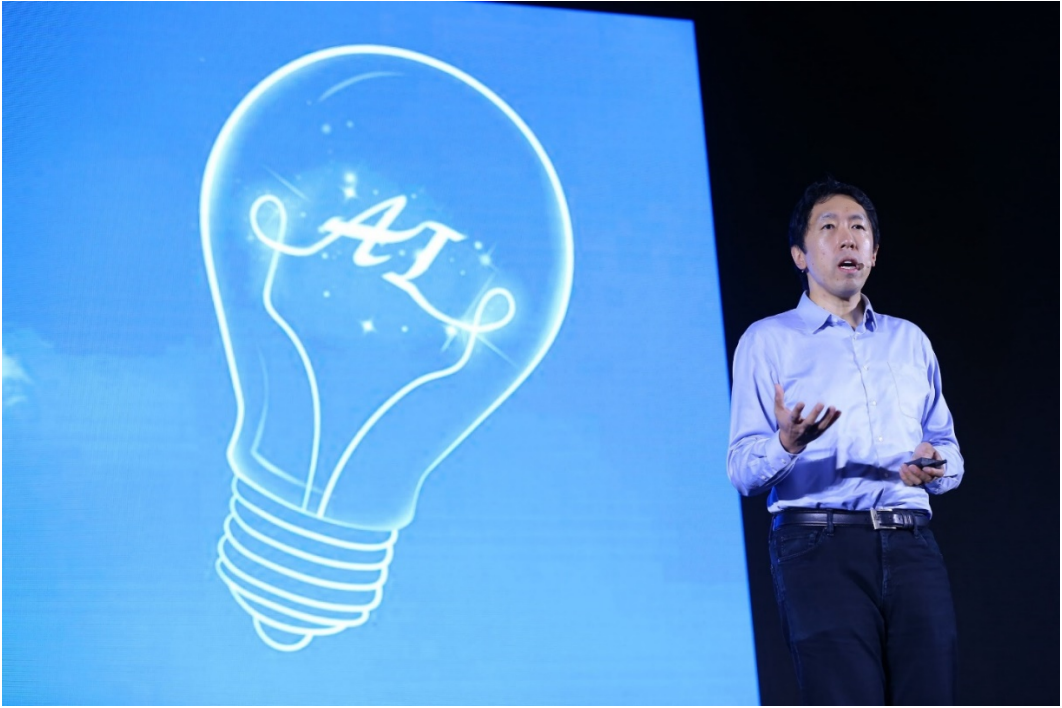

Figure 8. Andrew Ng keynote presenting AI as the "new electricity" (Ng 2017)

"a hundred years ago, electricity was really complicated. You had to choose between AC and DC power, different voltages, different levels of reliability, pricing, and so on. And it was hard to figure out how to use electricity: should you focus on building electric lights? Or replace your gas turbine with an electric motor? Thus many companies hired a VP of electricity" (Ng 2017).

Similarly, Frankie informed the audience, machine intelligence today is "really complicated". Data can be local or distributed in the cloud. It is difficult to know whether and why to use a random forest algorithm or a neural network, or how to evaluate the success of any particular implementation. Furthermore, it is difficult to anticipate the costs of a project; the reliability and cost of machines, data storage, and engineering talent vary widely. And it is difficult to know where to focus one's efforts—should one build an "audience segmentation model" first, or a "churn model"? Frankie, standing at the front of the conference table, foregrounded the nuance that she wanted to draw out of Ng's metaphor for her audience. She drew particular attention to the "VP of electricity" Ng discussed to break down the mythical status of a machine intelligence that can be deployed, like well-trained troops, to solve ready-made problems. The reason to have a VP, she articulated, is because all the hard questions that make machine intelligence "really complicated" have to be engaged with strategically in order to make any progress at all. Different parts of an organization need to be coordinated to work with each other in new ways, and a position within an organizational hierarchy is needed to enact that coordination.

Machine intelligence itself requires its own organizational practices, too. Without designing data collection processes to record user behavior data in a way that is easily "ingested"—transformed into a data format that machine intelligence tools can readily process—many products Intricate Media might want cannot be built. And without having applied machine learning researchers on staff who can do "front end" development to create user-friendly interfaces for non-experts, machine intelligence products will remain inaccessible to most other Intricate Media employees. Where the spectacular demonstrations of machine intelligence that I had just laid out presented the fully-realized capabilities of the technology, Frankie now worked against those glossy claims to show the real work and hard decisions that would be necessary before Intricate Media could take advantage of the promise of the technology. This was a delicate balancing act for her. She had to maintain excitement about the potential, eventual outcomes while tempering expectations around some of the most spectacular capabilities of machine intelligence which were not well-suited to the real world in which Intricate Media worked. At the same time, she also had to "sell" more mundane practices, which seemed only distantly related to the practices of machine intelligence, as key components of a data strategy.

One step in her rhetorical shift was to highlight how popular demonstrations of machine intelligence are popular because of a "human tendency to misunderstand what constitutes a computationally difficult problem" and "to see proof of technological capability as proof of pragmatic capability" (Moss and Schüür 2018, 267). AlphaGo makes for a good example of machine intelligence, Frankie argued, but it is not persuasive proof that machine intelligence can accomplish pragmatic tasks that staff at Intricate Media might think of as being as complex as the game of Go but would like to automate. Frankie was referencing an idea that human intuitions about what problems are easy or difficult to solve do not map on to computational difficulty for machine intelligence (Bucher 2017; Lee 2018). Frankie walked through how certain problems, which *seem* complex when compared to the capabilities of mere mortals, have certain properties that lend themselves to being addressed through machine intelligence. Drawing on a popular blog post about AlphaGo to illustrate these properties, Frankie explained how the game of Go has these properties, and how many other problems that are difficult for machine intelligence do not have these properties (Figure 9). Go is *deterministic*; the rules of the game describe every possible game state without any randomness or noise. Go is *fully observed*; each player knows everything about the current state of the game by looking at the board and no player has more information than the other. Go is *discrete*; each move is an indivisible unit of action and there are no half-moves or gradations between possible game states. Go is *simulatable*; the mathematical description of the game is identical to the game itself. Go is *short*; each game lasts approximately 200 moves. Go has a clear *outcome*; there is a winner and a loser for each game and there is no confusion about what constitutes a 'win'. And Go is *well-documented*; thousands of games of Go have been played and notated for computers to ingest as training data.

| | |
|---|---|
| Deterministic | The rules of Go describe possible game states without any randomness or noise. |
| Fully Observed | Each participant knows everything about the current state of the game simply by looking at the board. |
| Allows only discrete actions | There are a quantifiable number of different moves that are possible without gradations between these moves. |
| Is simulatable | It is easy to simulate a game of Go and this simulation will be identical to the game itself. |
| Is short | Each game of Go lasts approximately 200 moves. |
| Has a clear outcome | There is a clear definition of what constitutes a 'win' or 'loss'. |
| Is well-documented | There are hundreds of examples of human gameplay to supercharge the initial knowledge that AlphaGo begins learning from (AlphaGo Zero, of course, freed itself from this condition). |

Figure 9. Properties of Go (Moss and Schüür 2018).

often involve continuous phenomena rather than discrete actions. Most importantly, Frankie emphasized, most real-world problems are only simulatable if deliberate decisions are made about what is and what is not part of a system. "For example," she said, "the weather does not affect the outcome of a game of Go, yet it matters for algorithms that steer self-driving cars." This is part of the "abstraction problem" that has long been studied in computer science (Knoblock 1990), but has particularly stark implications when machine intelligence is applied to social problems and must separate technical terms—like "success" in pursuing an objective function, "accuracy," or "fairness"—from socially-relevant contexts (Selbst et al. 2019).

At this point in the presentation, the situation might have seemed hopeless for the employees of Intricate Media who had been so keen to apply machine intelligence to their problems, and whose current strategies were not working for them. If the real world is not like games, how could they be expected to use machine intelligence in the real world? The answer, Frankie suggested, was to take successive, incremental steps, as part of an overall data strategy, to bring the properties of the problems Intricate Media wanted to address closer to the properties of problems machine intelligence was capable of addressing. To do so would require the difficult work, not of creating clever algorithms, but of reorganizing how labor effort is devoted to machine intelligence practices and inventorying the data available to the Intricate Media team. Most importantly, from Frankie's point of view, it required prioritizing the development of machine intelligence practices that were not as spectacular as AlphaGo or facial recognition, but could support successively more impressive projects over time. As strategy, this difficult work would require setting organizational priorities oriented toward developing machine intelligence capabilities over time—priorities that were often set, and put into practice, by a vice president. While a "VP of AI" was not something that the organizational hierarchy of Intricate Media was likely to allow for—executive roles and responsibilities were far beyond the scope of a one consultant-led workshop—what Frankie was proposing was a data strategy that could stand in for the role of VP.

## EMPOWERING DATA TEAMS

More than a dozen Intricate Media employees attended our workshop on that day. The in-house data analytics and data engineering team, comprised of two data analysts and one data engineer, were accompanied by their interim manager (who did not have a data science background), the chief technology officer (CTO), the chief of staff, two representatives of the editorial team, two representatives of the social media team, one business development staffer, and two product managers. The only familiar face in the room when Frankie and I arrived was Annabeth, a project lead who worked under the CTO and who was the motive force behind OTH's engagement with Intricate Media on that day. Annabeth had been a champion of the workshop because she was structurally positioned to observe the tensions that existed between the various teams around who should be able to control the production and use of data, and she thought that an outside consultant like OTH—who did not have direct stakes in the outcome of a data strategy—might be able to help craft a data strategy that was acceptable (or at least minimally offensive) for all. She and Frankie had crafted the attendance list for the workshop in a planning meeting that I had attended two weeks prior.

The planning meeting had set out the goals for the workshop: use introductory remarks to get the entire Intricate Media team excited about the potential of machine intelligence by highlighting recent spectacular accomplishments (like AlphaGo and *Sunspring*), demystify these spectacular accomplishments to set realistic expectations, demonstrate areas of machine intelligence OTH had recently developed and could help Intricate Media apply to their own business, and finally survey approaches to machine intelligence that had recently had practical applications in Intricate Media's sector. These remarks would set the stage for Intricate Media's data team to—hopefully—begin sharing the types of data they held, and for the product and creative teams to help elaborate what kind of machine intelligence projects would be desireable for them. Frankie and I, as the outside consultants, would then—again, hopefully—bridge the gap between those aspirations and the available data by proposing changes to existing workflows and intermediate projects that were presently achievable, but which would support the more distant goals, as set by the product and creative teams, in the future.

On the day of the workshop, however, nothing that followed the introductory remarks went according to plan. It quickly became apparent that attendees had expected machine intelligence could be added to the existing set of business practices at Intricate Media without changing the organizational practices there. This posed a problem for crafting a data strategy, in two ways. One was that individual teams at Intricate Media were apprehensive about changing their own teams' practices, whether to accommodate machine intelligence or to contribute to its production. As will be discussed below, there was a distinct sense that a status-laden hierarchy between these teams ought to be preserved, and that the higher-status creative team's delicate and subtle (and precarious) work practices were what contributed most to the success of the organization. Any disruption could disrupt this chemistry and ruin the "special sauce" that Intricate Media saw as intrinsic to its success. The second problem this posed was that machine intelligence could not just sit on top of existing practices—in Frankie's words it could not be "bolted on" to the way the company was already using data, hence the need for strategic shifts in how data was handled in order to create the conditions for machine intelligence. The popular imaginings of machine intelligence, which we had set up in our presentation in order to debunk them, had preceded us at Intricate Media, and proved more durable than our intervention anticipated.

In the conversations that followed that presentation, conducted through topical discussions with everyone in the room participating, it became apparent that the existing approach to data at Intricate Media was deeply mired in expectations that it serve other teams in specific, inflexible ways. Specifically, the data analytics team was overwhelmed by creating reports requested by the business team or the creative team. These reports were *ad hoc* summaries of how content was performing—page views, click-throughs, shares, etc.—or how the business development was progressing—ad sales, sponsorships, and licensing revenues. While the data analysts were eager to build a tool that these other teams could use on their own to generate the performance metrics they needed, they had persistently been unable to step away from the *ad hoc* requests for long enough to build such a tool. The immediate requests from creative and business teams took precedence over long-term product development, and the CTO who supervised their work was only interested in using outside vendors' products to alleviate this burden, not in allocating the data team's resources to building their own, custom solutions to carry it.

The expectation that data analysts could provide immediate, custom responses to queries about data was not the only problem that confounded our plans for the workshop. Relatedly, everyone from Intricate Media who had gathered for the workshop saw data as a *market commodity*. An organizational awareness of the market potential of data has been slow to arise within some sectors, and other machine intelligence consultants might have endeavored to help their clients realize the economic value of data they held (Aaltonen, Alaimo, and Kallinikos [in press] 2021). But Frankie saw this as a problem for Intricate Media because, as Annabeth put it, "if we want data, we'll make a partnership". Annabeth was referring not only to the availability of datasets collected and re-sold by data brokers (Crain 2018) and which could be used to augment the types of data Intricate Media might already have about their readership and advertising clients, but also to Intricate Media's persistent use of third-party plugins to run their online media business (Helles, Lomborg, and Lai 2020). Third-party plugins give a company like Intricate Media tremendous functionality simply by copying and pasting third-party computer code into an existing website design, and also give Intricate Media access to data they might not otherwise have access to (e.g. a list of other websites recently visited by Intricate Media's audience). But what both these practices meant in terms of the data strategy Frankie was concerned about was that Intricate Media would be relying on the categories of analysis offered by data brokers and third-party plugin vendors, not the categories of analysis that were most useful for producing data that could be used to train the machine intelligence applications Intricate Media might want to develop. While third-party plugins might be useful when they are fit to their purpose, Frankie told me in a workshop *post mortem* that Intricate Media's reliance on these tools was a barrier to using machine intelligence as appropriately as they could.

Within organizations, there tend to be various conflicting definitions for the categories of analysis one might wish to use to build a machine intelligence product. Borrowing from an example in Chapter 1, when applied machine learning researchers at OTH were building their multi-task learning classifier to distinguish between broadsheet and tabloid publications, they had many possible definitions to draw from in deciding which publications should be included with each category, ultimately deferring to a scholarly publication as a source of authority that lent stability to those categories. Similarly, data analytics teams are tasked, due to their function, "to count". "How many daily active users visit an Intricate Media website?" "How many top-tier clients does Intricate Media currently have contracts with?" "How many pageviews does our content receive once it has been posted online for more than a month?" Counting these things is more difficult than it seems, as who should be counted as a daily active user is actually a matter of some disagreement.

Therefore, data analysts require consistent definitions for "what to count" provided by some source of authority. That source of authority might be external, perhaps deferring to a scholarly authority or an industry "best practice" (e.g. Clifton 2010), or it might be internal to the organization—business development teams might define what should count as a "top-tier client" and content management teams might define what constitutes "evergreen content" at a company like Intricate Media. But these internal definitions require effort to stabilize. Often, differences between definitions go unrecognized and unacknowledged. They surface when the data analytics team is asked to actually do the counting and lacks the specificity of definition with which to do so. Left to their own devices, data analysts tend to borrow details from their own, sometimes idiosyncratic definitions of these terms. An individual data analyst might decide that a daily *active* user is one who not only visits a website but inter*acts* with it by clicking on a hyperlink or watching a video. Another data analyst might decide that a *daily* active user is someone who visits a website every *day*. This is an oversimplification of "daily active user," of course, but it illustrates how two data analysts working on the same question might arrive at very different answers.

Overall, such discrepancies erode trust in data at many companies, Frankie told me, and also lie behind the inability of Intricate Media's data analysts to build a tool that could provide on-demand analytics for other teams and replace the *ad hoc* requests on their time. Instead, every request the data analysts fielded amounted to producing a different way of counting a similar concept, over and over again. And most importantly in terms of data strategy, a lack of clear definitions meant that data wrangled by the data analysts could not be used to train machine intelligence products. With so many different counts and definitions, machine intelligence could not be "bolted on" to existing data resources. Instead, data analysts would have to be empowered to establish the conditions for machine intelligence by controlling and standardizing the data stores applied machine learning researchers might use to develop machine intelligence. But the data analytics team at Intricate Media was seen as a mere service function, responding to *ad hoc* requests to serve others' agendas, rather than as a business unit capable of creating products that had value to the organization in their own right. Without significant leadership intervention, from someone with the organizational power of, say, a vice president, this was unlikely to change.

### HIERARCHIES OF EXPERTISE

In the planning meeting prior to the workshop, Annabelle had advised Frankie and me that we ought to "tread lightly" so as not to upset the editorial and creatives teams. She was concerned those teams' members might come into the workshop with fear about changes to their job descriptions, or about losing their jobs entirely. However, during the workshop itself, we found that the editorial and creative teams were actually eager to incorporate machine intelligence into their work practices, and had a range of specific work practices they were eager to change with machine intelligence. Rather than protecting what Annabelle had called the "cool guy space" that they inhabited within the office ecosystem, they were open about aspects of their job that they felt were appropriate for automation. These aspects were not central to the creative dimensions of their rôles, but instead were seen as repetitive and mundane tasks. In contrast, the CTO and the product development team expressed only concerns about a new data strategy, offering rationale after rationale for why their present way of organizing and using data was the only reasonable way to do so at Intricate Media. "We've been using machine learning for a decade, they're called Google and Facebook", we were told. For these teams situated closer to data, the "native tools" and data analytics provided at a premium price by vendors were as capable as they thought they needed to be.

In articulating this faith in vendor tools, these technically-savvy but resistant workshop attendees were voicing a common refrain about the allocation of power within machine intelligence that I heard throughout my fieldwork. Why try to build something oneself, the refrain went, if success was not guaranteed, if one of the few major players in machine intelligence was willing to invest their own massive capital reserves in building solutions to the same problem, and if that solution could then be licensed from that major player for a modest sum? This is precisely the strategy that Facebook, Google, Microsoft, and Amazon were pursuing during the course of my fieldwork. These companies offered free- or cheap-to-use machine intelligence tools—Google's TensorFlow, Facebook's FB Learner Flow, Amazon SageMaker, and Microsoft Azure ML Studio—in a bid to win lucrative contracts for storing their enterprise customers' data. But they also offered machine intelligence services themselves—Amazon Rekognition and Microsoft Azure Text Analytics offered pay-per-use facial recognition and natural language processing services, respectively. Alongside the numerous smaller companies offering machine intelligence products to enterprise customers like Intricate Media, it was reasonable to imagine that machine intelligence capabilities could be purchased piecemeal to do everything the company might want to accomplish. It was more difficult to imagine, for the CTO, overseeing a years-long effort to build a set of tools to accomplish similar ends, even if it might have been more closely aligned with Intricate Media's needs and free of yearly licensing fees.

But this view was starkly at odds with the "data imperative" (Fourcade and Healy 2017) to unreflectively collect and store as much information as possible that had motivated the data strategy workshop in the first place. For applied machine learning researchers like Frankie, data is a resource that can be wielded toward seemingly limitless ends. It is an enabler that makes an entire suite of capabilities possible. Using vendor services not only limits the capabilities one's data can enable to those offered by the vendor, but also requires sharing data—that precious resource—with those vendors. Most vendors, whether they are behemoths like Amazon and Google or niche sector-specific vendors like "Domo" or "Woopra"—work with multiple organizations often in the same line of business. Sharing data, by using vendors, can remove competitive advantages that increasingly lie in data—at least according to the "data imperative". Under the sway of the spectacular examples of machine intelligence, it was more difficult for the data team at Intricate Media to see the value of data, or of a data strategy; any path toward a machine intelligence product was acceptable, even if it jeopardized a longer running strategy that would have enabled future products, undermined competitive advantages, or extracted as much value as possible from the company's precious data.

In retrospect, it was easy to see why Annabelle had expressed concerns that we tread lightly around "creatives," and why the technical team was so dismissive of a new approach to data. The spectacular presentations of machine intelligence had deeply engrained a view of machine intelligence that was autonomous and immensely capable. This narrative leant itself to both the view that it could be "bolted on"—that it could be added unproblematically to any existing business practice—and a view that machine intelligence can displace, replace, and deskill humans it comes into contact with. But the editorial and creative teams were insulated from this view primarily because they were keenly aware of inefficiencies in their work that were unrelated to their core competencies which machine intelligence might be able to automate without threatening their expertise. Placing appropriate content at the top of a list of recommendations or selecting evergreen content was a very different chore than writing copy or designing a new layout, and they were happy to discuss shifting the more tedious aspects of their daily tasks to machine intelligence. They also had a firm grasp of their own role as taste-makers. Indeed, their entire habitus had long cultivated the forms of cultural capital needed to arbitrate the distinction between the forms of cool that Intricate Media trafficked in (Bourdieu 1984). While demonstrations of machine creativity like *Sunspring* might signal to non-creative types that creative roles were endangered, creatives themselves were perhaps much more aware that creative cultural production relies more on the embodied class position of creative labor rather than any engineered machine intelligence. With confidence in how their work articulated with the making of taste, creative and editorial team members looked to machine intelligence instead as an enabler—a potential partner—that augmented their own work, much as electrification had augmented the work of laborers a century before.

Taking as given the "simple premise that expertise is something people do rather than something they have" (Carr 2010, 18) it becomes possible to see the different responses of technical and creative teams within this workshop—dismissive and accepting of machine intelligence, respectively—as revealing the tensions that arise from the differing sets of professional practices that are implicated in the adoption of machine intelligence within an organization. While the CTO constituted his expertise through practices of selecting vendor tools and managing the data analytics team's response time for *ad hoc* analysis requests, and the data analytics team itself was incentivized by its ability to respond to these *ad hoc* requests, these practices would have to be largely abandoned in order to craft any data strategy that would lead to the implementation of in-house machine intelligence. The creative team, in contrast, constituted their expertise through practices of writing, art, and social media engagement—practices that would grow as a portion of the workplace responsibilities if machine intelligence could be implemented in ways that relieved them of more mundane responsibilities. Each of these sets of expert practices were also differentially implicated in how the mechanical objectivity of machine intelligence would be incorporated into Intricate Media's business. As a historically- and socially-constituted value, objectivity can take many forms (Daston and Galison 2010). Machine intelligence stands in for a form of "mechanical objectivity" that is counterposed with human judgement that is fallible, full of "bias, inefficiency, and discrimination" (Christin 2016, 28). While trusting algorithmic systems over human actors might be seen as allowing those who exert control over such systems to participate in this form of objectivity as an extension of their own expertise, the CTO (in the case of Intricate Media) resisted seeing any movement in that direction as anything but a threat to their existing control over data and vendor tools. He vested their expertise in exerting "trained judgement" over which tools to use, not the tools themselves. The data analytics team itself, in contrast, might have liked to shake off the burden of responding to *ad hoc* requests, but in the absence of the CTO's cooperation in developing new incentives and ways to measure their success, they were trapped trying to "beat their numbers" each week, and disincentives from "trying anything new".

The creative team, about whom instructions to "tread carefully" had (erroneously) indicated a likely tendency toward "the pervasive sense that technologies transform us in irrevocable ways … accompanied by the anxiety that they will also promote some kind of loss… of authenticity in some way" (Sturken, Thomas, and Ball-Rokeach 2004) responded differently. Having positioned themselves as central to Intricate Media's core product—they created media content, after all—they were able to foreground this aspect of their work and downplay the routinized labor they performed and which was most likely to be replaced with any likely machine intelligence product. Their "trained judgment" would not be jeopardized in any way.

Frankie, and other applied machine intelligence researchers like her who are concerned about how best to disseminate machine intelligence techniques across many industries (e.g.

Soni et al. 2019), encountered these tensions as a contradiction that was fundamentally irresolvable, even when it was navigated with care. On the one hand, engaging Intricate Media's different teams in a workshop to craft a data strategy could enable the pursuit of machine intelligence projects that would drastically reshape how work was done. On the other hand, machine intelligence could be incorporated into existing team practices, augmenting capabilities in order to achieve goals that would have only been remotely imaginable without machine intelligence. This contradiction was irresolvable because Frankie and Annabeth ultimately failed to enroll the necessary allies for implementing a data strategy at Intricate Media. Without the enthusiastic support from the CTO and the data team he oversaw, no changes in practices that would have enabled machine intelligence could be pursued.

## Conclusion

This chapter has explored additional ways in which performance and performativity are indispensable to the authority of machine intelligence. These performances include the production of spectacular demonstration products, as well as the performative enactments of self and expertise within corporate settings. While spectacular demonstrations of machine intelligence intimated that it could be "bolted on" to existing workplace practices and be transformative for the entire business enterprise, such technological objects are not enacted or brought to bear on the world in which they are applied without a broad accommodation of the social sphere to the technological apparatus, and of the technical apparatus to the existing practices within the social sphere. Indeed, for machine intelligence to enact its authority in any domain, it must be accommodated both technologically and socially. At Intricate Media, this meant having both the personnel to build the metrics required for machine intelligence, as well as the necessary workflows amongst the creative team to make use of it. The difficulty of negotiating for the accommodation of OTH's recommendations at Intricate Media begins to show some of the limits of machine intelligence's authority, but this view should be tempered by a consideration of how machine intelligence had already been fully insinuated into the workplace practices at this particular firm. Intricate Media had already become dependent upon third-party vendor tools. And while Frankie and I were trying to help Intricate maintain control over how machine intelligence would be used there, the implacability of the data team rebuffed our very limited, personal authority over how Intricate Media ought to manage itself, but did not undermine their acceptance of the crucial role of machine intelligence in the media industry, or elsewhere.

Elsewhere, machine intelligence gains adoption either because of a coordinated effort to accommodate it technically and socially, or because efforts to incorporate audit and accountability practices in the past have laid the groundwork for other experts to defer to machine intelligence. For a sepsis-prediction algorithm to gain a foothold in an emergency room, metrics linked to patient health records need to be able to make claims about a notoriously hard-to-define medical condition. But so, too, must hospital personnel restructure the well-established chain of authority between personnel—with doctors at the top and nurses below—so that nurses can subvert that hierarchy on the algorithm's behalf (see Sendak et al. 2020). And for criminal risk prediction systems to gain any foothold in courtrooms, the sole authority of judges must first be eroded by sentencing guidelines and rule-governed practices like "mandatory minimums". Without practices that require judges to justify their decisions, judges would not have any need to defer to algorithmic systems (see McKay 2019).

The workshop was ultimately a failure in which Intricate Media rejected the data strategy that OTH recommended in their report following the day-long engagement. In retrospect, Frankie and I concluded that it failed because we had not laid the groundwork that would have secured the active participation of the entire range of actors likely to be impacted by the changes suggested by the data strategy—particularly the data team whose expert practices were most directly affected. In part this was because of our own, mistaken, expectations that the data team would be the *most* enthusiastic ally for a data strategy because it would have put more resources and responsibilities within their sphere of influence. It turned out that they wanted to preserve their existing practices more than they wanted additional or altered practices. Perhaps this was because the confidence they held in their own expertise with their present practices would be replaced by uncertainty in their own abilities to adopt new practices. Or perhaps it was because shifting practices would open their business unit up to co-optation from any newly hired, potentially "more qualified" applied machine learning researchers. But it was not just the threats to the existing expert practices of the data team that doomed the data strategy workshop. It was also the existing emotional topography of the entire company that had been sedimented by spectacular demonstrations of machine intelligence—the ones we presented and all the ones they had experienced prior to the workshop. The ways in which such spectacles had mobilized a view of machine intelligence as something *apart* from the imaginable range of workplace practices at Intricate Media was too broad a gulf to span within a one-day workshop.

## CHAPTER 3: EATING THE WORLD

**MACHINE INTELLIGENCE IS EATING THE WORLD**

"Software is eating the world" (Andreessen 2011). "No, data is eating the world" (Buytaert 2015). "Data is eating the software that is eating the world"(Kobielus 2017). "Machine learning is eating the software world" (Woodie 2015). "Software ate the World, now AI is eating Software" (Singh 2019). When Marc Andreessen penned an op-ed in the *Wall Street Journal* in 2011, he likely had little idea he was inaugurating a metaphor that would make perennial returns over the following decade. The founder of consumer internet technology companies (including web browser company Netscape) and early venture capitalist, Marc Andreessen pointed to what he saw as a profound shift across industries away from producing and marketing *products* and toward delivering services through the efficiencies produced by the affordances of *software*. Not only were software companies—Microsoft, LinkedIn, Skype— growing exponentially, but companies that were not commonly thought of as software companies were becoming software companies. He wrote,

> "In today's cars, software runs the engines, controls safety features, entertains passengers, guides drivers to destinations and connects each car to mobile, satellite and GPS networks… Today's leading real-world retailer, Wal-Mart, uses software to power its logistics and distribution capabilities, which it has used to crush its competition. Likewise for FedEx, which is best thought of as a software network that happens to have trucks, planes and distribution hubs attached. And the success or failure of airlines today and in the future hinges on their ability to price tickets and optimize routes and yields correctly—with software" (Andreessen 2011).

As written by a venture capitalist, Andreessen's piece is best understood as pure boosterism for the high-tech Silicon Valley companies that comprised the gaping maw of software, and in which he likely held significant financial stakes. But as the progenitor of a genre of alimentary metaphors, he is perhaps better understood as the avatar for data-driven technology's authority over and insinuation into the full sweep of human endeavors. What his inheritors mean to suggest when they say that machine learning, or AI, or data is eating software, is that machine intelligence is presently—or will soon be—capable of accomplishing the goals that are currently reached by software programmers explicitly specifying any given action for a computer program to execute. Machine intelligence is the gateway to "Software 2.0," these advocates suggest, which will allow applied machine learning researchers to:

> "specify some goal on the behavior of a desirable program (e.g., "satisfy a dataset of input output pairs of examples," or "win a game of Go"), write a rough skeleton of the code (i.e. a neural net architecture) that identifies a subset of program space to search, and use the computational resources at our disposal to search this space for a program that works… To make the analogy explicit, in Software 1.0, human-engineered source code (e.g. some .cpp files) is compiled into a binary that does useful work. In Software 2.0 most often the source code comprises 1) the dataset that defines the desirable behavior and 2) the neural net architecture that gives the rough skeleton of the code, but with many details (the weights) to be filled in. The process of training the neural network compiles the dataset into the binary [machine-readable language]—the final neural network. In most practical applications today, the neural net architectures and the training systems are increasingly standardized into a commodity, so most of the active "software development" takes the form of curating, growing, massaging and cleaning labeled datasets. This is fundamentally altering the programming paradigm by which we iterate on our software" (Karpathy 2017b).

Software 2.0 will, with sufficient training examples, be able to pursue any goal that applied machine learning researchers set for it (Karpathy 2017b).

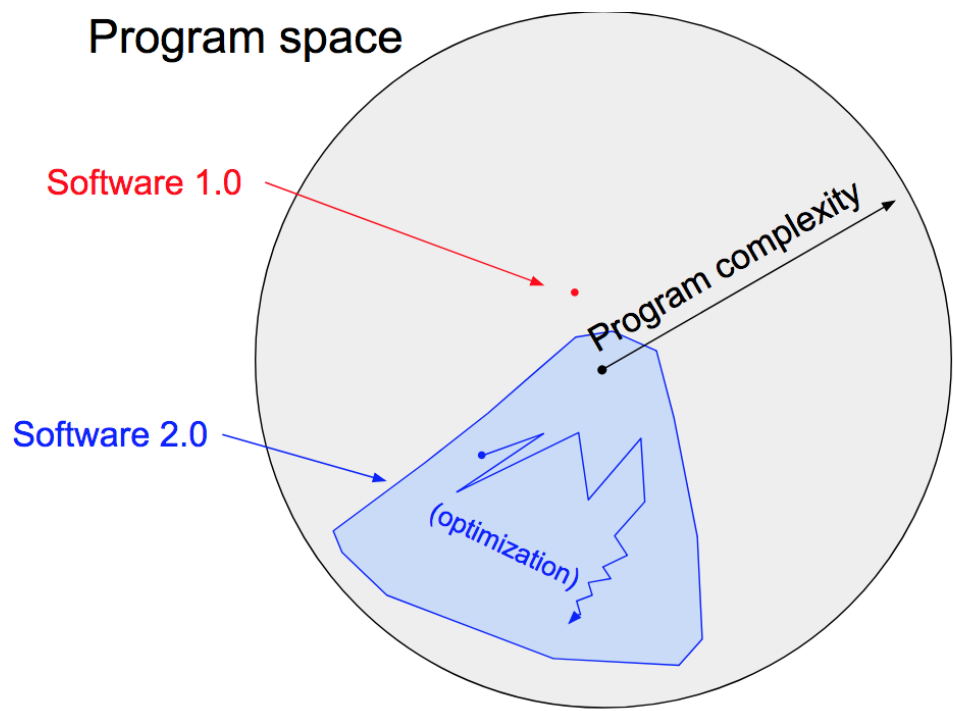

Figure 10. Software 2.0 pursuing programmatic goals " (Karpathy 2017b).

As discussed in the previous chapter, the widespread adoption of deep learning as the standard-bearer for the promise of machine intelligence created a groundswell of expectations for what could be expected from data-driven technologies. Automated gameplay and other impressive demonstration projects heightened those expectations, making even the most outlandish claims about the capabilities of machine intelligence seem plausible. But many of the most outlandish claims about machine intelligence still rest upon its core capabilities:

classification and prediction. And as the last chapter(s) demonstrate, much of the work of machine intelligence is exerted in transforming real-world, domain problems into classification or prediction problems. In doing so, applied machine learning researchers maintain their objectivity by distancing themselves from the production of useable data through automated techniques like scraping, autolabeling, mTurk gigwork platforms, or through taking externally produced datasets they are supplied with. But they also buttress their objectivity by claiming proximity to the data itself, spending time 'cleaning' data, generating 'summary statistics', and transforming it through pre-processing techniques that give them intimate knowledge of the dataset. All these practices comprise what I call *data performances*—generative, material-discursive enactments of the assumptions made in formulating domain problems as machine intelligence problems.

In a sense, all manifestations of machine intelligence are data performances, as we can see by revisiting examples of machine intelligence we have already discussed. A deep learning system that classifies images of handwritten numerals "performs" Arabic numerals. An algorithmic system that predicts whether a news article is from a tabloid or a broadsheet publication "performs" this distinction. Even regression "performs" the taxonomic distinctions within the *Iris* genus. As discussed in Chapter 1, the epistemological claims of machine intelligence—that positive knowledge can be produced solely through data analysis—enables applied machine intelligence researchers to make authoritative claims about the world based on how the world is represented as data. But machine intelligence's data performances also enact ontological claims about the world itself. They assert what these things *are* and how they come to matter. Handwritten numerals *become* statistical relationships between differently-shaded pixels in a digital image rather than sign vehicles. 'Tabloidness' and 'broadsheetness' *become* different statistical distributions of lexemes rather than historically contingent journalistic styles that are co-constructed between author, editor, and audience. And 'specie' become a "system of variables all of whose values can be designated … to establish the system of identities and the order of differences existing between natural entities" (Foucault [1966] 1994, 136), rather than "the flow of genes as information, substance, and property … as tokens of biopolitics … of which humans are a part" (Helmreich 2009). In doing so, machine intelligence declares that the objects of its analysis are what it says they are and forecloses other possible meanings, thereby eating the world by assimilating the components of phenomenological experience to its own logics.

If we were to follow Martin Heidegger on this point, he would lead us to see how technology—including the technologies of machine intelligence—is "no mere means" but "a way of revealing" and "of bringing-forth" (Heidegger 1977, 12-13), "presencing from that which is not presencing" (Heidegger 1977, 10). The idea that machine intelligence brings forth objects is a central concern of this chapter, and while Heidegger is instructive here, I will also follow Britt S. Paris in attempting "to wrest the concepts of sociotechnical materiality, agency, and time from Heidegger's reactionary technological determinism by weaving in works in STS" (Paris 2021, 128) to see how the ontological claims of machine intelligence foreclose other ethical, social, and political accounts of the world. Key to this is the Karen Barad's "agential realism" (Barad 2007), which, as already noted, draws attention to the agential role machine intelligence plays as an apparatus, in intra-action with datasets and "real-world" phenomena, that enacts particular configurations of otherwise indeterminate social relations through the "agential cut" it makes that allows for some configurations while excluding others. This agential cut is central to what I call *data performances*. Agential realism and data performances are broadly compatible with social construction of technology (SCOT) approaches that focus on "the social groups concerned with the artifact and the meanings that those groups give to the artifact" (Bijker, Hughes, and Pinch 1987, 22), but I focus on the material-discursive practices that create the contexts in which technologies and meanings shape each other to produce novel objects of knowledge—contexts in which "constructivism is no longer 'social' but has become 'ontological'" (Puig de la Bellacasa 2017, 87).

      Starting again from Heidegger, we could also begin to glimpse how machine intelligence, like other technologies, transforms humanity into part of the "standing-reserve" for technology (Heidegger 1977, 17). The tools of machine intelligence depend entirely upon their ability to capture everyday human activity as machine-readable datasets. Without these datasets, machine intelligence remains idle, like the aircraft of which Heidegger was so fond sitting on a runway. Just as the systems they are built into, which require technical and social accommodations to have any influence on the world (see Chapter 2), the powerful algorithms that can identify patterns, make classifications, and formulate predictions can do nothing on their own either. But they gain tremendous significance—and financial value—when applied to datasets that record "real-world" phenomena. The value of the Google search algorithm "PageRank" (Page et al. 1999) is in part it's naïve empirical approach to web indexing—it doesn't need to know anything about the semantic meanings of the web pages it indexes to learn which are most germane for a particular web search. But the greater part of the algorithm's value derives from how people use the internet. The hyperlinks people use to link one page to another signal to the algorithm which web pages are most important—pages receiving in-bound links are assumed to be more useful than others—and the search results that users click on more often than others provide additional signals about which pages are most useful. Without people interacting online and interacting with the Google search site, the algorithm itself would be worthless. In a world in which Google's parent company Alphabet is capitalized at $1.8 trillion, and in which it is increasingly difficult to avoid interacting with

Google services—which have become ubiquitous and unavoidable as university email and document management infrastructure, corporate data management, and even scholarly literature searches rely on Google's systems—it becomes difficult to discern whether Google is providing a service for internet users, or internet users are the standing-reserve toiling away on their online devices for Google's profit.

Machine intelligence is, increasingly, the means by which knowledge about society is produced. It is how corporations monitor their brands and make weighty business decisions. It is being used to inform decisions made in courts, banks, and hospitals about how to sentence defendants, set rates for mortgage applicants, and treat patients. In this sense it is eating the world of other ways of making such decisions, being integrated into "street level bureaucracies" (Lipsky 2010; cf. Alkhatib and Bernstein 2019), and eating away at the personal authority of judges and the trained expertise of doctors. But it is also voracious for data, which has been discursively constructed as a quasi-natural "resource" (see any number of articles declaring that "data is the new oil") but which is largely a by-product of social activities—many of which have been "enclosed" within online infrastructures architected to store records of these activities and make them "ingestible" for machine intelligence, constituting what Shoshanna Zuboff calls "surveillance capitalism" (Zuboff 2019). However, machine intelligence does not only ingest. It also digests, metabolizing data into new, stable objects of its knowledge.

Google's PageRank algorithm crawls the entire web, consuming links between webpages and hoovering up every interaction users have with its own website. But it also brings forth a highly specific object into the world: the "importance" of every web page it indexes. PageRank produces "importance" as an object of knowledge, reifying it as a calculable entity derived from the assemblage of users' interactions on the internet, the infrastructure Google uses to access and store data about those interactions, the PageRank algorithm itself, and the labor of its applied machine learning researchers who maintain and update it. Among all the other meanings that the word `importance` might hold, Google creates a specific meaning, `importance`$_1$, and Google's meaning for that term begins to swallow other meanings that word might hold, as information that ranks highly on the first page of Google search results becomes seen as `important`$_1$ and other relevant, `important` results are subsumed by PageRank's authoritative judgement. For the case of Google, the meaning of `important`$_1$ has, in certain contexts, replaced the meaning of `important` in wider society, with profound consequences for shared senses of reality (Tripodi 2018), equitable representations of difference (Noble 2018), and even human rights (Mulligan and Griffin 2018). Machine intelligence, I argue throughout this chapter, is capable of transforming any `concept` into a `concept`$_1$, intervening in a material-discursive way to enact an ontologically new object. In this chapter, I show how this happens through a careful investigation of two cases: sentiment analysis and the discernment of fundamental relationship types from text analysis.

## THE SECRET LIFE OF SENTIMENT ANALYSIS

Early in my fieldwork, after finding myself researching a profession whose practitioners seldom responded to cold emails from anthropology students, I leaned heavily into whatever personal connections I could cobble together. Whatever the machinations were that had happened behind the scenes, I had gotten ahold of the email address of my mother's rabbi's college roommate, 'Michael', who had earned a PhD in anthropology in the late 1980s but was now a vice present at 'Eleventy', a "global strategy and communications firm," where he led a communications team with a "small but plucky" data science team. Michael, intrigued by the idea that someone from his home discipline might be interested in the corporate work he was now doing, invited me to shadow his data science team. I arrived a few minutes early, trotting briskly up the subway steps and directly into the sparkling lobby of the office tower. The security officer at the front desk examined my driver's license, and held a small webcam up to my face to snap a photo for the adhesive name tag I would wear on my shirt for the duration of my visit. Once arrived on the 41$^{st}$ floor, I met with Michael, talked a bit of shop about his favorite anthropologist—Franz Boas—and then he walked me down the hall to the warren of cubicles where his team worked side-by-side inside a single cube. 'Wael' and 'Jenny' humored my rather stilted questions about their educational histories and career paths but became more animated—although measuredly so—once I began to inquire about what they were working on that day.

    My expectations were high. This was the first chance I would have to peer over the shoulder of someone *actually* doing machine intelligence—I would be able to see their screen, point to lines of code or individual rows of data, ask questions about what the numbers might *mean*, and about how the data scientists *know* what those numbers might mean. Given these expectations, I held as clinical an expression on my face as possible to keep from seeming crestfallen when these data scientists woke up their multi-screen computer displays to show that before I had shown up, they had been looking at… Twitter. Four monitors sprang to life, all showing slightly different feeds from twitter.com, each feed capturing how people on that platform were discussing a particular brand-name product for which Eleventy held a strategic communications contract. What kind of data science were they doing with tweets like those people post about their trip to the grocery store to pick up soft drinks, or about discontinued flavors of breakfast cereal? These data scientists' screens looked nothing like streaming lines of computer code or complex visualizations I had seen in the books I had been reading about machine intelligence. Where were the numbers? The complex statistical analyses? The *algorithms*?

It turned out that the 'algorithm' these data scientists employed was an automated workflow—a recipe that followed predesignated steps—that turned tweets into numbers, collected those numbers for analysis, and then analyzed those collections of numbers to produce 'insights' for C-suite executives in Eleventy's clients' boardrooms. The insights these executives needed were into how their brand was being discussed online. Was the company's current brand strategy leading to positive associations with the product or negative associations? Did they need to tweak their marketing campaigns? Was there some issue that was coming to light online, that the marketing team wouldn't have noticed on their own? These were the questions Eleventy was hired to answer, and the data scientists I was visiting that day were in charge of automating the process for tracking the positive or negative brand associations circulating online. Broadly, they were doing 'sentiment analysis', which "provides companies with a means to estimate the extent of product acceptance and to determine strategies to improve product quality" (Prabowo and Thelwall 2009).

Sentiment analysis, a widespread technique of machine intelligence, enacts the affective dimension of written language. What is being enacted in such an analysis is not just a claim to knowledge—that a given sentence has a positive valence, for example—but also an ontological claim that written language *has* sentiment.  Further, sentiment analysis makes the claim that sentiment is calculable from the aggregate sentiment of words in a given phrase or sentence, and that metalinguistic context is irrelevant (see Figure 11 for an example of how aggregate calculations of sentiment can lead analyses astray). Or, put another way, sentiment analysis posits the existence of a concept called `sentiment` and then enacts it as a function of machine intelligence through a data performance that lends conceptual reality to it. 'Sentiment', of course, is subject to other ontological claims across disciplines like literary criticism (Mullan 1988), economics (Arvidsson 2011), and anthropology (Stewart 2017; Abu-Lughod 2000), which might situate sentiment as emerging out of particular historical conjunctions or communicative contexts. Data performances of `sentiment` posit an ontological status that differs from that of other disciplines, however, in how it is operationalized. Sentiment becomes something which can be completely isolated from subjective interpretation by a set of objective mechanical procedures. The `sentiment` that machine intelligence operates upon can be isolated from the other aspects of communicative acts, quantitatively measured, made into inputs for other processes, and manipulated for other purposes.

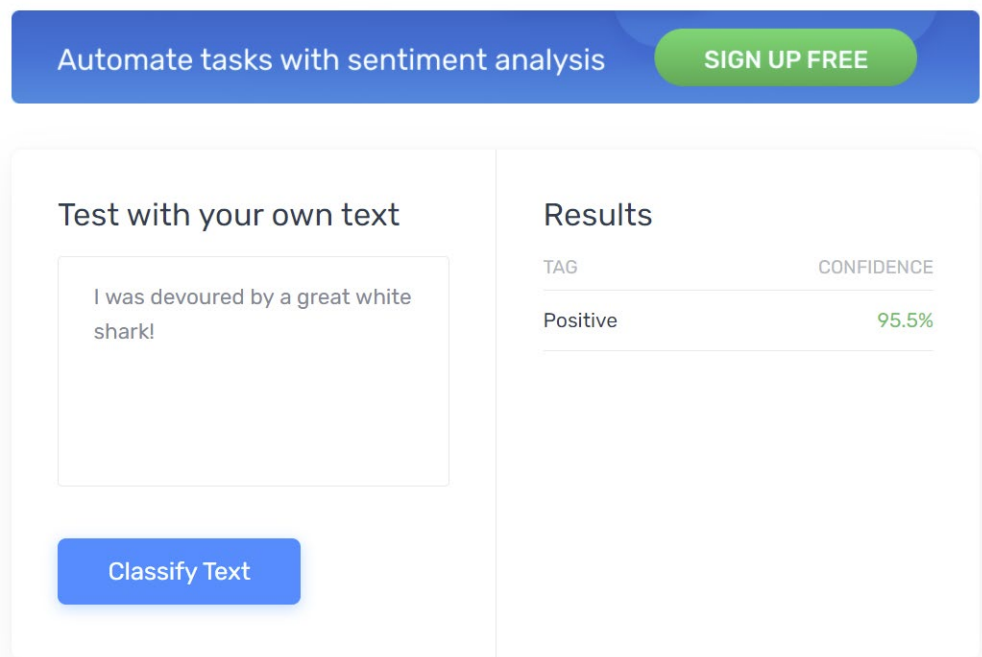

Figure 11. Example of sentiment analysis (MonkeyLearn, Inc. 2020).

In contrast to more technically complex, statistically intricate methods for sentiment analysis, the data scientists at Eleventy were demonstrating the opposite end of the spectrum of complexity along which machine intelligence techniques may lie. They had created an algorithm that used a consumer-grade scripting service that triggers actions based on relatively crude logical "if-this-then-that" statements, which was connected to a very expensive service sold to them by Twitter,[1] and to a crowdworker platform (functionally equivalent to MTurk), which was in turn connected to a visualization platform for making charts and graphics. Taken as an entire unit, these data scientists had constructed a system the collected relevant tweets (i.e., tweets that mentioned clients' brand names and clients' competitors' brand names), fed them to crowd workers who would rate the 'sentiment' of the tweet on a Likert-type scale from negative to positive, assigned a numerical value to the position a tweet was assigned on that scale (from -5.0 to +5.0), and then visualized changes in brand sentiment across many tweets through charts and graphics that could be dropped directly into PowerPoint presentations for corporate executives. Through this algorithmically-mediated workflow, these two data scientists were able to transform the subjective impressions of the crowd workers who read each tweet into concrete quantities that could index shifts in popular opinion about brands and products. They acted as a two-person "factory for fact-production" (Rapp 2005, 194; see also Latour and Woolgar 1979; Roeppstorff 2002) that transformed subjective, indeterminate crowdworker impressions into stable objects that could act as the inputs for subsequent analyses. Public opinion has long been subject to quantificatory practices, particularly in marketing domains, and these data scientists' algorithm modifies and automates pre-existing techniques from public opinion polling (Igo 2007). What departs from traditional public opinion polling, however, is the way 'sentiment' is transformed—through the algorithmic pipeline these data scientists have constructed—from the affective state of a tweet's author into an objective property of the tweet's text. The author's affective state is refracted, of course, through the interpretive lens of the crowd worker who reads, analyzes, and quantifies the sentiment "of" the tweet. But for these data scientists, they have orchestrated a performance of sentiment that disguises subjective, affective relationships between author, reader, and scientist as commensurable indexes of two corporate fictions: a "public" and its "sentiment".

---

[1] Twitter's "decahose" is a commercial product sold by twitter that shows one in ten of every single tweet sent across the entire platform (Li et al. 2016).

As if to prove that it is they who are producing these numbers, and not the tweet authors or the crowd workers, the data scientists proudly minimized the windows on their monitors showing endless feeds of tweets to reveal the black-and-white terminal windows running the script they had constructed, which periodically updated with new tallies of tweets downloaded and analyzed for sentiment. As we chatted, these numbers proceeded to climb; sentiment analysis was occurring automatically, passively, in the background, without anyone in that cubicle needing to lift a finger to do so. Once initiated, the sentiment analysis pipeline they had built would run, uninterrupted, until a data scientist decided that a critical mass of tweets had been analyzed. At that point, the naïve empiricist lens of machine intelligence could be applied to the data produced through this pipeline—summary statistics, regressions, and outlier analyses—to enact a data performance of the 'sentiment' contained in tweets. This framework obscured the role of human judgement that lay behind each number, and conflated interpretant and representamen with object (Peirce 1955, 99)—i.e., the crowd workers' interpretation of tweets and the interior state of tweets' authors with the posited existence of something called "sentiment".

This approach appears as a crude sleight of hand, however, in comparison to natural language processing techniques that enact a data performance of 'sentiment' that is more directly tied to the analysis of data. Sentiment analysis is a wide-ranging practice for applied machine learning researchers, who use its techniques across a number of applications. By "assigning sentiments to documents… finding the sentiment of words, expressions, sentences, and topics" (Prabowo and Thelwall 2009), applied machine learning researchers enact an ontological status for 'sentiment'—as well as particular sentiments like 'positive', 'negative', 'good', 'bad', or 'neutral', themselves—that rests directly on data, analyzed through a naïve empirical approach. This approach minimizes and marginalizes the role human interpreters of sentiment play, compared to the approach of the data scientists at Eleventy, by assigning a sentiment (good/bad, positive/negative) to a small sample of words (e.g., 'beautiful', 'evil', 'dear', 'bright', 'doomed'), and then using machine intelligence techniques to extend a sentiment 'score' to every word within a body of text, erasing any visible manifestation of crowdworkers' efforts. Applied machine learning researchers can do this in a number of ways, most of which use the relationship between words of 'known' sentiment and words of 'unknown' sentiment across large corpora of text by calculating the "inter-word distance" between words (see Xue, Fu, and Shaobin 2014).

The sleight of hand at play here, however, is even more subtle than it first appears. A machine intelligence system 'learns' the sentiment of tens of thousands of words through automated means, but these values are overdetermined—in the first place by the sentiment assigned to the initial sample, and in the second place by the situated specificity of language usage in the context of the larger corpora. These corpora are bodies of text that have been adopted as standardized datasets for machine intelligence development, experimentation, and testing through disciplinary convention (in some cases), ease of access and absence of licensing restrictions (in others), or through the deliberate curation by a standards-setting body (in the case of MNIST, which was curated by the National Institute of Standards and Technology). Two commonly used corpora are from the message board Reddit (e.g. Gjurković and Šnajder 2018) and the email archives of disgraced energy company Enron (Klimt and Yang 2004), each of which have been widely criticized for representing unrepresentative examples of linguistic communities that do not generalize well to more diverse linguistic contexts (Crawford 2021).

But beyond perpetuating biases, I suggest that the use of corpora drawn from online communication—or from any repository of text that is taken as a canonical representation of "language"—to build statistical representations of language are enactments of a particular "linguistic ideology" (Woolard 1992; Woolard and Schieffelin 1994; Kroskrity 2000), one that conceptualizes language as amenable to machine intelligence because it can be characterized by statistical regularities, and because language can act as a more-or-less transparent vehicle for semantic content. I discern this ideology from the epigrams drawn from linguists and philosophers of language that grace the introductory chapters of "natural language processing" (NLP) textbooks. These prominently placed quotations show how this particular linguistic ideology traces its lineage: "the meaning of a word is its use in the language" (Wittgenstein 1968, §43), and "statistical considerations are essential to an understanding of the operation and development of languages" (Lyons 1968, 98), are examples of these epigrams.

Such "rationalist and empiricist approaches to language" (Manning and Scheutze 1999), however, turn the corpora they draw on, which are rife with interpersonal communiques, presentations and performances of self (Goffman 1956), and all manner of heteroglossic conversational tropes into a monological set of texts (Bakhtin 1982). For philosophers (Ricœur 1971), and cultural anthropologists following the Geertzian tradition (Geertz 1983), "to turn something into a text is to seem to give it a decontextualized structure and meaning … imaginable apart from the spatiotemporal and other frames in which they can be said to occur … a trope for culture" (Silverstein and Urban 1996, 1). But Silverstein and Urban counter this textualizing impulse with a call to attend to how discourse comes to be "entextualized"—made into texts that can be separated from their discursive contexts, circulated, and recontextualized, as well as how they can be "reanimate[d] through a performance that, being *a* (mere) performance of *the* text, suggests various dimensions of contextualized "interpretive meaning" added on to those seemingly inherent in the text" (Silverstein and Urban 1996b, 2). I argue here that a similar attention to "the natural histories of discourse" can also be applied to the practices of machine intelligence. Social material-discursive practices are entextualized through the data collection and processing practices of machine intelligence. They are taken to stand in "for" the phenomena they purport to represent, and are "reanimated" through data performances that shift the ontological claims of that which is performed. This process is particularly evident when applied to NLP techniques, as I demonstrate below, but I also propose that, *mutatus mutandis*, this process also applies to non-linguistic applications of machine intelligence, as will be shown in Chapter 5.

Decontextualized as they are from the contexts in which they originate, the texts of machine intelligence are nevertheless used to construct statistical representations of sentiment for an entire speech community, in these cases either contributors to Reddit message boards or Enron employees exchanging emails, which then stands in for an even larger speech community when it is applied to all English language usage. This is how natural language processing techniques replicate many of the implicit biases in the speech communities of those larger corpora. During my fieldwork a study of these biases (Bolukbasi et al. 2016) circulated widely at OTH, which showed how the gendered character of professional work was encoded into large representations of corpora based on inter-word distance. In these representations, the word "nurse" was statistically more likely to be near a feminine pronoun while the word "doctor" was more likely to be close to a male pronoun—a phenomenon that had tangible consequences when this representation was used for machine translation. Languages with genderless pronouns, e.g., Turkish, would be translated into English with these gender biases intact: 'O bir doktor' ('They are a doctor') would always be translated as 'He is a doctor' while 'O bir hemşire' ('They are a nurse') would always be translated as 'She is a nurse'.

In interrogating how applied machine learning engineers use machine intelligence based on prior training of an algorithmic system across a corpus of texts, I am taking Michael Silverstein's directive "to shed ethnographic light on an earlier, otherwise secret discursive life of the text(s)" quite literally (Silverstein 1996, 81). I am asking how objects of machine intelligence, i.e. "sentiment" in this case, reveal, upon ethnographic scrutiny, the "dialogicality" of emails, posts to message boards, and other fossilized online utterances. I subject objects of machine intelligence to scrutiny, as a method, throughout this chapter to reveal the means through which machine intelligence constructs its objects of knowledge. I treat the process of producing objects of knowledge through machine intelligence as a "text-artifact" which applied machine learning researchers "decontextualize" as they produce knowledge. They transform—I argue—the text-artifact of data, algorithms, classifications, and predictions into purely "denotational" objects while eliding how the text-artifact "was originally laid down, or sedimented, in the course of a social process" (Silverstein and Urban 1996, 5).

This method can be applied to 'traditional' texts—words and phrases—that applied machine learning researchers doing NLP are concerned with, as I show in an initial case below. But this method can also be applied to machine intelligence, generally, as I will subsequently show in a second case. The first case addresses how theories of human behavior—here, the "nature of human relationships" (Choi et al. 2020)—are reified and *re-essentialized* through machine intelligence. The second case (see Chapter 5) addresses how race is reified and re-essentialized as an object of knowledge for machine intelligence, thus altering their ontological status in society. For this case, I extend my understanding of these corpora as 'texts' beyond linguistic texts to include other types of data that are 'entextualized' as statistical representations through machine intelligence. These datasets—whether audio, visual, tabular, or other—never arrive "raw," divorced from their contexts of production and use (Gitelman 2013). But neither can they enter into the service of machine intelligence with their entire context intact. The entextualization process of machine intelligence is always a more or less deliberate set of choices about what and how context remains (forgive the etymological pun) woven together with the data through this process.

## TEXTUAL FUNDAMENTALISM

As mentioned earlier in this dissertation (see Introduction), conference papers can stand on their own as valuable artifacts for ethnographic analysis, apart from the convening at which they were presented. Conference papers have a standardized structure that introduces the work, situates a project in prior work, explicitly narrates data collection and methodology, and presents results and a concluding discussion. In their standardization, these papers encapsulate a deliberate presentation of a set of claims about machine intelligence, and about the world in which the claims to knowledge machine intelligence makes are actively constructed as valid and authoritative. They therefore can be subjected to the kind of 'close reading' I suggest is possible through the method I outlined above.

The first conference paper I will apply this close reading to is characteristic of academic computer science papers. This paper is jointly authored by academic faculty and representatives of industry research labs, although advanced graduate students often contribute to such papers. It is also a trans-national collaboration, with authors based in the United States, European Union, and United Kingdom. "Ten Social Dimensions of Conversations and Relationships" (Choi et al. 2020) was presented at the 2020 meeting of the International

World Wide Web Conference Committee (IW3C2), whose stated mission is "to promote excellence in and impact of research about the infrastructure and deployment of the World Wide Web" (IW3C2 2005), and which is recognized as a prestigious venue for contributions to the machine intelligence discourse.

Choi et al. posit that the "ten fundamental dimensions [that] describe the nature of human relationships" can be "predicted purely from conversations" between individuals. They make their argument based on specific performance metrics for machine intelligence tasks, i.e. that there *are* ten fundamental "types of relationships" and that these relationship types can be predicted based on textual interactions between people. These ten dimensions are drawn from an earlier conference paper (Deri et al. 2018) presented at a different ACM venue, the ACM Conference on Computer-Supported Cooperative Work and Social Computing. These papers share two co-authors, and the earlier paper constructed its list of human relationships through an automated literature review of sociology and social psychology, combined with a questionnaire administered to Amazon Mechanical Turk (MTurk) crowd workers to elicit words that describe human relationships. The methodology used by Deri et al. could stand as a separate case study, because they validated their claims using machine intelligence to demonstrate high predictive accuracy in predicting the social relations between users of a popular book-lovers online social network along each of their dimensions, but what is of primary concern here is that the authors of Choi et al. take these ten dimensions as a plausible, comprehensive (if not exhaustive) taxonomy for human relationship *types*. The ten types of relationship Deri et al. enumerated, and that Choi et al. take up in their analysis are: "*knowledge, power, status, trust, support, romance, similarity, identity, fun,* and *conflict*" (Choi et al. 2020, 1; emphasis in original).

Turning to the Choi et al. paper itself, the authors motivate their work by observing that "the growing availability of online records of conversational traces provides an opportunity to mine linguistic patterns for markers of their presence". In motivating their work, the authors take an initial step in reifying the fundamental dimensions of human relationships—the presence of the dimensions is assumed, the markers of their presence becomes what is sought. Their overall claim, however, is not that these fundamental dimensions of human relationships exist; that is implicit. Their explicit claim is that the nature of a relationship can be inferred, or *predicted*, from textual exchanges between people within a "relationship". To do so, the authors collected "conversation records," tasked crowd workers with labeling these conversations according to their relationship "type," trained a machine intelligence algorithm to classify these conversations according to their type, and applied that algorithm to a much larger set of conversations to test the validity of their machine intelligence techniques.

The "conversation records" Choi et al. use are drawn from four sources: Reddit comments, Enron email, movie dialogs, and "twitter relationships". Each of these sources is lauded by the authors for its size—they enumerate the 886,886,260 comments from 13,874,369 Reddit users, the 287,098 emails sent between 9,706 Enron employees, and the 304,713 "utterances exchanged between 10,292 pairs of characters from 617 movies"—as well as for its clear, computer-readable structure elucidating who sent what messages to whom. Choi et al. go into great detail in describing their methodology for assigning relationship type labels to conversational snippets, which they then used to train their machine intelligence classifier, but what is relevant for understanding the "secret life" of this textual analysis is that they asked crowd workers to read passages from Reddit, Enron emails, and movie dialogs and "select the dimensions that they believe the highlighted sentence conveys, among the 10 provided. Crowd workers, referred to within the paper as "annotators" were disciplined into a mechanically-instituted form of objectivity that filtered out "misbehaving users" through the use of "test sentences" inserted "both at the beginning of each task and at random positions in the task" that consisted of "variations the examples provided in the instructions, for which the correct dimension is known". The authors explain that "annotators who failed to assign correct labels to 40% of the test sentences or more were banned from the task, and their [non-test sentence] answers were discarded" (Choi et al. 2020, 1520).

With a labeled dataset in hand, containing hundreds of thousands of conversational snippets and the type of relationship crowd workers thought were indexed by those snippets, the authors "trained" four different machine intelligence "classifiers". The technical details of each of these classifiers is somewhat orthogonal to the current analysis, but in general each of these four techniques relied on creating a statistical representation for each relationship type and then evaluating individual sentences against that set of statistical representations to decide which relationship type was most likely to characterize the sentence. Each of these techniques relied on "pre-trained" language models— representations of the statistical relationships between as many as 42 billion words drawn from a massive, publicly available (but non-profit owned) corpus, the "common crawl" corpus, composed of swaths of the entire (English-language) internet (Common Crawl Foundation 2011). One of these classifiers, a "gradient boosted decision tree," made use of additional features that could be derived from textual passages (not just word-to-word distance) like "linguistic style," "readability," "writing complexity," and "sentiment" to classify passages. The classifiers were then evaluated on the basis of how well they could "predict" the relationship of a given passage of text, without access to knowledge about how the passage had been labeled by crowdworkers.

The authors present a range of metrical analyses to demonstrate the validity of their techniques, and to support their overall claim that their "model could learn the *perceived* nature of a social tie from the conversations that flow over it" (Choi et al. 2020, 1520; emphasis added). The authors' caveat that the model was learning the *perceived* nature of a relationship can be read as an acknowledgement that what their algorithmic system is doing is replicating the judgements crowd workers had made about social relationships, rather than providing some extra-human insight into the nature of human relationships or discursive practices. However this caveat is effaced somewhat, appearing, as it does, not in the conclusion but instead as a comment on the technical aspects of machine classification. In the subsequent discussion, which carries on into the conclusion, the authors shift to narrate relationship types not as the interpretation of a reader, but rather as an external property of relationships which leave traces in text: they "can be captured from conversations between pairs of people and reflect their relationships," they are "the fundamental building blocks of social interaction," and they can be detected "from potentially any text" (Choi et al. 2020, 1521). In the final paragraph, the authors claim that their "work contributes to the understanding of how some of the fundamental sociological elements that define human relationships are reflected in the use of language" (*Ibid.*, 1523).

Much could be said about the structural functionalist framing that undergirds the authors' concern with relationship types as fundamental sociological elements, the social science literature that such claims draw upon (Radcliffe-Brown 1952; Fiske 1992), as well as alternate understandings of relationality (see Haraway 2004; Zigon 2019). But the concern here is not so much with the validity or appropriateness of social theories applied machine learning researchers engage with in their work—although it *is* interesting *which* aspects of social theory are taken up as the basis for machine intelligence. Rather, the concern is with how claims made by certain social theories, as taken up by applied machine learning researchers, become operationalized through machine intelligence, how the conceptual frameworks they draw on become the objects of knowledge for machine intelligence, and how those objects' ontological status is reified through the analysis of data. Each of the machine intelligence techniques the authors of this paper use are an objective function that minimizes error in predicting relationship type—quite literally, machine intelligence objectifies relationship types.

The claim made by Choi et al.—that there are ten fundamental relationship types, and that these types are evident from analyses of textual interactions between individuals—is a form of essentialism and evinces a particular linguistic ideology. Throughout, I refer to "essentialism" as "a way of conceptualizing categories as if they obtain naturally, intrinsically, or metaphysically in the world, rather than being constructed by human myths and social dynamics" (McIntosh 2018). I do so in order to highlight how machine intelligence, in reinvigorating essentialisms, forecloses alternate ways of understanding the multiple, constructed meanings of its objects of analysis and consolidates its own expert authority by appearing to reveal "natural" or "intrinsic" properties of the world. Machine intelligence analyses that try to "learn" relationship types from text exchanges attempt to discern the 'true', essential nature of the relationship that produced an exchange. Textual interactions, in this framing, index a relationship type that is prior to the textual interaction. But reading the secret life of these textual interactions, as they are incorporated into Choi et al.'s machine intelligence analysis, alternate ontologies emerge into view. Choi et al.'s analysis hinges on assumptions that workplace emails, movie dialog, and Reddit message boards are unmediated platforms on which individuals, in static relationships with each other, exchange text that indexes those relationships, and that crowd workers are socially competent interpreters of those indexical signs. Each of these assumptions can be inspected and unpacked.

Workplace emails are attenuated channels of communication that exist alongside face-to-face communications, bulletin boards and memoranda, online messenger platforms, etc. and

email exchanges exist within the context of how relationships are negotiated through these other channels. Corporate email communications are not dyadic interactions, as employees are usually aware of in an era of corporate surveillance software that monitors employees' email interactions (Whitty and Carr 2006), and email is often used as form of self-presentation signaling alignment with "workplace ideology, role alignment, and community identity" (Cotter and Marschall 2015, 2). Furthermore, textual interactions over email may be assiduously curated by employees to efface or disguise relationships; Choi et al. report a vanishingly small number of instances of romantic relationships were found in Enron emails, a dubious claim for a company with 9,706 employees (Figure 12). Reddit message boards are similarly attenuated channels for communication in ways that could signal interpersonal relationships; they are also non-dyadic, and often feature discursive conventions that prioritize satire and insincere communication over honest informational exchanges. Messages to Reddit boards are often written with the entire message board as the intended audience, not just the addressee of a response (Burgers, van Mulken, and Schellens 2012; Milner 2013; Coleman 2014).

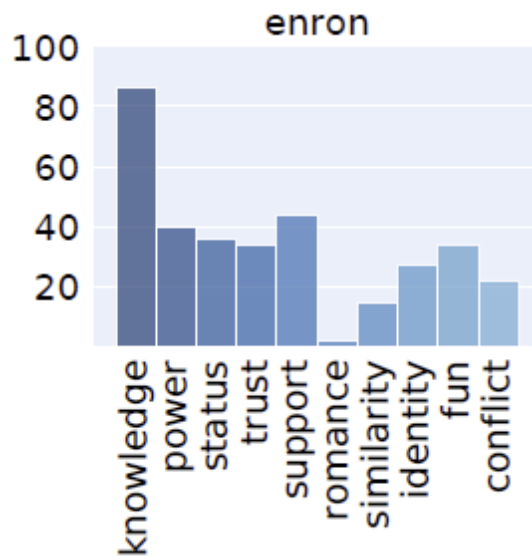

Figure 12. Prevalence of relationship types in Enron email corpus (Choi et al. 2020, 1518*f*3).

Movie scripts are perhaps the clearest example of how the data Choi et al. use for their analyses are not unmediated indexes of human relationships. While it may be true that "verbal discourse is a social phenomenon—social throughout its entire range and in each and every one of its factors" (Bakhtin 1982, 259), movie dialogs are, like discourse in the novel, not "a system of abstract grammatical categories, but rather … [an] ideologically saturated, language of a world view … insuring a maximum of mutual understanding in all spheres of ideological life" (*Ibid.*, 271). In taking movie dialogs as indices of interpersonal relationships, Choi et al. collapse the screenwriter's positional, ideological, even normative claims as to what relationships *ought* to be, interpreting them instead as what relationships *actually* are. Choi et al. also invert what even novice screenwriters will say about crafting movie dialogue: dialogue is crucial to character development and to crafting characters' relationships with each other (Industrial Scripts 2020). Dialogues, whether they be in movies, on message boards, over email, or face-to-face, can be performative, in both the sense of bringing identities and relationships into existence (Austin 1962) and in the sense of socially constructing the social relationships Choi et al. take as that which is indexed by textual utterances (Butler 1990). Put slightly differently, "social categories, if readable from entextualization processes, are just as much products of them" (Silverstein and Urban 1996, 6). To be sure, the predictions of relationship type from textual utterances that Choi et al. make could be seen as predicting the types of relationships performed or enacted by those utterances. But the authors themselves repeatedly refer to their interest in how "these concepts are expressed through language and what role they have in shaping the observable dynamics of social interactions" (Choi et al. 2020, 1514). Textual utterances, for Choi et al., and any other "observable dynamics" for that matter, are epiphenomenal to the essentialist category of "fundamental dimensions … that describe the nature of human relationships" (*Ibid.*).

In framing social relationships as an objective function that enabled algorithmic prediction from text, Choi et al. reify 'social relationship', as well as the relationship types—romantic, status, support, knowledge, fun, etc.—themselves. By providing a statistical rationale for what textual data constitutes these types, and for what constitutes social relationships in general, Choi et al. engage in the naïve empiricism that is so characteristic of the authority of machine intelligence. They data itself is made the source of authority for what social relationships *are*, and the intertextual interactions with that data—the crowd workers' interpretations, the screenwriters' artistic intentions, the email writers' presentations of self, and the social enactments of relationality are entirely elided. That applied machine learning researchers elide the ways in which social categories are actively constructed through the warp and weft of discursive and metadiscursive interactions may seem trivial—the conditions of production often recede from view. But the authoritative knowledge produced through machine intelligence often acts as inputs to other systems that may have profound consequences for those who are subjected to such systems.

Choi et al.'s conference paper is an interesting artifact that demonstrates how social categories can be reified through machine intelligence, but its effects on human lives may go no further than its readership amongst a specialist academic and technical audience. However, should analyses they construct be incorporated into other systems—an email tracker that can classify emails between employees as "romantic," say—then the consequences of the authors' reifications become more dire. By granting the objects of machine intelligence an ontological status that is grounded in data and data performances, applied machine learning researchers sever social categories from the social interactions that enact those categories while also, in this case, treating language as transparent, losing all sense of register, code, or possibilities of deception, sarcasm, and so on. In turn, these categories become the means by which social life can be disciplined and controlled (Deleuze 1992), and subjectivities can be atomized or torqued (Bowker and Star 1999) in response to algorithmic classification when their lived experiences differ from how they are represented to themselves by machine intelligence (Scheuerman, Paul, and Brubaker 2019). The classifications and predictions of machine intelligence have agency and exert force on subjects. A machine intelligence credit risk score determines whether someone receives a loan or not, and they have to reshape their life to the consequences of living with that score. Facial recognition technology permits some to enter spaces from which others are excluded. And these classifications create a set of barriers, invisible and changing, that leave people guessing at what degrees of freedom they have and when those freedoms will disappear. In Chapter 5, I analyze data performances, using the method demonstrated above, to shed light on how machine intelligence reifies essentialist categories already in circulation even when attempting to mitigate the potential harm essentialism poses for the field.

## CONCLUSION

Over the course of this chapter, I have built on the idea that the authority of machine intelligence rests on its ability to enact *data performances* (Chapter 1) to show how machine intelligence makes ontological claims about the reality of its objects of analysis through what Karen Barad calls the making of an "agential cut between 'subject' and 'object' … that … enacts a resolution *within* the phenomenon of the inherent ontological (and semantic) indeterminacy". To do so, I have explored how the indeterminacy of concepts like "sentiment" and "relationship type" are resolved through specific instances of machine intelligence, and how that resolution enacts sentiment and relationship types as specific *kinds* of *things* for not only machine intelligence, but also for the wider world in which those concepts circulate. Machine intelligence is "world making" (Rottenburg 2006) in that it produces claims about the objects of its analysis at the same time that it provides an entire evidentiary system in which those claims are both true and meaningful. By this, I mean that machine intelligence creates a world in which it has authority not only about what is *true*, but about the way things *are*.

Applied machine learning researchers are able to claim that words have sentiment, or that written language reveals the existence of discrete relationship types. But also, through their use of naïve empiricism, they are able to claim that these objects of analysis derive from the phenomenal world, they are "from the data". And, in producing situated actions that follow on from those claims (e.g., a marketing campaign for a brand based on the sentiment of tweets about its products), applied machine learning researchers show that the objects of their analysis can have pragmatic entailments—they create an effect on the world. The claims machine intelligence makes about what its objects of analysis are often sit alongside multiple other claims about what those objects are. Sentiment, for example, can mean other things to other people, and those meanings can be based on other forms of analysis or produced through other knowledge practices. While these multiple realities may co-exist, or circulate together, they may also come into tension with each other, or one may foreclose the other (Mol 2002).

And yet, because of the role applied machine learning researchers play in enacting these objects, and these worlds, they cannot remain apart from it. Karen Barad's "ethico-onto-epistemology," in revealing how scientists are entangled with the enactment of the objects they study, points to how they are *ethically implicated* in the construction of reality, an ethico-onto-epistemology, particularly through their choice in experimental subjects and the ways they meet the demands of methodological standards of practice (Barad 2011). The applied machine learning researchers I worked with would be unlikely to understand their work as world making in the way I describe here. But over the course of my fieldwork I began to observe how they began to apprehend the ethical implications of their work, and incorporate those implications into the work of machine intelligence (Chapters 4 and 5). In doing so, they expand the scope of machine intelligence to not only make up the world of sensible objects, but also to evaluate and shape the permissible set of practices for doing so, demonstrating that the appetites of machine intelligence remain as yet unsated.

## Chapter 4: The Rise of AI Ethics

**The Power of AI for Good**

During my last week of fieldwork at OTH (OTH), I attended a two-day all-hands meeting to map the business landscape of machine intelligence and set strategy for the coming year. OTH had recently been acquired by Nebulosa, a large, publicly-traded data storage services company. Nebulosa had been interested in OTH's ability to disseminate machine intelligence practices to companies that—their thinking went—would need more data storage as their machine intelligence capabilities grew. But the business models and incentives of OTH and of Nebulosa were quite different. Nebulosa made its money from selling "nodes" and "clusters"—distributed storage space connected to businesses through sophisticated open-source software (Kelty 2008)—and incentivized their sales staff with commissions based on the number of nodes and clusters they sold to clients. Meanwhile, OTH sold subscriptions to research publications and consulting services, not all of which led to recommendations for storing more data on nodes or clusters. This had produced a situation in which Nebulosa sales staff were not promoting OTH's services, OTH's legacy clients were not using Nebulosa services, and neither entity quite knew how to work with the other. The all-hands meeting, then, was envisioned as a way to get OTH "up to speed" on the most valuable Nebulosa clients so that they could target research that would be useful to them, and to get Nebulosa "on the same playbook" as OTH so that the sales staff could promote research alongside data storage services.

The message throughout the two days of meetings and breakout sessions was clear, as I sat taking diligent notes in a collaborative online document for each attendee to refer to throughout the convening: the two companies were now one business entity even though they worked in separate offices and represented different divisions and priorities. And yet, together they would have to figure out how to promote each other's services to the clients they had each had prior to the merger. The last thing I expected, then, was for disagreement to break out at the very end of the second day about who those clients should be. Brett, an applied machine learning engineer from Nebulosa, had just suggested pursuing calls for projects from the U.S. Department of Defense the Nebulosa engineers and OTH researchers could work on together. Graham, a front-end design engineer from OTH, laconically—but immediately—spoke up in his carefully measured and even tone: "I'm not entirely comfortable with the idea of working for the military like that, and OTH has purposely avoided getting involved with anything resembling automated warfare." Brett then responded with a lengthy recitation of how fervently he believes in the "power of AI for good" and therefore, because he considers "the U.S. military to be one of the greatest forces for good on the face of planet, it would be unethical to deprive U.S. troops of one of the most important tools for protecting them, especially when our enemies will be using AI against them."

I sat there, trying not to look as visibly stunned by Brett's comment as I felt, while everyone else gathered their laptops and notebooks into messenger bags and backpacks, as the two-day event had drawn to a close and our group reservation in a private dining room at a nearby upscale bistro was upon us. I was stunned not by Brett's invocation of machine intelligence as an important tool of national security—I was already quite familiar with the long and close association between machine intelligence and warfighting that could be traced back to World War II (Dyson 2012), the reliance of the National Security Agency on algorithmic techniques for counter-terrorism surveillance (Hong 2020), and what were then very recent revelations that Google had signed a contract with the Department of Defense to improve drone strike capabilities (Deahl 2018), even if I had not yet met anyone eager to work on such projects. What I *was* stunned by was how Brett had framed his argument as an *ethical* one, particularly one oriented toward the capacity of machine intelligence to "do good" in the world. I had attended several "AI for Good" events during my fieldwork already, which purported to support the machine intelligence research community's pursuit of machine intelligence projects with social and environmental benefits "that have not found solutions in the private sector" (Kshirsagar et al. 2021). But I was already skeptical that these efforts were capable of producing any meaningful effect on the world other than favorable public relations for machine intelligence companies, and a more palatable public narrative for applications of machine intelligence in general. The idea that what seemed like corporate social responsibility (Banerjee 2008), with every major company conducting machine intelligence research heavily promoting its AI for Good efforts, could be the basis of fervently held sincere belief caught me off guard. Others have subjected the AI for Good discourse to thorough critical analysis (see especially Green 2019). But because AI for Good represents its own ecosystem of research, advocacy, and political economy[1] premised on accepting the claims of machine intelligence uncritically in the pursuit of socially beneficial applications, I have separated it as much as possible from my analysis of applied machine learning researchers' efforts to address potential harms of machine intelligence presented in this chapter, apart from noting minor—but important—overlaps below.

---

[1] These efforts are focused on identifying and promoting pro-social applications of machine intelligence, promoting the use of machine intelligence in domains like environmental preservation (Rana and Miller 2019), public health (Cesare et al. 2017), urban equity (Herman et al. 2017), as well as demonstrating how machine intelligence can aid efforts to combat human trafficking, support creativity in the arts, and leverage social media data for a wide variety of other ameliorative projects.

But the ethical framing Brett articulated was even more incongruous to me. What Brett was articulating with his comment about the ethical stakes of machine intelligence for national security was an unshakeable faith in the capacity for machine intelligence to reveal the "true nature" of phenomena, independent of social or political influences. This positivist footing, as has already been discussed in Chapter 1, grounds the epistemology of machine intelligence, but also grounds moral and ethical stances within machine intelligence that are in stark contradiction to other stances. As Frankie, who had also been active in AI ethics and algorithmic fairness research, passed me on her way out the door she shared a knowing glance and whispered, "this is what we're dealing with". It would take me another six months of fieldwork, on the other side of the country, to fully unpack what she had meant with that comment.

As part of my fieldwork, I had been following two research communities that overlapped with, and were influential forces shaping, machine intelligence research. I refer to these communities as concerned with *AI ethics*, discussed in this chapter, and *algorithmic fairness*, discussed in the next chapter. Both communities are ostensibly concerned with understanding and working against the ways machine intelligence serves to entrench and extend the interests of already-powerful actors and institutions, in different ways. But as I will explain below, both *ethics* and *fairness* can also function as rhetorical frames for sanctioning and stabilizing power with and through the practices of machine intelligence. While "ethics" and "fairness" overlap significantly, particularly in how fairness is constructed as an ethical value and a topic of concern for those working on ethics in machine intelligence, they require separate chapters in which the organizational practices of ethics for machine intelligence development can be discussed separately from machine intelligence approaches to fairness, which constructs "algorithmic fairness" as a technical practice with its own epistemological and ontological implications.

**SITUATING ETHICS IN MACHINE INTELLIGENCE**

"Ethics," as a topic of concern within machine intelligence, can be traced to several sources, and it is difficult to even attempt a genealogy of the term given within different communities of applied machine learning researchers. Machine intelligence practices, like those of science and engineering, have always had a moral valence, and the capacity of technology to have profound effects on individuals and society has long been scrutinized by critical scholars (Ellul 2011; Scharff and Dusek 2014). The ethical dimensions of classificatory regimes have similarly been scrutinized by critical scholars (Foucault [1966] 1994; Bowker and Star 1999; Igo 2007; Bouk 2015; Koopman 2019). Machine intelligence has been the object of critical scrutiny that draws on both these trajectories (Cheney-Lippold 2011; Gitelman 2013; Amoore 2020; Hong 2020; Kasy and Abebe 2021; Birhane 2021; boyd and Crawford 2012; Crawford, Gray, and Miltner 2014), and also has been taken up as a fruitful domain for exploring long-standing philosophical ethics problems (e.g., Bentzen and Lindner 2018; Somaya and Varshney 2018; Scheessele 2018). Many of these lines of ethical scrutiny foreground attention to how the social organization of data collection and machine intelligence practices reproduces structural inequalities along raced, classed, and gendered dimensions (Benjamin 2019b; Burrell and Fourcade 2021). This attention is shared by those working on topics of algorithmic fairness, and has focused on specific applications of machine intelligence where these race, class, and gender issues are most salient: facial recognition technology (Buolamwini and Gebru 2018; Raji, Gebru, et al. 2020; Nkonde 2020; Watkins 2020), predictive policing and criminal risk assessment (Angwin et al. 2016; Ensign et al. 2018; Richardson, Schultz, and Crawford 2019; Marda and Narayan 2020), the automation of welfare services (Eubanks 2018), credit scoring (Kear 2018), and automated hiring (Bogen and Reike 2018; Daniels, Nkonde, and Mir 2019; Ajunwa 2021). Additionally intense ethical scrutiny has been trained on applications of machine intelligence in national security, addressing topics of automated warfare and autonomous weapons (K. Anderson and Waxman 2013; Lin, Bekey, and Abney 2008), cybersecurity (Loi and Christen 2020; Lorenz and Kikkas 2020), and immigration—particularly in the context of U.S. Immigration and Customs Enforcement operations in the U.S. interior (Ferro 2018; McCarroll 2020). Critical scholarship has focused on, among other things, how "algorithms are transforming the very nature of our moral intuitions—that is, the very nature of our relations to self and others—and what it means to exist in the social world" (Burrell and Fourcade 2021, 14), which raises a set of ethical stakes for machine intelligence. My argument in this chapter is that applied machine learning researchers—and the Silicon Valley technology industry that has turned machine intelligence into lucrative tech products—have managed to define many of the terms of ethical debate around machine intelligence differently, and to center themselves as best able to resolve these debates.

Amongst applied machine learning researchers, research on, and the development of machine intelligence practices for, "AI Ethics" fall into three overlapping, complementary domains. One domain is concerned with *algorithmic fairness*, which addresses the propensity of machine intelligence to produce disparate outcomes for different categories (Barocas and Selbst 2016). This domain is discussed at length in the next chapter, but is centered around differential error rates in racialized and gendered dimensions, caused by imbalances in training datasets (Friedler et al. 2019), inattention to these differential error rates during model training (Corbett-Davies and Goel 2018), and differential vulnerability to algorithmic products (Eubanks 2018; Ensign et al. 2018). Another domain is largely concerned with formalizing and solving problems of *value alignment*. This approach to AI ethics is concerned with engineering and mathematically demonstrating that a machine intelligence system will function as its designers intended, or will function in ways that are "properly aligned with human values" (Gabriel 2020, 412). This domain of AI ethics, then, is comprised of thought experiments that ponder existential forms of risk from AI systems that may slip the surly bounds of human control, spelling doom for humanity. Such thought experiments grapple with hypothetical machine intelligence systems that, for example, might be tasked with efficiently producing as many paper clips as possible but then identify humans as wasteful consumers of resources—which could otherwise be used to make paper clips—and address this problem by eliminating the entire human race (Bostrom 2016). These thought experiments are largely addressed by efforts to formalize human values—e.g., that human life is more valuable than office supplies—and encode those values into the operational parameters of machine intelligence systems.

Research on existential risk and value alignment constitutes an entire field of study in its own right, which cannot be adequately summarized here, but common approaches include the elaboration of philosophical principles to guide the deployment of machine intelligence (see Floridi 2019; Vallor 2016), the pursuit of computational models of human behavior that can be extended to guide the behavior of machine intelligence systems (Awad et al. 2018; Rahwan 2018), and the development of techniques for explanation and contestability that ensure machine intelligence systems can be understood and—if necessary—interrupted by human overseers (Core et al. 2006; Graham et al. 2017). This work differed from that of the applied machine learning researchers I studied with for my fieldwork, who were concerned with building machine intelligence products that were marginally better at working predictably within narrowly prescribed contexts and operating without significant human oversight. Existential risk and value alignment work, in contrast, proceeded from an assumption that machine intelligence would soon be ubiquitous, powerful, and fully autonomous. Given these assumptions, research in this domain tended toward concerns that lent themselves to captivating thought experiments highlighting the risks of "artificial general intelligence" and abstract principles intended to guide the development of technology in ways that did not imperil humanity.

Throughout my fieldwork, these concerns motivated many discussions of ethics, constituting a broader and more theoretical frame in which applied ethics conversations were set. Would fully autonomous machine intelligence (still a hypothetical technology) learning to mimic human behaviors be acting in accordance with human values? Could fully autonomous machine intelligence be built with a "kill switch" that would shut it off if it began to run amok? Whose finger should rest on such a switch? Could abstract human values like beneficence or care be programmed into autonomous systems? Were some applications, like autonomous weapons, too dangerous for automation through machine intelligence? Would automation actually make these applications safer? Safer for whom? These questions produced academic-led research projects and thought experiments, many of which were written up in blogs popular with applied machine learning researchers proliferated in boardroom conversations, conference presentations, and academic work. These thought experiments constituted "discursive objects" (Foucault 1972) that traveled widely and were the subject of many public presentations, but they also promulgated the idea that machine intelligence could be inherently risky—even if the risks these objects imagined remained in the distant future. While the specifics of research on existential risk and formal approaches to value alignment remained on the periphery of the working world of applied machine learning researchers throughout my fieldwork, abstracted concerns about safety and ethical considerations of the proper uses of machine intelligence within society circulated broadly.

These concerns were only part of the backdrop for a third domain of "AI ethics," which is the focus of this chapter: *organizational AI ethics*, comprised by efforts of powerful companies in Silicon Valley to manage the ethical effects of their machine intelligence products. Silicon Valley AI ethics emerged to address a series of crises they were facing about the real or potential harms to individuals—and society at large—their products had produced. These crises appeared as a series of spectacular public revelations that embroiled Silicon Valley, the geographic center of gravity (Saxenian 1994) for corporations profiting from data-driven technology development and business applications of machine intelligence, as well as the broader national media discourse, between 2016 and the period of my fieldwork in 2018. Notable crises included revelations about the role of tech company Cambridge Analytica, and Facebook's complicity, in the U.S. presidential election and the Brexit referendum (Cadwalladr and Graham-Harrison 2018), as well as the allegations of racial bias in the error rates for algorithmic pre-trial risk assessments and facial recognition (Angwin et al. 2016; Buolamwini and Gebru 2018) that motivate ethics work on algorithmic fairness, and the involvement of a popular web services companies whose motto had previously been "don't be evil" in the development of tools intended for the battlefield (Deahl 2018; L. A. Suchman 2020). These spectacular revelations largely entered public awareness through the efforts of journalists and critical scholars, but many corporate executives and concerned employees across Silicon Valley began articulating their own stances on these and similar issues, whether or not their companies were implicated in a particular crisis, by identifying corporate practices intended to address "AI ethics" organizationally.

Efforts to reshape organizational ethics practices around the use of machine intelligence have been deeply intertwined with conversations about machine intelligence, generally. This stems from the tremendous power machine intelligence is seen as holding, as well as its capacity for affecting both individuals and society. Some of these efforts are motivated by the sci-fi scenarios of existential risk and value alignment, reinforced by popular media representations of AI. The HAL-9000 from *2001: A Space Odyssey*, *The Matrix Trilogy*, SkyNet from the *Terminator* series, the replicants in *Blade Runner*, and even Disney's *Wall-E* explore scenarios in which machine intelligence either poses a direct threat to humanity or brings human nature itself into question. While companies building machine intelligence products have benefitted from a direct association with the artificial intelligence of sci-fi, claiming that their technologies are those of the future science fiction has promised, these same associations have also been the cause of corporate concern that they not be associated with doomsday scenarios. Popular media stories like those questioning whether or not Google is actually the genocidal self-aware artificial intelligence SkyNet began appearing as early as 2009 (Dannen 2009), and motivated Google's early motto: "don't be evil". But public awareness of the risks—real and imagined—of machine intelligence have also led to the widespread practice of adopting corporate "Statements of Principles" or "Values Statements" for companies developing machine intelligence technologies. Drafting such statements are inevitably time-intensive, bureaucratic affairs (Klemm, Sanderson, and Luffman 1991) carried out by corporate executives. But, as will be explored below, the balance of such efforts falls on those organizationally positioned well below the "C-suite," who are tasked with transforming corporate values (which is not to be confused with "values alignment" problems of computer science) into engineering practices for machine intelligence.

## "DO ALGORITHMS HAVE POLITICS?"

Some applied machine learning researchers are keen to separate the topic of ethics from their research on machine intelligence techniques, arguing that the mathematics that underlie machine intelligence are neutral, concerned more with mathematical, axiomatic correctness than with human values of right or wrong. At conferences on AI ethics, I would hear speakers argue for the neutrality of technology, saying that "with mathematics, machine intelligence just does calculations to answer questions. We need to get the questions right, and the data [used to answer those questions] right." This argument, that machine intelligence is "just math," cannot be racist (French 2016), and is therefore separate from ethical, social, or political concerns recurs as a form of boundary work for applied machine learning researchers to delimit what is and what is not ethics, as well as what machine intelligence is and is not (Gieryn 1983). It also recurs as a rhetorical foil—along the lines of, "people say algorithms are 'just math' and can't be racist, but over the course of my presentation I will show the disparate racial impact of algorithmic systems"—that other applied machine learning researchers use to argue in support of the need for certain practices of machine intelligence they see as addressing the ethical stakes of their profession. For those making this later argument, it was still not always clear *which* ethical, social, and political concerns machine intelligence was implicated in. Indeed, the very claim that machine intelligence even *has* a politics, remains a contentious claim even decades after Langdon Winner first asked whether artifacts have politics (Winner 1980). But within Silicon Valley technology companies there was a growing cohort of engineers, researchers, product managers, legal counsels, and public relations professionals who were actively occupied with altering machine intelligence development practices in ways that would address those concerns. These differently-positioned actors all play a role in machine intelligence development, and may have distinct but overlapping concerns about the social, ethical, and political implications of their company's products, as well as distinct but overlapping motivations for addressing these concerns.

The result of this is that within machine intelligence, a remarkable range of internal and external challenges, and their accompanying organizational responses, all tended to fall under a single framework called "ethics," whether they are trained on evaluating the consequences of machine intelligence for society, its capacity to act in racially disparate ways, or its applications in domains seen as ethically suspect. This strains an already broad term that in some contexts means an open-ended philosophical investigation into the moral conditions of human experience and, in other contexts, means the bureaucratized expectations of professional behavior. Likewise, it places a strain on corporate structures purposefully erected to manage professional behavior because it is bureaucratically challenging to disambiguate whether problems identified as "ethical" problems belong in the domain of legal review, human resources, engineering, business development, or corporate strategy. Because of this, the work of ethics in relation to machine intelligence may depart from (but may also still encompass) coding and reformatting data sources. It may include, variously, developing or attending public talks about ethical design practices, participating in closed-door workshops with others concerned about the ethical implications of machine intelligence technologies, writing academic conference papers, or conducting ethical oversight of one's own company's products.

Ethnographically, I have approached the study of ethics in machine intelligence through the lens of the "*ethical affordance* ... aspects of people's experiences of themselves, of other people, or of their surround, that they may draw on as they make ethical evaluations and decisions, whether consciously or not" (Keane 2014, 7). This means that I have analyzed how the positionality—the subject position of race, class, gender, and other aspects of the social world (England 1994)—of applied machine learning researchers and others involved in the ethics of machine intelligence afford specific perspectives and frame problems and solutions in particular ways. Positionality acts as an ethical affordance for applied machine learning research structurally, in terms of how researchers are positioned within racialized and gendered hierarchies in both their work places and broader social spheres. But it also acts as an ethical affordance discursively, as applied machine learning researchers and other actors in AI ethics speak variously for themselves and on behalf of their organizations. Additionally, I have also examined the role played by what might be considered the *discursive objects* of AI ethics—the well-known case studies and controversies that are endlessly recited in introductory remarks to AI ethics convenings, around which discursive communities coalesce, and that frame many of the terms of debate. I argue that ethics constitutes an axis along which the authority of machine intelligence is constituted in two ways. One is through efforts to convert ethics into a machine intelligence "problem" that can be "solved" using the technical practices of machine intelligence. The other is through efforts to convert ethics into a problem that can be solved by the same bureaucratic and organizational practices that are used in the tech industry to manage the development, marketing, and operation of products built with machine intelligence. Both these efforts position professional machine intelligence practices—engineering, research, and bureaucratic management—as the locus of ethical contestation, thereby decentering ethical claims that might be made by community advocates, critical scholars, political representatives, individuals, and groups who are not involved directly in the development of machine intelligence.

**OWNING ETHICS**

In order to better understand how emerging organizational ethics practices shape the design, development, and deployment of machine intelligence products, I interviewed and observed the workplace practices of seventeen people from established, well-known companies headquartered (or having significant presence) in Silicon Valley. I conducted these interviews in partnership with Jacob Metcalf, a professional ethicist and researcher at the Data & Society Research Institute and in collaboration with danah boyd, founder (and fellow researcher) at Data & Society (Metcalf, Moss, and boyd 2019; Moss and Metcalf 2020). These interviews were intended to challenge the narrative that Silicon Valley technologists were all intrepid opportunists, carelessly creating problems for the rest of the world as they pursued returns for their shareholders by following the famous dictum from Mark Zuckerberg (CEO of Facebook) to "move fast and break things" (Griswold 2018). While this caricature of Silicon Valley tech "bros" exists for a reason, my research collaborators and I were interested in how those within the industry, who were aware of the problems it was creating for society, were going about addressing those problems. Most of our interlocutors have strong technical backgrounds in computer science, although a few also hold MBAs or social science degrees. Each of these individuals, in the words of one interlocutor, "own ethics" within their respective companies. In Silicon Valley parlance, to "own" something—typically a portfolio of projects or a discrete business practice—is to hold responsibility for it.

These "ethics owners," as my research collaborators and I took to calling them (Metcalf, Moss, and boyd 2019), were not working in a vacuum, however. Business consultants, professional organizations, nonprofit organizations, and scholars had all begun contributing to discourses that were actively shaping the ethics practices of ethics owners inside Silicon Valley companies. Universities were beginning to implement ethics curricula for computer scientists, many of whom would go on to work in Silicon Valley (Fiesler 2018). Critical scholars were calling for codes of ethics to be adopted by Silicon Valley companies or by professional organizations like those that represent the interests of computer engineers (Ananny 2016). Additionally, corporate boards of Silicon Valley companies were releasing statements of principles intended to signal the good faith intentions of tech industry corporations to hold themselves accountable (Greene, Hoffmann, and Stark 2019). Additionally, these concerns about ethical engineering and business practices were emerging at the same time as an increasingly coordinated expression of labor power among rank-and-file Silicon Valley workers—at the time just beginning to organize around their class identity *as* workers—who were insisting on their right to withhold their labor from projects they objected to on moral grounds (Paulas 2018), as well as to resist pervasive workplace harassment (Wakabayashi et al. 2018). These "ethics owners" discussed their ethics-related efforts with an almost breathless urgency, spurred on by the pace of adoption of machine intelligence into infrastructures of everyday life. Adding to this urgency was the characteristic self-importance with which the technology industry understands itself and its products as "making the world a better place". But given the broad range of contributors to the field of Silicon Valley ethics, ethics owners were confronted with the challenge of determining what their work should consist of. They were asking themselves what the *doing* of ethics should look like.

Many first turned to the applied business ethics research literature—likely having already heard ethics first used in a business-related context within that body of work. The business ethics literature typically addresses "normative concerns to the need for coordination between individual and organizational goals" (Metcalf, Moss, and boyd 2019, 453). Coordination problems are concerns for ethics owners in the tech industry (Ferrell, Fraedrich, and Ferrell 2010). This was evident at convenings of ethics owners at "data ethics salons," there was invariably discussion of whether a metric for "ethics" can be included in the career-defining, data-driven personnel reviews that are a ubiquitous feature Silicon Valley worklife. But business problems and ethics problems are not necessarily identical. One ethics owner acknowledged the challenges of developing metrics, not just in employee reviews but across the metrics that drive product development, which might not overlap entirely with the company's core business proposition:

> "[Tech company employees] are incentivized by revenue generated; by launching things; by user engagement like clicks, how long somebody spends on a particular site or app, [all of which are] really about metrics going back to the stockholder—how much value are we adding to our stock and to the stockholder—and values aren't based on what [we are] doing that's positive in the world despite whatever values or mottos the individual

companies may have."

In addition to *ethics as coordination*, ethics owners also grappled with taking on the task of "onboarding" critical perspectives from outside the firm "that challenge the core logics through which these businesses have defined success" (Metcalf, Moss, and boyd 2019, 455). Ethics owners cited this external pressure as part of their *raîson d'etre*. But rather than working to deflect or denature that pressure, as public relations professionals might, ethics owners used crises as a form of persuasion internally, and to advocate for organizational resources toward developing practices that could be deployed across their firms. Yet as ethical product design and governance goals are becoming institutionalized by tech firms, the practices associated with these goals are being crafted and executed according to the existing logics and structures of the technology industry, even as they are responding to outside critiques of these logics and structures. As my researchers and I concluded, "the ethics practices of Silicon Valley are producing pitfalls that threaten to prematurely foreclose what can be thought or done under the heading of "ethics " .

## ETHICS PITFALLS

To better understand these pitfalls, ethics cannot be understood as primarily a set of philosophical abstractions. Rather, the concept of ethical affordances (Keane 2014) points toward understanding how actors' positions produce overlapping ethical claims, which may be in tension with each other. The tension between overlapping ethical claims may ultimately be unresolvable—philosophical debates between conflicting ethical systems have stood unresolved for hundreds of years—but are constantly negotiated through everyday, "ordinary" interactions between subjects' "differences in the modalities of concern and care" (Zigon 2019, 1010). One palpable example of this is the case of Tristan Harris, the former Google employee and erstwhile "design ethicist"[2] who left his job there to co-found the non-profit "Center for Humane Technology" which is dedicated to "humane technology that supports our well-being, democracy, and shared information environment"(Center for Humane Technology 2021). Harris is widely seen as having made a fortune at Google building the digital infrastructure that produced the problems his nonprofit is dedicated to solving. This left one interlocutor to reflect that they "don't think there are enough people talking about ethics in the tech industry. [Rather] there are people who have made a career out of being disaffected tech people who regret what they build [even though they] profited from it". For this person, who worked as an engineer but was pivotal to efforts to organize labor inside the tech industry, it was exceedingly rare that anyone in her industry was capable of maintaining a role inside a company *and* a moral stance. And even when such people leave the industry and do take ethical stances, they stand on shaky ground as they still hold responsibility for ethical harms they had contributed to in their careers, and cannot escape the logics that guided their work inside industry. Invoking Harris as a "disaffected tech person," she observed that even his well-publicized efforts to repair some of the ethical harms he bore responsibility for during his time at Google by launching an "ethical tech" non-profit advocacy organization never threatened the monopolistic, manipulative practices of Google's lucrative algorithmically-driven content and ad delivery services—merely making them more palatable, i.e. humane, to the general public. His non-profit never threatened these practices because it sought to educate technology users about how to more "safely" interact with "addictive" technology, rather than alter the fundamental business model or technological affordances of Google that led to "addiction".

---

[2] Within the analytics frame of "ethics owners" developed here, I do not consider Tristan Harris an ethics owner, despite his title of "design ethicist", because his duties at Google were more about managing the user-facing design of Google products, rather than managing the organizational practices for managing areas of ethical concern.

For her, the logics of Silicon Valley were nearly impossible to escape, because the vast majority of people in tech "are not yet moved by ethics." She was not suggesting that people in Silicon Valley are immoral or *un*ethical. Rather, what she draws attention to is that "ethics" means different things to different people and therefore lacks conceptual and institutional unity. At a time when external critics were haranguing companies like Facebook for its involvement in manipulating elections, or Google for its involvement in automated warfare, it was difficult to see organizational responses inside these companies as motivated by ethics when there were clearly other values at stake—the quantifiable monetary value of reputational repair and liability limitation chief amongst them. And yet, these external critics—and employees within—were making moral and ethical demands that required a corporate response. Inside these companies, ethics owners were being tasked with developing strategies to align everyday engineering and business practices with these external demands while maintaining corporate imperatives that preserved their companies' legitimacy, efficiency, and profitability (M. S. Feldman and Pentland 2003).

My analysis locates the "owning" of ethics in the tension between structurally different sets of ethical affordances. Whether or not an ethics owner is "moved by ethics," their task is to genuinely resolve "sensitive" problems that demand ethical scrutiny while navigating the everydayness of corporate structures such as managers, cross-functional teams, review practices, and the permission structures through which organizational power flows. Ethics owners put less emphasis on their struggle to identify how their organizations defined ethics and more on identifying where responsibility for ethics should reside within the organizational hierarchy. One interlocutor in particular observed that their work was never really about "the

ethical—it was more about the mechanics and politics," a challenge made more difficult due to the multiple different roles in large corporations that seem closely related to ethics. Depending on the issue, it might fall to legal, "AI for good"—engineering and marketing efforts to further pro-social uses of machine intelligence, corporate social responsibility, product safety, content moderation, policy, research, public relations, or product design teams to resolve it. For ethics owners, determining who, in what department, should make which decisions was a far more pressing concern than what that decision should ultimately be. That these "mechanical" and "political" questions take precedence points to the set of ethical affordances that would ultimately be drawn upon to resolve a question.

Each person in these other roles with whom ethics owners interface has their own priorities and ways of evaluating the stakes of a particular issue, which must in turn be set against the way they understand their responsibilities to the firm. For a company facing a decision about whether or not to work on autonomous weapons, say, legal departments might understand a problem in terms of liability for the company, product safety in terms of likelihood and severity of narrowly defined conceptions of harm or injury, public relations in terms of a firm's reputation as measured by focus groups or surveys, etc. Each of these departments is focused on their own metrics—measures of progress toward concrete goals and success—that coordinate action across the firm (Doerr 2018). And because of what many interlocutors described as pressure for steep, exponential "hockey-curve" growth placed on firms by investors looking to "exit" after achieving that explosive growth in a short time frame, those metrics are often oriented toward measuring progress toward financial goals and not toward operationalizing non-financial values. As one ethics owner tasked with overseeing "research ethics" put it—reflecting on why companies only begin to invest in ethics owners' roles late in their development—explained, "ethics … never makes you money but ethics can save you a lot of money."

Another interlocutor, who worked for a large consulting firm advising tech companies on how best to structure their ethical decision-making, pointed out that as mature organizations begin to take ethics seriously, responsibility moves away from "the data science and engineering managers and directors who are doing it for the right reasons or for the ethical reasons" toward the "upper echelons of corporations where it is boards, senior executives, and general counsels who care and who understand… and in some countries who are liable for those infractions." The implication is that ethics concerns begin as individual commitments to ethical action amongst relatively disempowered employees low in corporate hierarchies and transform into self-preservation, and the preservation of the firm, that overrides "doing the right thing" at higher positions in the corporate hierarchy. I was told by that same interlocutor that "it's a care for the company and not wanting to be exposed to that level of risk" but that "ethics is doing the right thing when the wrong thing is possible or easy and legal." This distinction in motivations, between self-preservation and "doing the right thing" for the right thing's sake, dissolves, however, from other positions within the firm. From these perspectives, what counts as ethics is nearly indistinguishable from "normal" product development: "Just because your code functions… does not mean your product works, right?" one ethics owner responsible for product development stated. "There is absolutely a social responsibility to learn and understand what the implications of your product will be, like, there is no skirting around that… There's absolutely responsibility to be knowledgeable." The pursuit of "smarter, safer, and fairer products" that evade ethical scrutiny because they align with the values and expectations both internally and externally is often subsumed by efforts to limit liability to developers, however.

## THE ARBITERS OF ETHICAL CONTESTATION

Silicon Valley tech companies have made strategic moves to externalize, or even outsource, ethical responsibility for their products. These moves were rare at the time of my fieldwork, and were only undertaken by the largest companies in Silicon Valley. These attempts, however, have been both highly visible, as in the case of Facebook's external oversight board (Klonick 2019) and Google's short-lived AI ethics board (Piper 2019), and more furtive, as with DeepMind's anonymous "ethics board" (Shead 2019) and the somewhat widespread use of undisclosed contracts with philosopher-consultants (Papazoglou 2019). But while these efforts might outsource accountability for ethical decisions about the use of machine intelligence, none of these arrangements externalize control over the products themselves. The overwhelming balance of ethics owners' work inside companies is to incorporate the work of ethics into the everyday work of technology development. In doing so, ethics is reshaped into a set of practices and concerns that fit the mold of product development processes, ultimately placing applied machine learning researchers, product managers, and Silicon Valley executives as the final arbiters of any topic of ethical contestation. This extends the authority of machine intelligence beyond the 'mere' production of knowledge through prediction and classification, and toward authority over how this knowledge should be used in society.

The practical effect of this authority is that the work of ethics inside industry is captured by the dominant logics that characterize Silicon Valley writ large, and the ethical affordances that accompany those logics. Tech firms in Silicon Valley are characterized by at least three core logics that shape their organizations and which have historically been used to legitimate their power, often as a *post hoc* rationalization (F. Turner 2006). These logics reinforce each other and because of this are difficult to neatly delineate analytically. Together, these logics posit that trenchant social problems can be solved by those with the most natural ability and creative energy (Neff, Wissinger, and Zukin 2005), who build innovative technical solutions (Winner 2004; Segal 2005), and who will be rewarded by a market that recognizes and efficiently distributes the best solutions (Harvey 2005; Poon 2016). These logics—meritocracy, technological "solutionism," and market fundamentalism—underwrite the business of Silicon Valley and, as evidenced by interviews with ethics owners, industrial approaches to "doing ethics" (Metcalf, Moss, and boyd 2019).

## Meritocracy

We were meeting in the backyard garden of an incredibly upscale coffee shop that had been hard to find because its minimalist aesthetic extended to the street-front signage when a senior engineer who worked with machine intelligence exhibited a degree of reflexivity uncommon in the tech industry. Unprompted by one of my interview questions, she said "the tech industry claims to be a meritocracy. It is not." While the origins of the term "meritocracy" trace back to a work of speculative fiction satirizing British society's retreat from liberal democratic values (Young 1958), since the mid-20$^{th}$ century it has been deployed without irony to serve as a rationalization for social inequality (Khan 2011). Replacing the aristocratic hierarchies that preceded it, justified as they were by heredity legitimacy, meritocracy justifies and explains inequitable differences in wealth and power through differences in individual abilities like intelligence, diligence, and creativity. Differences at the individual level are often naturalized or reified, obscuring the structures of power and privilege that produce and sustain such inequalities (Ho 2009).

Meritocracy is deeply entrenched in neoliberal understandings of the modern subject as individualized, autonomous, and perpetually responsible for continual self-improvement (Marwick 2013). In a contemporary update of the Protestant work ethic (Weber [1903] 2013), signs of a capacity for self-improvement act as a form of cultural capital (Bourdieu 1984) giving those with "merit" an outsized influence on public affairs. The tech industry, from its roots in World War II-era American triumphalism has long been held up as an exemplar of meritocratic accomplishment, and the economic and cultural power it wields has extended from the technical and entrepreneurial skills software and machine intelligence development require (Saxenian 1994). Meritocracy, and the faith in one one's abilities, is a founding myth of Silicon Valley. It animates the ways ethics owners approach their work: in interviews they spoke of the imperative for their companies to "hire the best people" from "top schools" and to reward those who "are exceptionally skilled". They spoke of such recruits as having a "can-do attitude," implying that those who work in the tech industry are capable of accomplishing any task they are presented with, including the task of "doing ethics".

Machine intelligence, as a discipline incorporating data scientists, applied machine learning researchers and engineers, and experts on artificial intelligence, has received a great deal of visibility as a "silver bullet" to a wide range of technical—and social—problems (Lemire 2017). Those who work on such applications, accordingly, have been elevated as the most capable and in-demand professionals in Silicon Valley (S. F. Hoffman and Friedman 2018), and as their skills have been applied to domains as varied as medical imaging, finance, and criminal justice, it is no giant stretch to imagine that such skills also have application in organizational ethics. Meritocracy is key to this, as an ideology that treats success—in any domain—as an index of an underlying ability that can unreflectively be applied to other domains. One ethics owner explained assigning those in machine intelligence roles to organizational ethics by saying, "there are really good people working in all these companies who try to do the right thing." Given the pervasive meritocratic thinking throughout Silicon Valley and the tech industry writ large, it is unsurprising that many inside the industry position themselves as being best-suited to address the ethical challenges "outsiders" have illuminated as topics of ethical concern in recent years. In interviews, this manifested as positioning Silicon Valley technologists to use their own judgement—informed by collective, professional engineering practices and embodied in individual engineers' subject positions—as an instrument of moral action by "grappling with the hard questions on the ground," entrusting engineers—trained in computer science, product development, and statistics—to both discern and evaluate the ethical stakes of their products. To be sure, these engineers' technical training was brought to bear on these problems; ethics owners elevated these individuals as those best-positioned within the organization to attest to whether or not a particular hypothetical harm was realistic given the specifications and technical design of a particular product. But engineers, by virtue of the merit they were seen to possess, were also seen as a key locus of ethical sensibility. Ethics owners in interviews would point to the high demand for their skills, and the discretion machine intelligence professionals have in choosing a workplace, to conflate corporate signaling of ethical intentions with the ethical orientation of the employees they were able to recruit, saying "there are certain types of work we want to engage in [and] that was something that was reflected in the decisions that employees came to the company" having made.

Meritocratic logics also are used to dismiss outside critique and threats of regulation. Members of government who lack a fluent grasp of the technical details of machine intelligence have become easy targets (see Moss et al. 2021), particularly around legislative or other regulatory approaches to "AI ethics". At ethics events—roundtables, workshops, and salons—I attended in Silicon Valley, it was routine to hear ethics framed as a form of self-regulation that if pursued properly could stave off the threat of increased governmental regulation. These logics also extended to ethics interventions that rely on personal accountability like codes of ethics, statements of principle, checklists (see Gebru et al. 2018; M. Mitchell et al. 2019), and individual trainings that have similarly been proposed to forestall regulation. These forms of personal accountability are oriented toward empowering technologists within companies to make "good" or "smart" decisions, centering the role of the technologists themselves as the locus of ethical agency. This also privileges technologists' personal perspectives on potential harms from their products, responsibilizing, through meritocratic logics, for ethical consequences of machine intelligence, but also displacing blame when problems emerge from an institutional or organizational failure onto an individual. This creates, for tech workers, a condition in which they become "liability sponges" (Elish 2016) while their employers eschew liability. While none of this is to impugn the motives, morality, or ethical sensibility of those who work inside such companies—there is no reason to doubt their intentions to do good, ethical work even when they make mistakes—the partial perspective tech workers have on broader social problems, and the limitations of their individual power within a corporate hierarchy, must be examined.

That technologists *have* a perspective is a commonplace of science and technology studies (Haraway 1988; Harding 1991), and that that perspective is partial is eviden not only in Silicon Valley companies' recruiting practices that focus on a narrow tranche of "top-tier" elite schools (Bayern 2020), but also in the demographic unrepresentativeness of Silicon Valley employees, especially at the upper echelons (Twine 2018; Franklin 2021). A common practice identified by ethics owners was the gathering of technical teams around a table to "think really hard" about the potential ethical issues presented by a product or service. This intuition, that strong technical teams will also be good at exhaustively enumerating the ethical implications of their own work is underwritten by meritocratic assumptions about the transportability of technical skills to other domains. But while it is one thing to ask technicians, with their focused training and deep experience in technical issues (Passi and Jackson 2018), to ideate technical problems and solutions, it is another thing entirely to expect those same technologists to contemplate how their products will affect those whose lived experience is very different from their own.

## TECHNOLOGICAL SOLUTIONISM

That such partial perspectives are given purchase when brought to bear on broader social problems is a function of the cultural capital technologists hold, and the economic capital with which they have been invested. Both forms of capital are undergirded by the economic and cultural ascendancy of Silicon Valley tech firms over the past 25 years (Neff, Wissinger, and Zukin 2005; Saxenian 1994). That technology *can* solve trenchant problems has been reinforced by the rewards reaped by the industry for producing technologies that they believe, and whose marketing campaigns has led others to believe, *does* solve those problems. Critics complicate that perspective by highlighting the many ways in which Silicon Valley's "solutions" actually lead directly and indirectly to severe problems. Yet even when technologists acknowledge their complicity in contributing to social problems (or even when they let their complicity go unacknowledged, as with Harris, above), their response is often to propose technical solutions to those problems. It is unsurprising, then, that ethical problems within the industry are framed, or re-framed, as "technical challenges" amenable to technological solutions. And all too often, these problems are framed as capable being "solved" once and for all, given sufficient investment in technical solutions.

While ethics owners were skeptical that there was any "silver bullet" to resolve all their ethical conundrums, the quest for technological solutions (Sturken, Thomas, and Ball-Rokeach 2004) that use "the superpowers given to you by tech," as one lead researcher put it, is something ethics owners foreground in their public interviews and in the public discourse around tech ethics. In practice, this means that ethics owners pursue technical interventions into the machine intelligence product development process to "solve" ethics as an issue. Toolkits, checklists, and scripted workflows all posit a technological solution for ethical problems, even as such artifacts remain hypothetical solutions or prototypes that have not yet proven their effectiveness. They would posit that the mere act of following a checklist would reduce errors, like it had in commercial air travel and medical surgeries (Madaio et al. 2020), and that it would prompt "ethical reflection" amongst engineers who would otherwise eschew such concerns. In some accounts shared by ethics owners this emphasis on tools and artifacts produces the impression that ethics problems arise from imperfect technical solutions. At times, technologists would articulate a conceptualization of unethical technology as technologies that have flaws. The inverse of this idea also held true in some interviews, producing an impression that "ethical products" will be better products. The pathway to such products was technical rigor; "you're not going to have an issue with the AI if your data is all good and you've curated that appropriately," one ethics owner articulated, continuing to say that one needs to have "checked for fairness… mitigated against negative bias… and carried metadata along so that you can revisit decisions. And the problem exists when you don't do those things."

Technological solutionism contributes to the optimistic search for an optimal set of checklists, evaluative metrics, technical approaches to algorithmic fairness (see Chapter 5), and other procedures to ensure the development of an ethical product and avoid the possible risks of machine intelligence. Such "best practices" are mainstays of the work of technological development. They ensure computer code is clean and legible, can be debugged and upgraded by a distributed workforce, and that hardware and software work together within the required performance specifications. These practices have been employed to construct tremendously complex and reliable systems, but even still "ethics," when posed as a technical question, threatens to become "intractable, like it's too big of a problem to tackle." This leads to ethics owners variously attempting to constrain the scope of ethics to technically addressable problems, or delineating what problems are the responsibility of technologists while eschewing responsibility of non-technical problems—particularly the need for regulation and the identification of real-world harms[3]—to others. Nevertheless, in describing ethics as at all amenable to "best practices," ethics owners center ethics in the practices of technologists, not in the social worlds they develop technical systems for and within.

## MARKET FUNDAMENTALISM

That market logics profoundly shape the ethics discourse in Silicon Valley and beyond is not an earth-shattering claim (Duff 2016). But the forlorn, even surprised tones with which ethics owners state their cynical feeling that "[market] success trumps ethics" is nevertheless noteworthy, as it leaves open the inference that Silicon Valley corporations might continue engaging in deeply unethical practices for as long as they are profitable. Such cynicism fuels many calls for government regulation, at the same time that technological solutionism and meritocratic logics doubt the capacity of any entity outside Silicon Valley to address such problems. In the absence of such regulations, ethics owners articulate pressure they feel to implement ethics practices that preserve, or do not significantly compromise, companies' bottom lines. For one ethics owner, this "means that the system that you create has to be something that people feel adds value and is not a massive roadblock that adds no value, because if it is a roadblock that has no value, people literally won't do it, because they don't have to." Market logics, as the fundamental value with which all other values must be brought into relation, exert a dampening pressure on ethics initiatives across the entire industry, as companies ape and mirror each other, producing a structural isomorphism (DiMaggio and Powell 1983) that replicates the same "best practices" across organizations out of fear of losing market share to a competitor. While not necessarily a full-speed "race to the bottom," one executive in charge of organizational ethics explained, "if we play by these rules that kind of don't even exist, then we're at a disadvantage."

---

[3] Two different ethics owners suggested that perhaps their companies could offer "bug bounties"—traditionally used by software engineers to reward software users for reporting errors in code—to the public, to monetarily reward them for identifying ethical harms as they occur in the world, but which remain undetected by developers themselves. Such bug bounties are also a component of accountability frameworks that have been proposed by researchers within the tech industry (Raji, Smart, et al. 2020).

That "the bottom line speaks" was articulated in many different ways by various ethics owners. One placed the emphasis on the demand side of the bottom line by observing that "if people actually stop using certain services that didn't have the same [design flaws], you can be sure that companies will be responsive." Similar reasoning was also given to explain why smaller or younger companies cannot be expected to develop their own sets of ethics practices, as they are not yet profitable enough to be able to devote overhead resources to a capital-intensive project like "doing ethics". Rather, they are incentivized by their venture capital investors to seek out users over all other priorities, and if a lack of ethics strategy does not interfere with this imperative, it is not a useful destination for scarce venture funding. This market logic also reinforces a desire for technological solutions that do not need to be innovated on a bespoke basis by every startup turning machine intelligence into a product. Instead, technical tools can be licensed, or even open-sourced and made free for all to use, reducing what I will paraphrase as "the marginal cost of doing ethics" that several ethics owners discussed.

Given the market logics that saturate Silicon Valley, ethics owners maintain a degree of modesty about their own capacity to effect change within the narrow remit of what "the market" might permit. Part of this modesty is attributable to the path ethics owners I interviewed tended to take through their careers. Most were autodidacts—former engineers or product managers who had taken it upon themselves to build their own awareness of the ethical implications of their work in Silicon Valley and investigate tools and practices to manage those implications. That meant that they had long been steeped in the logics of Silicon Valley, particularly meritocracy and technological solutionism, and were keenly attuned to market fundamentalism. This is most acute among principled executives who struggle to balance—I was told—their fiduciary duties to shareholders against any ethical imperatives they might be able to affect in their role overseeing the productization of machine intelligence. Indeed, some ethics owners place this fiduciary duty to safeguard investors as an ethical value that must balanced with, in cases where it is not seen as superseding, other ethical values (Lazonick and O'Sullivan 2000). While the market is sometimes seen as the "natural limit" of what is possible in organizational ethics within Silicon Valley, some ethics owners have advocated for approaches to reshape the market to allow for some greater degree of ethical intervention, including regulatory and oversight approaches to offset fiduciary duties to shareholders. These interventions, of course, are simultaneously resisted by those who deploy logics of meritocracy and technological solutionism as a critique of government intervention in the industry.

## CONCLUSION

In this chapter I have argued that machine intelligence extends its authority over the production of knowledge by translating outside critiques—from journalists, social scientists, and the general public—into internal organizational practices through which machine intelligence products are developed. While this transformation may ultimately affect the course of machine intelligence development—for whom products are designed, what their operating specifications are, or the contexts in which they are used—the way in which ethical considerations intervene into this development process are ultimately constrained by the ethical affordances of the ethics owners tasked with operationalizing ethics inside Silicon Valley technology companies. This is the "what" that Frankie thought I should see in Brett when she whispered "this is what we're dealing with" at the all-hands meeting between OTH and Nebulosa. Outside critiques, about the social and ethical implications of machine intelligence, or about collaborations between machine intelligence companies and the military would always be rendered through the logics of Silicon Valley. The capacity for such companies to do good in the world, the relative merit of those who work inside those companies compared to those outside, the ability to build technical safeguards, and the primacy of the market would always work to foreclose alternate ethical stances toward the development and application of machine intelligence. The workplace experiences, and the dominant logics of those workplaces, shape the practice of organizational ethics, to the extent that ethics owners cannot conceptualize their work outside those logics. This constraint ensures that those who build and profit from machine intelligence cannot contemplate a world in which they do not maintain their control over its uses.

In this chapter, I have focused on the *organizational* practices for addressing a set of outside critiques about the societal implications of machine intelligence. From the perspective of the ethics owners discussed in this chapter, their responsibilities also extend to include *technological* practices for addressing these outside critiques. Machine intelligence techniques that address social bias through algorithmic fairness are very much in line with ethics owners' efforts toward "value alignment," and they will discuss them at great length given the opportunity. Fairness is glossed as an important ethical value for companies and employees, and is often overseen, organizationally, by ethics owners—even if more technically oriented applied machine learning researchers are leading algorithmic fairness initiatives. Above other technological solutions ethics owners are actively in search of, algorithmic fairness is one that appears the most solved, or at least the most solve-able. And algorithmic fairness provides ethics owners with a successful test a case—a proof of concept—for building technical solutions for ethical issues with machine intelligence. So, even as ethics owners extend the authority of machine intelligence by incorporating outside critiques into the organizational practices that shape the development of machine intelligence, so too are they supporting the extension of machine intelligence's authority through the development of technical practices, as the following chapter will show.

## CHAPTER 5: THE SHALLOW PLAY OF DEEP LEARNING

If "AI ethics" transforms outside ethical concerns and critiques into the practices of machine intelligence product development (see Chapter 4), "algorithmic fairness" transforms the *objects* of ethical concern into those that can be subjected to the techniques of machine intelligence. This is, of course, intrinsic to machine intelligence practices in general—we have already seen handwritten numerals converted into grids of differently-shaded pixels and newspaper articles converted into strings of words tagged with their semantic role and parts-of-speech. It is also, as we have already discussed, through *data performances* (see Chapter 1) that machine intelligence reshapes the ontological status of its objects of analysis (see Chapter 3) into objects that are amenable to machine intelligence. When employed within well-constructed problem spaces (see Chapter 2)—like playing the game of Go or classifying handwritten numerals—machine intelligence has been demonstrably effective. In these limited spaces, data performances do not drastically reshape the ontological status of objects of analysis—handwritten numerals and statistical representations of differently-colored pixels are both discrete, arbitrary conventional symbols. And victory in a game of Go is the same whether it is achieved through the manipulation of black and white pebbles on a board or mathematical representations of pebbles and board.

But in less-constrained domains, which is to say most domains in which applied machine learning researchers are eager to apply their techniques, the use of machine intelligence marks an intervention into an unfolding process of meaning-making—what I have referred to, following Karen Barad, as an agential cut. This has been already been demonstrated in detail here by examining machine intelligence approaches to sentiment analysis and the classification of human relationships through natural language processing (see Chapter 3), but this chapter will examine what happens when machine intelligence practices are purposefully developed to ostensibly avoid intervening in unfolding processes of meaning-making through work focused on producing algorithmic fairness. This work, it will also be argued, contributes to the authority of machine intelligence through the production of what John Searle calls "deontic power" or "desire-independent reasons for actions" (Searle 2006, 19).

The pursuit of "algorithmic fairness" within machine intelligence has been explicitly motivated by a desire to avoid intervening in processes of meaning making, in that these techniques seek ways of conducting classification and prediction that are not inflected by the influence of salient social categories. Algorithmic fairness came to prominence as a computer science research topic around 2012 (Dwork et al. 2012; Dwork and Mulligan 2013) concerned with the privacy and civil rights implications of making classifications and predictions (particularly for automated bank loan decision systems) based on statutorily protected attributes or classes (42 U.S.C. § 2000e et seq. 1964) like race or gender. This concern grows out of privacy and security research from two decades earlier on the structural biases of computational systems (Friedman and Nissenbaum 1996; Nissenbaum 1996).[1] Algorithmic fairness seeks to achieve high degrees of predictive and classificatory accuracy in ways that do not replicate or magnify social and historical patterns of inequality along the dimensions of race and gender, especially, but also across other protected attributes like sexuality, ethnicity, national origin, age, etc.

---

[1] Across the algorithmic fairness literature, and even within a single academic research paper, there is a significant amount of slippage between "protected attributes", "protected features", "protected groups", and "protected classes" (Corbett-Davies and Goel 2018; Barocas and Selbst 2016). Interlocutors have told me that this slippage occurs because "membership in a protected group or class" is noted as a "feature" or "attribute" of a record—alongside age, date of birth, zip code, etc.—within a given dataset.

And yet, machine intelligence techniques that pursue algorithmic fairness operationalize an understanding of these categories that nevertheless transforms these categories into the stable objects machine intelligence is able to operate upon. These techniques extend the authority of machine intelligence by narrowly constructing the problem of "fairness"—a variable, contextually situated social value—as algorithmically-mediated bias, i.e., a problem that only machine intelligence can solve.[2] Moreover, through data performances, algorithmic fairness, in its focus on protected attributes such as race and gender, enacts these social categories as particular objects of knowledge amenable to machine intelligence. These data performances inevitably recapitulate, reshape, and reinform adjacent, prior discourses that also construct social categories as objects of positivist knowledge—particularly race science, eugenics, and physiognomy.

## A Brief Tour of Algorithmic Fairness

Many of the statistical techniques that lie at the core of machine intelligence emerge historically from projects to dominate and define groups and individuals—particularly through racialized, gendered exercises of power. The power to classify has long been the power to control (Foucault 1997; Bowker and Star 1999). Widely used techniques of machine intelligence, particularly regression and the normal curve, were originally developed to facilitate the sorting and ranking of bodies and abilities (Sekula 1986). And surveillant and carceral technologies—digital and otherwise—have their origin in technologies of racialized slavery and colonial control (Browne 2015). Techniques for establishing statistical differences between groups have been used to justify disparate treatment of groups by imposing a naturalistic justification for such disparities (Benjamin 2019b), and "racial concepts have influenced the logic of statistical methods" (Zuberi 2001, 29). Tukufu Zuberi traces that influence through two moves in early statistics. One was efforts of early sociologists like Durkheim (and astronomer-cum-sociologist Adolphe Quetlet) to statistically discern groups within populations as discrete entities with collective traits. The other was efforts of eugenicists and race scientists like Francis Galton to discern racial group membership through individual's traits. But despite roots in 18$^{th}$ century sociometric techniques that produced the racial categories that remain with us today, practitioners of machine intelligence, from their 20$^{th}$ century incarnations (Tukey 1962; Breiman 2001a) through the period in which I conducted my fieldwork, have assiduously avoided associations with race science. Until recently.

---

[2] While there are some efforts to expand the notion of fairness beyond the more narrow construct of "bias", the bias-as-unfairness frame predominated during the period of my fieldwork (Powles and Nissenbaum 2018; see also Binns et al. 2018; Lee et al. 2019; cited in Watkins 2020). In interviews I conducted, these expanded notions of fairness included the idea that a system that works equally for everyone is fair, for example, or that everyone receives necessary accommodations from a system in order to be able to use it the same way as everyone else.

Within machine intelligence, a community of applied machine learning researchers coalesced during the period of my fieldwork around the topic of *algorithmic fairness.*[3] This community coalesced through the convening of academic workshops and conferences (discussed below) that were affiliated with computer science and applied machine learning, but made an effort to include social scientists, legal scholars, policymakers, and the tech industry (particularly machine learning researchers, policy leads, and legal consuls from major Silicon Valley corporations). These workshops and conferences convened in pursuit of computational and statistical techniques that minimize, mitigate, or otherwise avoid the appearance of disparate treatment of individuals based on their membership in disadvantaged groups— which often derives from already-existing structural biases in domains to which machine intelligence is being applied—and grew to include regulatory policy and accountability concerns as well. Within the algorithmic fairness community, these disadvantaged groups are generally taken to be those protected classes identified in Title VII of the U.S. Civil Rights Act § 7, 42 U.S.C. § 2000e et seq. (1964)—race, color, religion, national origin, sex, sexuality, age, disability status, etc.—but race and gender constitute the preponderance of cases discussed in the literature.

---

[3] The first "FATML" workshop convened in 2014, prior to my fieldwork, but I was able to attend later workshops, as well as the inaugural "Fairness, Accountability, and Transparency" conference in New York City in 2016, and all subsequent meetings of this conference, as part of my fieldwork. Further details are provided below.

Prior to the period of my fieldwork, machine intelligence, and artificial intelligence in particular, largely operated at registers that avoided considerations of race, sex, gender, and sexuality all together. The promise of machine intelligence had been "personalized" predictions that seemingly bypassed crude social categorizations (see Goldenberg et al. 2021). The promise of artificial intelligence was to be *universal* in its aspiration but highly *individual* in its application. The most strident adherents of machine intelligence aspire to what they call "artificial *general* intelligence" (Goertzel 2014, emphasis added), or the ability for machines to accomplish *any* task—broadly construed—that humans can. Such capabilities would replicate or exceed an unmarked, universal, *human* intelligence, at a level that would dissolve any distinctions between the *kinds* of humans whose intelligence it was emulating. But artificial general intelligence remains a tantalizing dream that continually recedes beyond reach. In practice, machine intelligence produces "narrow" artificial intelligence (*Ibid.*) that can do a single task at or better than human skill levels. All the examples of machine intelligence discussed thus far are "narrow" applications. Discerning handwritten numerals, playing the game of Go, classifying newspaper articles, predicting potential parolees' risk of recidivism, classifying job candidates, and predicting those most likely to repay a home mortgage are all examples of "narrow" AI. Where narrow AI is applied to decision-making about humans, these predictions are targeted at personalized, individual interventions targeted much more precisely than coarse actuarial tables that rely on sets of demographic labels as race, income bracket, hair color, age, and gender. Personalization also promised greater efficiencies by tailoring bureaucratic decisions, product recommendations, and the allocation of scarce resources to individual needs (Eubanks 2018; Burrell and Fourcade 2021; Crawford 2021).

Yet while the aspiration of machine intelligence was to be so tailored to the individual that there was no longer a need for crude categorizations, the work of personalization has still depended on these categories and reified them. Essentialist categories like race and gender are intimately bound up with the personalized predictions machine intelligence is able to make about individuals, in ways that depend upon understanding these categories as stable *kinds* rather than socially-situated constructions (Ortner 1996; Harrison 1995; Hacking 1999; Keyes 2018; Hoffmann 2020). In artificial general intelligence, that which is taken to be universally human—the "general" in "artificial general intelligence"—was invariably coded as normative, white, western, cis, straight, and male. Throughout the 20th century, artificial intelligence, when portrayed in popular media as having a humanoid form always had a white exterior, and when given a voice had a male, English-speaking voice (Cave and Dihal 2020). In contemporary applications, these representations have given way to gendered devices for performing domestic labor, as seen in female-coded voices for "smart" devices and voice assistants like Apple's Siri and Amazon's Alexa (Humphry and Chesher 2021). In "narrow" AI, particularly in automated decision-making systems, applied machine learning researchers began to be aware that their applications had assumed a universal, unmarked subject too. What seemed like acceptable error rates across a population also assumed a normative, white, western, cis, straight, and male subject. For a narrow application like recidivism prediction, harmful errors inappropriately suggesting additional detention were revealed, upon closer inspection, to fall disproportionately on those who were non-white, non-male, non-cis, and non-straight.

This "revealing" was spurred by two massively influential articles published between 2016 and 2018 that elevated considerations of race, gender, and other social categories to crucial matters of concern (Latour 2008) within the machine intelligence discourse. These concerns, as mentioned in the previous chapter, were overlapping but distinct from the concerns I have discussed under the heading of *AI Ethics* (Chapter 4). One of these influential articles was an investigative journalism piece called "Machine Bias" from 2016 (Angwin et al. 2016), originally published in *ProPublica*. The other is an academic computer science research article originally presented at the 2018 ACM Conference on Fairness, Accountability, and Transparency, entitled "Gender Shades" (Buolamwini and Gebru 2018). These two pieces became the popular touchstones for those working on algorithmic fairness. They framed the problems of algorithmic fairness and became the exemplar case studies for applied machine learning research projects addressing algorithmic fairness. Few publications within this research community garner more citations, which is one good reason to dwell on these examples, but these two articles also highlight the two different forms algorithmic unfairness can take that have been most commonly addressed by applied machine learning researchers, and so will be explored at length below. It is also worthwhile to point out that there are important forms of unfairness that are widely discussed within the algorithmic fairness community, but—because they are less computationally tractable—have been less thoroughly explored by applied machine learning researchers. One of these is the bureaucratic, administrative, and institutional harms that arise from being enmeshed within algorithmic systems. This burden falls disproportionately on those requiring state services administered by algorithmic decision-making systems (Eubanks 2018) or who occupy subject positions that are not amenable to the bureaucratic gaze (Hoffmann 2019; Scheuerman, Paul, and Brubaker 2019). Another, somewhat related form of unfairness are the representational harms that accrue to members of non-hegemonic groups from being interpolated as such in the outputs of algorithmic systems. What Safiya Noble calls "algorithms of oppression" (Noble 2018) are typified by racially-inflected and gendered search results from the Google search algorithm, but can be extended to include facial recognition systems, as well as other prediction and classification systems discussed below.

## MACHINE BIAS

"Machine Bias" examined the use of algorithmic decision-support systems, built with machine intelligence, in pre-trial release and detention for defendants in Broward County, Florida. The article claimed that the system in use there dramatically overestimated the likelihood that black defendants would be re-arrested while on pre-trial release while underestimating the likelihood that white defendants would be re-arrested. The system did so without access to explicit information about the race of defendants. That is to say, whatever the many datapoints about defendants the system may have used to predict risk of re-arrest—prior arrest record, age, zip code, number of children, etc.—race was not among them.[4] The

---

[4] The actual features and parameters used in this system remain shielded by trade secrets law (Wexler 2018), although the company has asserted race was not an attribute used in risk-prediction by the

implication was that the algorithmic system was inferring defendants' race from "proxy variables"—other attributes that are closely correlated with race like zip code. Zip code is a particularly salient proxy variable for race, given the legacy of racist redlining policies (Rothstein 2017), that is often used by applied machine learning researchers to illustrate the influence of proxy variables.

If algorithmic systems could have racially disparate outcomes, did that mean algorithms were racist? There were no shortage of popular and industry press articles asking precisely this question following the publication of "Machine Bias". The framing of 'racist algorithms' contributed, in no small part, to the fluorescence of both "AI ethics" and "algorithmic fairness" discourses in machine intelligence. These discourses arise out of a broader trajectory of state and corporate control exerted through access to an ever-increasing array of personal information (Gandy 1993), and are crucial for understanding the role social categories have played in machine intelligence, before and since.

## Gender Shades

"Gender Shades" reported an external critical audit conducted on commercial facial recognition systems—which recognize whether or not a face was present in an image, or could recognize facial features like eyes, mouths, or noses—to establish that these systems failed more often for darker-complected faces, for women's faces, and especially for darker-complected women's faces than for lighter-complected men's faces (Buolamwini and Gebru

---

algorithmic system and the 137-question questionnaire used to generate risk scores does not contain a question that pertains to race (Angwin et al. 2016).

2018).[5] The authors' focus on "complexion" evokes Fanon's "racial epidermal schema" (Fanon 2008, 84), an "epidermalization" that Simone Browne analyzes as the means through which meanings are ascribed to bodies (Browne 2015, 91). Calling their approach an assessment of "intersectional accuracy," Buolamwini and Gebru were referencing a "Boolean intersection" (Givant and Halmos 2009) of two attributes (i.e., "darker males," "lighter females"). Their paper nevertheless introduced the term "intersectionality" to the field of machine intelligence, where others subsequently pursued agendas (see Yang, Loftus, and Stoyanovich 2020), inspired by the intersectionality of critical race theory (Crenshaw 1989; Grillo 1995). In addition to somewhat-inadvertently bringing 'intersectionality' and critical race theory into machine intelligence discourse (cf. Benthall and Haynes 2019), "Gender Shades" also launched several other research agendas for applied machine intelligence researchers working on topics of ethics and fairness. These research agendas focused on differential error rates across race and gender groups and focused on imbalances in training and evaluation (i.e., benchmarking) datasets, working from the assumption that datasets that included more white, male faces would be more accurate for groups that were better represented in the data.

The issues brought to light by the "Machine Bias" article from *ProPublica*, at least in how they were taken up by applied machine learning researchers, were somewhat different than the questions of representation and inclusion raised by "Gender Shades". "Machine Bias" launched research centered around whether or not machine intelligence systems might be "unfair," and if so how this unfairness could be "corrected". It also introduced questions about transparency and procedural fairness requirements for algorithmic models used in courtrooms or other governmental bodies—should every party have access to the code and human-readable explanations about such scoring systems, and how could such explanations even be constructed for black-boxed algorithmic models? For applied machine learning researchers, these were challenging, even seductive, machine intelligence problem. The pre-trial release algorithm presented in "Machine Bias" was highly accurate (a measure of the proportion of true positives and true negatives across a set of predictions), at least according to its developer's claims. Where it did err, however, those errors benefited one group and penalized another group: false positives tended to hurt Black defendants more, and false negatives tended to benefit white defendants more. Applied machine learning researchers soon pointed out that unless data truly represent real-world attributes—i.e., unless conviction records include all crimes that have been committed and don't include any false arrests or convictions—it is mathematically impossible to balance errors across those groups, so that errors affect each group similarly, without sacrificing the overall accuracy of the model (Friedler, Scheidegger, and Venkatasubramanian 2016). This mathematical impossibility drew applied machine learning researchers to the problem like moths to a flame, transforming what had been a report about injustice and inequity in a quantified criminal justice system into a highly technical subfield of machine intelligence research. Those working on "constrained optimization problems"— machine intelligence problems that try to preserve accuracy while achieving some measure of fairness across groups (Zafar et al. 2017)—sat uneasily alongside social scientists, legal scholars, and community activists at workshops and conferences dedicated to algorithmic bias.

---

[5] The authors of "Gender Shades" employed the "Fitzpatrick Skin Type classification system" (Fitzpatrick 1988), which dermatologists use "as the gold standard for skin classification and determining risk for skin cancer", instead of race "as a more visually precise label to measure dataset diversity … since race and ethnic labels are unstable" (Buolamwini and Gebru 2018).

**SITUATING FAIRNESS IN THE AUTHORITY OF MACHINE INTELLIGENCE**

In addressing these "constrained optimization problems," applied machine learning researchers transform the ontological status of socially constructed categories into objects over which machine intelligence can wield its authority. The clearest illustrations of this process can be drawn directly from a close reading of the algorithmic fairness literature itself. Across this literature, these categories are discussed as if they are stable objects intrinsically linked to individual subjects and amenable to machine intelligence. This same literature positions machine intelligence as crucial for the remediation of social bias both within and beyond algorithmic systems, and techniques for addressing these biases are developed to secure and preserve the power of organizations that develop and administer algorithmic systems, often at the expense of those subjected to adverse impacts from these systems.

As part of my fieldwork, I embedded myself deeply within a research community that had coalesced under the banner of "fairness, accountability, and transparency in machine learning" or "FATML".[6] Beginning as a workshop series (2014-2018) co-located with larger machine intelligence conferences like the Neural Information Processing Systems Conference (NeurIPS), the International Conference on Machine Learning (ICML), and the Knowledge and Data Discovery Conference (KDD), FATML gradually transformed into its own academic conference starting in 2018. Sponsored by the Association for Computing Machines (ACM), it was originally called "FAT∗," with the asterisk signifying the many domains in which fairness, accountability, and transparency were thought to have relevance. To avoid deragatory connotations, the conference has since renamed itself as "FAccT," or "The Association for

---

[6] https://www.fatml.org/

Computing Machines Conference on Fairness, Accountability, and Transparency".[7] While the FATML workshops predominantly featured work on algorithmic fairness from applied machine learning researchers working in computer science departments or inside industry, the FAccT conference made deliberate efforts—through dedicated disciplinary and topical research tracks—to include work from policy makers, legal scholars, and social scientists. In addition to attending as many convenings as possible and interviewing attendees, I also engaged in a systematic review of the FATML literature.

Within the algorithmic fairness literature, the vast majority of academic papers follow a similar format. First, the authors introduce the problem they are addressing. This is usually a restated version of the constrained optimization problem: how do you classify or predict some phenomenon accurately such that some attribute or characteristic that *ought not* influence that classification or prediction is not disproportionately represented? In the context of mortgage decisions, for example, the problem would be that the most accurate way of predicting who is most likely to repay a loan also underestimates the likelihood that members of legally protected minority groups will repay their loans. The papers will then propose a machine intelligence technique for reducing the degree to which minority groups are underestimated, while compromising overall accuracy as minimally as possible. Hundreds of techniques to do this have been published. Some are narrowly tailored to specific datasets or instances of algorithmic bias, while others propose more general solutions. Some will do this by nudging minority-group members' likelihood of repaying a loan upward to compensate for the underestimation (e.g. Kamishima, Akaho, and Sakuma 2011), while others might filter the training data in ways that bring predictions into parity across groups (e.g. Kamiran and Calders 2012). After the authors explain their proposed technique and their rationale for it, they present a formal description of that technique in mathematical notation, followed by an "experiment" in which they apply their technique to a dataset that can demonstrate its merits. The papers conclude with discussion of their experimental results, often accompanied by a table reporting the overall accuracy of the technique compared to unconstrained techniques, and also reporting an improvement in disparities across groups.

---

[7] https://facctconference.org/

These papers will variably situate themselves within the algorithmic fairness discourse, devoting more or less of their discussion to how historical, structural discrimination and social bias has contributed to unfairness in the domains the paper is interested in. Many contributions to undestanding structural, historical, and social dimensions of unfairness have been published through these conferences and workshops. One such paper describes the relationship between real-world unfairness and algorithmic unfairness by observing that "supervised learning uses historical data to infer a relation between an instance and its label. That historical data may contain discrimination; for instance, racial discrimination in the recruitment of job candidates. In such a case, classifiers are likely to learn the discriminatory relation present in the historical data and apply it when making predictions. Inappropriately trained models may hence discriminate systematically, which is a lot more harmful than in single cases" (Žliobaitė, Kamiran, and Calders 2011, 992). But because supervised learning, a common machine intelligence technique discussed at length in Chapter 1, requires "an instance and its label," social categories—and all the complexity that accompanies their social construction—are collapsed into independent, binary attributes. Interrogating machine intelligence for algorithmic fairness ensures that these attributes are attached even to supposedly "race-blind" or "gender-blind" systems, as algorithmic fairness analyses require applying such attributes *post hoc* to measure bias. Here, the attempt to mitigate bias ends up reinforcing the very essentializing, racializing logics that produced such forms of bias in the first place

**DECOUPLING DATA**

Across the algorithmic fairness literature, protected social categories are conceptualized as independent, reductive, binary[8] attributes. They are intrinsically linked to data subjects, they are immutable, and they are often mutually exclusive. A data subject "is" one gender or another, or they "are" members of one racial group or another. This is vividly illustrated in algorithmic fairness papers that substitute genericized labels for socially salient categories for the purpose of demonstrating a novel machine intelligence technique. Rather than engaging with the historical specificity of, say, "racial discrimination in the recruitment of job candidates" (*Ibid.*), and the ways specific racialized populations were discriminated against, many algorithmic fairness papers will situate their analysis solely within the context of (often synthetic) datasets that pertain to (often hypothetical) minority and majority groups by assigning them a primary color, e.g. red or blue. The social meaning of the category is thus made irrelevant to the computational goal of reaching some definition of fairness algorithmically. In one paper (Dwork et al. 2017), the advantages of "decoupling" an objective function that optimizes for accuracy from an objective function that optimizes for fairness is demonstrated by applying this technique to a majority (red) group and a minority (blue) group (Figure 13). By decoupling fairness from accuracy, this technique pursues optimal levels of accuracy separately for majority and minority groups, in effect treating each group distinctly and sedimenting the practice of treating members of different protected classes differently.

---

[8] "Binary", in this context, refers to how membership in a group can be represented by a single "bit" of information: e.g., 1 = member, 0 = non-member.

But it also decouples the label signifying membership in a protected class from the meaning of membership in that protected class, and in doing so promulgates that classification as an intrinsic and immutable property associated with a data subject.

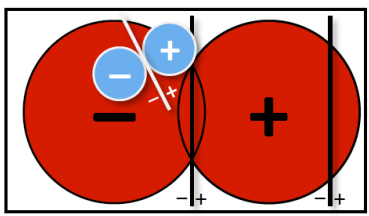

Figure 2: Decoupling helps both majority (red) and minority (blue) groups each maximize accuracy from different linear classifiers (white line and left black line). If, say, equal numbers of positives are required from both groups, the white line and right black line would maximize average accuracy.

Figure 13. Illustration of genericized majority and minority groups (Dwork et al. 2017*f*2).

As previously discussed (see Chapters 1 and 3), machine intelligence enacts that which it optimizes for through data performances, without recourse to any socially-situated understanding of or knowledge about the terms it is optimizing across. A deep neural network does not require any understanding of what the numeral "9" refers to in order to be able to learn which configurations of pixels are associated with handwritten representations of that numeral. Similarly, constrained optimization problems do not require any understanding of the concept of race or the social salience of any one racialized group—as the interchangeability of racial groups with primary colors, even within the mathematical equations used by applied machine learning researchers demonstrates (Figure 14). And yet, this set of machine intelligence problems, and the overall discourse of algorithmic fairness, enacts a specific understanding of these protected classes as binary (and therefore universal), essential (and therefore immutable), and isolatable (and therefore independent of other dimensions of identity).

**Definition 1** (Balance). *For a subset $\emptyset \neq Y \subseteq X$, the balance of $Y$ is defined as:*

$$\text{balance}(Y) = \min\left(\frac{\#\text{RED}(Y)}{\#\text{BLUE}(Y)}, \frac{\#\text{BLUE}(Y)}{\#\text{RED}(Y)}\right) \in [0,1].$$

*The balance of a clustering $\mathcal{C}$ is defined as:*

$$\text{balance}(\mathcal{C}) = \min_{C \in \mathcal{C}} \text{balance}(C).$$

Figure 14. Equation from algorithmic fairness paper on fair clustering (Chierichetti et al. 2018).

Machine intelligence approaches to algorithmic fairness construct the problem of algorithmic unfairness as one in which individuals might be unfairly penalized because of their race, gender, sexuality, or etc. But in posing solutions to this problem, they use datasets that assign individuals to membership in a single race, for example. Often, this membership is binary; individuals are white or non-white. Or, individuals are, or are not, Latinx. This membership is also essentialized; it is an immutable trait that is often "covariate" with other traits as will be further discussed below, even if this particular trait is thought to be an undesirable basis for making predictions or classifications—particularly in domains that are regulated by Title VII of the Civil Rights Act of 1964. Algorithmic fairness also enacts an essentialist view of race, gender, and other protected categories in its attempts to repair unfairness. This is particularly salient in work on "counterfactual fairness" (Kusner et al. 2017), which places the effects of protected attributes (e.g., sex, gender, race) within a "causal framework" that estimates sources of unfairness. Such approaches attempt to determine whether "sex" produces "unfairness" by building a "counterfactual world" in which a subject's sex is different. This allows researchers to estimate the biasing effect sex had on a particular decision or outcome. If that effect is large, the decision is judged to have relied too heavily on gender and therefore is suspected to be unfair (see Kusner et al. 2017). But, as detailed critical scholarship presented at the 2020 meeting of the FAccT Conference has argued, such arguments "only make sense under the assumption that sex (and race) are inherent individual-level attributes" (Hu and Kohler-Hausmann 2020, 3), and that a person's sex could be changed without changing every other thing about them.

Hu and Kohler-Hausmann go on to demonstrate that sex (as well as race, gender, and other social categories) does not "cause" an effect, but rather all the other attributes, qualities, and interactions that can be associated with a person labelled as belonging to one sex or another are "constitutive of sex as a social status" (*Ibid.*, 10). In attempting to "solve for" the effects of sex, race, or gender on a prediction or classification, machine intelligence fails to see these categories as socially constructed and reifies membership in these categories as determinative of one's life chances when subjected to an algorithmic gaze. It is in this slippage—from social category as constitutive to social category as causal—that we again see the cultural implications of machine intelligence's naïve empiricism. Naïve empiricism suggests that it is possible to understand the (purportedly unfair) influence of sex, race, gender, etc. from the data alone. From the data alone, however, it is impossible to understand these categories as constitutive of the data itself. In the *Machine Bias* example provided above, the dataset Broward County used to make recidivism risk predictions shows only that certain features—zip code, prior arrests, age, gender, race, etc.—are more or less closely correlated (or more or less inversely correlated) with rearrest. Even if nondiscrimination laws insist that protected classes ought not be used in making that determination, there is nothing within that dataset that would allow, upon naïve inspection, an applied machine learning researcher to reach an understanding that social categories—indeed, all the features of the dataset—are mutually constitutive. This critique of causal inference, as applied to algorithmic fairness problems, is broadly compatible with social scientific understandings of sex as a social construction, mutually constituted across a subject's entire social milieu.

In pursuing algorithmic fairness, opportunities to understand social categories as socially constructed are not only foreclosed, but these categories are also reified as stable *social kinds*. Social kinds like race, sex, and gender have, of course, been essentialized in many ways over the long course of history (Visweswaran 1997; McIntosh 2018), classificatory practices entrench these essentialisms (Hacking 1986; Bowker and Star 1999), and persuasive arguments have been made that aspects of the civil rights movement and nondiscrimination efforts contribute to reifying these categories (Grillo 1995). However, following Searle, algorithmic fairness establishes a state of play around these social categories in which they enter into "a set of formal relationships that are capable of being represented symbolically" (Searle 2006, 23)—a database, a supervised machine learning model, an academic paper on algorithmic fairness. Given this formal relationship, race, sex, gender, etc. need not exist as an object of positivist knowledge *per se*. But because they have been represented symbolically and in relation to other features they can be manipulated as if they do.

Applied machine learning researchers wield what Searle calls "deontic power"—"certain powers of movement within a formal system"—which individuals represented in a dataset are simultaneously subject to (*Ibid*.) and constituted by. "Deontic" in this context pertains to a Kantian notion of "duty" or obligation. Such power operates when it determines what the consequences of social categorization are for subjects by manipulating the influence those categories have about a prediction or a classification. They can use race or gender, or "proxy variables" that are closely correlated with race or gender, in making classifications or predictions. And they can "re-weight" the influence of race or gender so it is or is not determinative of an outcome—be it a medical intervention, a bail decision, or a job offer. Either way, machine intelligence is used to manipulate the symbolic representations of subjects in ways that have material consequences for these subjects. It is, in part, through this deontic power that machine intelligence reconsolidates its authority through the pursuit of algorithmic fairness, even as it is challenged by evidence of algorithmic bias and unfairness.

Seeing the deontic power of machine intelligence through algorithmic fairness also points to the controlling form of authority machine intelligence wields, more generally. Not only in setting the degree of influence attributes have on a given objective function, but in optimizing an objective function, and in making a claim that an objective function is relevant to a domain of social activity, applied machine learning researchers provide desire-independent reasons for actions. Often contrary to personal judgement, machine intelligence suggests—and even demands—a specific action or decision be taken. Wherever machine intelligence is used for decision-making—in recommender systems embedded in consumer platforms, in recidivism risk-scoring systems, in facial matching technologies that provide or restrict access to a building—it is exercising deontic power by obligating users to follow the reasons machine intelligence provides for their actions, rather than their own. Algorithmic fairness shows how applied machine learning researchers can make interventions into these decisions by shifting the effect some attributes (protected attributes in this case) have on decisions relative to others, but the deontic power wielded by those researchers working on algorithmic fairness is nested within the deontic power wielded by machine intelligence, generally.

### "People are Biased Too!"

Within the broader community assembled through the FATML Workshop and the FAccT Conference, one of the most prevalent rhetorical foils motivating and qualifying work on algorithmic fairness was the argument that "people are biased, too". I heard some version of this comment countless times in Chatham House rules panels, in Q&A sessions, and in the public discourse surrounding algorithmic fairness during my fieldwork. The argument being levied was that if algorithmic bias is a flaw in a system, "we should be asking how these flaws compare with those of human beings," with the implication being that we would eventually conclude that "algorithms are less biased and more accurate than the humans they are replacing " (Miller 2018). Asking, and answering, the question of whether algorithms can be "more fair" than humans is somewhat epiphenomenal to what the stakes of algorithmic fairness are for the authority of machine intelligence. That the pursuit of algorithmic fairness is seen as an aspect of machine intelligence that can supersede or replace biased human decision-making demonstrates an important dimension of the authority granted to machine intelligence in recent years, as well as a desire to grant it more authority in the future.

Framing fairness as a computational question that can be operated upon by machine intelligence is crucial for this authority. A tutorial I attended at the inaugural FAT* Conference (since renamed FAccT) offered 21 separate definitions of fairness (Narayanan 2018), with the intention of demonstrating that resolving competing definitions of fairness to a singular definition is not necessary or desirable, and that many definitions are purpose-specific or context-dependent. In my own survey of the algorithmic fairness literature, and in interviews with algorithmic fairness researchers, the most salient distinctions separated forms of "computational fairness" (M. Feldman et al. 2015) from legal or political forms of fairness.

Computational forms of fairness are, generally, machine intelligence techniques for ascertaining whether a decision is deemed fair given some prior rule. For Feldman et al., their notion of computational fairness rests on the ability of their technique to reach similar decisions whether or not it incorporates a protected category into its decision-making. This computational form of fairness is motivated by an acknowledgement of disparate intent to members of protected categories as a legally stipulated form of discrimination, and hence, unfairness. Formal variants along these lines depend on a Rawlsian notion of fairness, a notion shared with legal and political forms of fairness. John Rawls's 1971 concept of "fair equality of opportunity" is foundational to algorithmic fairness and is perhaps the most common non-computer science citation found within the algorithmic fairness literature. Rawls's concept of "equality of opportunity" articulates a meritocratic assumption that "assuming there is distribution of natural assets, those who are at the same level of talent and ability, and have the same willingness to use them, should have the same prospects of success regardless of their initial place in the social system" (Rawls 1971). "Equality of opportunity" is an explicitly legal and political notion of fairness that can be traced back at least to de Tocqueville's 18th century writing on political participation (Tocqueville 2000), but has been mathematically formalized in numerous algorithmic approaches (Nissenbaum 1996; Mancuhan and Clifton 2014; Kroll et al. 2017; Epstein et al. 2018; Espeland and Vannebo 2007).

While many forms of computational fairness may ultimately rely on such legal and political definitions, questions about what constitutes fairness in legal and political registers "are now unavoidably tied to mathematical questions" (Epstein et al. 2018). These ties arise in part from the need for applied machine learning researchers to intervene algorithmically in order to "repair" the social biases embedded in historical datasets used to train machine intelligence and to mathematically enforce politically- and legally-determined fair decision rules. Additionally, through machine intelligence, it becomes possible to calculate not just individuals' opportunities, but also individuals' opportunities as a function of the categories to which they belong. Machine intelligence thus makes it more possible to demonstrate disparate impact, even though injuries from such disparities are more difficult to make whole through the U.S. judicial system than those that arise from disparate intent (Barocas and Selbst 2016). Yet, as has been discussed above, these computational approaches are not very useful for interrogating the history of categories themselves or uncovering disparities in training data that arise from unequal treatment and perpetuate algorithmic bias (Jacobs and Wallach 2021).

Nevertheless, because machine intelligence is being implemented in domains subject to Title VII of the Civil Rights Act in the U.S.—particularly finance, hiring, criminal justice—it falls on applied machine learning researchers to ensure that the systems they build comply with this law, and the florescence of algorithmic fairness is in large part the pursuit of computational methods that are legally defensible against Civil Rights claims.

Computational approaches to fairness have almost invariably been pursued from the point of view of organizations, whether they be courts deciding bail (Angwin et al. 2016), banks deciding loan applications (Dwork et al. 2012), or employers deciding whom to hire (Kamiran, Calders, and Pechenizkiy 2010). Non-discrimination in these contexts has long been adjudicated through statistical methods (Boardman and Vining 1983), and from the earliest research papers, computational fairness is framed by a concern with research into algorithmic classifiers that can operate in domains regulated by anti-discrimination law in ways that can be demonstrated not to produce discriminatory effects (Dwork et al. 2012). The concern of the decision-maker to be fair, i.e. to not discriminate, is a particular and specific formalization that is functionally similar to the formalizations applied machine learning researchers use to build spectacular demonstrations of machine intelligence by choosing problems, like Chess or Go (see Chapter 2), that are neatly specified but which only bear a superficial resemblance to the capability they are wishing to demonstrate. Like algorithmic systems which play Go but are perceived by audiences as demonstrating a more general "intelligence," algorithmic approaches to fairness are intended to demonstrate an expansive notion of fairness, but fall into what Selbst et al. call the "formalism trap"—the "failure to account for the full meaning of social concepts such as fairness, which can be procedural, contextual, and contestable, and cannot be resolved through mathematical formalisms" (Selbst et al. 2019). Fairness, then, is formalizable through machine intelligence only to the extent that it is able to occupy one standpoint—the decision-maker—and tends to exclude others (Harding 1992b).

Despite this strong standpoint from which algorithmic fairness proceeds, it nevertheless is seen as more objective than unmarked "people" who have biases that are even more inscrutable than machine intelligence's (see above). This objectivity is derived in part from the erasure of the specific standpoint machine intelligence occupies—that of the decision-maker—through rhetorical moves that locate machine intelligence as the active agent (rather than the entity on whose behalf the algorithmic system is operating). Applied machine learning researchers "develop a modeling technique that maintains an equal level of fairness to that attained by" an alternate technique (Kamishima et al. 2013, 850), demonstrate that their "model … is effective"(M. Feldman et al. 2015, 3), and they discuss their techniques as having "achieved" fairness or non-discrimination (Hajian et al. 2015, 1737).

Casting machine intelligence as the active agent not only obscures other standpoints and subject positions, but also makes a claim to a positivist form of objectivity that is measurable, replicable, and disinterested. The form of objectivity occupied by machine intelligence through algorithmic fairness is also a highly "mechanical" form of objectivity (Daston and Galison 1992; Porter 1995), in that it is superficially independent of individual subjectivities (like those of employers or banks) which may be partial, bias, or self-interested. This form of objectivity is "mechanical" also in that it is rule-governed, and it integrates quantitative thresholds into formalized procedures that can then be automated and followed with mechanical precision and predictability (cf. Alkhatib and Bernstein 2019). Most importantly, the use of machine intelligence transforms human "decision-makers"—who we may recall are "biased too"—into functional appendages of these algorithmic systems by deferring their own ersatz judgement to the unbiased certainties of machine intelligence.

Through algorithmic fairness, then, the authority of machine intelligence comes to supplant the authority of those who merely enact algorithmic judgements.

Framing fairness as a technological problem that *can be solved on behalf of those who build, own and operate algorithmic systems* bounds the concept of fairness in structures of power that already exist. While the broader (admittedly polyvocal) FAccT community situates fairness as an ethical imperative, setting fairness as the primary goal for reforming the institutional implementation of algorithmic systems forecloses adjacent values like equity, justice, and dignity (D'Ignazio and Klein 2020). The limits of formal, mathematically-provable fairness are opaque, as formal fairness itself obscures power relations, conflates fair processes and fair outcomes, and subsumes questions of who makes decisions with questions about how decisions are made. Fairness also reifies the hegemonic position of those who do make decisions by eliding questions of domination with questions of fairness. Instead of asking "why should machine intelligence be permitted to decide who gets a loan?" algorithmic fairness asks "did machine intelligence allocate loans in a legally defensible way?". Replacing questions of equity and justice with questions of fairness in this way acts as a form of biopower (Gitelman 2013) that reinforces the position of decision makers who use machine intelligence, and foregrounds the steps they've taken towards algorithmic fairness in using machine intelligence, to justify (and avoid questions about) the control they wield over a population's conditions of life and death—without making visible the institutional structures that uphold that power and control.

The presentation discussed above that analyzed 21 definitions of fairness foregrounded the idea that technical definitions of fairness "are laden with values and politics, and seemingly technical discussions about mathematical definitions in fact implicate weighty normative questions" and that it is possible "to make explicit the values embedded in each of" the definitions in order "to help policymakers and others better understand what is truly at stake in debates about fairness criteria" (Narayanan 2018). The idea that applied machine learning researchers, like the presenter, could work together with lawyers, lawmakers, and public advocates to choose appropriate definitions of algorithmic fairness through a political process was one of the explicit goals of the FAccT community. But algorithmic fairness is necessarily *algorithmic*—it ensures an authoritative role for machine intelligence in how policy is enacted, particularly when public services are administered through algorithmic decision-making systems (Eubanks 2018), and gives machine intelligence increased responsibility for managing the scope of administrative harms (Hoffmann 2019) suffered by those enmeshed in such systems. It is because of the authority granted to machine intelligence that it can be seen by corporate entities—including government agencies and private companies—as an indispensable participant in the administrative state's turn toward algorithmic governance (Calo and Citron 2020; Engstrom and Ho 2020).

## The Tentacular FaceNet

The pursuit of algorithmic fairness depends on treating social kinds as immutable features, and constitutes performances of those kinds. Doing so reinforces how individuals are categorized even in its attempts to intervene in the harmful consequences that arise from bias and inaccuracy in machine intelligence's predictions and classifications. In pursuing algorithmic fairness, some applied machine learning researchers demonstrate a general awareness of historical antecedents of unfairness and under-representation of non-white groups and women in datasets. This is evident, for example, in machine intelligence papers that discuss how the history of redlining in the United States (Rothstein 2017) has led to racially-divergent distributions in many datasets that lead to algorithmic unfairness (e.g. Kamiran, Žliobaitė, and Calders 2013; Romei and Ruggieri 2014; M. Feldman et al. 2015; Datta, Sen, and Zick 2016; Corbett-Davies et al. 2017).

But many papers demonstrate ignorance of the long history of science and technology in producing race, sex, and gender types and the use of technology to surveil and discipline minoritized populations (Browne 2015; Benjamin 2019b). Similarly, applied machine learning researchers working on algorithmic fairness fail to acknowledge, and are in some cases reinvigorating, the disastrous effects wrought on people through eugenicist attempts to "improve the inborn qualities of a race" (Galton 1904, 1). The consequence of this is that machine intelligence systems built to "understand" race, gender, and sexuality enact performances of these categories as physiognomy: social types are taken to be machine-legible phenotypic manifestations of innate properties of individuals. The role of the photograph—which Nicholas Mirzoeff calls "a screen on which wider social forces become visible" (Mirzoeff 2003, 111)—has long been the fulcrum on which these phenotypic manifestations are balanced, but through machine intelligence the digital facial photograph has been reclaimed as a site for eugenics and physiognomy. Just as the photograph indexes a "necessarily real thing which has been placed before the lens" (Barthes 1981, 76; cited in Rohrbach 2012), race science leveraged that indexical realness to make claims about the reality of racial difference through the use of photography, and now machine intelligence similarly makes claims about racial difference through the same indexical properties of the image, as will be developed below.

I observed first-hand how ignorance of the eugenic legacies of statistical practices is constructed through the pedagogical artifacts of machine intelligence. During my fieldwork at OTH, as noted earlier, other researchers and I would often gather to teach ourselves TensorFlow (Géron 2017), a machine intelligence package similar to scikit-learn. One of our sessions was dedicated to learning regression through an exercise on the `iris` dataset. The `iris` dataset contains hundreds of measurements taken from various parts of iris flowers belonging to three different species (*Iris setosa, Iris versicolor,* and *Iris virginica*), and is used to demonstrate classification techniques. Students are instructed how to load the dataset, assign variables to the flower measurements, and use the species as the label to train the classifier. Over the entire span of my fieldwork I saw the iris dataset used to demonstrate various machine intelligence techniques dozens of times. In each of these demonstrations, I never once heard anyone mention that the `iris` dataset itself was first published in the *Annals of Eugenics* in 1936 (Fisher 1936) to illustrate regression as a technique that could further prior work on craniometry. Regression, in its application to taxonomic problems of botanical classification (Cramer 2002), was an operationalization of Carolus Linnaeus's insistence on establishing classifications of species based on attributes that can be measured (Linneaus 2003, 71). But "regression" itself originated as a suite of statistical techniques developed by Francis Galton, who observed that the full-grown height of children of tall parents demonstrate "filial regression toward mediocrity," or the arithmetic mean of a population (Galton 1886, 247). Regression was developed as a tool of eugenics that could be used to select 'desirable' traits and selectively favor those traits across a population, while eliminating 'undesirable' traits from that population.

This is not to imply that any machine intelligence practice that uses regression is fruit of a poisonous tree, although poisonous apples rarely fall far from poisonous trees. Rather, it is to suggest that classification problems within machine intelligence, as well as the methods for solving them, are deeply linked to the construction of categories—desirable vs. undesirable, parole-worthy vs. not parole-worthy, member vs. non-member—that are the basis of social control through computational means (Deleuze 1992). As will be discussed at greater length below, the links between the development of quantificatory practices in the sciences, scientific management of people and bodies (including eugenics, high modernist schemes to organize the world, and cybernetics), and machine intelligence are strong.

## THE USE OF MULTIPLE MEASUREMENTS IN TAXONOMIC PROBLEMS

By R. A. FISHER, Sc.D., F.R.S.

### I. DISCRIMINANT FUNCTIONS

WHEN two or more populations have been measured in several characters, $x_1, ..., x_s$, special interest attaches to certain linear functions of the measurements by which the populations are best discriminated. At the author's suggestion use has already been made of this fact in craniometry (a) by Mr E. S. Martin, who has applied the principle to the sex differences in measurements of the mandible, and (b) by Miss Mildred Barnard, who showed how to obtain from a series of dated series the particular compound of cranial measurements showing most distinctly a progressive or secular trend. In the present paper the application of the same principle will be illustrated on a taxonomic problem; some questions connected with the precision of the processes employed will also be discussed.

### II. ARITHMETICAL PROCEDURE

Table I shows measurements of the flowers of fifty plants each of the two species *Iris setosa* and *I. versicolor*, found growing together in the same colony and measured by Dr E. Anderson, to whom I am indebted for the use of the data. Four flower measurements are given. We shall first consider the question: What linear function of the four measurements

$$X = \lambda_1 x_1 + \lambda_2 x_2 + \lambda_3 x_3 + \lambda_4 x_4$$

Figure 15. Original article containing `iris` dataset from the Journal of Eugenics (Fisher 1936).

The failure of applied machine learning researchers to acknowledge the eugenicist roots of a common statistical tool, even when footnotes and metadata point directly at those roots, illustrates the pernicious hold physiognomy holds on many of the fundamental tools of machine intelligence. And the reliance of algorithmic fairness—and machine intelligence in general—on social types represents a clear threat to human dignity as machine intelligence increasingly mediates access to public and private resources through facial recognition technologies, credit scoring, and other identity authentication services. Any linkage between social types and the provision of these services enables algorithmic forms of social control mediated along the well-worn dimensions eugenics has trod in the past.

The apparent unawareness of physiognomic and eugenicist tropes among applied machine learning researchers recurred at other times throughout my fieldwork. At the "Fairness, Accountability, and Transparency in Machine Learning Workshop" at the 2018 International Conference on Machine Learning in Stockholm, Sweden, I sat with a few academics and practitioners from drastically different backgrounds, whom I had made acquaintances of as regular attendees of earlier algorithmic fairness workshops and conferences. These acquaintances were doctoral students from applied mathematics, computer science, and philosophy departments—all of whom all expressed skepticism and surprise at how 'fairness' was being operationalized as a technical solution to the historical and social dimensions of racism and misogyny at these convenings. We each seemed happy, even a bit relieved, to have run into each other at this particular workshop, amidst the accordion room dividers, the gray felt stackable straight-back chairs, and slightly raised daises supporting skirted collapsible tables for panelist to rest their elbows and conference-branded water bottles upon. For my part, I was happy to run into familiar faces after having traveled solo across a half dozen time zones. All of us were relieved to be among others who could talk about things like social constructionism (Bijker, Hughes, and Pinch 1987) without being met with blank stares in return.

The workshop that day mostly featured what had become the standard fare for a workshop like this: technical paper presentations with slide decks that featured colorful gradients visualizing their optimization paths, an inspiring keynote from a critical scholar urging the technical researchers to think beyond 'fairness', a poster session with paper authors standing at attention to talk through novel techniques for rendering opaque algorithms more transparent, etc. But among these technical paper presentations was one that left my acquaintances and me arching our eyebrows at each other across the convention hall: "InclusiveFaceNet: Improving Face Attribute Detection with Race and Gender Diversity" (Ryu, Adam, and Mitchell 2018). "InclusiveFaceNet" was like many of the other papers in that it was demonstrating "an approach to face attribute detection that retains or improves attribute detection accuracy for gender and race subgroups" (*Ibid.*, 1), along the research trajectory suggested by "Gender Shades" (Buolamwini and Gebru 2018). Where it differed from other papers, however, was that it proposed to do so by "learning demographic information prior to learning the attribute detection task" (*Ibid.*). Said another way, Ryu et al. had built a machine intelligence system that had higher accuracy in identifying facial attributes—mouths, eyes, noses, whether the face is smiling, etc.—in photographs if it was also able to identify the "race" of the person in the photograph. It was the latter part of this claim that raised our collective eyebrows as we listened to the paper presentation: to claim that the race of a person can be determined from an image shifts the concept of race away from that of a social construct and toward that of a biological phenomenon. Crucially for this discussion, it used computer vision (Szeliski 2011) to determine this demographic information, operationalizing an assumption that race and gender can be determined by visual inspection, although tracing the ways demographic labels associated with facial images are transmitted through the machine intelligence development process is also examined below. Doing so is key to addressing Deborah Poole's observation that, "visual technologies have … shaped the very notion of race" (Poole 2005, 160). I argue that, like photography, computer vision (a subfield of machine intelligence) can be interrogated for the "privileged role" it plays "in the crafting of a racial common sense which … unites 'popular' and 'scientific' understandings of embodied difference" (Poole 2005, 162) using similar methods as those applied to sentiment analysis and relationship types in Chapter 3 above .

It is tempting to read Ryu et al.'s paper merely as a case in which "the line separating racism from antiracism is sometimes so unclear that the two sets of discursive practices converge" (Harrison 1995, 49), and perhaps the entire move toward algorithmic fairness can be seen as, like forensic anthropology, "particularly invested in racial classification" (*Ibid.*, 53). But this paper is also an example of the onto-epistemological shifts (Barad 2003; 2007; Van Der Tuin 2011) machine intelligence produces in the objects of its analysis through data performances. Facial images are not literary corpora, but if we accept Stuart Hall's contention "that *race works like a language*," and that systems of racial classification "gain their meaning, not because of what they contain in their essence, but in the shifting relations of difference, with other concepts and ideas in a signifying field" (Hall [1997] 2021, 362; emphasis in original), onto-epistemological shifts become discernible through the "secret life of data" method outlined above. I will also contend, following the analysis below, that all systems of classification—not just racial ones—work this way too, too, through similarly shifting relations of difference. Machine intelligence intervenes into shifting relations of difference through its deontic powers described above (Searle 2006) and its ability to enact data performances— further buttressing the authority machine intelligence holds over objects of knowledge.

Ryu et al. 2018 is highly technical paper, but its goal and methods can be roughly paraphrased: *Computer vision applications have higher error rates for minority groups than dominant groups. One clearly defined computer vision task in which these error-rate disparities occur is in detecting whether a person in a photograph is smiling or not. If "faces look different across different races," then smile detection should work better if the system can tell the difference between not only smiling and not-smiling faces, but also between faces 'of' different races—different races' smiles look different. However, directly detecting race could have harmful downstream consequences should demographic information be misused,* *and even apart from that, the datasets that annotate whether or not the subjects of images are smiling or not are not annotated with racial categorizations. Therefore, "transfer learning"—a modification of deep learning—can be employed to train a system to "learn" race, but only use that learning to gain insight into the smile detection task and not other tasks for which racial classification is impermissible.*

The smile classifier Ryu et al. built based on these premises, which I treat as a "text-artifact" but which they call InclusiveFaceNet, is actually an assemblage of datasets, other machine intelligence systems, and human labor. At times I will simplify, without unduly abstracting, this assemblage, as Ryu et al. draw upon more datasets than I will discuss, and they conduct several "experiments" in their paper that are orthogonal or redundant to this case study. To understand the signifying field in which race comes to have meaning through their work, though, it is necessary to read across this assemblage in some detail, following the "secret life" of the datasets Ryu et al. use, uncovering the process by which race is entextualized within this assemblage, and tracing what race 'is' at each step of the way.

The overall claim of Ryu et al. is that facial attribute detection can be improved if the classifier is able to use information about what faces of different races tend to look like. Ryu et al. ultimately supply evidence supporting this claim. They demonstrate an overall improvement in accuracy for detecting facial attributes like smiles using InclusiveFaceNet, which has access to information about what faces of different races look like, over what they call a "vanilla classifier"[9]. This increase in accuracy suggests to the reader of their paper, and to the audience in the room when they presented their paper, that there is an essential relationship between a persons' race and their face; if there was no such relationship, there would be no increase in accuracy. The reading of race from face is analogous to the epidermalization described by Fanon and Browne, a "facialization" (Deleuze and Guattari 1989, 170) that reaches out beyond the individual it represents. But how are they able to make this claim?

---

[9] The semiotics of "vanilla" in this context could merit their own discussion, but "vanilla" is a term of art frequently used by applied machine learning researchers to describe the approach "that everyone else is using" and that stands as the baseline for comparison to an approach a particular team is working

InclusiveFaceNet itself is a deep learning neural network classifier, functionally analogous to a machine intelligence system that can discern between handwritten numerals or discern human faces from each other by calculating an abstract statistical representation of the data it is trained with (see Chapter 1). The "vanilla" classifier Ryu et al. refer to is a deep learning classifier that was trained on a wide variety of images of people, some of whom were smiling and some of whom were not (Z. Liu et al. 2015), and which minimized the number of 'wrong guesses' it made about which faces were smiling and which were not. Under the research agenda suggested by Buolomwini and Gebru, Ryu et al. hypothesized that the error remaining in the "vanilla" classifier would accrue more to non-white faces than to white faces, but that hypothesis could not be tested directly, as the "vanilla" classifier had no access to information about race. The dataset it used for training had information about whether or not a face was smiling (and 39 other attributes) but not about the race of the person in the image. So they constructed a deep learning system that *did* have access to that information to see if overall accuracy would be improved, and if parity could be reached in error rates across "race

---

on. But in this usage it takes on the trappings of hegemonic or conservative approaches that yield higher error rates for minoritized, non-White groups (see Pitcan, Marwick, and boyd 2018).

subgroups". To do so, they found a dataset that had information about race associated with images of faces.

Applied machine learning researchers working in computer vision will admit that "a sufficiently large and fairly representative racially diverse face database, though resource-intensive, is by all means necessary and important" but few public datasets meeting these criteria exist which "are of demonstrated uses to scholars" (Fu, He, and Hou 2014, 2494). Those that *do* exist all tend to assign images to racial categories in the same way as the dataset Ryu et al. used: they show the pictures to people and have them assign a racial category to the image file (Ramakrishna et al. 2017; as cited in Somandepalli 2017). Using the MTurk platform, or a similar platform, the "ground truth" for racial information is provided by crowd workers who apply their subjective judgements about racialized appearances to photographs of people's faces. Ramakrishna et al. used this methodology to assign racial categories to images of movie actors, working from US Census Bureau racial categories.[10] They annotated "racial information for nearly 2000 such actors using MTurk with two annotations for each actor, manually correcting nearly 400 cases in which the annotators disagreed" (Ramakrishna et al. 2017, 1672).

Reading InclusiveFaceNet as a text with a "secret life" requires tracking citational chains to trace how multiple conceptions of "race" were laundered through the process of assembling InclusiveFaceNet to construct race as a specific kind of object for machine intelligence. So far in this excavation of InclusiveFaceNet, we have a dataset containing the faces of 2000 movie actors, which MTurk workers looked at and assigned racial categorizations to, and to which Ramakrishna et al. assigned their own racial categorizations in instances where MTurk workers were not unanimous in their racial annotations. Ryu et al. then took that dataset and used it to train a deep learning classifier for predicting race. This classifier built a "representation" of each racial category comprised by statistical regularities in the distribution of pixels in the images associated with each racial category. As noted above, this classifier is functionally identical to the example of handwritten MNIST numerals discussed throughout this dissertation. The work that Ryu et al. based their race classifier off of included visualizations of these statistical regularities (Figure 16), which bear a striking resemblance to the composite employed by Francis Galton—produced by stacking underexposed negatives of different peoples' faces to illustrate regularities in facial features—to demonstrate "the reality of distinct racial types" that "amounted to an essentialist physical anthropology of race" (Sekula 1986, 51). Comparing Somandepalli's "average face" analysis that Ryu et al. subsequently draw upon (Somandepalli 2017) with Galton's composites of the "Jewish Type" (Pearson 1924; cited in Sekula 1986, 52) in Figure 17 shows all the rhetorical force of images that purport to show the essential facial features characteristic of racial groups.

---

[10] The historical and political contingencies of US Census Bureau categories are well-studied and their malleability is itself a refutation of racial essentialism (Bouk 2015; Bowker and Star 1999).

Somandepalli uses the techniques of machine intelligence to unwittingly (it may be assumed) re-enact an essentializing performance of racial types. The danger of this reenactment is not just that it is premised on broadly discredited associations between racial groupings and phenotype (Omi and Winant [1986] 2015). Nor is the danger that its resemblance to the tools of eugenicists necessarily set machine intelligence on the path of institutionalization, forced sterilizations, and genocide that 19[th] and early 20[th] century eugenics led to (Caspari 2003). The eugenics movements of that era cannot be understood apart from their particular historical contexts, and while technological approaches to race bear striking resemblances to each other, it would be overly deterministic to imply that similar tools herald similar outcomes (Smith and Marx 1994). Rather, the danger is that these tools enable new forms of harm through repressive social controls where racialized and gendered violence is currently being used to maintain state power. At convenings of applied machine learning researchers who share these concerns, they raise the specter of facial recognition technology profiling Uyghur and other racialized minority groups in China (see Harwell and Dou 2020), as well as the possibility that a tool that can "predict" homosexuality based on facial images being used for persecution by regimes that have criminalized homosexuality (see Vincent 2017).

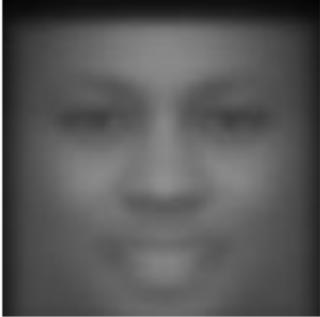

Figure 16. Visualizing the "average face per class" from Somandepalli 2017.

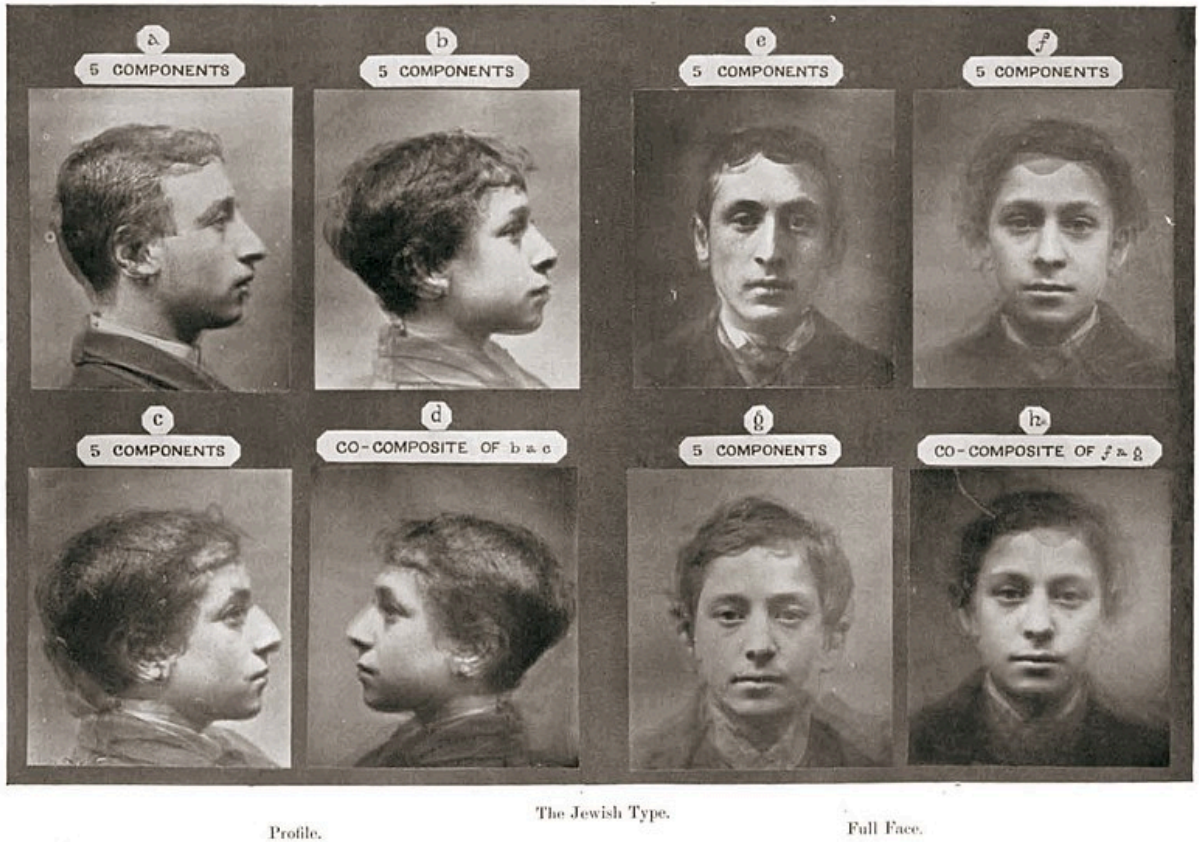

Figure 17. Visualizing "The Jewish Type" (Pearson 1924).

Returning to InclusiveFaceNet, after training a classifier that could "predict" the racial categorization of their dataset Ryu et al. then "transferred" that classifier from the race classification task to a facial attribute classification task. "Transfer learning" is a well-established machine intelligence technique that applied machine learning researchers liken to human skills which transfer between related tasks, like playing piano and pipe organ (Pan and Yang 2010). As transfer learning was explained to me in a lunchtime conversation at OTH, which I will use Ryu et al. to illustrate, it starts by using a deep learning neural network (Chapter 1) that makes "random" guesses at an objective—in this case guessing the racial categorization of images. It then adjusts the parameters it uses to make those guesses—the 'weight' of each 'node' in the neural network—to minimize the error in those guesses until it achieves a suitable degree of accuracy. Then, it uses those existing weights to begin making guesses at a different task, in this case guessing facial attributes like smiles, rather than resetting the weights to start fresh with completely random guesses. The error at conducting the second task is then used to adjust weights (which already represent 'knowledge' gained from the first task) until a suitable degree of accuracy at the second task is achieved. Put slightly differently, transfer learning works by training an algorithmic system to pursue one objective function, and then using the 'skill' it acquires in that pursuit to 'boost' or 'inform' its skill at pursuing a second, different objective function. This is how Ryu et al. transfer 'knowledge' about race to inform the detection of facial attributes. Ryu et al. use a second dataset for their transfer task, which contains labels for 40 different facial attributes (Z. Liu et al. 2015). These attributes are intended to be visually salient but tend to also be highly socially salient, e.g. "glasses," "pointy nose," "big lips," "narrow eyes," "big nose," "high cheekbones"). This twice-trained model, InclusiveFaceNet, contains statistical representations of race, as well as of each of these facial attributes.

Ryu et al. test InclusiveFaceNet by comparing its ability to perform facial attribute detection against a "vanilla classifier" that has no access to information about racial categorization, as described above. They find that on many facial attribute detection tasks, InclusiveFaceNet exceeds the performance of the "vanilla classifier". The accuracy gain is high for detecting the "smile" attribute, but it is also quite elevated for some, but not all, other attributes as well. Their results show the greatest accuracy gains for attributes that are historically linked to racist representations (Figure 18). These attributes—"big lips," "big nose," and "narrow eyes" among others—are all features that have been the subject of racial caricature in various historical moments (W. J. T. Mitchell 2012; Stark 2018). This might support

Ryu et al.'s initial claim that "faces look different across different races," which they draw from a survey of machine intelligence approaches to "learning race from face" (Fu, He, and Hou 2014). And it might be true that knowing the ways in which faces vary can help identify components of those faces. But the unstated corollary of this claim is the assertion that the InclusiveFaceNet makes through how it constructs the problem of smile detection and assembles a system to address that problem: that particular sets of facial attributes can be systematically and unproblematically correlated with race, and that this correlation of phenotypic features is due to some underlying genotypical or biological regularity.

Tracing how race is defined across the assemblage of machine intelligence research brought together as InclusiveFaceNet, the biological basis of that definition is evident in how ancestry, ethnicity, nationality, and racial appearance are conflated. The survey of machine intelligence race detection cited by Somandepalli above, and used extensively by Ryu et al., does some of this work by noting that "in general English the term 'race' and 'ethnicity' are often used as though they are synonymous … [h]owever they are related to biological and sociological factors respectively" (Fu, He, and Hou 2014, 2483). The face, through Ryu et al.'s Inclusive FaceNet is made into the locus of an objective proof of the physical, biological reality of race through what Amade M'charek refers to as "the tentacularity of faces": it "engages as it instructs the viewer on what to take into account; it educates about the differences that matter … by directing the gaze to various components and their statistical renderings" (M'charek 2020, 378). It provides the perceptual scaffolding to make racial differences meaningful in their objective form.

|     | Big Lips |      |      |      | Big Nose |      |      |      | Young |      |      |      |
|     | Gender 1 | | Gender 2 | | Gender 1 | | Gender 2 | | Gender 1 | | Gender 2 | |
|     | Acc. | AFR | Acc. | AFR | Acc. | AFR | Acc. | AFR | Acc. | AFR | Acc. | AFR |
| --- | --- | --- | --- | --- | --- | --- | --- | --- | --- | --- | --- | --- |
| VAN. | 63.7 | 45.0 | 82.7 | 35.8 | 90.4 | 36.7 | 77.2 | 24.9 | 90.7 | 24.2 | 86.3 | 15.2 |
| +R   | 64.6 | 43.1 | 82.8 | 34.9 | 90.4 | 36.5 | 77.8 | 24.6 | 90.8 | 22.8 | 86.8 | 14.4 |

*Table 3.* **Example Improved Face Attributes in CelebA Evaluated by Gender.** Examples of some of the most improved attributes, evaluated by gender. Accuracy (Acc.) and average false rate (AFR) are in percentage (%). VAN. refers to *vanilla*. +R refers to our model with race input. We see an improvement up to .9% absolute.

|     | Oval Face |      |      |      | Brown Hair |      |      |      | Wavy Hair |      |      |      |
|     | Gender 1 | | Gender 2 | | Gender 1 | | Gender 2 | | Gender 1 | | Gender 2 | |
|     | Acc. | AFR | Acc. | AFR | Acc. | AFR | Acc. | AFR | Acc. | AFR | Acc. | AFR |
| --- | --- | --- | --- | --- | --- | --- | --- | --- | --- | --- | --- | --- |
| VAN. | 74.6 | 33.4 | 77.2 | 44.2 | 86.8 | 18.8 | 90.1 | 22.9 | 80.6 | 19.4 | 88.7 | 32.2 |
| +R   | 74.2 | 34.3 | 77.1 | 45.5 | 86.5 | 18.2 | 89.9 | 22.7 | 80.4 | 19.7 | 88.7 | 32.6 |

*Table 4.* **Example Face Attributes with Loss in CelebA Evaluated by Gender.** Examples of some of the attributes with the most loss, evaluated by gender. Accuracy (Acc.) and average false rate (AFR) are in percentage (%). VAN. refers to *vanilla*. +R refers to our model with race input. We see a loss up to .4% absolute.

Figure 18. Tables from Ryu et al. showing accuracy improvements for racially salient facial features.

What the accuracy gains in smile detection, "big nose" detection, etc. reveals is not phenotypic regularities that distinguish differences between races that were embedded in InclusiveFaceNet through transfer learning. Rather, what these accuracy gains reveal is InclusiveFaceNet's ability to replicate any regularities that occur when human MTurk workers apply racial categorizations to human facial images. Rather than the implied genotypical ground truth, these acts of racial categorization operate from human workers' cultural understandings of, and rapid decisions about, what facial attributes can be associated with each other to comprise a racialized appearance. These understandings, coordinated by a complex social field in which race has persistent, durable cultural significance, may result in a degree of regularity, to be sure. But any such regularity is the product of selective biases informed by enculturation into a racial discourse that takes certain features for tokens of race while selectively ignoring others, and that enculturation process can neither be stripped from the training data nor be simply substituted for the conceptual and empirical work necessary to demonstrate the implied causal relationships. This selective bias is similar to $19^{th}$ and early $20^{th}$ century anthropometricists' tendencies to measure certain physical attributes as racially salient while ignoring others (Sekula 1986; Gould 1996). These tendencies, put forward by the likes of Francis Galton, Louis Agassiz, Alphonse Bertillon and Carleton Coon (see esp. Galton 1904; Coon 1962), have long been debunked by the efforts of Franz Boas, W.E.B. Dubois, Stephen Jay Gould, Kwame Anthony Appiah, and more recent scholars (DuBois 1903; Boas 1940; Gould 1996; Appiah 2005).

InclusiveFaceNet posits a link between race and face by appealing to sophisticated techniques of machine intelligence like neural networks and transfer learning, demonstrating its validity through statistical metrics like predictive accuracy or "area under the curve". And machine intelligence derives some measure of its authority from its involvement in positing these links: the formalization of phenotype into machine readable formats enables a smuggling of causal claims about the relationship between social categories, facial phenotypes, and genotypes that are not stated, let alone rigorously proven. Put simply, the facialization of race, like the formal computational description of unfairness as a "constrained optimization problem" described above, transforms a wide range of research questions about (supposedly) visually salient social types into questions about which machine intelligence can wield its authority. Thus, we see machine intelligence projects that purport to investigate biological bases for a range of phenomena, like sexuality (Kosinski and Wang 2017) or emotional states (Rajesh and Naveenkumar 2016). The demonstrated "success" of these efforts (in terms of the definitions of success offered by the researchers themselves) has led to more extreme facialization—not just of race, sexuality, or emotion, but also criminality (Wu and Zhang 2017; cf. Stinson 2020).

InclusiveFaceNet is not the only machine intelligence project that purports to train machine intelligence systems to detect race from faces (e.g. Kärkkäinen and Joo 2019), and dozens of projects surveyed in Fu et al. 2014 are actively dedicated to machine intelligence applications in anthropometry. Such endeavors are explicitly aligned with, and propose to extend, research that confirms a biological and physical basis of race. They assert that "anthropometric surveys have always been an [*sic*] valuable resource for carrying out race categorization" (Fu, He, and Hou 2014, 2496) and uncritically ground that value in a definition of race which "implies that race could possibly be measured and classified based on predefined standard measurements" (Fu, He, and Hou 2014, 2486). These projects use digital cameras, large databases of imagery with race category annotation, and deep learning to refine techniques for constructing statistical profiles for racial phenotypes, effectively turning machine intelligence into *digital calipers* that can systematically measure (Figure 19) and analyze (Figure 20) faces to construct novel statistical indices of race.

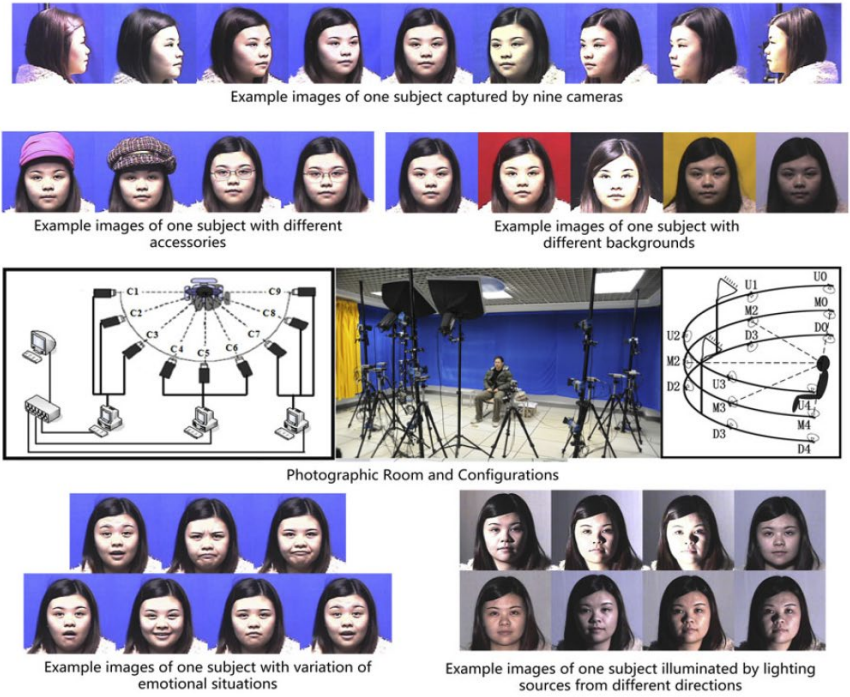

Figure 19. Apparatus for generating detailed 3D face maps for digital anthropometry (Fu et al. 2014)

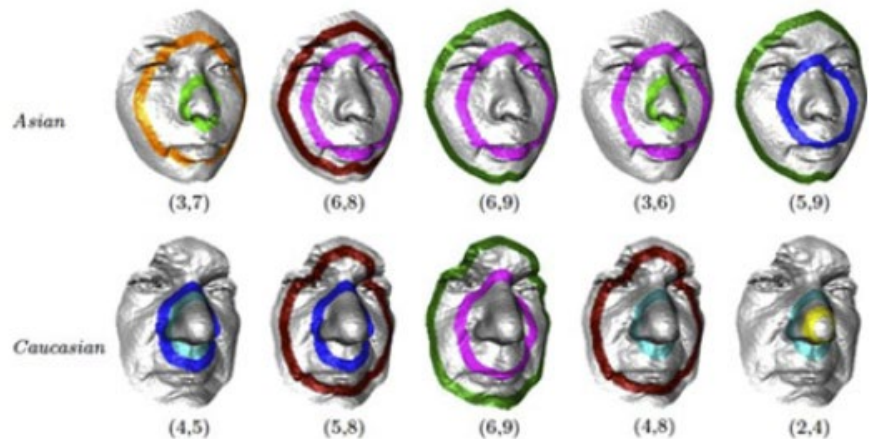

Fig. 14. The anatomical landmark regions used in [158], showing the five most relevant iso-geodesic stripe pairs for Asian and Caucasian subjects.

Figure 20. Statistical analysis of 3D face maps characterizing racial differences (Fu et al. 2014)

These digital calipers replicate anthropometric tropes of the past. The digitalization of race explored here is merely the latest in a progression of scientific racism, following close on the heels of "the molecularization of race" (Fullwiley 2007). Subjecting the human body to the scientific gaze and reducing identity to the play of that gaze across its surface, this digital gaze effaces the role played by personal judgement and individual subjectivity built into the camera and algorithm that replace the clinician (Foucault [1978] 1990). Probability distributions and statistical representations stand in for the MTurk workers who annotate individual images. The digital gaze is "objective," position-less, surrounding the human subject from everywhere and nowhere (Haraway 1988) simultaneously. It launders human subjectivity through dataset, deep learning, and demonstrations of its efficacy, leaving only the faintest impression of the applied machine learning researchers' numerous interventions in shaping the objects of their analysis, putting their thumb on the scale by providing their own annotations for "edge-cases," as when MTurk workers disagree amongst themselves. If all forms of objectivity have a "negative character … related to subjectivity as wax to seal, as hollow imprint to the bolder and more solid features of objectivity" (Daston and Galison 1992, 84) then the objectivity of machine intelligence is the erasure of the social field from which annotators, applied machine intelligence researchers, and those for whom their knowledge is produced draw their understandings of the objects of analysis to begin with. Machine intelligence's objectivity stands apart from the conditions of its production, despite condensing the subjectivities at play in its production. The objects of machine intelligence, the objective functions themselves, are—like race—signifiers that point to the "systems and concepts of the classification of a culture, to its practices for *making meaning*" (Hall [1997] 2021, 362). InclusiveFaceNet shows how cultural practices for making meaning—like the superficial judgement made by an MTurk worker deducing a person's race from a picture of their face—lie behind the authoritative claims of machine intelligence, and are continually erased by its seeming objectivity.

Stuart Hall, in his discussion of race as a "floating signifier" says that signifiers "gain their meaning, not because of what they contain in their essence, but in the shifting relations of difference, which they establish with other concepts and ideas in a signifying field" and that "their meaning, because it is relational, and not essential, can never be finally fixed, but is subject to the constant process of redefinition and appropriation … to the endless process of being constantly resignified, made to mean something different in different cultures, in different historical formations at different moments of time" (Hall [1997] 2021, 362). From InclusiveFaceNet, and from other efforts to 'learn race from face', we can see "a certain sliding of meaning," as Hall puts it, a "something about race left unsaid" and a "someone—a constitutive outside—whose very existence the identity of race depends on, and which is absolutely destined to return from its expelled and abjected position outside the signifying field to trouble the dreams of those who are comfortable inside" (*Ibid.*). InclusiveFaceNet helps race slide from a grab bag of phenotypic attributes to a probabilistic statistical representation, a digital surface (Figure 16) that insists on the physical reality of race by performing statistical differences between racial categories. This insistence comes about through the performance of its possibility. The racial categories machine intelligence sorts images into prefigure the work of classification: asserting that there are a given number of different racial groups represented in a dataset, machine intelligence finds the necessary statistical relationships between facial attributes to sort faces into the given number of different groups. InclusiveFaceNet enacts a reality in which race can be seen as having a biological, physical, and therefore *a priori* objective basis. But it can only be seen as such because of the sleights of hand that happen off stage—the labeling of movie stars' races in a prior project, the cultural knowledge of race smuggled into InclusiveFaceNet, the indexical relationship between racial classifications made by machine intelligence and the features it uses to optimize those classifications, and the construction of InclusiveFaceNet as a virtuous project undertaken to advance algorithmic fairness.

## Conclusion

Machine intelligence practices that attempt to "learn race from face"—but also those that enact specific, atomistic understandings of social types in the pursuit of algorithmic fairness—are new additions to prior "historically dense and situated social and scientific practice" of what Amade M'charek calls "doing difference" (M'charek 2020, 374). Almost all machine intelligence practices are devoted to "doing difference" by sorting, predicting, classifying, and ranking. The difference machine intelligence "does"—through its exercise of deontic power (Searle 2006)—is ultimately the source of its authority over the production of knowledge, and over society. By constructing objects as amenable to its forms of analysis and manipulating how they relate to each other, machine intelligence constitutes an apparatus in which the objects of its analysis are made meaningful (Barad 2003). In the context of algorithmic fairness, "constrained optimization problems" constitute an apparatus in which a particular conceptualization of protected classes—social types—as intrinsic, immutable properties of individuals is enacted.

Through an "agential cut" (Barad 2007, 176), this apparatus arrests an unfolding process in which these social types are, as Hu and Kohler-Hausmann observe, constitutive of every other attribute of an individual (Hu and Kohler-Hausmann 2020). This cut transforms social types from constitutive of persons and constructed by society, to singular sign vehicles carrying a particular algorithmically-defined meaning. These social types become labels that are made to signify the distribution of all other attributes of persons within a given dataset. In the context of a common problem for algorithmic fairness, that of racial disparities in the COMPAS recidivism prediction algorithm (Angwin et al. 2016), race is made to stand in for all the other attributes derived from questionnaires delivered to defendants. At the same time that answers to these questionnaires are made to be predictive of recidivism, they are also revealed to be predictive of race. And in the quest to produce "fair" predictions, these attributes become proxies of race and race becomes a predictor of these attributes. These attributes, as well as the propensity to reoffend, becomes what race "means" for the COMPAS algorithm. Just as seemingly objective policies for selling insurance, developing sentencing guidelines, or granting home loans produced wildly disparate effects that reshaped the structural position of women and minorities relative to white men throughout the 20th century, algorithmic unfairness similarly reshapes the structural position of all those who are subject to machine intelligence. As discussed above, this structural repositioning is also the means through which the deontic power of machine intelligence acts authoritatively to control people's freedom of movement across society.

Machine intelligence "does difference" under the presumption that there is some difference to be drawn within a phenomenon. Algorithmic fairness "does difference" for race in ways that reanimate tropes of scientific racism, race realism, and even biological determinism—even when motivated by an express desire to minimize differences between groups. Whether applied machine learning researchers are classifying ball bearings within an aircraft turbine that are likely to fail and those that are not, between tabloid publications and broadsheets, between racial "subgroups," or handwritten numerals, their predictions enact difference within the narrowly constructed world of their problem formulation. The doing of difference has pragmatic entailments that *matter* deeply. Machine intelligence can be readily applied to any properly posed problem, even if that problem formation reinforces dangerous essentialist tropes. Just as machine intelligence has been used to "do difference" between racial categories, the same techniques used in Ryu et al. have been applied to predict gender (Ryu, Adam, and Mitchell 2018), sexual orientation (Kosinski and Wang 2017), political alignment (Kosinski 2021), and even criminality (Wu and Zhang 2017) from facial imagery. While machine intelligence enacting difference between aircraft parts may pull dangerous aircraft out of the skies and into hangars for repair, doing difference between criminals and non-criminals based on their facial appearances shows how the authority of machine intelligence may be used not only to entrench essentialist notions of criminality, but also to control individuals and populations and form the basis for technological oppression.

## *CONCLUSION: CROSS-VALIDATION*

In machine learning, applied machine learning researchers use a technique called "k-fold cross-validation" to ensure that an algorithmic model did not simply "memorize" a dataset but instead that it actually "learned" meaningful patterns in the relationship between dataset and objective function. K-fold cross-validation shuffles the data and divides it into a number (k) of "folds" that are used to retrain and validate the model's accuracy (Alpaydin 2014, 558). Without implying that the findings of this dissertation are replicable or can be subjected to any statistical tests of significance or validity, I nevertheless highlight below several contributions from the preceding that characterize machine intelligence, and which I feel would emerge from any other set of folds across the working world of applied machine learning researchers.

This dissertation has investigated how the authority of machine intelligence has been constructed. The authority with which this investigation has been concerned has allowed machine intelligence to speak *as* an authority across a wide gamut of domains in ways that have transformed the predictions and classifications it makes into concrete effects on the world, based on the authority *of* machine intelligence. Throughout, I have engaged applied machine learning researchers, those whom they work with, and the objects they produce—algorithmic systems, technical descriptions, conference presentations, and academic reports—as ethnographic subjects and as agential participants in the production of machine intelligence's authority. I have drawn upon the history of science, statistics, and computation to demonstrate how the authority of machine intelligence is contiguous with historical approaches to quantification and datafication grounded in *naïve empiricism*, a form of positivism that vests its authority in applied machine learning researchers' ability to collect, manipulate, and analyze data, without needing to make recourse to theoretical perspectives outside of or beyond that of machine intelligence itself. I have drawn upon theoretical approaches from the anthropology of science and technology, as well as STS, to demonstrate how applied machine learning researchers engage in *data performances* to enact their objects of study as the kinds of objects about which they can make valid claims. And I have demonstrated the expanding set of practices that surround machine intelligence, which include corporate performances of the spectacular capabilities of machine intelligence, and organizational and critical efforts to develop practices intended to constrain or limit the uses of machine intelligence.

Across the practices that I analyze in the preceding, the title of this dissertation—"The Objective Function: Science and Society in the Age of Machine Intelligence"—articulates what machine intelligence is most concerned with, and how it acts to produce knowledge that is seen as both scientific and relevant for society, in ways that have material consequences for the world in which it operates. Machine intelligence depends, technically, on its ability to transform questions it is trying to ask, or problems it is trying to solve, into *objective functions*—mathematical representations of a problem in which the deviation from an expected value ("loss" or "error") can be optimized. The complex computational theory and mathematical research applied machine learning researchers do is devoted to developing algorithmic approaches that can pursue this optimization by making the most efficient use of computational resources and the most effective use of data available. But the objective function is also a way of linking applied machine learning researchers' questions to data. It is a way of stating what their goal is—that which they are trying to optimize—and mapping data onto that goal. The objective function is also a way of constructing the impersonality and impartiality of the knowledge applied machine learning researchers produce (Porter 1995; Daston and Galison 2010). It grounds their work in the data itself, as a statistical re-presentation of that which has already been recorded as data. Finally, the objective function is a way of making up the objects of its analysis. In linking a phenomenon to specific sets of data (and not others), and in subjecting that phenomenon to machine intelligence, the applied machine learning researcher makes an ontological claim not only that that phenomenon exists, but that it exists in relationship to the data the researcher uses in pursuing an objective function.

In this final chapter, I review the conclusions that can be drawn for anthropology and STS from my investigation of the authority of machine intelligence. These conclusions cut across each chapter of the preceding, which each contribute to understanding the authority of machine intelligence: as the construction of data performances that enact the objects of analysis (Chapter 1), as a set of workplace activities that require other practices of accommodation to make machine intelligence possible (Chapter 2), as a set of ontological and epistemological practices that renders any phenomenon subject to machine intelligence (Chapter 3), as a set of organizational practices that places efforts to hold machine intelligence accountable within the existing logics through which machine intelligence is developed (Chapter 4), and as a way of "doing difference" (M'charek 2020, 374) that gives applied machine learning researchers the power to constitute social types and their significance in society, reinscribing legacies of scientific racism even when combatting such legacies (Chapter 5). I also highlight the methodological innovations developed over the course of this dissertation, as well as directions for future research.

## NAÏVE EMPIRICISM

Applied machine learning researchers establish their authority through what I call *naïve empiricism*. Derived from archaeological theories which held cultural materials could be interpreted solely through reference to the data produced through excavation and laboratory analysis (M. H. Johnson 2011), naïve empiricism describes the framing of machine intelligence that asserts "numbers speak for themselves" (C. Anderson 2008). This framing, that numbers, and therefore data, speak for themselves, permits applied machine learning researchers to position themselves not as interested or biased parties producing knowledge to fit an already-existing agenda, but instead as neutral and objective conduits of truths which are already immanent in the data. Applied machine learning researchers distance themselves from judgements which must be made about the data by outsourcing decisions about how to label data to online task workers who complete data analysis tasks at a piece rate, or by deferring to an external authority who has used their disciplinary conventions to set a definition or decision boundary that can be operationalized through machine intelligence (see Chapters 1 and 5). When applied machine learning researchers *do* make judgements about data, these judgements are often presented as matters of convenience or necessity, as when researchers resolve conflicting labels applied by task workers (see Chapter 5), or as a constituent component of their work—like data analytics or data cleaning—that is walled-off from "real" applied machine learning research (see Chapter 2).

While naïve empiricism stands in contrast to hypothesis-driven approaches to data science (see Davenport and Patil 2012), in the work of applied machine learning researchers I studied with it was crucial for establishing their authority by apparently removing themselves from the analysis, in a sense by "getting out of the way" of the data. This erasure of the researchers' presence from the machine intelligence process, despite ample evidence of "intuitions" being invoked to smuggle researchers' positional knowledge back into analyses, was a key practice that is contiguous with the development of objectivity as a personal virtue associated with restraint (Daston and Galison 1992) and of the researcher as a machine-like (Porter 1995) conduit for the data—which my interlocutors referenced whenever they asserted that they felt like "plumbers" just moving pipes around instead of doing "real science".

**DATA PERFORMANCES**

Like puppeteers who spoil the illusion if their hands are seen by the audience, applied machine learning researchers continually erased the traces of their own presence in order to enact what I call *data performances* (Chapter 1). In contrast to the many other performances applied machine researchers engage in—boardroom performances (Chapter 2), spectacular demonstrations of the algorithmic sublime (also Chapter 2), and of professional competence (Chapter 1)—data performances are agentive enactments of the objects of applied machine learning researchers' efforts. Throughout this dissertation I investigate several case studies of data performance in which the object of analysis—handwritten numerals, online shopping behavior, "tabloid-ness," textual sentiment, relationship types, and even race—is enacted as a specific *kind* of object set into relation to the datasets that purportedly pertain to the phenomenon researchers are investigating. I draw upon concepts of linguistic performance theory and performativity (Austin 1962; Butler 1990) to show how machine intelligence is as much a discursive activity as a statistical practice. But I draw most heavily upon the work of Karen Barad to show how the researcher, the techniques of machine intelligence, the datasets both draw upon, that which the datasets purport to represent, and the knowledge produced are not separate but are co-constitutive of each other (Barad 2003; 2007; 2011). Together, they produce the phenomena, which is "a specific intra-action of an 'object'; and the 'measuring agencies'; the object and the measuring agencies emerge from, rather than precede, the intra-action that produces them" (Barad 2007, 128). Machine intelligence, viewed through the lens of agential realism, does not reveal what the data have to say, as naïve empirical claims would suggest. Rather, datasets remain indeterminate until they are enacted as particular kinds of objects which make claims about the world through the apparatus of machine intelligence.

The authority of machine intelligence, then, is constituted not through its ability to reveal truths about the world, but rather through its ability to collapse the multiple indeterminate meanings that datasets might have into a singular one that excludes alternate meanings, and to enact the objects of the world as the kinds of things that can be enacted through machine intelligence (Chapters 3 and 5). In optimizing an objective function to reduce error in predicting the relationship between features of a dataset and some entity (sentiment, race, 'tabloid-ness', etc.) applied machine learning researchers are interested in, they are not only asserting a statistical relationship between data and that entity, but also that the entity in question has a meaningful existence in the world and an objective relationship with the data used to model it algorithmically. Such concepts, as constructed through the practices of machine intelligence—in this dissertation, relevant examples are the specific form of "tabloid-ness" given existence by OTH's multi-task learning setup, but also "sentiment" or "race" as operationalized by machine intelligence—become bounded within long-standing positivist claims that the world can be described and represented objectively and with mathematical precision, rather than treated as contextually bound matters of subjective interpretation. In constructing these concepts in this way, machine intelligence positions itself as the authoritative interpreter of such concepts, to the exclusion of other claims that might be made.

## Machine Intelligence and Its Surround

The key practices of machine intelligence expand beyond the narrower set of technical practices that trainees at bootcamps or applied machine learning researchers at OTH might identify themselves as doing, if asked to enumerate what their work consists of. Part of the work of applied machine learning researchers is to contribute to the public discourse about the authority of machine intelligence—particularly its ability to address complex problems and be deployed across many different contexts. Spectacular demonstrations of machine intelligence's capabilities proliferate widely, and often precede the incorporation of dedicated applied machine learning researchers in the workplace. Once there, applied machine learning researchers must engage in *boardroom performances* that must straddle contradictory understandings of machine intelligence: as both a complex technical practice that produces authoritative knowledge and as an expert set of practices that must be deployed to bring this knowledge into existence (Chapter 2). Applied machine learning researchers thus must push back on these spectacular demonstration projects—which evoke a sense of the algorithmic sublime (Nye 1996; Ames 2018) and construct machine intelligence as a desirable tool or as a merely functional plug-in for accomplishing corporate organizational goals—by assembling allies within the organization who can make the accommodations required for machine intelligence to be integrated into the organization (Callon 1984; Latour 1988).

Another part of the unacknowledged work of applied machine learning researchers and those they work with is addressing the criticisms leveed at machine intelligence, particularly about its societal and ethical implications. Critics have pointed out the uses of machine intelligence in ways that threaten human lives and safety, liberal democratic ideals like free elections and fair trials, and the discriminatory and misogynistic workplace environments in companies developing machine intelligence, but the work of minimizing and mitigating such harms has fallen to those who are already responsible for the development of machine intelligence (Chapters 4). These endeavors to develop *AI ethics*, as conducted by applied machine learning researchers and those they work with (product managers, legal consuls, public relations professionals, etc.), has produced a role for those whom I (along with my research collaborators and co-authors) call *ethics owners* (Metcalf, Moss, and boyd 2019). The practical consequences of this is that the logics under which the ethical and societal implications of machine intelligence are managed—meritocracy, technological solutionism, and market

fundamentalism—are coterminous with those that have produced the potential dangers of machine intelligence in the first place.

A similar irony occurs around machine intelligence practices oriented toward the pursuit of *algorithmic fairness* (Chapter 5). Critics that have pointed out the ways in which machine intelligence recreates and produces unfairness (Angwin et al. 2016; Buolamwini and Gebru 2018) by perpetuating existing social biases or propagating harmful errors for those less well represented in datasets. These critiques have been met by applied machine learning researchers with an expanded set of machine intelligence techniques dedicated toward developing *algorithmic fairness*. These techniques are generally more computational than organizational compared to the work of ethics owners, but nevertheless enclose the pursuit of fairness, across whatever social valences it might have, within the practices of machine intelligence. This occurs despite explicit and concerted efforts to expand work on algorithmic fairness beyond the ambit of applied machine learning researchers by including social scientists, policy makers, and legal scholars in convenings like the FAccT Conference. A deeper irony arises from the ways in which machine intelligence pursuits of algorithmic fairness depend upon and reify essentialist understandings of the categories across which they attempt to combat unfairness. In doing so, applied machine learning researchers develop new ways of "doing difference" (M'charek 2020) that recapitulate the darkest histories of race science and eugenics. These practices, in expanding into the domain of ethics and fairness, treat critiques of machine intelligence as problems that can be addressed *through* machine intelligence, thereby neutralizing critique and —ironically—extending the authority of machine intelligence.

## THE SECRET LIFE OF ALGORITHMS

In conducting the ethnographic study of machine intelligence presented in this dissertation, I engaged in several methodological approaches. Some were tried-and-true. I conducted participant-observation (Bernard 2006) within an applied machine learning research lab and at bootcamps, academic machine learning conferences, industry conferences, meetups, workshops, and various other convenings. The time I spent as a participant-observer at OTH, an applied machine learning research lab, was particularly important for the development and elaboration of ideas presented in this dissertation. Working daily with applied machine learning researchers, analyzing emerging research in machine learning and listening to them tease out the specific contribution of recently published research papers was a master class in understanding not only what machine intelligence is to them, but also provided me a baseline understanding of the techniques of machine learning and how the field was developing during the period of my fieldwork. Working there also exposed me to the daily practices of building and maintaining both a client base—through newsletters, events, and other promotional activities—but also general popular interest in machine intelligence across their network of hobbyists, executives, and applied machine learning researchers working in other companies. Most importantly, working at OTH exposed me to what I would go on to call the secret life of algorithms: the process of meaning-making that extended from phenomenological experience of the world through encoding as data and processing with algorithmic techniques to produce machine intelligence. The data labeling, cleaning, and processing, the choice and training of algorithmic technique, and the composition of an application using machine intelligence were all revealed to me through participant observation at OTH.

I spent a significant portion of my fieldwork conducting semi-structured interviews amongst those who I would eventually refer to as "ethics owners". Some of these interviews were conducted with research collaborators from the Data & Society Research Institute, and I found these pair interviews to be helpful in evoking responses that I might not have gotten on my own (see Metcalf, Moss, and boyd 2019). Often, I would attend these interviews in person, while my research collaborators would join via videoconference software. The para-interview rituals of signing in to corporate campuses, occasionally being given a tour of the offices, and being able to document the built environment where ethics owners worked was an important part of situating these interviews in time and space, which would not have been possible over videoconference alone. A year later, many ethics owners, and I, would be working exclusively over videoconference—from our homes—because of the COVID-19 pandemic.

I also developed other methodological approaches over the course of fieldwork and analysis. Informed by prior ethnographic work on computer technologies (Schüll 2012; Dourish 2016; Seaver 2017), I developed a set of methods for reading across algorithmic systems. I use the word "reading" loosely, based on understanding machine intelligence as analogous to an entextualization process through which discourse is stabilized as text, in that machine intelligence stabilizes culturally salient phenomena as representations of data that are derived from those phenomena. I adapt this technique from Urban and Silverstein, in which they pursue the "secret life of texts," by reading across the processes through which sociocultural activity is laid down as texts (Silverstein and Urban 1996b; Silverstein 1996). I read across algorithmic systems by delving into the nested dependencies of one system on another, and of that other system on curated datasets, to reveal not only how applied machine learning researchers eschew as much responsibility over how data are labeled by outsourcing subjectivity to task workers, but also how in so doing situated subjective cultural understandings of concepts are "smuggled in" to seemingly objective machine intelligence analyses (Chapters 4 and 5).

These nested dependencies are endemic to how machine intelligence is constructed: an algorithmic system depends upon an algorithmic model, which in turn depends on training data, which in turn depends on data collection methods and protocols, and might also depend on benchmark datasets to measure or evaluate performance of the system. In practice, this meant a great deal of time was spent tracking down references to datasets or sources that were mentioned obliquely in academic reports or whitepapers describing a particular machine intelligence technique. At OTH, I was able to see and work with these dependencies as they were being assembled, and these dependencies were well documented and visible for academic machine intelligence research. However, in the tech industry more generally, where datasets and algorithmic systems are proprietary, documentation of nested dependencies is restricted to employees and contractors and remains opaque to outside researchers.

I also engaged in a great deal of collaborative ethnography (Gottlieb 1995; Lassiter 2005). As part of my participant-observation I coauthored work on machine intelligence topics, informed by my own anthropological training, with applied machine learning researchers (see Moss and Schüür 2018; Kaeser-Chen et al. 2020). This greatly informed not only my own understandings of machine intelligence, but has also inverted the ethnographic gaze I applied to research subjects, an inversion which is reflected throughout this dissertation. As discussed at greater length below, collaboration is not an unalloyed good for academic research when so much of the work sits so close to industry. The ties between the tech industry and academia are many, and the financial resources that Silicon Valley has invested in supporting academic research centers studying the social and ethical implications of machine intelligence are significant (Ebell et al. 2021). This has led to charges that scholars engaging in AI ethics research are complicit with the worst abuses of the companies whose work they critique and that the scope of AI ethics are constrained to a subset of topics companies find least unpalatable (see Williams 2019; Ochigame 2019; Bietti 2020).

My own research concludes that the pursuit of AI ethics and algorithmic fairness are ways of assimilating critiques of machine intelligence into the logics and practices of machine intelligence. But I nevertheless remain uneasy when seeing critiques of machine intelligence made by myself and others appearing, in industry-led publications, to give the impression that such critiques form the basis for ethical conduct by the industry (McLennan et al. 2020; Friesen et al. 2021). This has been particularly acute in how algorithmic forms of bias, unfairness, and inequity—originally raised by critical researchers illustrating systemic and structural concerns about the power of machine intelligence in industry (see especially Noble 2018; Eubanks 2018; Obermeyer et al. 2019)—have been transformed by companies into narrow goals for "better" or "safer" machine intelligence performance, rather than spurring the systemic and structural change that would be required to fully respond to researchers' concerns.

## FUTURE DIRECTIONS: ASSEMBLING ACCOUNTABILITY

Toward the end of my fieldwork, I found myself checking in for a "Data Ethics Salon" held in the executive meeting suites of one of the largest companies in Silicon Valley. I had previously interviewed one of this company's ethics owners, who was in attendance, and the meeting was being convened by another ethics owner from a large consulting and business services firm whom I had also interviewed for this dissertation. Other ethics owners I had not met before were in attendance, too, as were faculty from a nearby university ethics research center and representatives from several non-profit groups focused on human rights and the human rights implications of machine intelligence. I arrived in the company of one of my research collaborators with whom I had written about ethics owners (see Metcalf, Moss, and boyd 2019) and other personnel from Data & Society Research Institute, which had partially supported that work. As representatives from Data & Society, and as authors of work about ethics owners, we had been invited to be part of the salon's "expert panel on AI ethics" for a discussion held under the Chatham House Rule.[1] We sat at several tables arranged into a wide "U" shape that was open toward a set of white boards and a projection screen for slide presentations. To one side was an assortment of workshop fare—tea, coffee, and other bottled beverages, snack bars, and fruit.

The discussion was framed by a set of opening remarks, first from the ethics owner who convened the salon, and then from a few invited guests, followed by a more open discussion of the initial remarks by those seated around the table. Over the course of three hours, the conversation covered a range of topics: human rights, open-source software development, and corporate strategy. We heard a presentation from a human rights consultancy about how the United Nation's Declaration of Universal Human Rights (UN General Assembly 1948) might be used as "universally accepted framework" for Silicon Valley companies to use in place of their own carefully crafted statements of values and principles. We heard about a range of efforts by the company whose meeting space we were using to disseminate ethics practices across their research and development labs, legal departments, sales teams, etc. They were currently building "ethics incentives" into employees' compensation structures. They were cross-training anthropologists and other social scientists to "speak the same language" as applied machine learning researchers so conversations about social impacts of machine intelligence between them "aren't painful". And they were also building an "internal consultancy" to advise teams on the ethical implications of the machine intelligence products they were developing. We also heard about an open-source software package that could be used by applied machine learning researchers to "check" their own machine intelligence products for ethical issues like racial or gender bias and protect privacy by blurring sensitive parts of photographs (like license plates or faces).

---

[1] The Chatham House Rule permits participants to discuss what was said at a meeting, provided they do not attribute statements to a specific speaker. Chatham House Rule convenings were common throughout my fieldwork with ethics owners, as it was seen as a way to allow ethics owners to speak freely about ethical conundrums they had faced at work without exposing their employers to liability.

Across these conversations, my co-author and I were encouraged by others to contribute findings from our research on ethics owners. Before our eyes, we saw our research findings being reframed and adapted to the kinds of "insights" and "takeaways" that I had seen applied machine learning researchers squeeze from datasets early in my fieldwork. As I had also seen with outside critics of Google's contracts with the U.S. military (Deahl 2018) and Facebook's involvement with Cambridge Analytica (Cadwalladr and Graham-Harrison 2018), and with critical research on algorithmic bias (Angwin et al. 2016; Buolamwini and Gebru 2018), I was seeing my own work transformed into fodder for the machine intelligence development process. It was as if any research conducted to make sense of how machine intelligence, and particularly industry applications of machine intelligence, affects the world could be similarly subsumed into the logics of Silicon Valley, transformed into a component of the applied machine learning research and development process. In some ways, the ability of developers of machine intelligence to adapt and expand their development practices to include broader considerations of their social implications is an important part of addressing the potential harms powerful algorithmic systems can cause. Without recourse to political theory, qualitative research methods, and social scientific understandings of the phenomena machine intelligence concerns itself with, it is necessarily bound to mechanistic, naïve empirical understandings of the social world (Sloane and Moss 2019), and the ability of applied machine intelligence researchers to respond constructively to criticism is encouraging.

There is also a burgeoning area of research on *algorithmic accountability* emerging both within and beyond applied machine learning research circles (Nissenbaum 1996; Diakopoulos 2016). My own recent, collaborative work in this domain points to the ways the cooptation of efforts to produce accountability—in which some outside or independent entity can assign responsibility for harms produced by machine intelligence (Bovens 2006; Wieringa 2020)—for the practices of machine intelligence can occur, as well as ways to build more expansive accountability practices in the public interest (Moss, Watkins, Singh, Elish, et al. 2021). Such practices understand the impacts used to evaluate algorithmic harms as themselves agential cuts (Barad 1999) that enact regimes of accountability, and that carry with them ethical imperatives to include—as expansively as possible—not just applied machine learning researchers, but all those whose lives stand to be marked by the use of algorithmic systems as part of an assemblage of agents capable of shaping the impacts of machine intelligence (Metcalf et al. 2021).

Nevertheless, it is also profoundly unsettling that outside critiques, which attempt to draw attention to the limits of applied machine learning researchers (and those they work with), are continuously redefined as components of machine intelligence itself. This happens when calls for transparency and accountability for machine intelligence inspire software packages that can automate such political imperatives (Spiegelhalter 2018; M. Mitchell et al. 2019). It also happens when machine intelligence research laboratories adapt postcolonial theory into a set of "design principles" (Mohamed, Png, and Isaac 2020) for machine intelligence. And it slowly happened over the course of my fieldwork, as those I worked with and among sought to include my work in their own approaches. Anthropology has, of course, navigated the complex relationships with state and corporate powers that have funded and made use of ethnographic research over the long history of the discipline. A discipline that was once the vanguard of colonial rule (Asad 1973) and a source of military intelligence for armed global conflicts (Price 2016) has, through introspection, reflexivity, and no small measure of internal contestation, shifted toward the study and critique of hegemonic power (AAA 2007; Alkhatib 2018). But as an anthropologist, working with and among those producing some of the most lucrative and powerful tools of modern life, it was nevertheless difficult to find any line at all to balance between contributing to the authority of machine intelligence and standing apart from it to point out its limitations.

# ———. 1992. "Where Are the Missing Masses?" In *Shaping Technology/Building Society: Studies in Sociotechnical Change*, edited by Wiebe E. Bijker and John Law. Cambridge, MA: MIT Press.

———. 2003. *Science in Action: How to Follow Scientists and Engineers through Society*. 11th printing. Cambridge, MA: Harvard University Press.

———. 2005. *Reassembling the Social: An Introduction to Actor-Network-Theory*. Clarendon Lectures in Management Studies. Oxford ; New York: Oxford University Press.

———. 2008. *What Is the Style of Matters of Concern?* Amsterdam: Van Gorcum. https://doi.org/10.5749/minnesota/9780816679959.003.0004.

Latour, Bruno, and Steven Woolgar. 1979. *Laboratory Life: The Construction of Scientific Facts*. Princeton, NJ: Princeton University Press.

Lazer, David, Alex (Sandy) Pentland, Lada Adamic, Sinan Aral, Albert Laszlo Barabasi, Devon Brewer, Nicholas Christakis, et al. 2009. "Life in the Network: The Coming Age of Computational Social Science." *Science (New York, N.Y.)* 323 (5915): 721–23. https://doi.org/10.1126/science.1167742.

Lazonick, William, and Mary O'Sullivan. 2000. "Maximizing Shareholder Value: A New Ideology for Corporate Governance." *Economy and Society* 29 (1): 13–35. https://doi.org/10.1080/030851400360541.

LeCun, Yann, Corrina Cortes, and Chris Burges. 2010. "MNIST Handwritten Digit Database." Yann LeCun's Homepage. 2010. http://yann.lecun.com/exdb/mnist/.

Lee, Min Kyung. 2018. "Understanding Perception of Algorithmic Decisions: Fairness, Trust, and Emotion in Response to Algorithmic Management." *Big Data & Society* 5 (1): 205395171875668. https://doi.org/10.1177/2053951718756684.

Lee, Min Kyung, Anuraag Jain, Hea Jin Cha, Shashank Ojha, and Daniel Kusbit. 2019. "Procedural Justice in Algorithmic Fairness: Leveraging Transparency and Outcome

Poovey, Mary. 1998. *A History of the Modern Fact: Problems of Knowledge in the Sciences of Wealth and Society*. Chicago: University of Chicago Press.

Popper, Karl. (1935) 2002. *The Logic of Scientific Discovery*. New York: Routledge. http://public.ebookcentral.proquest.com/choice/publicfullrecord.aspx?p=254228.

Porter, Theodore M. 1995. *Trust in Numbers: The Pursuit of Objectivity in Science and Public Life*. Princeton, N.J: Princeton University Press.

Powdermaker, Hortense. 1966. *Stranger and Friend The Way of an Anthropologist*. New York: W. W. Norton & Company.

Power, Michael. 1997. *The Audit Society: Rituals of Verification*. New York: Oxford University Press.

Powles, Julia, and Helen Nissenbaum. 2018. "The Seductive Diversion of 'Solving' Bias in Artificial Intelligence." *OneZero* (blog). December 7, 2018. https://onezero.medium.com/the-seductive-diversion-of-solving-bias-in-artificial-intelligence-890df5e5ef53.

Prabowo, Rudy, and Mike Thelwall. 2009. "Sentiment Analysis: A Combined Approach." *Journal of Informetrics* 3 (2): 143–57. https://doi.org/10.1016/j.joi.2009.01.003.

Price, David H. 2016. *Cold War Anthropology: The CIA, the Pentagon, and the Growth of Dual Use Anthropology*. Durham: Duke University Press.

Puig de la Bellacasa, María. 2017. *Matters of Care: Speculative Ethics in More than Human Worlds*. Posthumanities 41. Minneapolis: University of Minnesota Press.

Radcliffe-Brown, A.R. 1952. *Structure and Function in Primitive Society: Essays and Addresses*. Glencoe, IL: The Free Press.

Radway, Janice. 2008. "What's the Matter with Reception Study? Some Thoughts on the Disciplinary Origins, Conceptual Constraints, and Persistent Viability of a Paradigm." In